\documentclass[sort&compress,preprint,3p]{elsarticle}

\makeatletter
\@addtoreset{equation}{section}
\makeatother

\newcommand{\gea}{\raisebox{-.3ex}{\small $ \
\stackrel{\textstyle >}{\sim} $ }}
\newcommand{\bbox}[1]{\mbox{\boldmath $#1$}}
\def\be{\begin{eqnarray}}
\def\ee{\end{eqnarray}}

\usepackage{graphicx}
\usepackage{amssymb}
\journal{Physics Reports [published: Phys. Rep. 503 (2011) 1-75]}

\begin{document}

\begin{frontmatter}



\title{\bf Chiral effective field theory and nuclear forces}


\author[Idaho]{R. Machleidt\corref{cor}}
\cortext[cor]{Corresponding author.}
\ead{machleid@uidaho.edu}
\author[Salamanca]{D. R. Entem}
\ead{entem@usal.es}

\address[Idaho]{Department of Physics, University of Idaho, 
Moscow, Idaho 83844, USA}
\address[Salamanca]{Grupo de Fisica Nuclear and IUFFyM, University of Salamanca, 
E-37008 Salamanca, Spain}

\begin{abstract}
We review how nuclear forces emerge from low-energy QCD via chiral
effective field theory. The presentation is accessible to the non-specialist.
At the same time, we also provide considerable detailed information
(mostly in appendices)
for the benefit of researchers who wish to
start working in this field.
\end{abstract}


\end{frontmatter}

\tableofcontents

\newpage

\section{Introduction and historical perspective} 

The theory of nuclear forces has a long history (cf.\ Table~\ref{tab_hist}).
Based upon the seminal idea by Yukawa~\cite{Yuk35}, 
first field-theoretic attempts
to derive the nucleon-nucleon ($NN$) interaction
focused on pion exchange.
While the one-pion exchange turned out to be very useful
in explaining $NN$ scattering data and, in particular, the properties
of the deuteron~\cite{Sup56}, 
multi-pion exchange was beset with
serious ambiguities~\cite{TMO52,BW53}
that could not be resolved in a satisfactory way.
Thus, the ``pion theories'' of the 1950s
are generally judged as failures---for reasons
we understand today: pion dynamics is constrained by chiral
symmetry, a crucial point that was unknown in the 1950s.

Historically, the experimental discovery of heavy 
mesons~\cite{Erw61} in the early 1960s
saved the situation. The one-boson-exchange (OBE)
model~\cite{Sup67,BS64,Erk74,HM75,NRS78,Mac89} emerged, which still today is the most economical
and quantitative
phenomenology for describing the 
$NN$ interaction~\cite{Sto94,Mac01,GS08}.
The weak point of this model, however, is the scalar-isoscalar
``sigma'' or ``epsilon'' boson, for which empirical
evidence remains controversial. Since this boson is associated
with the  correlated (or resonant) exchange of two pions,
a vast theoretical effort 
was launched to derive the 2$\pi$-exchange contribution
of the nuclear force, which creates the intermediate-range attraction.
During this effort, which occupied more than a decade,
dispersion theory 
(Stony Brook~\cite{JRV75,BJ76} and Paris~\cite{Vin79,Lac80} potentials) 
as well as field theory 
(Partovi-Lomon model~\cite{PL70}, Bonn potential~\cite{Mac89,MHE87})
were invoked.

The nuclear force problem appeared to be solved; however,
with the discovery of quantum chromo-dynamics (QCD), 
all ``meson theories'' were
relegated to the status of models and the attempts to derive
the nuclear force had to start all over again.

The problem with a derivation of nuclear forces from QCD is that
this theory is non-perturbative in the low-energy regime
characteristic of nuclear physics, which makes direct solutions
very difficult.
Therefore, during the first round of new attempts, QCD-inspired quark 
models~\cite{Tar80,Har81,OY81,Fae83,Tho83,MW88,TSY89,Bar93,EFV00,Wu00,Dow10} 
became popular. 
The positive aspect of these models is that they try to explain hadron structure
and hadron-hadron interactions on an equal footing and, indeed, 
some of the gross features of the $NN$ interaction are explained successfully.
However, on a critical note, it must be pointed out
that these quark-based
approaches are nothing but
another set of models and, thus, do not represent
fundamental progress. 
For the purpose of describing hadron-hadron interactions, 
one may equally well stay
with the simpler and much more quantitative meson models.

A major breakthrough occurred when 
the concept of an effective field theory (EFT) was introduced
and applied to low-energy QCD.
As outlined by Weinberg in a seminal paper~\cite{Wei79},
one has to write down the most general Lagrangian consistent
with the assumed symmetry principles, particularly
the (broken) chiral symmetry of QCD. At low energy, the effective degrees of freedom are pions 
(the Goldstone bosons of the broken symmetry) and
nucleons rather than quarks and gluons; heavy mesons and
nucleon resonances are ``integrated out''.
So, the circle of history is closing and we are {\it back to Yukawa's meson (pion) theory},
except that we have finally learned how to deal with it:
broken chiral symmetry is a crucial constraint that generates
and controls the dynamics and establishes a clear connection
with the underlying theory, QCD.

\begin{table}[t]
\caption{The Theory of Nuclear Forces:
Seven Decades of Struggle
\label{tab_hist}} 
\smallskip
\begin{tabular*}{\textwidth}{@{\extracolsep{\fill}}ll}
\hline 
\hline 
\noalign{\smallskip}
\bf 1935   &
\bf Yukawa: Meson Theory
\\
\noalign{\smallskip}
\hline
\noalign{\smallskip}
\bf 1950's &            
{\it The ``Pion Theories''.}
\\
          &
One-pion exchange: good;
multi-pion exchange: disaster.
\\
\noalign{\smallskip}
\hline
\noalign{\smallskip}
\bf 1960's & 
Many pions $\equiv$ multi-pion resonances:
{\boldmath $\sigma$, $\rho$, $\omega$, ...}
\\
           & 
The One-Boson-Exchange Model: success.
\\
\noalign{\smallskip}
\hline
\noalign{\smallskip}
\bf 1970's & 
Divers two-pion-exchange models: 
\\
   &
Partovi-Lomon, Stony Brook, Paris, Bonn.
\\
\noalign{\smallskip}
\hline
\noalign{\smallskip}
\bf 1980's &  
Nuclear physicists discover {\bf QCD}:
\\
           & Quark Models.
\\
\noalign{\smallskip}
\hline
\noalign{\smallskip}
\bf 1990's & 
Nuclear physicists discover {\bf EFT}; Weinberg, van Kolck, \ldots
\\
\bf and beyond &
{\bf Back to Meson (Pion) Theory!} {\it But, constrained by Chiral Symmetry.}
\\
\noalign{\smallskip}
\hline
\hline
\end{tabular*}
\vspace*{0.5cm}
\end{table}

The idea of chiral symmetry has an interesting history of its own.
The modern understanding is that this symmetry arises because the up and
down quarks happen to have relatively small masses.
However, chiral symmetry and its significance for low-energy hadron (pion)
physics was discovered long before QCD.
In 1960, based upon concepts proposed by Schwinger~\cite{Sch57},
Gell-Mann and Levy~\cite{GL60} developed the sigma model, which is
a linear realization of chiral symmetry.\footnote{For a pedagogical introduction into chiral symmetry
and the sigma model, see~\cite{EW88}.}
One major problem researchers had been struggling with in the 1950's was that the pion-nucleon
scattering length came out two orders of magnitude too
large when the (renormalizable) pseudo-scalar ($\gamma_5$) $\pi N$ interaction was used.
This unrealistic prediction was due to very large contributions from
virtual anti-nucleon states (the so-called ``pair terms'' or ``Z-graphs''). Similar problems
occurred in the $2\pi$-exchange contribution to the $NN$ interaction. 
In the sigma model, the large pair terms are canceled by processes
involving the (fictitious) $\sigma$ boson. In this way, the linear sigma model
demonstrates how imposing chiral invariance fixes the problem with low-energy
$\pi$-$N$ scattering. However, the fictitious character of the $\sigma$ particle as well as 
the reliance on the perfect
cancelation of huge terms are uncomfortable features.
In 1967, motivated by the current algebra approach to soft pion physics,
Weinberg~\cite{Wei67} worked out what has become known as the non-linear sigma model, which
does not include a $\sigma$ anymore and has pions and nucleons interact via pseudo-vector
(derivative, $\gamma_5 \gamma^\mu \partial_\mu$) 
coupling besides a new (non-linear) $\pi\pi NN$ term also involving a derivative
(``Weinberg-Tomozawa term''~\cite{Wei66,Tom66}).
The derivative (equivalent to momentum) guarantees that the interaction vanishes when the momentum goes to zero providing a natural explanation for the weakness of the interaction by
soft pions  which does not rely on the cancelation of large terms. 
Following suggestions by Schwinger, Weinberg~\cite{Wei68} developed, soon after, 
a general theory of non-linear realizations
of chiral symmetry, which was further generalized in an elegant way by
Callan, Coleman, Wess, and Zumino~\cite{CCWZ}.

Even though the original work on chiral symmetry was obviously all performed by particle physicists, 
it must be stated---to the honor of nuclear physics---that 
there have been some far-sighted nuclear physicists who early on
understood and appreciated the significance of chiral symmetry for low-energy hadron
interactions. One of them was Gerry Brown, who as early as 1968 published with two co-workers
a paper~\cite{BGG68} on three-nucleon forces, where the consequences of chiral symmetry are fully
exploited.  In 1970, Brown wrote a remarkable Comment~\cite{Bro70} and, in 1979, he published
a book chapter entitled ``Chiral symmetry and the nucleon-nucleon
interaction''\cite{Bro79}.
Moreover, in the more sophisticated relativistic meson models of the 
past~\cite{MHE87,FT80,BM84,GOH92}
the pseudo-vector coupling was applied in $\pi NN$ vertices (instead of the simpler pseudo-scalar
one that was commonly in use) in recognition of chiral symmetry.
However, this chiral patch work, even though it points into the right direction,
cannot be perceived as a serious chirally invariant theory.
Moreover, one has to face the problem that the derivative coupling is not renormalizable
in the conventional sense. 

Therefore, ideas were still needed for how
to implement chiral symmetry consistently in the theory of pionic and nuclear interactions
and how to deal with the renormalization issue.
In his contribution to the `Festschrift' in honor of Schwinger of 1979~\cite{Wei79,Wei09}, Weinberg proposed
to consider the most general possible Lagrangian including all higher-derivative 
terms that are consistent
with chiral symmetry (besides the other commonly accepted symmetry principles).
For this theory to be manageable, one needs to assume some sort of perturbative
expansion such that only a finite number of terms contribute at a given order.
This expansion is provided by powers of small external momenta over the chiral symmetry
breaking scale, $\Lambda_\chi \sim 1$ GeV. The higher-derivative terms supply the counterterms
that make possible an order-by-order renormalization, which is the appropriate renormalization 
procedure for an effective field theory.
Weinberg's suggestions were soon picked up  
by Gasser, Leutwyler, and associates who worked out, to one loop, the cases of 
$\pi\pi$~\cite{GL84} and $\pi N$ scattering~\cite{GSS88} with great success.

But there was still the problem of the nuclear force which is more difficult, 
since nuclear interactions do not vanish in the chiral limit ($q\rightarrow 0$; $m_{u/d}, m_\pi \rightarrow 0$)
and require a non-perturbative
treatment because of the existence of nuclear bound states.
In a series of papers published around 1990~\cite{Wei90,Wei91,Wei92}, 
Weinberg picked up the nuclear force issue and suggested to calculate the $NN$ potential
perturbatively in the chiral expansion and then iterate it to all orders in a 
Schroedinger or Lippmann-Schwinger
equation to obtain the nuclear amplitude. 
Here, the introduction of four-nucleon contact terms is crucial for renormalization.

Following the Weinberg proposal, pioneering
work was performed by Ord\'o\~nez, Ray, and
van Kolck \cite{ORK94,ORK96} who applied time-ordered perturbation theory to
construct a $NN$ potential up to next-to-next-to-leading order (NNLO).
The results were encouraging and nuclear EFT quickly developed into one of the most popular
branches of modern nuclear physics.
The Munich group used covariant perturbation theory and dimensional regularization
to calculate the perturbative
$NN$ amplitude without~\cite{KBW97} and 
with $\Delta(1232)$-isobar degrees of freedom~\cite{KGW98} at NNLO.
Besides this, the Munich group worked out important loop contributions of
higher order~\cite{Kai00a,Kai00b,Kai01,Kai01a,Kai01b}.
A relativistic approach was also taken by the Brazil group~\cite{RR94,RR03}.
The Bochum-J\"ulich group devised
a method of unitarity transformations to eliminate the energy-dependence of 
time-ordered perturbation theory amplitudes and calculated 
the $NN$ potentials up to NNLO~\cite{EGM98,EGM00}.
The Idaho group managed to construct a chiral $NN$ potential
at next-to-next-to-next-to-leading order (N$^3$LO) and showed that only at
this order can one achieve the precision necessary for reliable few-nucleon
and nuclear structure calculations~\cite{EM02a,EM02,EM03,ME05}.
Progress extended beyond the $NN$ interaction, as nuclear many-body forces based 
upon chiral perturbation theory were also developed~\cite{Wei92,Kol94,Epe02b,IR07,Ber08}.

During the past decade or so,
chiral two-nucleon forces have been used in many microscopic calculations of
nuclear reactions and structure~\cite{DF07,Cor02,Cor05,Cor10,NC04,FNO05,Kow04,DH04,Wlo05,Dea05,Gou06,Hag08,Hag10,FOS04,FOS09} 
and the combination of chiral two- and three-nucleon forces has been applied in
few-nucleon reactions~\cite{Epe02b,Erm05,Kis05,Wit06,Ley06,Ste07,KE07,Mar09,Kie10,Viv10},
structure of light- and medium-mass nuclei~\cite{Nog06,Nav07,Hag07,Ots09},
and nuclear and neutron matter~\cite{Bog05,HS09}---with a great deal of success.
The majority of nuclear structure calculations is nowadays based upon chiral forces.

Consequently, it may be of interest to the community
to have a good understanding of these forces
and their background.
It is therefore the purpose of this report to provide an accessible review 
on how nuclear forces emerge from low-energy QCD via chiral effective field theory.
A pedagogical introduction into the phenomenology and the traditional
view of nuclear forces can be found in Refs.~\cite{Mac89,MS01}.
Alternative reviews on various aspects of the modern perspective are published in Refs.~\cite{Kol99,BK02,Epe06,EHM09}.

This article is organized as follows.
In Section~\ref{sec_EFT}, we sketch the foundations of an EFT for low-energy
QCD including the effective Lagrangians. Section~\ref{sec_overview} provides an overview
on nuclear forces derived from chiral EFT. The two-nucleon force is then discussed in detail in Section~\ref{sec_2NF}
and many-body forces in Section~\ref{sec_manyNF}. 
The extension of the theory through the introduction of $\Delta(1232)$-isobar degrees of freedom
and higher order contributions to nuclear forces are considered in Section~\ref{sec_delta}.
Finally, Section~\ref{sec_concl} contains our conclusions.
The appendices provide many mathematical details which may be useful to researchers
who wish to start working in the field.

\section{Effective field theory for low-energy QCD
\label{sec_EFT}}

Quantum chromodynamics (QCD) is the theory of strong interactions.
It deals with quarks, gluons and their interactions and is
part of the Standard Model of Particle Physics.
QCD is a non-Abelian gauge field theory
with color $SU(3)$ the underlying gauge group.
The non-Abelian nature of the theory has dramatic
consequences. While 
the interaction between colored objects is weak 
at short distances or high momentum transfer
(``asymptotic freedom'');
it is strong at long distances ($\gea 1$ fm) or low energies,
leading to the confinement of quarks into colorless
objects, the hadrons. Consequently, QCD allows for a 
perturbative analysis at large energies, whereas it is
highly non-perturbative in the low-energy regime.
Nuclear physics resides at low energies and
the force between nucleons is
a residual color interaction
similar to the van der Waals force between neutral molecules.
Therefore, in terms of quarks and gluons, the nuclear force
is a very complicated problem that, nevertheless, can be attacked
with brute computing power on a discretized, Euclidean space-time lattice
(known as lattice QCD). In a recent study~\cite{Bea06}, the neutron-proton scattering lengths
in the singlet and triplet $S$-waves have been determined in fully dynamical
lattice QCD, with a smallest pion mass of 354 MeV. This result is then extrapolated
to the physical pion mass with the help of chiral perturbation theory. The pion mass
of 354 MeV is still too large to allow for reliable extrapolations, but the feasibility has been
demonstrated and more progress can be expected for the near future.
In a lattice calculation of a very different kind, the nucleon-nucleon ($NN$) potential
was studied~\cite{IAH07,Ino10}.
The $NN$ potential is extracted from the equal-time Bethe-Salpeter
amplitude with local interpolating operators for the nucleons. The central part of the
potential shows a repulsive core plus attraction of intermediate range. This is a very promising
result, but it must be noted that also in this investigation still rather large pion masses are being used.
In any case, advanced lattice QCD calculations are under way and continuously improved. 
However, since these calculations are very time-consuming
and expensive, they can only be used to check a few representative key-issues. For everyday
nuclear structure physics, a more efficient approach is needed. 

The efficient approach is an effective field theory.
For the development of an EFT, it is crucial to identify a separation of
scales. In the hadron spectrum, a large gap between the masses of
the pions and the masses of the vector mesons, like $\rho(770)$ and $\omega(782)$,
can clearly be identified. Thus, it is natural to assume that the pion mass sets the soft scale, 
$Q \sim m_\pi$,
and the rho mass the hard scale, $\Lambda_\chi \sim m_\rho$, also known
as the chiral-symmetry breaking scale.
This is suggestive of considering an expansion in terms of the soft scale over the hard scale,
$Q/\Lambda_\chi$.
Concerning the relevant degrees of freedom, we noticed already that,
for the ground state and the
low-energy excitation spectrum of
an atomic nucleus as well as for conventional nuclear
reactions,
quarks and gluons are ineffective degrees of freedom,
while nucleons and pions are the appropriate ones.
To make sure that this EFT is not just another phenomenology,
it must have a firm link with QCD.
The link is established by having the EFT observe
all relevant symmetries of the underlying theory.
This requirement is based upon a `folk theorem' by
Weinberg~\cite{Wei79}:
\begin{quote}
If one writes down the most general possible Lagrangian, including {\it all}
terms consistent with assumed symmetry principles,
and then calculates matrix elements with this Lagrangian to any given order of
perturbation theory, the result will simply be the most general possible 
S-matrix consistent with analyticity, perturbative unitarity,
cluster decomposition, and the assumed symmetry principles.
\end{quote}
In summary, the EFT program consists of the following steps:
\begin{enumerate}
\item
Identify the soft and hard scales, and the degrees of freedom appropriate
for (low-energy) nuclear physics.
\item
Identify the relevant symmetries of low-energy QCD and
investigate if and how they are broken.
\item
Construct the most general Lagrangian consistent with those
symmetries and symmetry breakings.
\item
Design an organizational scheme that can distinguish
between more and less important contributions: 
a low-momentum expansion.
\item
Guided by the expansion, calculate Feynman diagrams
for the problem under consideration
to the desired accuracy.
\end{enumerate}
In the following (sub)sections,
we will elaborate on these steps, one by one.
Since we discussed the first step already, we will address now step two.

\subsection{Symmetries of low-energy QCD}

In this section, we will give a brief introduction into 
(low-energy) QCD,
its symmetries and symmetry breakings.
More detailed presentations of this topic are provided
in Refs.~\cite{Wei05,Sch03,SS05}.

\subsubsection{Chiral symmetry}

The QCD Lagrangian reads
\begin{equation}
{\cal L}_{\rm QCD} = 
\bar{q} (i \gamma^\mu {\cal D}_\mu - {\cal M})q
 - \frac14 
{\cal G}_{\mu\nu,a}
{\cal G}^{\mu\nu}_{a} 
\label{eq_LQCD}
\end{equation}
with the gauge-covariant derivative
\begin{equation}
{\cal D}_\mu = \partial_\mu -ig\frac{\lambda_a}{2}
{\cal A}_{\mu,a}
\end{equation}
and the gluon field strength tensor\footnote{For $SU(N)$ group indices, we use 
Latin letters, $\ldots,a,b,c,\ldots,i,j,k,\dots$,
and, in general, do not distinguish between subscripts and superscripts.}
\begin{equation}
{\cal G}_{\mu\nu,a} =
\partial_\mu {\cal A}_{\nu,a}
-\partial_\nu {\cal A}_{\mu,a}
 + g f_{abc}
{\cal A}_{\mu,b}
{\cal A}_{\nu,c} \,.
\end{equation}
In the above, $q$ denotes the quark fields and ${\cal M}$
the quark mass matrix. Further, $g$ is the
strong coupling constant and ${\cal A}_{\mu,a}$
are the gluon fields. The $\lambda_a$ are the
Gell-Mann matrices and the $f_{abc}$
the structure constants of the $SU(3)_{\rm color}$
Lie algebra $(a,b,c=1,\dots ,8)$;
summation over repeated indices is always implied.
The gluon-gluon term in the last equation arises
from the non-Abelian nature of the gauge theory
and is the reason for the peculiar features
of the color force.

The masses of the up $(u)$, down $(d)$, and
strange (s) quarks are~\cite{PDG}:
\begin{eqnarray}
m_u &=& 2.5\pm 0.8 \mbox{ MeV} ,
\label{eq_umass} \\
m_d &=& 5\pm 0.9 \mbox{ MeV} ,
\label{eq_dmass} \\
m_s &=& 101\pm 25 \mbox{ MeV} .
\label{eq_smass}
\end{eqnarray}
These masses are small as compared to
a typical hadronic scale, i.e., a scale of low-mass
hadrons which are not Goldstone bosons, e.g., 
$m_\rho=0.78 \mbox{ GeV} \approx 1 \mbox{ GeV}$.

It is therefore of interest to discuss the
QCD Lagrangian in the limit of vanishing quark
masses:
\begin{equation}
{\cal L}_{\rm QCD}^0 = \bar{q} i \gamma^\mu {\cal D}_\mu
q - \frac14 
{\cal G}_{\mu\nu,a}
{\cal G}^{\mu\nu}_{a} \,.
\end{equation}
Defining right- and left-handed quark fields,
\begin{equation}
q_R=P_Rq \,, \;\;\;
q_L=P_Lq \,,
\end{equation}
with 
\begin{equation}
P_R=\frac12(1+\gamma_5) \,, \;\;\;
P_L=\frac12(1-\gamma_5) \,,
\end{equation}
we can rewrite the Lagrangian as follows:
\begin{equation}
{\cal L}_{\rm QCD}^0 = 
\bar{q}_R i \gamma^\mu {\cal D}_\mu q_R 
+\bar{q}_L i \gamma^\mu {\cal D}_\mu q_L 
- \frac14 
{\cal G}_{\mu\nu,a}
{\cal G}^{\mu\nu}_{a} \, .
\end{equation}
Restricting ourselves now to
up and down quarks,
we see that
${\cal L}_{\rm QCD}^0$ 
is invariant under the global unitary transformations
\begin{equation}
q_R =
\left( \begin{array}{c}
u_R \\ d_R
\end{array} \right)
\longmapsto
g_R \, q_R =
\exp
\left(
-i \Theta_i^R \frac{\tau_i}{2}
\right)
\left( \begin{array}{c}
u_R \\ d_R
\end{array} \right) 
\label{eq_gR}
\end{equation}
and
\begin{equation}
q_L =
\left( \begin{array}{c}
u_L \\ d_L
\end{array} \right)
\longmapsto
g_L \, q_L =
\exp
\left(
-i \Theta_i^L \frac{\tau_i}{2}
\right)
\left( \begin{array}{c}
u_L \\ d_L
\end{array} \right) 
\,,
\label{eq_gL}
\end{equation}
where $\tau_i \; (i=1,2,3)$ 
are the generators of $SU(2)_{\rm flavor}$,
the usual Pauli spin matrices with commutation relations
\begin{equation}
\left[\frac{\tau_i}{2},\frac{\tau_j}{2} \right]
= i\epsilon^{ijk}\frac{\tau_k}{2}
\,,
\end{equation}
and $g_R$ and $g_L$  are elements of 
$SU(2)_R$ and $SU(2)_L$, respectively. 
In conclusion: {\it The right- and left-handed components of
massless quarks do not mix.} 
This is
$SU(2)_R\times SU(2)_L$ 
symmetry, also known as {\it chiral symmetry}.
Noether's Theorem implies the existence of
six conserved currents;
three right-handed currents
\begin{equation}
R^\mu_i = \bar{q}_R \gamma^\mu \frac{\tau_i}{2} q_R 
\;\;\;\; \mbox{\small\rm with} \;\;\;\; \partial_\mu R^\mu_i = 0
\end{equation}
and three left-handed currents
\begin{equation}
L^\mu_i = \bar{q}_L \gamma^\mu \frac{\tau_i}{2} q_L 
\;\;\;\; \mbox{\small\rm with} \;\;\;\; \partial_\mu L^\mu_i = 0
\,.
\end{equation}
It is useful to consider the following linear
combinations; namely,
three vector currents
\begin{equation}
V^\mu_i = R^\mu_i + L^\mu_i = \bar{q} \gamma^\mu \frac{\tau_i}{2} q
\;\;\;\; \mbox{\small\rm with} \;\;\;\; \partial_\mu V^\mu_i = 0
\label{eq_Vmu}
\end{equation}
and three axial-vector currents
\begin{equation}
A^\mu_i = R^\mu_i - L^\mu_i = \bar{q} \gamma^\mu \gamma_5 
\frac{\tau_i}{2} q
\;\;\;\; \mbox{\small\rm with} \;\;\;\; \partial_\mu A^\mu_i = 0
\,,
\label{eq_Amu}
\end{equation}
which got their names from the fact that they
transform under parity as vector and axial-vector current densities, respectively.
The vector
transformations are given by
\begin{equation}
q =
\left( \begin{array}{c}
u \\ d
\end{array} \right)
\longmapsto
\exp
\left(
-i \Theta_i^V \frac{\tau_i}{2}
\right)
\left( \begin{array}{c}
u \\ d
\end{array} \right) ,
\end{equation}
which are isospin rotations
and, therefore, invariance under vector transformations can be
identified with isospin symmetry.

There are six conserved charges,
\begin{equation}
Q^R_i = \int d^3x \; R^0_i = \int d^3x \; q_R^\dagger (t,\vec x) 
\frac{\tau_i}{2} q_R(t,\vec x)
\;\;\;\; \mbox{\small\rm with} \;\;\;\; \frac{dQ^R_i}{dt} = 0
\end{equation}
and
\begin{equation}
Q^L_i = \int d^3x \; L^0_i = \int d^3x \; q_L^\dagger (t,\vec x) 
\frac{\tau_i}{2} q_L(t,\vec x)
\;\;\;\; \mbox{\small\rm with} \;\;\;\; \frac{dQ^L_i}{dt} = 0
\, ,
\end{equation}
or, alternatively,
\begin{equation}
Q^V_i = \int d^3x \; V^0_i = \int d^3x \; q^\dagger (t,\vec x) 
\frac{\tau_i}{2} q(t,\vec x)
\;\;\;\; \mbox{\small\rm with} \;\;\;\; \frac{dQ^V_i}{dt} = 0
\end{equation}
and
\begin{equation}
Q^A_i = \int d^3x \; A^0_i = \int d^3x \; q^\dagger (t,\vec x) 
\gamma_5 \frac{\tau_i}{2} q(t,\vec x)
\;\;\;\; \mbox{\small\rm with} \;\;\;\; \frac{dQ^A_i}{dt} = 0
\, .
\label{eq_charges}
\end{equation}
The $Q^L_i$ and $Q^R_i$ satisfy the commutation relations 
of the Lie algebra of $SU(2)_R\times SU(2)_L$ (chiral algebra),
\begin{equation}
\left[ Q^R_i, Q^R_j \right] = i \epsilon^{ijk} Q^R_k, \quad \quad
\left[ Q^L_i, Q^L_j \right] = i \epsilon^{ijk} Q^L_k, \quad \quad
\left[ Q^R_i, Q^L_j \right] = 0
\,.
\end{equation}
For $Q^V_i$ and $Q^A_i$, the commutation relations read,
\begin{equation}
\left[ Q^V_i, Q^V_j \right] = i \epsilon^{ijk} Q^V_k, \quad \quad
\left[ Q^A_i, Q^A_j \right] = i \epsilon^{ijk} Q^V_k, \quad \quad
\left[ Q^V_i, Q^A_j \right] = i \epsilon^{ijk} Q^A_k
\,.
\end{equation}
For reasons of completeness, we mention that massless $u$ and $d$ quarks satisfy
an even larger symmetry group, namely,  $SU(2)_R\times SU(2)_L\times U(1)_V\times U(1)_A$. 
While the $U(1)_V$ symmetry corresponds to quark number conservation, the
$U(1)_A$ is broken on the quantum level (``$U(1)_A$ anomaly'') and is not a symmetry
of the system.

\subsubsection{Explicit symmetry breaking}

The mass term  
 $- \bar{q}{\cal M}q$
in the QCD Lagrangian Eq.~(\ref{eq_LQCD}) 
breaks chiral symmetry explicitly. To better see this,
let's rewrite ${\cal M}$ for the two-flavor case,
\begin{eqnarray}
{\cal M} & = & 
\left( \begin{array}{cc}
            m_u & 0 \\
              0  & m_d 
           \end{array} \right)  \nonumber \\
  & = & \frac12 (m_u+m_d) 
\left( \begin{array}{cc}
            1 & 0 \\
              0  & 1 
           \end{array} \right) 
+ \frac12 (m_u-m_d) 
\left( \begin{array}{cc}
            1 & 0 \\
              0  & -1 
           \end{array} \right)  \nonumber \\
 & = & \frac12 (m_u+m_d) \; I + \frac12 (m_u-m_d) \; \tau_3 \,.
\label{eq_mmatr}
\end{eqnarray}
The first term in the last equation in invariant under $SU(2)_V$
(isospin symmetry) and the second term vanishes for
$m_u=m_d$.
Thus, isospin is an exact symmetry if 
$m_u=m_d$.
However, both terms in Eq.~(\ref{eq_mmatr}) break chiral symmetry.
Since the up and down quark masses
[Eqs.~(\ref{eq_umass}) and (\ref{eq_dmass})]
are small as compared to
the typical hadronic mass scale of $\sim 1$ GeV,
the explicit chiral symmetry breaking due to non-vanishing
quark masses is very small.
Note also that, as a consequence of the non-vanishing quark masses,
the axial-vector current, Eq.~(\ref{eq_Amu}),
is not conserved anymore.

\subsubsection{Spontaneous symmetry breaking}

A (continuous) symmetry is said to be {\it spontaneously
broken} if a symmetry of the Lagrangian 
is not realized in the ground state of the system.
There is evidence that the (approximate) chiral
symmetry of the QCD Lagrangian is spontaneously 
broken---for dynamical reasons of nonperturbative origin
which are not fully understood at this time.
The most plausible evidence comes from the hadron spectrum.

Since the conserved quantity $Q^A_i$, Eq.~(\ref{eq_charges}),
commutes with the Hamiltonian and has negative parity,
one naively expects the existence of degenerate hadron
multiplets of opposite parity, i.e., for any hadron of positive
parity one would expect a degenerate hadron state of negative 
parity and vice versa. However, these ``parity doublets'' are
not observed in nature. For example, take the $\rho$-meson which is
a vector meson of negative parity ($J^P=1^-$) and mass 
776 MeV. There does exist a $1^+$ meson, the $a_1$, but it
has a mass of 1230 MeV and, therefore, cannot be perceived
as degenerate with the $\rho$. On the other hand, the $\rho$
meson comes in three charge states (equivalent to
three isospin states), the $\rho^\pm$ and the $\rho^0$,
with masses that differ by at most a few MeV. Thus,
in the hadron spectrum,
$SU(2)_V$ (isospin) symmetry is well observed,
while axial symmetry is broken:
$SU(2)_R\times SU(2)_L$ is broken down to $SU(2)_V$.
As a consequence of this,
the vacuum (QCD ground state) is invariant under 
vector transformations, i.~e., $Q^V_i|0\rangle = 0$, while this is not the case for axial
transformations, 
$Q^A_i|0\rangle \neq 0$,
where $|0\rangle$ denotes the vacuum.

A spontaneously broken global symmetry implies the existence
of (massless) Goldstone bosons with the quantum numbers
of the broken generators~\cite{Gol61,GSW62}. The broken generators are the $Q^A_i$
of Eq.~(\ref{eq_charges}) which are pseudoscalar.
The Goldstone bosons are identified with the isospin
triplet of the (pseudoscalar) pions, 
which explains why pions are so light.
The pion masses are not exactly zero because the up
and down quark masses
are not exactly zero either (explicit symmetry breaking).
Thus, pions are a truly remarkable species:
they reflect spontaneous as well as explicit symmetry
breaking.
Goldstone bosons interact weakly at low energy.
They are degenerate with the vacuum 
and, therefore, interactions between them must
vanish at zero momentum and in the chiral limit
($m_\pi \rightarrow 0$).

\subsection{Chiral effective Lagrangians 
\label{sec_Lpi} }

\subsubsection{Relativistic formulation}

The next step in our EFT program is to build the most general
Lagrangian consistent with the (broken) symmetries discussed
above.
An elegant formalism for the construction of such Lagrangians
was developed by 
Callan, Coleman, Wess, and Zumino (CCWZ)~\cite{CCWZ}
who worked out
the group-theoretical foundations 
of non-linear realizations of chiral symmetry.\footnote{An accessible introduction into the rather
involved CCWZ formalism can be found in Ref.~\cite{Sch03}.}
It is characteristic for these non-linear realizations that, whenever functions of the Goldstone bosons
appear in the Lagrangian, they are always accompanied with at least one
space-time derivative.
The Lagrangians given below are built upon the CCWZ
formalism. 

As discussed, the relevant degrees of freedom are
pions (Goldstone bosons) and nucleons.
Since the interactions of Goldstone bosons must
vanish at zero momentum transfer and in the chiral
limit ($m_\pi \rightarrow 0$), the low-energy expansion
of the Lagrangian is arranged in powers of derivatives
and pion masses.
The hard scale is the chiral-symmetry breaking
scale, $\Lambda_\chi \approx 1$ GeV. Thus, the expansion is in terms
of powers of $Q/\Lambda_\chi$ where $Q$ is a (small) momentum
or pion mass.
This is chiral perturbation theory (ChPT).

The effective Lagrangian can formally be written as,
\begin{equation}
{\cal L_{\rm eff}} 
=
{\cal L}_{\pi\pi} 
+
{\cal L}_{\pi N} 
 + \, \ldots \,,
\end{equation}
where ${\cal L}_{\pi\pi}$
deals with the dynamics among pions, 
${\cal L}_{\pi N}$ 
describes the interaction
between pions and a nucleon,
and the ellipsis stands for
terms that involve pions and two or more nucleons.
The individual Lagrangians are organized as follows:
\begin{equation}
{\cal L}_{\pi\pi} 
 = 
{\cal L}_{\pi\pi}^{(2)} 
 + {\cal L}_{\pi\pi}^{(4)} 
 + \ldots 
\end{equation}
and
\begin{equation}
{\cal L}_{\pi N} 
= 
{\cal L}_{\pi N}^{(1)} 
+
{\cal L}_{\pi N}^{(2)} 
+
{\cal L}_{\pi N}^{(3)} 
+ \ldots ,
\end{equation}
where the superscript refers to the number of derivatives or 
pion mass insertions (chiral dimension)
and the ellipsis stands for terms of higher dimensions.

To construct chiral Lagrangians,
we introduce the following $SU(2)$ matrix $U$ in flavor space
which collects the Goldstone pion fields,
{\boldmath $\pi$}:
\begin{equation}
U  = 1 + 
\frac{i}{f_\pi}
\mbox{\boldmath $\tau$} \cdot 
\mbox{\boldmath $\pi$}
-\frac{1}{2f_\pi^2} 
\mbox{\boldmath $\pi$}^2
-\frac{i\alpha}{f_\pi^3}
(\mbox{\boldmath $\tau$} \cdot \mbox{\boldmath $\pi$})^3
+\frac{8\alpha-1}{8f_\pi^4} 
\mbox{\boldmath $\pi$}^4
+ \ldots  \,,
\label{eq_U}
\end{equation}
where $f_\pi$ denotes the pion decay constant.
In this expansion, the coefficient of the 
term linear 
in the pion field 
{\boldmath $\pi$}
is fixed such as to produce the correct kinetic
term in the $\pi\pi$ Lagrangian, below,
and the coefficient of the
quadratic term is chosen 
to satisfy the unitary condition $U^\dagger U = 1$,
at second order in the pion field.
However, the coefficient $\alpha$
of the third order is arbitrary
because of our freedom of choice for the interpolating
pion fields
(constrained only by $U$ unitary and $\det U = 1$).
The coefficient of the fourth order is then
dictated by the unitarity condition,
$U^\dagger U = 1$, at fourth order.
Note that (on-shell) observables must not depend
on the choice for the pion fields 
or, in other words, they must not depend
on the (unphysical) parameter $\alpha$ (and there are more
such parameters as you continue to higher orders
in the above expansion of $U$).
Therefore, diagrams with vertices that involve
three or four pions must always be grouped together
such that the $\alpha$ dependence drops out.
For more on this issue, see the calculation of the
two-loop $2\pi$ and $3\pi$ contributions, below,
where those vertices enter.
Popular choices for the pion fields are 
the exponential parametrization
$U=\exp
(i\mbox{\boldmath $\tau$} \cdot \mbox{\boldmath $\pi$}
/f_\pi)$ which corresponds to 
$\alpha=1/6$
and the so-called sigma representation
$U = ( \sigma +
i\mbox{\boldmath $\tau$} \cdot \mbox{\boldmath $\pi$} )
/f_\pi$
with
$\sigma=\sqrt{f_\pi^2- \mbox{\boldmath $\pi$}^2}$
which is equivalent to
$\alpha=0$.

The leading order (LO) 
$\pi\pi$ Lagrangian is now given by~\cite{GL84}
\begin{equation}
{\cal L}_{\pi\pi}^{(2)} =
\frac{f^2_\pi}{4} \, {\rm tr} \left[ \,
\partial_\mu U 
\partial^\mu U^\dagger 
+ m_\pi^2 ( U + U^\dagger)\, \right] \,,
\label{eq_Lpipi}
\end{equation}
where tr denotes the trace in flavor space and $m_\pi$ is the pion mass.
Since Goldstone bosons can interact only when they
carry momentum, the interaction between pions comes
in powers of $\partial_\mu U$.
Only even powers are allowed because of Lorentz invariance. 
Note that the $U$ field transforms under global chiral rotations via
\begin{equation}
U  \longmapsto  g_L \, U \, g_R^\dagger
\end{equation}
with $g_R$ and $g_L$  elements of 
$SU(2)_R$ and $SU(2)_L$, respectively,
cf.\ Eqs.~(\ref{eq_gR}) and (\ref{eq_gL}).
Since we consider global chiral rotations, 
$g_R$ and $g_L$ do not depend on space-time and, therefore,
$\partial_\mu U$ transforms in the same way as $U$.
Thus, the first term in Eq.~(\ref{eq_Lpipi}) is clearly chiral invariant.
The second term
breaks chiral symmetry explicitly with
the coefficient 
chosen such as to reproduce
the correct mass term, which can be seen by
inserting $U$ and expanding in numbers of
pion fields:
\begin{eqnarray}
	{\cal L}_{\pi\pi}^{(2)} 
     &=&
	\frac{1}{2} 
	\partial_\mu \mbox{\boldmath{$\pi$}} \cdot 
	\partial^\mu \mbox{\boldmath{$\pi$}}
     -  \frac{1}{2} m_\pi^2 \mbox{\boldmath{$\pi$}}^2
\label{eq_Lpipi2a}\\ &&
     +  \frac{1-4\alpha}{2f_\pi^2} 
        (\mbox{\boldmath{$\pi$}} \cdot \partial_\mu \mbox{\boldmath{$\pi$}})
        (\mbox{\boldmath{$\pi$}} \cdot \partial^\mu \mbox{\boldmath{$\pi$}})
     -  \frac{\alpha}{f_\pi^2} 
        \mbox{\boldmath{$\pi$}}^2
	\partial_\mu \mbox{\boldmath{$\pi$}} \cdot 
	\partial^\mu \mbox{\boldmath{$\pi$}}
     +\;  \frac{8\alpha-1}{8f_\pi^2} 
        m_\pi^2 \mbox{\boldmath{$\pi$}}^4
     + \; {\cal O} (\mbox{\boldmath $\pi$}^6)
\,,
\label{eq_Lpipi2}
\end{eqnarray}
where we dropped the constant term $f_\pi^2m_\pi^2$,
since it does not contribute to the dynamics.

Baryon fields can also be incorporated into the effective
field theory in a chirally consistent manner.
Gasser {\it et al.}~\cite{GSS88} have derived
the LO relativistic $\pi N$ Lagrangian to be
\begin{equation}
{\cal L}^{(1)}_{\pi N}  =  
 \bar{\Psi} \left(i\gamma^\mu {D}_\mu 
 - M_N
 + \frac{g_A}{2} \gamma^\mu \gamma_5 u_\mu
  \right) \Psi  
\label{eq_L1rel}
\end{equation}
with
\begin{equation}
{D}_\mu  =  \partial_\mu + \Gamma_\mu 
\end{equation}
the chirally covariant derivative 
which introduces the so-called
chiral connection
(an analogy to a gauge term)
\begin{eqnarray}
\Gamma_\mu & = & 
                 \frac12      \left[
                           \xi^\dagger, \partial_\mu \xi
                              \right]
=                \frac12      \left(
                           \xi^\dagger \partial_\mu \xi
                        +  \xi \partial_\mu \xi^\dagger
                               \right)
\label{eq_gamma}
\\
&=& 
\frac{i}{4f^2_\pi} \,
\mbox{\boldmath $\tau$} \cdot 
 ( \mbox{\boldmath $\pi$}
\times
 \partial_\mu \mbox{\boldmath $\pi$})
 \; + \; {\cal O} (\mbox{\boldmath $\pi$}^4) \,,
\end{eqnarray}
representing a vector current
that leads to a coupling of even numbers of pions with the nucleon.
Besides this, the Lagrangian includes a
coupling term which involves the axial vector
\begin{eqnarray}
u_\mu & = & 
                i \left\{
                           \xi^\dagger, \partial_\mu \xi
                  \right\}
              = i \left(
                           \xi^\dagger \partial_\mu \xi
                        -  \xi \partial_\mu \xi^\dagger
                  \right)  
\label{eq_umu}
\\
       &=&
      - \frac{1}{f_\pi} \mbox{\boldmath{$\tau$}} 
	\cdot \partial_\mu \mbox{\boldmath{$\pi$}}
        + \frac{4\alpha-1}{2f_\pi^3} 
	(\mbox{\boldmath{$\tau$}} \cdot \mbox{\boldmath{$\pi$}})
	(\mbox{\boldmath{$\pi$}} \cdot \partial_\mu \mbox{\boldmath{$\pi$}})
        + \frac{\alpha}{f_\pi^3} 
        \mbox{\boldmath{$\pi$}}^2 \,
	(\mbox{\boldmath{$\tau$}} \cdot \partial_\mu \mbox{\boldmath{$\pi$}})
 \; + \; {\cal O} (\mbox{\boldmath $\pi$}^5)  \,,
\end{eqnarray}
which couples an odd number of pions to the nucleon.
The definition of $\xi$ used in the above is
\begin{equation}
\xi  = \sqrt{U} = 1 +
\frac{i}{2f_\pi}
\mbox{\boldmath $\tau$} \cdot \mbox{\boldmath $\pi$}
-\frac{1}{8f_\pi^2} 
\mbox{\boldmath $\pi$}^2
-\frac{i(8\alpha-1)}{16f_\pi^3}
(\mbox{\boldmath $\tau$} \cdot \mbox{\boldmath $\pi$})^3
+ \ldots \,.
\label{eq_xi}
\end{equation}
Note that 
$\xi\xi^\dagger = 1$; therefore,
$\partial_\mu (\xi\xi^\dagger) =  
 (\partial_\mu \xi ) \xi^\dagger 
+\xi \partial_\mu \xi^\dagger =0$,
which is applied in Eqs.~(\ref{eq_gamma}) and (\ref{eq_umu}).

Thus, more explicitly, the LO relativistic $\pi N$
Lagrangian, Eq.~(\ref{eq_L1rel}), reads
\begin{equation}
{\cal L}^{(1)}_{\pi N}  =  
 \bar{\Psi} \left(i\gamma^\mu \partial_\mu - M_N
-\frac{1}{4f^2_\pi} \, \gamma^\mu
\mbox{\boldmath $\tau$} \cdot 
 ( \mbox{\boldmath $\pi$}
\times
 \partial_\mu \mbox{\boldmath $\pi$})
      - \frac{g_A}{2f_\pi} \gamma^\mu \gamma_5
        \mbox{\boldmath{$\tau$}} 
	\cdot \partial_\mu \mbox{\boldmath{$\pi$}}
  + \ldots
  \right) \Psi  
\,.
\label{eq_L1relalt}
\end{equation}
In this equation,
$\Psi$ is the relativistic four-component Dirac spinor field
representing the nucleon,
$M_N$ denotes the nucleon mass and
$g_A$ the axial-vector coupling constant.
Numerical values will be given later.
The term proportional to $g_A/2f_\pi$ is the familiar axial-vector
coupling of one pion to the nucleon,
while the nonlinear term proportional to $1/4f_\pi^2$ is 
known as the Weinberg-Tomozawa coupling (a $2\pi$ contact term)~\cite{Wei66,Tom66},
which is crucial for $\pi$-$N$ $s$-wave scattering at threshold.

\subsubsection{Heavy baryon formalism}

The relativistic treatment of baryons in chiral perturbation
theory leads to problems. The reason for the problems is
the fact that the time-derivative of a relativistic baryon field generates
a factor $E\approx M$ (where $M$ denotes the baryon mass)
which is not small as compared to the chiral-symmetry breaking
scale $\Lambda_\chi \approx 1$ GeV; in fact,
$M_N/\Lambda_\chi \approx 1$.
Note also that the nucleon mass does not vanish in the chiral limit.
The consequence of all this is that the one-to-one correspondence
between the expansion in pion loops, on the one hand,
and the expansion in terms of small external momenta and pion masses,
on the other side, is destroyed~\cite{GSS88,Ber92,BKM95}.

A solution to these problems has been proposed by Jenkins and
Manohar~\cite{JM91} using effective field theory techniques
originally developed by Georgi~\cite{Geo90}
for the study of heavy quark systems.
The basic idea is to treat the baryons
as heavy static sources (``extreme non-relativistic limit'')
such that the momentum transfer between
baryons by pion exchange is small as compared to the baryon mass.
The expansion is performed in terms of these small momenta
over the baryon mass and has become known as
heavy baryon (HB)  
chiral perturbation theory (HBChPT).
We will now briefly sketch this approach.

The four-momentum of the heavy baryon is parametrized as
\begin{equation}
p^\mu=Mv^\mu+l^\mu
\label{eq_pmu}
\end{equation}
where $v^\mu$ is the four-velocity satisfying $v^2=1$ and
$l^\mu$ is a small residual momentum, $v\cdot l \ll M$.
Defining projection operators
\begin{equation}
P_v^\pm=\frac{1\pm \gamma_\mu v^\mu}{2} \,,
\quad \quad \quad
P_v^+ + P_v^- = 1 \,,
\end{equation}
we introduce the so-called velocity-dependent fields
\begin{equation}
N=e^{iMv\cdot x} P_v^+ \Psi \,,
\quad \quad \quad
h=e^{iMv\cdot x} P_v^- \Psi \,,
\label{eq_Nh}
\end{equation}
such that the relativistic four-component 
Dirac spinor field representing the baryon can be written as
\begin{equation}
\Psi=e^{-iMv\cdot x}(N+h) \,.
\label{eq_PsiNh}
\end{equation}
The exponential factor in Eq.~(\ref{eq_Nh}) eliminates the kinematical dependence
on the baryon mass.
To make the meaning of the fields $N$ and $h$ more
transparent, consider a positive-energy plane wave solution of
the Dirac equation
\begin{equation}
\psi_p(x)=u(\vec p,s) e^{-i p\cdot x} \,,
\quad \quad
	u(\vec p,s) = \sqrt{\frac{E+M}{2M}} 
        \left(
	\begin{array}{c}
	  I   \\
	\frac{\vec \sigma \cdot \vec p}{E+M}
	\end{array}
	\right)
        \chi_s
\,,
\end{equation}
where $I$ is the two-dimensional identity matrix, 
$p^0=E=\sqrt{{\vec p}~^2+M^2}$, and $\chi_s$ a Pauli 
spinor describing the spin state of the baryon.
Assuming for the four-velocity $v^\mu = (1,0,0,0)$,
which is what we have in the rest frame of the baryon,
the explicit expressions for the wave function components
$N_p$ and $h_p$ are
\begin{eqnarray}
N_p&=&\sqrt{\frac{E+M}{2M}}
        \left(
	\begin{array}{c}
	 \chi_s\\
	   0
	\end{array}
	\right)
  e^{-i(E-M)t+i\vec p \cdot \vec x} 
\longmapsto 
        \left(
	\begin{array}{c}
	 \chi_s\\
	   0
	\end{array}
	\right)
  e^{-i(E-M)t+i\vec p \cdot \vec x} 
\,,
\\
h_p&=&\sqrt{\frac{E+M}{2M}}
        \left(
	\begin{array}{c}
	   0   \\
	\frac{\vec \sigma \cdot \vec p}{E+M}\chi_s
	\end{array}
	\right)
  e^{-i(E-M)t+i\vec p \cdot \vec x} \,,
\end{eqnarray}
where the $0$ represents a column vector that consists
of two zeros.
The arrow indicates the properly normalized spinor at leading order.
So, for $v^\mu=(1,0,0,0)$, 
$N_p$ represents the large/upper component
and $h_p$ the small/lower component of the 
Dirac wave function $\psi_p$.
Note also that the energy of 
$N_p$ and $h_p$ 
is different
from $\psi_p$; namely,  
they carry the small
residual energy
[cf.\ Eq.~(\ref{eq_pmu})]
\begin{equation}
l_0=E-M\approx 0 + \frac{{\vec p}~^2}{2M} + \ldots \,.
\end{equation}
Thus, $l_0$ is zero in leading order and ${\vec p}~^2/2M$ at 
next-to-leading order (NLO).

To derive the leading-order heavy-baryon Lagrangian,
we start from the leading-order relativistic one,
\begin{equation}
{\cal L}^{(1)}_{\pi N}  =  
 \bar{\Psi} \left(i\gamma^\mu {D}_\mu 
 - M
 + \frac{g_A}{2} \gamma^\mu \gamma_5 u_\mu
  \right) \Psi  
\,,
\end{equation}
and insert 
Eq.~(\ref{eq_PsiNh}),
assuming $v^\mu=(1,0,0,0)$ for simplicity\footnote{For a 
full-fledged derivation, see Ref.~\cite{Sch03}}.
One immediate result is that the $e^{-iMt}$ factor of
Eq.~(\ref{eq_PsiNh})
generates a term $\gamma_0 M$ which kills the
troublesome baryon mass term $(-M)$ in the upper component
projection yielding
\begin{equation}
{\cal L}^{(1)}_{\pi N}  =  
 \bar{N} \left(i\gamma^\mu {D}_\mu 
 + \frac{g_A}{2} \gamma^\mu \gamma_5 u_\mu
  \right) N  
 + \ldots
\label{eq_L1relN}
\end{equation}
where the ellipsis stands for additional expressions 
involving the lower-component
field $h$. Since $N$ contains only upper components, 
Eq.~(\ref{eq_L1relN})
collapses to
\begin{equation}
\widehat{\cal L}^{(1)}_{\pi N} 
 =  
\bar{N} \left(
 i {D}_0 
 - \frac{g_A}{2} \; 
\vec \sigma \cdot \vec u
\right) N  
\label{eq_L1HB}
\end{equation}
plus $1/M$ corrections which are generated by expressing $h$
in terms of $N$ via the equations of motion for $N$ and $h$. 
The first such $1/M$ corrections are shown in
Eq.~(\ref{eq_L2fixed}), below.

Equation~(\ref{eq_L1HB}) is the leading order $\pi N$
Lagrangian in the HB formalism (as indicated by the hat).
Expanding in numbers of pion fields, 
this Lagrangian reads
\begin{eqnarray}
\widehat{\cal L}^{(1)}_{\pi N} 
& = & 
\bar{N} \left\{ 
i \partial_0 
- \frac{1}{4f_\pi^2} \;
\mbox{\boldmath $\tau$} \cdot 
 ( \mbox{\boldmath $\pi$}
\times
 \partial_0 \mbox{\boldmath $\pi$})
- \frac{g_A}{2f_\pi} \;
\mbox{\boldmath $\tau$} \cdot 
 ( \vec \sigma \cdot \vec \nabla )
\mbox{\boldmath $\pi$} 
\right. 
\nonumber \\ && 
\left.
+ \frac{g_A(4\alpha-1)}{4f_\pi^3} \;
(\mbox{\boldmath $\tau$} \cdot 
\mbox{\boldmath $\pi$}) 
\left[ \mbox{\boldmath $\pi$} \cdot 
 ( \vec \sigma \cdot \vec \nabla )
\mbox{\boldmath $\pi$} \right]
+ \frac{g_A\alpha}{2f_\pi^3} \;
\mbox{\boldmath $\pi$}^2 
\left[ \mbox{\boldmath $\tau$} \cdot 
 ( \vec \sigma \cdot \vec \nabla )
\mbox{\boldmath $\pi$} 
\right]
\right\} N 
+ \ldots
\label{eq_L1}
\end{eqnarray}
where the ellipsis stands for terms involving four
or more pions.

{\it At dimension two},
the relativistic $\pi N$ Lagrangian reads
\begin{equation}
{\cal L}^{(2)}_{\pi N} 
= \sum_{i=1}^{4} c_i \bar{\Psi} O^{(2)}_i \Psi \, .
\label{eq_L2rel}
\end{equation}
The various operators $O^{(2)}_i$ are given in Ref.~\cite{Fet00}.
The fundamental rule by which this Lagrangian---as well as all 
the other
ones---are assembled is that they must contain {\it all\/} terms
consistent with chiral symmetry and Lorentz invariance 
(apart from other trivial
symmetries) at a given chiral dimension (here: order two).
The parameters $c_i$ are known as low-energy constants (LECs)
and are determined empirically from fits to $\pi N$ data.

The HB projected $\pi N$ Lagrangian at order two 
is most conveniently broken up into two pieces,
\begin{equation}
\widehat{\cal L}^{(2)}_{\pi N} \, = \,
\widehat{\cal L}^{(2)}_{\pi N, \, \rm fixed} \, + \,
\widehat{\cal L}^{(2)}_{\pi N, \, \rm ct} \, ,
\label{eq_L2}
\end{equation}
with
\begin{eqnarray}
\widehat{\cal L}^{(2)}_{\pi N, \, \rm fixed}  
 & = &  
 \bar{N} \left[
\frac{1}{2M_N}\: \vec D \cdot \vec D
+ i\, \frac{g_A}{4M_N}\: \{\vec \sigma \cdot \vec D, u_0\}
 \right] N
\nonumber \\ & = & 
 \bar{N} \left\{
 \frac{{\vec \nabla}^2}{2M_N} 
-\frac{ig_A}{4M_Nf_\pi} 
\mbox{\boldmath $\tau$} \cdot 
\left[
\vec \sigma \cdot
\left( \stackrel{\leftarrow}{\nabla} 
\partial_0 \mbox{\boldmath $\pi$}
 -
\partial_0 \mbox{\boldmath $\pi$}
\stackrel{\rightarrow}{\nabla} \right)
\right]
\right.
\nonumber \\ &&
\left.
- \frac{i}{8M_N f_\pi^2}
\mbox{\boldmath $\tau$} \cdot 
\left[
\stackrel{\leftarrow}{\nabla} 
\cdot
( \mbox{\boldmath $\pi$} \times \vec\nabla \mbox{\boldmath $\pi$} )
   -   
( \mbox{\boldmath $\pi$} \times \vec\nabla \mbox{\boldmath $\pi$} )
\cdot
\stackrel{\rightarrow}{\nabla} 
\right]
\right\} N 
 + \ldots
\label{eq_L2fixed}
\end{eqnarray}
and
\begin{eqnarray}
\widehat{\cal L}^{(2)}_{\pi N, \, \rm ct}
& = & 
 \bar{N} \left[
 2\, c_1
\, m_\pi^2\, (U+U^\dagger)
\, + \, 
\left( c_2 - \frac{g_A^2}{8M_N}\right) u_0^2
 \, + \,
c_3
\, u_\mu  u^\mu
%\right.  \nonumber \\ && \left.
+ \, \frac{i}{2} 
\left( c_4 + \frac{1}{4M_N} \right) 
  \vec \sigma \cdot ( \vec u \times \vec u)
 \right] N 
\nonumber \\ & = & 
 \bar{N} \left[
 4c_1m_\pi^2
-\frac{2 c_1}{f_\pi^2} \, m_\pi^2\, \mbox{\boldmath $\pi$}^2 
\, + \, 
\left( c_2 - \frac{g_A^2}{8M_N}\right) 
\frac{1}{f_\pi^2}
(\partial_0 \mbox{\boldmath{$\pi$}} \cdot 
 \partial_0 \mbox{\boldmath{$\pi$}})
%\right.  \nonumber \\ && \left.
 + \, \frac{c_3}{f_\pi^2}\,
(\partial_\mu \mbox{\boldmath{$\pi$}} \cdot 
\partial^\mu \mbox{\boldmath{$\pi$}})
\right.  
\nonumber \\ && \left.
 - \, \left( c_4 + \frac{1}{4M_N} \right) 
\frac{1}{2f_\pi^2}
\epsilon^{ijk} \epsilon^{abc} \sigma^i \tau^a
(\partial^j \pi^b) (\partial^k \pi^c)
 \right] N 
\, + \, \ldots ,
\label{eq_L2ct}
\end{eqnarray}
where we neglected the isospin-breaking $c_5$-term
proportional to $(m_u-m_d)$;
the ellipsis represents terms involving 
more pions.

Note that 
$\widehat{\cal L}^{(2)}_{\pi N, \, \rm fixed}$  
is created entirely from the HB expansion of the relativistic
${\cal L}^{(1)}_{\pi N}$ and thus has no free parameters (``fixed''),
while 
$\widehat{\cal L}^{(2)}_{\pi N, \, \rm ct}$
is made up by the new $\pi N$ contact terms proportional to the
$c_i$ parameters (plus those $1/M_N$ corrections
which happen to have the same mathematical structure as
$c_i$ terms).

{\it At dimension three},
the relativistic $\pi N$ Lagrangian can be formally written as
\begin{equation}
{\cal L}^{(3)}_{\pi N} 
= \sum_{i=1}^{31} d_i \bar{\Psi} O^{(3)}_i \Psi \, ,
\label{eq_L3rel}
\end{equation}
with the operators, $O^{(3)}_i$, listed in Refs.~\cite{Fet00,FMS98}; 
not all 31 terms will be of interest here.
The new LECs that occur at this order are the $d_i$.
Similar to the order two case,
the HB projected Lagrangian at order three can be broken into two pieces,
\begin{equation}
\widehat{\cal L}^{(3)}_{\pi N} \, = \,
\widehat{\cal L}^{(3)}_{\pi N, \, \rm fixed} \, + \,
\widehat{\cal L}^{(3)}_{\pi N, \, \rm ct} \, ,
\label{eq_L3}
\end{equation}
with
$\widehat{\cal L}^{(3)}_{\pi N, \, \rm fixed}$
and
$\widehat{\cal L}^{(3)}_{\pi N, \, \rm ct}$
given in Refs.~\cite{Fet00,FMS98}.

\subsection{Nucleon contact Lagrangians \label{sec_Lct}}

Two-nucleon contact interactions consist of four nucleon
fields (four nucleon legs) and no meson fields.
Such terms are needed to renormalize
loop integrals, to make results reasonably independent
of regulators, and to parametrize the 
unresolved short-distance dynamics of the nuclear
force. For more on the role of contact terms,
see Section~\ref{sec_ct}.

Because of parity, nucleon contact interactions come only
in even powers of derivatives, thus,
\begin{equation}
\label{eq_LNN}
\widehat{\cal L}_{NN} =
\widehat{\cal L}^{(0)}_{NN} +
\widehat{\cal L}^{(2)}_{NN} +
\widehat{\cal L}^{(4)}_{NN} + \ldots
\end{equation}

The lowest order (or leading order) $NN$ Lagrangian has no derivatives 
and reads~\cite{Wei90,Wei91}
\begin{equation}
\label{eq_LNN0}
\widehat{\cal L}^{(0)}_{NN} =
-\frac{1}{2} C_S \bar{N} N \bar{N} N 
-\frac{1}{2} C_T (\bar{N} \vec \sigma N) \cdot (\bar{N} \vec \sigma N) \, ,
\end{equation}
where $N$ is the heavy baryon nucleon field ($\bar{N}=N^\dagger$).
$C_S$ and $C_T$ are unknown constants which are determined
by a fit to the $NN$ data.
The second order $NN$ Lagrangian can be stated as
follows~\cite{ORK96}
\begin{eqnarray}
\label{eq_LNN2}
\widehat{\cal L}^{(2)}_{NN} &=&
-C'_1 \left[(\bar{N} \vec \nabla N)^2+ (\overline{\vec \nabla N} N)^2 \right]
-C'_2 (\bar{N} \vec \nabla N)\cdot (\overline{\vec \nabla N} N)
%\nonumber \\ &&
-C'_3 \bar{N} N \left[\bar N \vec \nabla^2 N+\overline{\vec \nabla^2 N} N \right]
\nonumber \\ &&
-i C'_4 \left[
\bar N \vec \nabla N \cdot 
(\overline{\vec \nabla N} \times \vec \sigma N) +
\overline{(\vec \nabla N)} N \cdot 
(\bar N \vec \sigma \times \vec \nabla N) \right]
\nonumber \\ &&
-i C'_5 \bar N N(\overline{\vec \nabla N} \cdot \vec \sigma \times \vec \nabla N)
-i C'_6 (\bar N \vec \sigma N)\cdot (\overline{\vec \nabla N} \times \vec \nabla N)
\nonumber \\ &&
-\left(C'_7 \delta_{ik} \delta_{jl}+C'_8 \delta_{il} \delta_{kj}
+C'_9 \delta_{ij} \delta_{kl}\right)
%\nonumber \\ && \times
\left[\bar N \sigma_k \partial_i N \bar N \sigma_l \partial_j N +
\overline{\partial_i N} \sigma_k N \overline{\partial_j N} \sigma_l N \right]
\nonumber \\ &&
-\left(C'_{10} \delta_{ik} \delta_{jl}+C'_{11} \delta_{il} \delta_{kj}+C'_{12} \delta_{ij} \delta_{kl}\right)
\bar N \sigma_k \partial_i N \overline{\partial_j N} \sigma_l N
\nonumber \\ &&
-\left(\frac{1}{2} C'_{13} (\delta_{ik} \delta_{jl}+
\delta_{il} \delta_{kj}) 
%\nonumber \\ &&
+C'_{14} \delta_{ij} \delta_{kl} \right)
\left[\overline{\partial_i N} \sigma_k \partial_j N + \overline{\partial_j N} \sigma_k \partial_i N\right]
\bar N \sigma_l N \, .
\end{eqnarray}
For a thorough discussion of second order contact Lagrangians, see Refs.~\cite{Epe00,Gir10}. 

Similar to $C_S$ and $C_T$ of Eq.~(\ref{eq_LNN0}), the $C'_i$ of Eq.~(\ref{eq_LNN2}) are unknown constants 
which are fixed in a fit to the $NN$ data.
Obviously, these contact Lagrangians blow up quite a bit
with increasing order, which is why we do not give
$\widehat{\cal L}^{(4)}_{NN}$
explicitly here.
The $NN$ contact potentials that emerge from these Lagrangians
are given in Section~\ref{sec_ct}.

Besides the above contact Lagrangians involving two nucleons,
there exist contact interactions among three or more nucleons
representing nuclear many-body forces 
(cf.\ last term of Eq.~(\ref{eq_LD1}), below, and Section~\ref{sec_manyNF}).

\subsection{Summary: Effective Lagrangians organized by 
interaction index $\Delta$}

To summarize, the effective Lagrangian needed to derive nuclear
forces includes the following parts:
\begin{equation}
{\cal L_{\rm eff}} 
=
{\cal L}_{\pi\pi} 
+
{\cal L}_{\pi N} 
+
{\cal L}_{NN} 
+ \ldots ,
\end{equation}
where the ellipsis stands for terms that involve two nucleons plus
pions and three or more
nucleons with or without pions, relevant for nuclear
many-body forces 
(cf.\ last two terms of Eq.~(\ref{eq_LD1}), below, and Section~\ref{sec_manyNF}).

In previous sections, we organized the Lagrangians by the number
of derivatives or pion-mass insertions. This is 
the standard way, appropriate particularly for
considerations of $\pi$-$\pi$ and $\pi$-$N$ scattering.
As it turns out (cf.\ Section~\ref{sec_chpt}), 
for interactions among nucleons,
it is sometimes useful to also consider the so-called
index of the interaction,
\begin{equation}
\Delta  \equiv   d + \frac{n}{2} - 2  \, ,
\label{eq_Delta}
\end{equation}
where $d$ is the number of derivatives or pion-mass insertions 
and $n$ the number of nucleon field operators (nucleon legs).
We will now re-write the HB Lagrangian in terms
of increasing values of the parameter $\Delta$.

The leading-order Lagrangian reads,
\begin{eqnarray}
\widehat{\cal L}^{\Delta=0} &=&
	\frac{1}{2} 
	\partial_\mu \mbox{\boldmath{$\pi$}} \cdot 
	\partial^\mu \mbox{\boldmath{$\pi$}}
     -  \frac{1}{2} m_\pi^2 \mbox{\boldmath{$\pi$}}^2
\nonumber \\ &&
     +  \frac{1-4\alpha}{2f_\pi^2} 
        (\mbox{\boldmath{$\pi$}} \cdot \partial_\mu \mbox{\boldmath{$\pi$}})
        (\mbox{\boldmath{$\pi$}} \cdot \partial^\mu \mbox{\boldmath{$\pi$}})
     -  \frac{\alpha}{f_\pi^2} 
        \mbox{\boldmath{$\pi$}}^2
	\partial_\mu \mbox{\boldmath{$\pi$}} \cdot 
	\partial^\mu \mbox{\boldmath{$\pi$}}
     +\;  \frac{8\alpha-1}{8f_\pi^2} 
        m_\pi^2 \mbox{\boldmath{$\pi$}}^4
\nonumber \\ &&
+ \bar{N} \left[ 
i \partial_0 
- \frac{g_A}{2f_\pi} \; \mbox{\boldmath $\tau$} \cdot 
 ( \vec \sigma \cdot \vec \nabla ) \mbox{\boldmath $\pi$} 
- \frac{1}{4f_\pi^2} \; \mbox{\boldmath $\tau$} \cdot 
 ( \mbox{\boldmath $\pi$}
\times \partial_0 \mbox{\boldmath $\pi$})
\right] N
\nonumber \\ &&
+ \bar{N} \left\{
 \frac{g_A(4\alpha-1)}{4f_\pi^3} \;
(\mbox{\boldmath $\tau$} \cdot 
\mbox{\boldmath $\pi$}) 
\left[ \mbox{\boldmath $\pi$} \cdot 
 ( \vec \sigma \cdot \vec \nabla )
\mbox{\boldmath $\pi$} \right]
+ \frac{g_A\alpha}{2f_\pi^3} \;
\mbox{\boldmath $\pi$}^2 
\left[ \mbox{\boldmath $\tau$} \cdot 
 ( \vec \sigma \cdot \vec \nabla )
\mbox{\boldmath $\pi$} 
\right]
\right\} N 
\nonumber \\ &&
-\frac{1}{2} C_S \bar{N} N \bar{N} N 
-\frac{1}{2} C_T (\bar{N} \vec \sigma N) \cdot (\bar{N} \vec \sigma N) 
\; + \; \ldots \,,
\label{eq_LD0}
\end{eqnarray}
and subleading Lagrangians are,
\begin{eqnarray}
\widehat{\cal L}^{\Delta=1} &=&
 \bar{N} \left\{
 \frac{{\vec \nabla}^2}{2M_N} 
-\frac{ig_A}{4M_Nf_\pi} 
\mbox{\boldmath $\tau$} \cdot 
\left[
\vec \sigma \cdot
\left( \stackrel{\leftarrow}{\nabla} 
\partial_0 \mbox{\boldmath $\pi$}
 -
\partial_0 \mbox{\boldmath $\pi$}
\stackrel{\rightarrow}{\nabla} \right)
\right]
\right.
\nonumber \\ &&
\left.
- \frac{i}{8M_N f_\pi^2}
\mbox{\boldmath $\tau$} \cdot 
\left[
\stackrel{\leftarrow}{\nabla} 
\cdot
( \mbox{\boldmath $\pi$} \times \vec\nabla \mbox{\boldmath $\pi$} )
   -   
( \mbox{\boldmath $\pi$} \times \vec\nabla \mbox{\boldmath $\pi$} )
\cdot
\stackrel{\rightarrow}{\nabla} 
\right]
\right\} N 
\nonumber \\ &&
+ \bar{N} \left[
 4c_1m_\pi^2
-\frac{2 c_1}{f_\pi^2} \, m_\pi^2\, \mbox{\boldmath $\pi$}^2 
\, + \, 
\left( c_2 - \frac{g_A^2}{8M_N}\right) 
\frac{1}{f_\pi^2}
(\partial_0 \mbox{\boldmath{$\pi$}} \cdot 
 \partial_0 \mbox{\boldmath{$\pi$}})
\right.  \nonumber \\ &&  \left.
 + \, \frac{c_3}{f_\pi^2}\,
(\partial_\mu \mbox{\boldmath{$\pi$}} \cdot 
\partial^\mu \mbox{\boldmath{$\pi$}})
%\right.  \nonumber \\ &&  \left.
 - \, \left( c_4 + \frac{1}{4M_N} \right) 
\frac{1}{2f_\pi^2}
\epsilon^{ijk} \epsilon^{abc} \sigma^i \tau^a
(\partial^j \pi^b) (\partial^k \pi^c)
 \right] N 
\nonumber \\ &&
- \frac{D}{4f_\pi} (\bar{N}N) \bar{N} \left[ 
\mbox{\boldmath $\tau$} 
\cdot ( \vec \sigma \cdot \vec \nabla )
\mbox{\boldmath $\pi$} 
\right] N
-\frac12 E
(\bar{N}N)
(\bar{N}
\mbox{\boldmath $\tau$} 
N)
\cdot
(\bar{N}
\mbox{\boldmath $\tau$} 
N)
\; + \; \ldots \,,
\label{eq_LD1}
\\
\widehat{\cal L}^{\Delta=2} &=&
\; {\cal L}^{(4)}_{\pi\pi} \; +
\; \widehat{\cal L}^{(3)}_{\pi N} \; + \; \widehat{\cal L}^{(2)}_{NN}
\; + \; \ldots \,,
\label{eq_LD2}
\\
\widehat{\cal L}^{\Delta=4} &=&
\; \widehat{\cal L}^{(4)}_{NN}
\; + \; \ldots \,,
\label{eq_LD4}
\end{eqnarray}
where the ellipses represent terms that are irrelevant
for the derivation of nuclear forces up to fourth
order.

\section{Nuclear forces from EFT: Overview
\label{sec_overview}}

In the beginning of Section~\ref{sec_EFT},
we listed the steps we have to take for carrying out
the EFT program of a derivation of nuclear forces.
So far, we discussed steps one to three.
What is left are steps four (low-momentum expansion) and
five (Feynman diagrams).
In this section, we will say more about the expansion
we are using and
give an overview of the Feynman diagrams that arise
order by order.

\subsection{Chiral perturbation theory and power counting
\label{sec_chpt}}

Effective Lagrangians have infinitely
many terms, and an unlimited number of Feynman graphs can be calculated
from them. Therefore, 
we need a scheme that makes the theory manageable and calculable.
This scheme
which tells us how to distinguish between large
(important) and small (unimportant) contributions
is chiral perturbation theory (ChPT).

In ChPT, 
graphs are analyzed
in terms of powers of small external momenta over the large scale:
$(Q/\Lambda_\chi)^\nu$,
where $Q$ is generic for a momentum (nucleon three-momentum or
pion four-momentum) or a pion mass and $\Lambda_\chi \sim 1$ GeV
is the chiral symmetry breaking scale (hadronic scale, hard scale).
Determining the power $\nu$ 
has become known as power counting.

For the moment, we will consider only so-called irreducible
graphs; the problem of reducible or iterative
diagrams and their relevance for the $NN$ system
will be discussed later (cf.\ Section~\ref{sec_reno}).
By definition, an irreducible graph is a diagram that
cannot be separated into two
by cutting only nucleon lines.
Following the Feynman rules of covariant perturbation theory,
a nucleon propagator is $Q^{-1}$,
a pion propagator $Q^{-2}$,
each derivative in any interaction is $Q$,
and each four-momentum integration $Q^4$.
This is also known as naive dimensional analysis.
Applying then some topological identities, Weinberg obtained
for the power of an irreducible diagram
involving $A$ nucleons~\cite{Wei90,Wei91,Wei92}
\begin{equation}
\nu_W = 4 - A - 2C + 2L + \sum_i \Delta_i 
\label{eq_nuw}
\end{equation}
with
\begin{equation}
\Delta_i  \equiv   d_i + \frac{n_i}{2} - 2  \, ,
\label{eq_Deltai}
\end{equation}
where $C$ denotes the number of separately connected pieces and
$L$ the number of loops in the diagram;
$d_i$ is the number of derivatives or pion-mass insertions 
and $n_i$ the number of nucleon fields (nucleon legs)
involved in vertex $i$;
the sum runs over all vertices $i$ contained in the diagram 
under consideration.
Note that $\Delta_i \geq 0$
for all interactions allowed by chiral symmetry.
Purely pionic interactions have at least two derivatives
($d_i\geq 2, n_i=0$);
interactions of pions with a nucleon have at least one
derivative
($d_i\geq 1, n_i=2$);
and nucleon-nucleon contact terms ($n_i=4$)
have $d_i\geq0$.
This demonstrates how chiral symmetry
guarantees a low-energy expansion.

The Weinberg formula 
Eq.~(\ref{eq_nuw})
works well for connected diagrams
with $A \leq 2$ nucleons and any number of pions, 
but there are problems when applied in 
systems with $A\geq 3$.

To illustrate the problem, consider
one-pion exchange 
($L=0$, $\Delta_i=0$) 
between two nucleons
in an $A=2$ system ($C=1$) which yields 
$\nu_W=0$.
Now, when the same interaction occurs
in an $A=3$ environment, then $C=2$, since one
nucleon is not interacting, and 
Eq.~(\ref{eq_nuw}) 
produces $\nu_W=-3$.
The reason for this result is that, by widespread convention~\cite{PDG},
particle states are normalized to Dirac $\delta$-functions,
\begin{equation}
\langle p' | p \rangle = 
\delta^3(\vec p - {\vec p}~') 
\,,
\label{eq_norm}
\end{equation}
which carry dimension $Q^{-3}$.
A nucleon line 
passing through a diagram
without interaction is represented by such a 
three-momentum-conserving
$\delta$-function. 

The above-illustrated $A$-dependence of the power $\nu_W$
is undesirable.
As indicated, the reason for the problem is the dimension $(-3)$
introduced for each nucleon through the normalization Eq.~(\ref{eq_norm}).
Therefore, to fix the problem,
we add $(+3A)$ to Eq.~(\ref{eq_nuw}) and subtract $6$ to have the $A=2$ case
unaltered.
Thus, we introduce the new power $\; \nu=\nu_W+3A-6$, which reads explicitly,
\begin{equation} \nu = -2 +2A - 2C + 2L 
+ \sum_i \Delta_i \, .  
\label{eq_nu} 
\end{equation}
This definition of the power $\nu$ will be applied
throughout this report.
It is also what is used in other works~\cite{Epe06}.
Another way to get the formula Eq.~(\ref{eq_nu}) is to define $\nu$
as the difference between $\nu_W$ and the smallest possible power,
$\nu_{\rm min}$: $\nu=\nu_W-\nu_{\rm min}$.
The minimal power $\nu_{\rm min}$ is obtained from Eq.~(\ref{eq_nuw})
for the case of no loops ($L=0$), $\Delta_i=0$ for all vertices, and the 
maximal number of separately connected pieces, which is $C=A-1$
if at least two nucleons interact;
thus, $\nu_{\rm min}=6-3A$.

Notice that, even though $A$ still appears in 
Eq.~(\ref{eq_nu}), 
the formula is essentially $A$ independent:
when a non-interacting nucleon is added,
$A$ and $C$ go up by one, which cancels.
From the above derivation, 
it should also be clear that Eq.~(\ref{eq_nu})
is suitable only for systems with $A \geq 2$ nucleons
while 
Eq.~(\ref{eq_nuw})
may be used for connected graphs with $A \leq 2$.

An alternative method for developing a reasonable
power formula has been presented by Friar~\cite{Fri97},
who considers the expectation value of an $m$-nucleon operator 
($m\leq A$) in $A$-nucleon space.
Working in configuration space generates additional
phase-space factors of power $3(A-1)$
to be added to Eq.~(\ref{eq_nuw}).
Thus, Friar's power is $\nu_F =\nu_W +3A-3=\nu+3$.
The additional power of three by which Friar differs from
Eq.~(\ref{eq_nu})
is due to an additional momentum-space integration
that converts the units of a momentum-space potential
into units of energy, which is reasonable.
Finally, we note that
normalizing the
nucleon states in a box to a Kronecker-$\delta$ (instead of
using the Dirac $\delta$-function normalization of the continuum)
should lead to Eq.~(\ref{eq_nu}) 
in a straightforward fashion.

In any case,
the most important observation from power counting is that
the powers are bounded from below and, 
specifically, $\nu \geq 0$. 
This fact is crucial for the convergence of 
the low-momentum expansion.

Moreover, the power formula 
Eq.~(\ref{eq_nu}) 
allows to predict
the leading orders of connected multi-nucleon forces.
Consider a $m$-nucleon irreducibly connected diagram
($m$-nucleon force) in an $A$-nucleon system ($m\leq A$).
The number of separately connected pieces is
$C=A-m+1$. Inserting this into
Eq.~(\ref{eq_nu}) together with $L=0$ and 
$\sum_i \Delta_i=0$ yields
$\nu=2m-4$. Thus, two-nucleon forces ($m=2$) start 
at $\nu=0$, three-nucleon forces ($m=3$) at
$\nu=2$ (but they happen to cancel at that order),
and four-nucleon forces at $\nu=4$ (they don't cancel).
More about this in the next sub-section and 
Section~\ref{sec_manyNF}.

For later purposes, we note that for an irreducible 
$NN$ diagram ($A=2$, $C=1$), the
power formula collapses to the very simple expression
\begin{equation}
\nu =  2L + \sum_i \Delta_i \,.
\label{eq_nunn}
\end{equation}

In summary, the chief point of the ChPT expansion is
that,
at a given order $\nu$, there exists only a finite number
of graphs. This is what makes the theory calculable.
The expression $(Q/\Lambda_\chi)^{\nu+1}$ provides a rough estimate
of the relative size of the contributions left out and, thus,
of the accuracy at order $\nu$.
In this sense, the theory can be calculated to any
desired accuracy and has
predictive power.

\subsection{The hierarchy of nuclear forces}
\begin{figure}[t]\centering
\vspace*{-0.25cm}
\scalebox{0.65}{\includegraphics{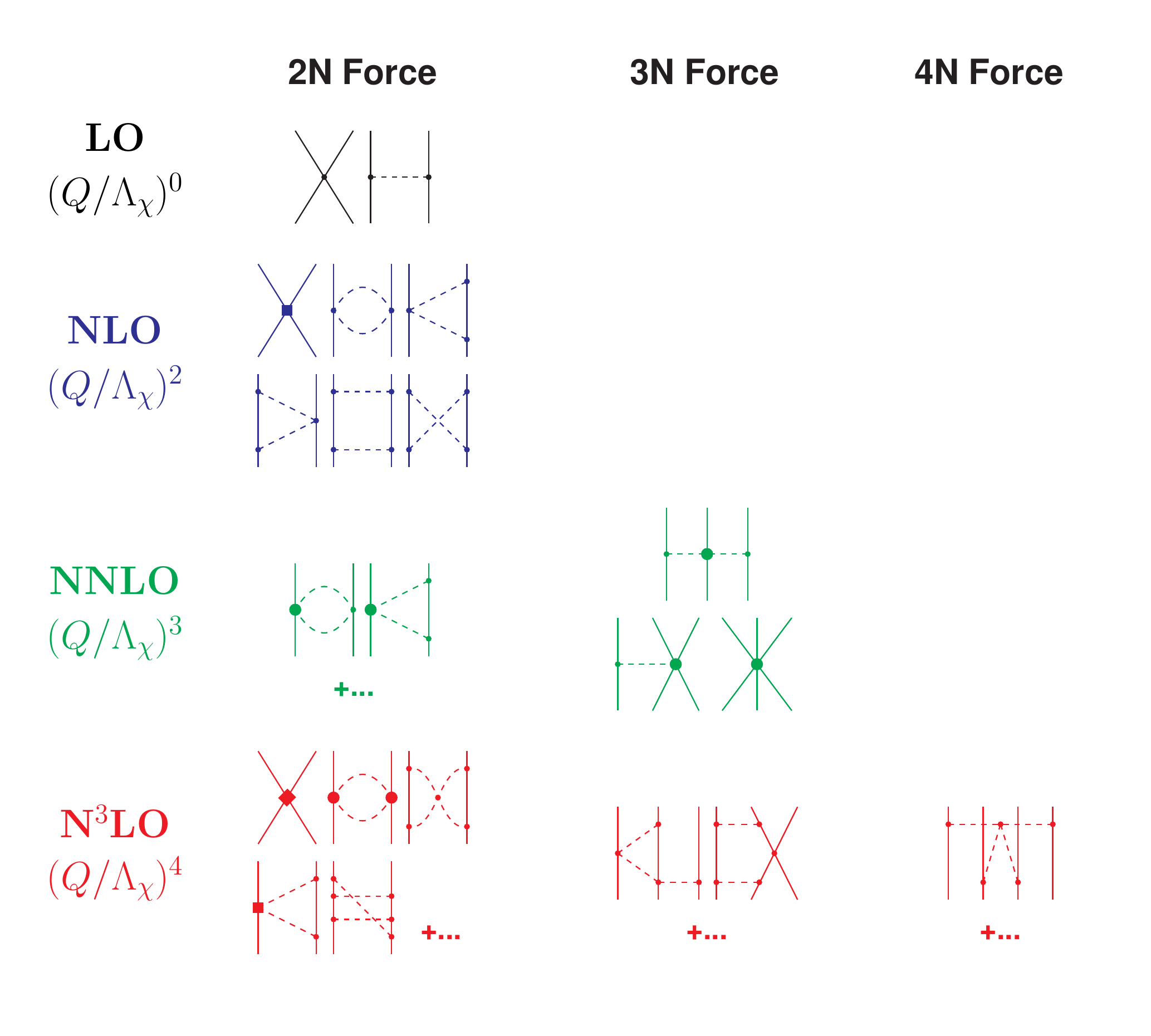}}
\vspace*{-1.0cm}
\caption{Hierarchy of nuclear forces in ChPT. Solid lines
represent nucleons and dashed lines pions. 
Small dots, large solid dots, solid squares, and solid diamonds
denote vertices of index $\Delta= \, $ 0, 1, 2, and 4, respectively. 
Further explanations are
given in the text.}
\label{fig_hi}
\end{figure}

Chiral perturbation theory and power counting
imply that nuclear forces emerge as a hierarchy
controlled by the power $\nu$, Fig.~\ref{fig_hi}.

In lowest order, better known as leading order (LO, $\nu = 0$), 
the $NN$ amplitude
is made up by two momentum-independent contact terms
($\sim Q^0$), 
represented by the 
four-nucleon-leg graph
with a small-dot vertex shown in the first row of 
Fig.~\ref{fig_hi},
and
static one-pion exchange (1PE), second
diagram in the first row of the figure.
This is, of course, a rather crude approximation
to the two-nucleon force (2NF), but accounts already for some
important features.
The 1PE provides the tensor force,
necessary to describe the deuteron, and it explains
$NN$ scattering in peripheral partial waves of very high
orbital angular momentum. At this order, the two contacts 
which contribute only in $S$-waves provide
the short- and intermediate-range interaction which is somewhat
crude.

In the next order,
$\nu=1$, all contributions vanish due to parity
and time-reversal invariance.

Therefore, the next-to-leading order (NLO) is $\nu=2$.
Two-pion exchange (2PE) occurs for the first time
(``leading 2PE'') and, thus, the creation of
a more sophisticated
description of the intermediate-range interaction
is starting here. 
Since the loop involved in each pion-diagram implies
already $\nu=2$ [cf.\ Eq.~(\ref{eq_nunn})],
the vertices must have $\Delta_i = 0$.
Therefore, at this order, only the lowest order
$\pi NN$ and $\pi \pi NN$ vertices are allowed which
is why the leading 2PE is rather weak.
Furthermore, there are 
seven contact terms of 
${\cal O}(Q^2)$, 
shown by
the four-nucleon-leg graph with a solid square,
 which contribute
in $S$ and $P$ waves. The operator structure of these
contacts include a spin-orbit term besides central,
spin-spin, and tensor terms. Thus, essentially all spin-isospin
structures necessary to describe the two-nucleon
force phenomenologically have been generated at this order.
The main deficiency at this stage of development 
is an insufficient intermediate-range attraction.

This problem is finally fixed at order three 
($\nu=3$), next-to-next-to-leading order (NNLO).
The 2PE involves now the two-derivative
$\pi\pi NN$ seagull vertices (proportional to
the $c_i$ LECs) denoted by a large solid dot
in Fig.~\ref{fig_hi}.
These vertices represent correlated 2PE
as well as intermediate $\Delta(1232)$-isobar contributions.
It is well known from the meson phenomenology of 
nuclear forces~\cite{MHE87,Lac80}
that these two contributions are crucial
for a realistic and quantitative 2PE model.
Consequently, the 2PE now assumes a realistic size
and describes the intermediate-range attraction of the
nuclear force about right. Moreover, first relativistic 
corrections come into play at this order.
There are no new contacts.

The reason why we talk of a hierarchy of nuclear forces is that 
two- and many-nucleon forces are created on an equal footing
and emerge in increasing number as we go to higher and higher orders.
At NNLO, the first set of
nonvanishing three-nucleon forces (3NF) occur~\cite{Kol94,Epe02b},
cf.\ column `3N Force' of
Fig.~\ref{fig_hi}. 
In fact, at the previous order, NLO,
irreducible 3N graphs appear already, however,
it has been shown by Weinberg~\cite{Wei92} and 
others~\cite{Kol94,YG86,CF86} that these diagrams all cancel.
Since nonvanishing 3NF contributions happen first
at order 
$(Q/\Lambda_\chi)^3$, 
they are very weak as compared to 2NF which start at
$(Q/\Lambda_\chi)^0$.

More 2PE is produced at $\nu =4$, next-to-next-to-next-to-leading
order (N$^3$LO), of which we show only a few symbolic diagrams in 
Fig.~\ref{fig_hi}. 
Two-loop 2PE
graphs show up for the first time and so does
three-pion exchange (3PE) which necessarily involves
two loops.
3PE was found to be negligible at this order~\cite{Kai00a,Kai00b}.
Most importantly, 15 new contact terms $\sim Q^4$
arise and are represented 
by the four-nucleon-leg graph with a solid diamond.
They include a quadratic spin-orbit term and
contribute up to $D$-waves.
Mainly due to the increased number of contact terms,
a quantitative description of the
two-nucleon interaction up to about 300 MeV
lab.\ energy is possible, 
at N$^3$LO 
(for details, see below).
Besides further 3NF,
four-nucleon forces (4NF) start
at this order. Since the leading 4NF 
come into existence one
order higher than the leading 3NF, 4NF are weaker
than 3NF.
Thus, ChPT provides a straightforward explanation for
the empirically known fact that 2NF $\gg$ 3NF $\gg$ 4NF
\ldots.

\section{Two-nucleon interactions
\label{sec_2NF}}

The last section was just an overview.
In this section, 
we will fill in all the details involved in the
ChPT development of the $NN$ interaction; and 3NF and 4NF
will be discussed in Section~\ref{sec_manyNF}.
We start by talking about the various pion-exchange
contributions.

\subsection{Pion-exchange contributions in ChPT \label{sec_pi}}

Based upon the effective pion Lagrangians of Section~\ref{sec_Lpi},
we will now derive
the pion-exchange contributions to the $NN$ 
interaction order by order.

As noted before, there are infinitely many pion-exchange 
contributions to the $NN$ interaction and, thus, we need
to get organized.
First, we arrange the various pion-exchange contributions 
according to the number of pions being exchanged between the two
nucleons:
\begin{equation}
V_\pi = V_{1\pi} + V_{2\pi} + V_{3\pi} + \ldots \,,
\end{equation}
where the meaning of the subscripts is obvious
and
the ellipsis represents $4\pi$ and higher pion exchanges.
Second, for each of the above terms, we assume a 
low-momentum expansion:
\begin{eqnarray}
V_{1\pi} & = & V_{1\pi}^{(0)} + V_{1\pi}^{(2)} 
+ V_{1\pi}^{(3)} + V_{1\pi}^{(4)} + \ldots 
\label{eq_1pe_orders}
\\
V_{2\pi} & = & V_{2\pi}^{(2)} + V_{2\pi}^{(3)} + V_{2\pi}^{(4)} 
+ \ldots \\
V_{3\pi} & = & V_{3\pi}^{(4)} + \ldots \,,
\end{eqnarray}
where the superscript denotes the order $\nu$
and the ellipses stand for contributions of fifth
and higher orders.
Due to parity and time-reversal,
there are no first order contributions.
Moreover, since $n$ pions create $L=n-1$ loops, 
the leading
order for $n$-pion exchange occurs at $\nu=2n-2$
[cf.\ Eq.~(\ref{eq_nunn})].

In the following subsections, we will discuss
$V_{1\pi}$, $V_{2\pi}$, and $V_{3\pi}$,
one by one and order by order.

\begin{figure}[t]\centering
%\hspace{1.25cm}
\scalebox{0.9}{\includegraphics{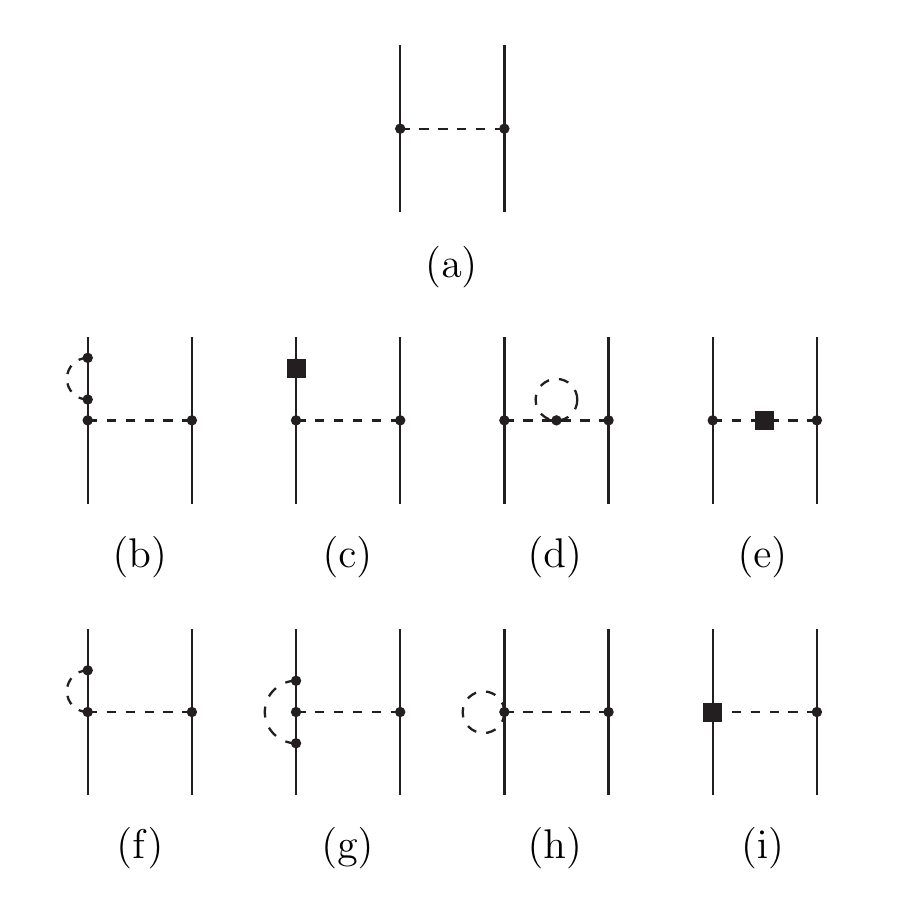}}
\vspace{-0.30cm}
\caption{One-pion exchange contributions.
Diagram (a) represents the leading-order, 
while graphs (b)-(i) contribute to renormalization.
Notation as in Fig.~\ref{fig_hi}. Diagrams that result from 
interchange of nucleon lines and/or time reversal are not shown.}
\label{fig_1pe}
\end{figure}

\subsubsection{One-pion exchange \label{sec_1pe}}

One-pion exchange (1PE) diagrams are
shown in 
Fig.~\ref{fig_1pe}. 
At leading order (LO, $\nu=0$), 
we have the well-known static 1PE, diagram (a) of
Fig.~\ref{fig_1pe}, 
which is given by [for notation, see Eq.~(\ref{eq_defqk})
below]:
\begin{equation}
V_{1\pi} ({\vec p}~', \vec p) = - 
%\frac{1}{(2\pi)^3} 
\frac{g_A^2}{4f_\pi^2}
\: 
\bbox{\tau}_1 \cdot \bbox{\tau}_2 
\:
\frac{
\vec \sigma_1 \cdot \vec q \,\, \vec \sigma_2 \cdot \vec q}
{q^2 + m_\pi^2} 
\,.
\label{eq_1pe}
\end{equation}
On-shell and in the center-of-mass system (CMS), 
there are no relativistic
corrections at any order.
Off-shell corrections come into play when the relativistic 1PE
is iterated, i.e., in four-dimensional planar box diagrams.
These corrections will be taken into account
in the evaluation of those box diagrams up to the
given order (see below).

At second order (NLO), 
the 1PE gets renormalized due to one-loop graphs 
and counter term insertions, shown in
the second and third row of Fig.~\ref{fig_1pe}.
Graphs (b) and (c) renormalize the nucleon and
graphs (d) and (e) the pion lines. Diagrams (f)-(i)
renormalize the pion-nucleon coupling.
In the one-loop graphs,
all vertices are from the leading
order Lagrangian
$\widehat{\cal L}^{\Delta=0}$, Eq.~(\ref{eq_LD0}),
while counter term insertions stem from
$\widehat{\cal L}^{\Delta=2}$.
Note that graph (f) vanishes because it involves
an odd power of the loop momentum that is integrated over.
In graph (i), the solid square includes
the $d_{18}$-vertex from
$\widehat{\cal L}^{(3)}_{\pi N}$ [which is part of 
$\widehat{\cal L}^{\Delta=2}$, Eq.~(\ref{eq_LD2})].
This correction, which is known as the 
Goldberger-Treiman discrepancy, 
can be taken care of by replacing
\begin{equation}
g_A \longrightarrow g_A - 2d_{18} m_\pi^2 \,.
\end{equation}
For more details on the NLO corrections to the 1PE
see Ref.~\cite{EMG03}.
At NNLO, there are further one-loop corrections, but
with one subleading vertex,
which renormalize $g_A$.

Finally, at N$^3$LO, there are two-loop corrections with
leading vertices only, as well as one-loop and
tree diagrams including (sub-)subleading vertices.
It has been shown~\cite{KBW97,Kai00a} that these contributions
renormalize various LECs and the pion mass $m_\pi$,
but do not generate any $\pi N$ form-factor-like functions.
Furthermore, a correction to the 
Goldberger-Treiman discrepancy arises at N$^3$LO.

We use $g_A=1.290$ (instead of $g_A=1.276$~\cite{Liu10})
to account for the Goldberger-Treiman discrepancy. 
Via the Goldberger-Treiman relation,
$g_{\pi NN} =
 g_A  M_N/f_\pi$, 
our value for $g_A$ 
together with $f_\pi=92.4$ MeV and $M_N=938.918$ MeV
implies
$g_{\pi NN}^2/4\pi = 13.67$
which is consistent with the empirical value
$g_{\pi NN}^2/4\pi = 13.65\pm 0.08$
obtained from $\pi N$ and $NN$ 
data analysis~\cite{AWP94,STS93}.
The renormalizations of $f_\pi$, $m_\pi$, and $M_N$
are taken care of by working with their physical
vales.

In summary, the familiar expression for 1PE,
Eq.~(\ref{eq_1pe}),
is appropriate to at least fourth order.

\subsubsection{Two-pion exchange \label{sec_2pi}}

\begin{figure}[t]\centering
%\vspace{-1.00cm}
%\hspace{1.25cm}
\scalebox{0.75}{\includegraphics{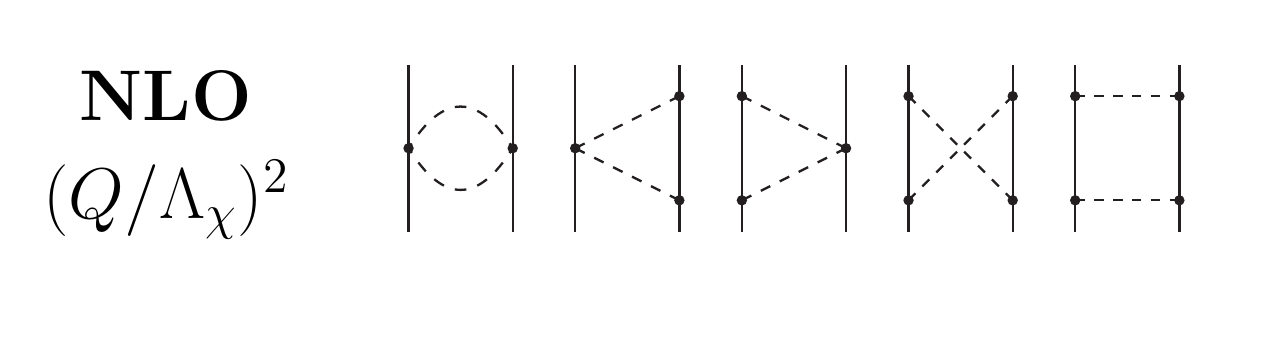}}
\vspace{-0.50cm}
\caption{Leading two-pion exchange contributions to the $NN$
interaction.
Notation as in Fig.~\ref{fig_hi}.}
\label{fig_nlo}
\end{figure}

The exchange of two or more pions always
involves loop diagrams,
which implies that we are faced with a non-trivial problem.
To conduct such calculations, a scheme of field-theoretic
perturbation theory as well as regularization and renormalization
methods for dealing with divergent loop integrals
must be adopted.
Ord\'o\~nez, Ray, and van Kolck~\cite{ORK94,ORK96},
who were the first to calculate chiral $2\pi$-exchange
to NNLO, used 
time-ordered perturbation theory 
and a Gaussian
cutoff function for regularization.
The Juelich group~\cite{Epe06}
developed a method involving
unitary transformations (to get rid of the energy-dependence
of time-ordered perturbation theory) and applied
so-called spectral function regularization.
Both, the Brazil~\cite{RR94,RR03} and the 
Munich~\cite{KBW97,Kai01a,Kai01b}
groups
use covariant perturbation theory
and dimensional regularization, but nevertheless, their works
differ substantially in detail.
We will follow here the method chosen by the Munich group
since we believe it to be the most efficient and elegant one.
In this approach, one starts
with the relativistic versions of the $\pi N$ Lagrangians
(cf.\ Section~\ref{sec_Lpi}) and sets up four-dimensional
(covariant) loop integrals.
Relativistic vertices and nucleon propagators are then 
expanded in powers of $1/M_N$. 
The divergences that occur in conjunction with the four-dimensional
loop integrals are treated by means of dimensional regularization,
a prescription which is
consistent with chiral symmetry and power counting.
The results derived in this way
are the same obtained when starting right away
with the HB versions of the $\pi N$ Lagrangians.
However, as it turns out, the method used by
the Munich group is more efficient in dealing 
with the rather tedious calculations and particularly
useful in conjunction with the planar box diagram.
To give the reader a taste of the rather involved calculations, 
we present in \ref{app_NLO} the explicit evaluation of
the NLO diagrams shown in Fig.~\ref{fig_nlo}.

The results will be stated in terms of contributions to the 
momentum-space $NN$ amplitude
in the CMS, 
which takes the general form
\begin{eqnarray} 
V({\vec p}~', \vec p) &  = &
%\frac{1}{(2\pi)^3} \;
% \bigg\{ 
% &&
 \:\, V_C \:\, + 
\bbox{\tau}_1 \cdot \bbox{\tau}_2 
\, W_C 
\nonumber \\ &+&  
\left[ \, V_S \:\, + \bbox{\tau}_1 \cdot \bbox{\tau}_2 \, W_S 
\,\:\, \right] \,
\vec\sigma_1 \cdot \vec \sigma_2
\nonumber \\ &+& 
\left[ \, V_{LS} + \bbox{\tau}_1 \cdot \bbox{\tau}_2 \, W_{LS}    
\right] \,
\left(-i \vec S \cdot (\vec q \times \vec k) \,\right)
\nonumber \\ &+& 
\left[ \, V_T \:\,     + \bbox{\tau}_1 \cdot \bbox{\tau}_2 \, W_T 
\,\:\, \right] \,
\vec \sigma_1 \cdot \vec q \,\, \vec \sigma_2 \cdot \vec q  
\nonumber \\ &+& 
\left[ \, V_{\sigma L} + \bbox{\tau}_1 \cdot \bbox{\tau}_2 \, 
      W_{\sigma L} \, \right] \,
\vec\sigma_1\cdot(\vec q\times \vec k\,) \,\,
\vec \sigma_2 \cdot(\vec q\times \vec k\,)
%\bigg\}
\, ,
%\nonumber \\ && 
\label{eq_nnamp}
\end{eqnarray}
where ${\vec p}~'$ and $\vec p$ denote the 
final and initial nucleon momenta 
in the CMS, 
respectively; moreover,
\begin{equation}
\begin{array}{llll}
\vec q &\equiv& {\vec p}~' - \vec p &  \mbox{\rm is the 
momentum transfer},\\
\vec k &\equiv& \frac12 ({\vec p}~' + \vec p) & \mbox{\rm the 
average momentum},\\
\vec S &\equiv& \frac12 (\vec\sigma_1+\vec\sigma_2) & 
\mbox{\rm the total spin},
\end{array}
\label{eq_defqk}
\end{equation}
and $\vec \sigma_{1,2}$ and $\bbox{\tau}_{1,2}$ are 
the spin and isospin 
operators, respectively, of nucleon 1 and 2.
For on-energy-shell scattering, $V_\alpha$ and $W_\alpha$ 
($\alpha=C,S,LS,T,\sigma L$) can be expressed as functions of 
$q$ and $k$ (with
$q\equiv |\vec q|$ and $k\equiv |\vec k|$), only.

Our notation and conventions are similar to the ones used by the Munich 
group~\cite{KBW97,Kai01a,Kai01b} except for two
differences: 
our spin-orbit potentials,
$V_{LS}$ and $W_{LS}$, differ by a factor of $(+2)$
and all other potentials 
differ by a factor of $(-1)$ 
from the Munich amplitudes.
Our definitions are more
in tune with what is commonly used in nuclear physics.

In all expressions given below, we will state only 
the {\it nonpolynomial} contributions to the $NN$ amplitude.
Note, however, that dimensional regularization typically 
generates also polynomial terms which are, in part,
infinite or scale dependent (cf.\ \ref{app_NLO}).
These polynomials are absorbed by the contact interactions 
to be discussed in a later section.

\paragraph{Next-to-leading order (NLO)}
\label{sec_NLO}

The leading two-pion exchange
appears at second order
($\nu=2$, next-to-leading order, NLO)
and is shown in 
Fig.~\ref{fig_nlo}.
Since a loop creates already $\nu=2$ [cf.\ Eq.~(\ref{eq_nunn})],
the vertices involved at this order have index 
$\Delta_i=0$, i.e., they are
from the leading order
Lagrangian $\widehat{\cal L}^{\Delta=0}$, Eq.~(\ref{eq_LD0}),
where the $\pi N$ vertices carry only one derivative.
These vertices are denoted by small dots in the figures.
The rather complicated evaluation of these diagrams is 
presented in \ref{app_NLO}.
 
Concerning the planar box diagram in 
Fig.~\ref{fig_nlo},
we should note that
we include only the non-iterative part
of this diagram which is obtained by subtracting 
the iterated 1PE contribution 
Eq.~(\ref{eq_2piitKBW}) or
(\ref{eq_2piitEM}), 
below, but using
$M_N^2/E_p 
\approx M_N^2/E_{p''} 
\approx M_N$
at this order (NLO).
Summarizing all contributions from irreducible two-pion exchange
at second order, one obtains:
\begin{eqnarray} 
W_C &=&-{L(q)\over384\pi^2 f_\pi^4} 
\left[4m_\pi^2(5g_A^4-4g_A^2-1)
+q^2(23g_A^4 -10g_A^2-1) 
+ {48g_A^4 m_\pi^4 \over w^2} \right] \,,  
\label{eq_2C}
\\   
V_T &=& -{1\over q^2} V_{S} 
    \; = \; -{3g_A^4 L(q)\over 64\pi^2 f_\pi^4} \,, 
\label{eq_2T}
\end{eqnarray}  
where
\begin{equation} 
L(q)  \equiv  {w\over q} \ln {w+q \over 2m_\pi}
\end{equation}
and
\begin{equation} 
 w  \equiv  \sqrt{4m_\pi^2+q^2} \,. 
\end{equation}

As will be demonstrated in Sec.~\ref{sec_peri}, below;
this part of the 2PE is rather weak and insufficient to properly
describe the $NN$ interaction at intermediate range.

\paragraph{Next-to-next-to-leading order (NNLO)}
\label{sec_NNLO}

\begin{figure}[t]\centering
%\vspace{-1.00cm}
%\hspace{2.50cm}
\scalebox{0.75}{\includegraphics{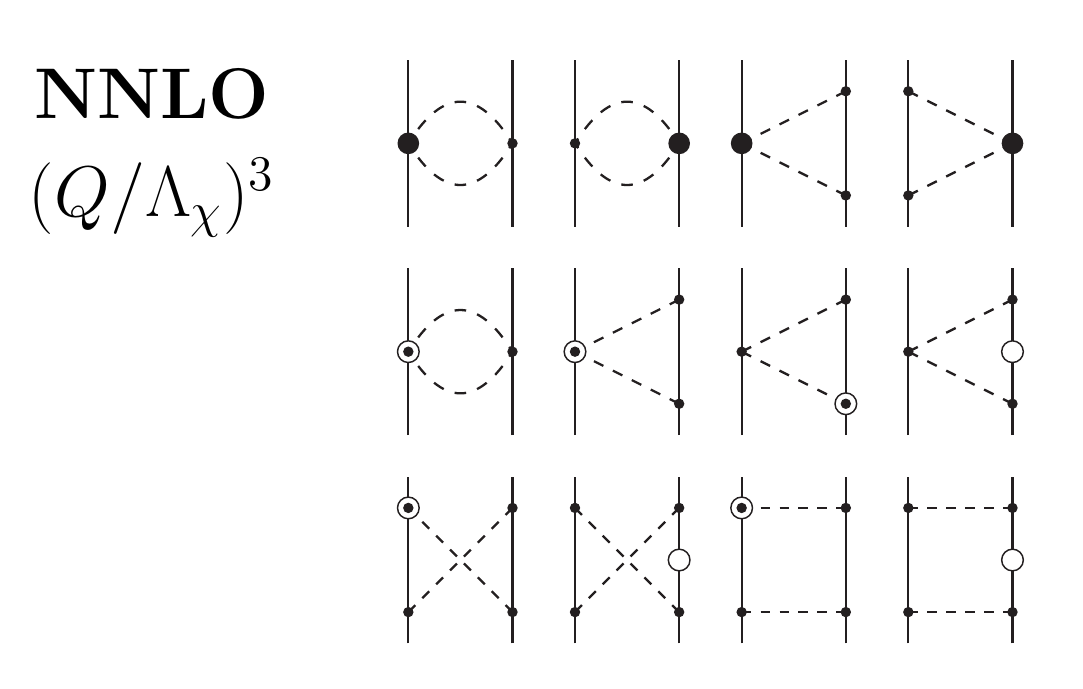}}
\vspace{-0.25cm}
\caption{Two-pion exchange contributions to the $NN$
interaction at order three in small momenta (NNLO).
Basic notation as in Fig.~\ref{fig_hi}.
Large solid dots denote vertices
from the Lagrangian
$\widehat{\cal L}^{\Delta=1}$,
Eq.~(\protect\ref{eq_LD1}),
proportional to the LECs $c_i$.
Symbols with an open circles are relativistic $1/M_N$ corrections
which are also part of 
$\widehat{\cal L}^{\Delta=1}$.
Only a few representative examples of $1/M_N$ corrections are shown.
Note that all football diagrams shown in this figure vanish.
\label{fig_nnlo}}
\end{figure}

The two-pion exchange diagrams of order three ($\nu=3$, 
next-to-next-to-leading order,
NNLO) are very similar to the ones of order two, except that 
they contain one insertion from 
$\widehat{\cal L}^{\Delta=1}$,
Eq.~(\protect\ref{eq_LD1}).
The resulting contributions are typically either 
proportional to one of the 
low-energy constants $c_i$ or they contain a factor $1/M_N$.
Notice that relativistic $1/M_N$ corrections derive
from vertices and nucleon propagators.
In Fig.~\ref{fig_nnlo}, we show in row one the diagrams with
one vertex proportional to $c_i$ (large solid dot), 
and in row two and three a few representative graphs with a $1/M_N$ 
correction (symbols with an open circle). The number 
of $1/M_N$ correction graphs
is large and not all are shown.
Note that all football diagrams vanish at this order,
because the loop integrals involve odd powers of the time-component
of the loop momentum.
Again, the planar box diagram is corrected for a contribution from
the iterated 1PE. If the iterative 2PE of 
Eq.~(\ref{eq_2piitKBW}), below, is used, the expansion of
the factor 
$M^2_N/E_p = M_N - p^2/2M_N + \ldots$ 
is applied and
the term proportional to $(-p^2/2M_N)$ is subtracted from
the third order box diagram contribution.
Then, one obtains for the full third order~\cite{KBW97}:
\begin{eqnarray} 
V_C &=&{3g_A^2 \over 16\pi f_\pi^4} 
\left\{ 
 {g_A^2 m_\pi^5  \over 16M_N w^2}  
-\left[2m_\pi^2( 2c_1-c_3)-q^2  \left(c_3 +{3g_A^2\over16M_N}\right)
\right]
\widetilde{w}^2 A(q) \right\} \,, 
\label{eq_3C}
\\
W_C &=& {g_A^2\over128\pi M_N f_\pi^4} \left\{ 
 3g_A^2 m_\pi^5 w^{-2} 
 - \left[ 4m_\pi^2 +2q^2-g_A^2(4m_\pi^2+3q^2) \right] 
\widetilde{w}^2 A(q)
\right\} 
\,,\\ 
V_T &=& -{1 \over q^2} V_{S}
   \; = \; {9g_A^4 \widetilde{w}^2 A(q) \over 512\pi M_N f_\pi^4} 
 \,,  \\ 
W_T &=&-{1\over q^2}W_{S} 
    =-{g_A^2 A(q) \over 32\pi f_\pi^4}
\left[
\left( c_4 +{1\over 4M_N} \right) w^2
-{g_A^2 \over 8M_N} (10m_\pi^2+3q^2)  \right] 
\,,
\label{eq_3T}
\\
V_{LS} &=&  {3g_A^4  \widetilde{w}^2 A(q) \over 32\pi M_N f_\pi^4} 
 \,,\\  
W_{LS} &=& {g_A^2(1-g_A^2)\over 32\pi M_N f_\pi^4} 
w^2 A(q) \,, 
\label{eq_3LS}
\end{eqnarray}   
with
\begin{equation} 
A(q) \equiv {1\over 2q}\arctan{q \over 2m_\pi} 
\end{equation}
and
\begin{equation} 
\widetilde{w} \equiv  \sqrt{2m_\pi^2+q^2} \,. 
\end{equation}

This contribution to the 2PE is the crucial one, because it provides
an intermediate-range attraction of proper strength (Sec.~\ref{sec_peri}).
The iso-scalar central potential, $V_C$, 
is strong and attractive due to the LEC $c_3$, which is negative and of large magnitude 
(cf.\ Table~\ref{tab_LEC}, below). Via resonance saturation, $c_3$ is associated with 
$\pi$-$\pi$ correlations (`$\sigma$ meson') and virtual $\Delta$-isobar excitations,
which create the most crucial contributions to 2PE in the frame work of conventional 
meson theory~\cite{Mac89,MHE87}.
The configuration-space expressions, which correspond to the above momentum-space
potentials, are given in Ref.~\cite{KBW97}, where also a detailed comparison with
meson-exchange potentials is conducted.

\begin{figure}[t]\centering
%\vspace{-1cm}
%\hspace{2.0cm}
\scalebox{0.70}{\includegraphics{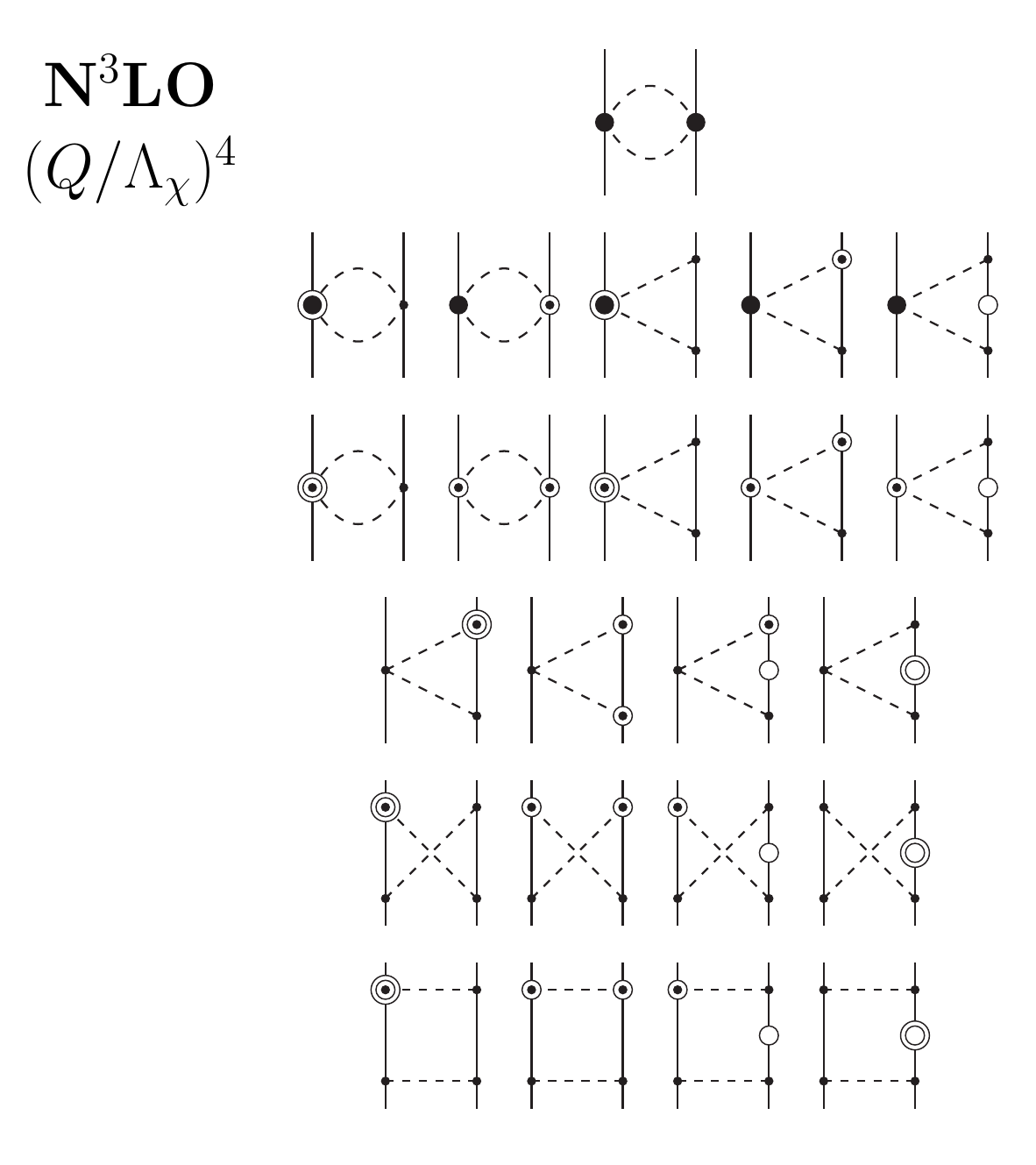}}
\vspace{-0.25cm}
\caption{One-loop $2\pi$-exchange contributions to the $NN$
interaction at order four. Notation as in Fig.~\ref{fig_nnlo}.
Moreover, symbols with a large solid dot and an open circle denote 
$1/M_N$ corrections of vertices
proportional to $c_i$.
Symbols with two open circles mark 
relativistic $1/M^2_N$ corrections.
Both corrections are part of the Lagrangian
$\widehat{\cal L}^{\Delta=2}$, Eq.~(\ref{eq_LD2}).
Representative examples for all types of one-loop graphs that occur 
at this order are shown.
\label{fig_n3lo1}}
\end{figure}

If the iterative 2PE defined in Eq.~(\ref{eq_2piitEM}), below, is applied,
the $1/M_N$ terms are slightly different. 
As derived in \ref{app_NNLO},
the changes are taken care of
by adding to Eqs.~(\ref{eq_3C})-(\ref{eq_3T}) the
following terms:
\begin{eqnarray}
V_C &=& -\frac{3 g_A^4}{256 \pi f_\pi^4 M_N} 
(m_\pi w^2 + \widetilde w^4 A(q) )
\,,
\label{eq_3EM1}
\\
W_C &=& \frac{g_A^4}{128 \pi f_\pi^4 M_N} 
(m_\pi w^2 + \widetilde w^4 A(q) )
\,,
\\
V_T &=& -\frac{1}{q^2} V_S = \frac{3 g_A^4}{512 \pi f_\pi^4 M_N} 
(m_\pi + w^2 A(q) )
\,,
\\
W_T &=& -\frac{1}{q^2} W_S = -\frac{g_A^4}{256 \pi f_\pi^4 M_N} 
(m_\pi + w^2 A(q) )
\,.
\label{eq_3EM4}
\end{eqnarray}

\paragraph{Next-to-next-to-next-to-leading order (N$^3$LO)}
\label{sec_N3LO}

\begin{figure}[t]\centering
%\vspace{-1.0cm}
%\hspace{2.8cm}
\scalebox{0.70}{\includegraphics{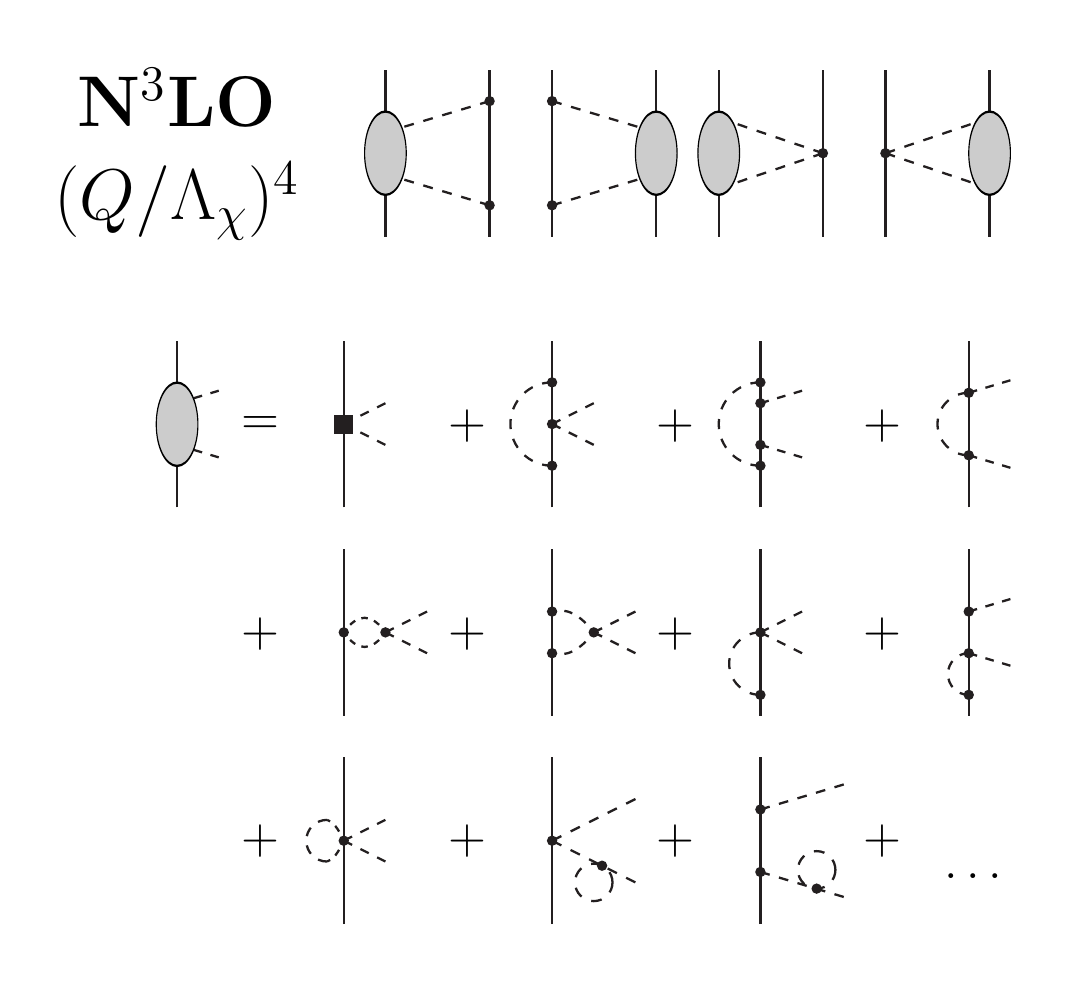}}
\vspace{-0.25cm}
\caption{Two-loop $2\pi$-exchange contributions at order four.
Basic notation as in Fig.~\ref{fig_hi}.
The oval stands for all one-loop $\pi N$ graphs
some of which are shown in the lower part of the figure.
The solid square represents vertices proportional to the LECs
$d_i$ introduced in
${\cal L}^{(3)}_{\pi N}$
which is part of
$\widehat{\cal L}^{\Delta=2}$, Eq.~(\ref{eq_LD2}).
More explanations are given in \ref{app_N3LO}.
\label{fig_n3lo2}}
\end{figure}

This order ($\nu=4$) is very involved.
The contributions can be subdivided into two groups,
one-loop graphs, 
Fig.~\ref{fig_n3lo1}, 
and
two-loop diagrams, 
Fig.~\ref{fig_n3lo2}.
Applying Eq.~(\ref{eq_nunn}),
it is easy to verify that all contributions result in $\nu=4$.
We have relegated the comprehensive mathematical expressions 
of this order to \ref{app_N3LO}.
The net effect of this contribution is small indicating a trend towards convergence
(cf.\ Sec.~\ref{sec_peri}).

\paragraph{Iterated one-pion-exchange}
\label{sec_2piit}

Besides all the irreducible 2PE contributions presented above,
there is also the reducible 2PE which is generated
from iterated 1PE. This 
``iterative 2PE'' is the only 2PE contribution
which produces an imaginary part. Thus, one wishes
to formulate this contribution such that relativistic
elastic unitarity is satisfied.
There are several ways to achieve this.

Kaiser {\it et al.}~\cite{KBW97} define the iterative
2PE contribution as follows,
\begin{equation}
V_{2\pi, \rm it}^{\rm(KBW)} ({\vec p}~',{\vec p})  = 
\:
\frac{M_N^2}{E_{p}} 
\:
\int 
\frac{d^3p''}{(2\pi)^3} 
\:
\frac{V_{1\pi}({\vec p}~',{\vec p}~'')\,
V_{1\pi}({\vec p}~'',{\vec p})} 
{{ p}^{2}-{p''}^{2}+i\epsilon}
\label{eq_2piitKBW}
\end{equation}
with $V_{1\pi}$ given in Eq.~(\ref{eq_1pe}).

As it will turn out (cf.\ Section~\ref{sec_pot}),
in the development of a chiral $NN$ potential,
it is useful
to adopt the relativistic scheme 
by Blankenbecler and Sugar~\cite{BS66} (BbS).
In this approach, the iterated 1PE
has the following form:
\begin{equation}
V_{2\pi, \rm it}^{\rm(EM)} ({\vec p}~',{\vec p})  = 
\int 
\frac{d^3p''}{(2\pi)^3} 
\:
\frac{M_N^2}{E_{p''}} 
\:
\frac{V_{1\pi}({\vec p}~',{\vec p}~'')\,
V_{1\pi}({\vec p}~'',{\vec p})} 
{{ p}^{2}-{p''}^{2}+i\epsilon}
\,.
\label{eq_2piitEM}
\end{equation}

Equations 
(\ref{eq_2piitKBW})
and (\ref{eq_2piitEM})
state the iterative 2PE to all orders.
On the other hand, the covariant box diagram is
calculated order by order and, therefore,
requires a subtraction of the iterative 2PE order
by order.
For this, the expansion
$M^2_N/E_p = M_N - p^2/2M_N + \ldots$ 
is applied in Eq.~(\ref{eq_2piitKBW}) and
$M^2_N/E_{p''} = M_N - {p''}^2/2M_N + \ldots$ 
in Eq.~(\ref{eq_2piitEM}).
At NLO, both choices for the iterative 2PE
collapse to the same, while at NNLO there are 
differences which give rise to the correction terms
Eqs.~(\ref{eq_3EM1})-(\ref{eq_3EM4})
when Eq.~(\ref{eq_2piitEM}) is used (see \ref{app_NNLO}).

\subsubsection{Three-pion exchange}
\begin{figure}[t]\centering
\vspace*{-0.3cm}
%\hspace*{1.10cm}
\scalebox{0.70}{\includegraphics{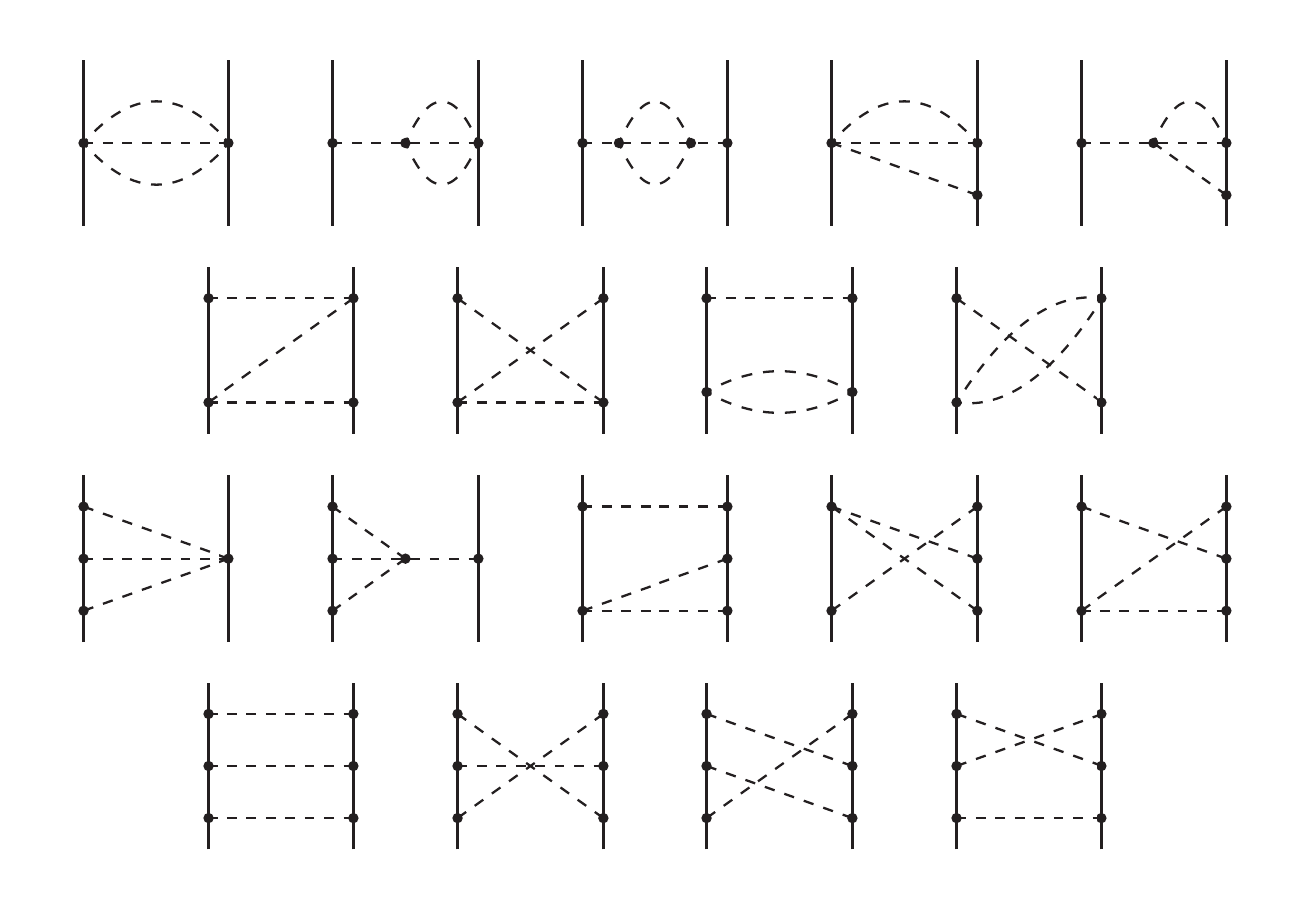}}
\vspace{-0.30cm}
\caption{Leading three-pion exchange contributions
to the $NN$ interaction.
Notation as in Fig.~\ref{fig_hi}.
Diagrams that result from interchange of nucleon lines
and/or time reversal are not shown.
\label{fig_3pe}}
\end{figure}

Since the exchange of three pions involves
two loops,
three-pion exchange (3PE) starts
at order four (N$^3$LO).
Two loops generate $\nu=4$ and, therefore, at this order,
all vertices
have to be from the leading order Lagrangian
$\widehat{\cal L}^{\Delta=0}$, Eq.~(\ref{eq_LD0}).
One can distinguish between three groups of diagrams,
namely, diagrams proportional to
$g_A^2$ (first and second row 
of Fig.~\ref{fig_3pe}), $g_A^4$ (third
row of the figure), 
and $g_A^6$ (fourth row).
The graphs in the first row
involve the $3\pi NN$ contact vertex from
the leading $\pi N$ Lagrangian 
and the $4\pi$ vertex from
the leading $\pi\pi$ Lagrangian
(both belong to $\widehat{\cal L}^{\Delta=0}$).
Note that these vertices contain the unphysical
parameter $\alpha$
which was introduced to parametrize the interpolating pion
fields, Eq.~(\ref{eq_U}). 
Therefore, this group of graphs has to
be calculated together such that the parameter $\alpha$
drops out, as it should, since measurable
quantities must not depend on $\alpha$.

The 3PE contributions at N$^3$LO have been
calculated by the Munich group and found to be
negligible~\cite{Kai00a,Kai00b}. 
This is not surprising since only the leading 
vertices are involved which are known to be weak.
As discussed, for similar reasons, the leading
2PE also turned out to be rather small (even though
not negligible). Note, however, that the subleading
3PE contributions which involve one vertex from 
the Lagrangian
$\widehat{\cal L}^{\Delta=1}$,
Eq.~(\ref{eq_LD1}), proportional the LEC $c_i$,
have been found to be sizable~\cite{Kai01}.
They contribute at fifth order (N$^4$LO).
Our current analysis is restricted to N$^3$LO
where only the leading 3PE contributes which
we will ignore because of 
its negligible strength.

\subsection{Perturbative $NN$ scattering in peripheral partial waves
\label{sec_peri}}

\begin{table}[t]
\caption{Hadron masses and
low-energy constants (LECs).
$m_{\pi^\pm}$ and $m_{\pi^0}$ denote the charged- and neutral-pion masses,
and $M_p$ and $M_n$ are the proton and neutron masses, respectively, in units of MeV.
$g_A$ is the axial-vector coupling constant (dimensionless)
and $f_\pi$ the pion decay constant (in units of MeV).
The $c_i$ belong to the dimension-two $\pi N$ Lagrangian,
Eqs.~(\ref{eq_L2ct}) and (\ref{eq_LD1}), and are in units of GeV$^{-1}$,
whereas the $\bar{d}_i$ are associated with the dimension-three
Lagrangian, Eqs.~(\ref{eq_L3rel}) and (\ref{eq_LD2}), and
are in units of GeV$^{-2}$.
Column ``$NN$ Potential'' lists the values
used for a $NN$ potential
at N$^3$LO presented in Section~\ref{sec_potn3lo}, while
column ``Peripheral perturbative $NN$'' shows the parameters applied in the 
peripheral $NN$ scattering calculations of Section~\ref{sec_peri}.
Finally, the last column displays values from empirical
determinations (see text for comments).
\label{tab_LEC}}
\smallskip
\begin{tabular*}{\textwidth}{@{\extracolsep{\fill}}cccc}
\hline
\hline
\noalign{\smallskip}
  &                           & Peripheral           &                \\
  & $NN$ Potential & perturbative $NN$    & Empirical \\
     & (Sec.~\ref{sec_potn3lo}) & (Sec.~\ref{sec_peri}) & \\
\hline
\noalign{\smallskip}
$m_{\pi^\pm}$ & 139.5702 & 139.5702 & 139.57018(35)~\cite{PDG} \\
$m_{\pi^0}$ & 134.9766 & 134.9766 & 134.9766(6)~\cite{PDG} \\
$M_p$    & 938.2720 & 938.2720   & 938.272013(23)~\cite{PDG} \\
$M_n$    & 939.5653 & 939.5653   & 939.565346(23)~\cite{PDG} \\
$g_A$ & $1.29^a$ & $1.29^a$ & $1.2759(45)$~\cite{Liu10} \\
$f_\pi$ & 92.4 & 92.4 & $92.2\pm 0.2$~\cite{PDG} \\
$c_1$ & --0.81 & --0.81 & $-0.81\pm 0.15^b$ \\
$c_2$ & 2.80 & 3.28 & $3.28\pm 0.23^c$ \\
$c_3$ & --3.20 & --3.40 & $-4.69\pm 1.34^b$ \\
$c_4$ & 5.40 & 3.40 & $3.40\pm 0.04^b$ \\
$\bar{d}_1 + \bar{d}_2$ & 3.06 & 3.06 & $3.06\pm 0.21^c$ \\
$\bar{d}_3$ & --3.27 & --3.27 & $-3.27\pm 0.73^c$ \\
$\bar{d}_5$ & 0.45 & 0.45 & $0.45\pm 0.42^c$ \\
$\bar{d}_{14} - \bar{d}_{15}$ & --5.65 & --5.65 & $-5.65\pm 0.41^c$ \\
\hline
\hline
\noalign{\smallskip}
\end{tabular*}
$^a$Confer discussion of Goldberger-Treiman discrepancy in
Section~\ref{sec_1pe}.\\
$^b$Table~1, Fit~1 of Ref.~\cite{BM00}.\\
$^c$Table~2, Fit~1 of Ref.~\cite{FMS98}.
\end{table}

Nucleon-nucleon scattering in peripheral partial waves
is of special interest---for several reasons.
First, these partial waves probe the long- and 
intermediate-range of the nuclear force. Due to the centrifugal
barrier, there is only small sensitivity to short-range
contributions and, in fact, the N$^3$LO contact terms
make no contributions for orbital angular momenta
$L\geq 3$ (cf.\ Section~\ref{sec_ct}).
Thus, for $F$ and higher waves and energies below the pion-production
threshold, we have a window in which the $NN$ interaction
is governed by chiral symmetry alone (chiral one-
and two-pion exchanges), and we can 
conduct a relatively clean test of how well 
the theory works.
Using values for the LECs from $\pi N$ analysis,
the $NN$ predictions are even parameter free.
Moreover, the smallness of the phase shifts in peripheral
partial waves suggests that the calculation can
be done perturbatively. This avoids the complications
and the possible model-dependence
that the non-perturbative treatment of the
Schroedinger equation, necessary for low partial
waves (Section~\ref{sec_reno}), is beset with.
Because of the importance of peripheral $NN$ scattering,
many calculations
based upon chiral dynamics can be found in the
literature~\cite{RR94,KBW97,KGW98,EM02,Ric99,BM04,Hig04}.
We will follow Ref.~\cite{EM02}.

Since we will compare the predictions with neutron-proton ($np$) phase shifts,
we will specifically calculate $np$ scattering in this sub-section.
Defining,
\begin{equation}
V_{1\pi} (m_\pi) \equiv -\, 
%\frac{1}{(2\pi)^3} 
\frac{g_A^2}{4f_\pi^2}\, \frac{
\vec \sigma_1 \cdot \vec q \,\, \vec \sigma_2 \cdot \vec q}
{q^2 + m_\pi^2} 
\,,
\label{eq_1pe_bare}
\end{equation}
the correct 1PE for $np$ scattering is given by
\begin{equation}
V_{1\pi}^{(np)} ({\vec p}~', \vec p) 
= -V_{1\pi}(m_{\pi^0}) + (-1)^{I+1}\, 2\, V_{1\pi} (m_{\pi^\pm})
\,,
\label{eq_1penp}
\end{equation}
where $I$ denotes the isospin of the two-nucleon system.
We use the pion masses given in Table~\ref{tab_LEC} and
\begin{equation}
M_N  =  \frac{2M_pM_n}{M_p+M_n} = 938.9182 \mbox{ MeV}
\,.
\end{equation}
Also in the iterative 2PE, we apply the correct $np$
1PE, i.e., in Eq.~(\ref{eq_2piitKBW}) we replace $V_{1\pi}$
with $V_{1\pi}^{(np)}$.
Thus, the perturbative relativistic $T$-matrix for $np$ scattering,
taking the exchange of up to two pions into account,
is calculated in the following way,
\begin{eqnarray}
T({\vec p}~',\vec p) &=& 
 V_{1\pi}^{(np)} ({\vec p}~', \vec p) +
 V_{2\pi}^{(np)} ({\vec p}~', \vec p) 
\nonumber \\
      &=&
 V_{1\pi}^{(np)} ({\vec p}~', \vec p) +
 V_{2\pi, \rm it}^{({\rm KBW},np)} 
({\vec p}~',{\vec p}) +
 V_{2\pi}'({\vec p}~',{\vec p}) 
\,.
\label{eq_tall}
\end{eqnarray}
As discussed, the expression for $V_{1\pi}^{(np)}$ is good to any order
we will consider in this article (cf.\ 
Section~\ref{sec_1pe}) and 
$V_{2\pi, \rm it}^{({\rm KBW},np)}$ 
includes all orders.
There is no need to break the latter term
up into orders, because
the admixture of (very small) higher order contributions from 
$V_{2\pi, \rm it}^{({\rm KBW},np)}$ 
will not affect the accuracy we are working at.
The most important term in the above equation is
$V_{2\pi}'$, 
{\it the irreducible 2PE contributions}, 
for which we have the low-momentum expansion,
\begin{equation}
V_{2\pi}' = 
V_{2\pi}^{'(2)} +
V_{2\pi}^{'(3)} +
V_{2\pi}^{'(4)} 
+ \ldots 
\,,
\end{equation}
with the various 
$V_{2\pi}^{'(\nu)}$ 
given in Section~\ref{sec_2pi}.
In the calculation of 
$V_{2\pi}'$, 
we use the average pion mass $m_\pi = 138.039$ MeV and, thus,
neglect the charge-dependence due to pion-mass splitting
in irreducible diagrams.
The charge-dependence that emerges from irreducible $2\pi$ 
exchange was investigated in Ref.~\cite{LM98b} and found to be
negligible for partial waves with $L\geq 3$.

\begin{figure}[t]\centering
\vspace*{-2.8cm}
%\hspace*{-0.7cm}
\scalebox{0.55}{\includegraphics{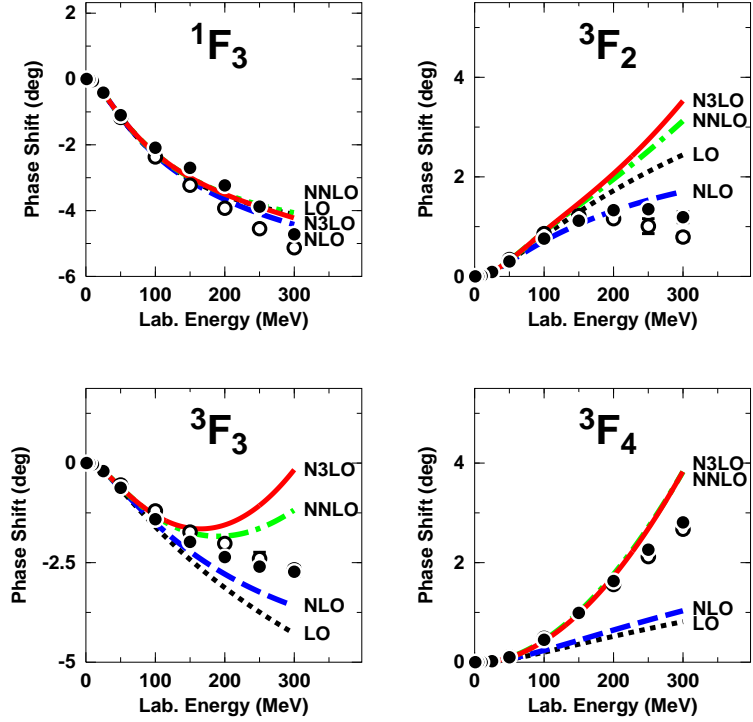}}
\vspace*{-3.0cm}
\caption{$F$-wave phase shifts of neutron-proton scattering
for laboratory kinetic energies below 300 MeV.
We show the predictions from chiral pion exchange
to leading order (LO), next-to-leading order (NLO),
next-to-next-to-leading order (NNLO), and
next-to-next-to-next-to-leading order (N3LO).
Note that in $^3F_4$, the NNLO and N3LO curves
cannot be distinguished on the scale of the figure.
The solid dots and open circles are the results from the Nijmegen
multi-energy $np$ phase shift analysis~\protect\cite{Sto93} 
and the VPI/GWU
single-energy $np$ analysis SM99~\protect\cite{SM99}, respectively.}
\label{fig_f}
\end{figure}

For the $T$-matrix given in Eq.~(\ref{eq_tall}),
we calculate phase shifts for
partial waves with $L\geq 3$ and 
$T_{lab}\leq 300$ MeV (see Ref.~\cite{EM02} for the details of
this calculation).
The LECs used in this calculation are shown in Table~\ref{tab_LEC},
column ``Peripheral perturbative $NN$''.
Note that many determinations of the LECs, $c_i$ and $\bar{d}_i$,
can be found in the literature.
The most reliable way to determine the LECs from empirical
$\pi N$ information is to extract them from the $\pi N$ amplitude
inside the Mandelstam triangle (unphysical region) which can
be constructed with the help of dispersion relations from empirical
$\pi N$ data. This method was used by B\"uttiker and Mei\ss ner~\cite{BM00}.
Unfortunately, the values for $c_2$ and all $\bar{d}_i$ parameters
obtained in Ref.~\cite{BM00} carry uncertainties,
so large that the values cannot provide any guidance.
Therefore, in Table~\ref{tab_LEC}, only  $c_1$, $c_3$, and $c_4$
are from Ref.~\cite{BM00}, while the other LECs
are taken from Ref.~\cite{FMS98} where the $\pi N$ amplitude in the
physical region was considered.
To establish a link between $\pi N$ and $NN$, we apply 
the values from the above determinations in our calculations
of the $NN$ peripheral phase shifts.
In general, we use the central values;
the only exception is $c_3$, where we choose
a value that is, in terms of magnitude, about one standard
deviation below the one from Ref.~\cite{BM00}.
With the exception of $c_3$,
phase shift predictions do not depend sensitively on
variations of the LECs within the quoted uncertainties.

\begin{figure}[t]\centering
\vspace*{-2.8cm}
%\hspace*{-0.7cm}
\scalebox{0.55}{\includegraphics{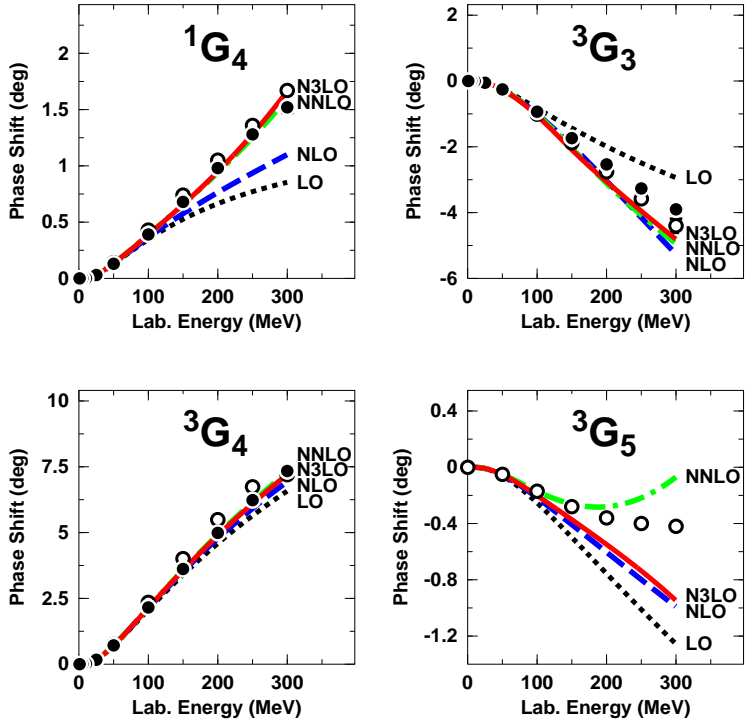}}
\vspace*{-3.0cm}
\caption{Same as Fig.~\ref{fig_f}, but for $G$ waves.}
\label{fig_g}
\end{figure}

In Figs.~\ref{fig_f} and \ref{fig_g},
we show the phase-shift
predictions for neutron-proton scattering in $F$ and $G$
waves, respectively, for laboratory kinetic energies below 300 MeV.
The orders displayed are defined as follows:
\begin{itemize}
\item
Leading order (LO) is just 1PE, 
first term on the r.h.s.\
of Eq.~(\ref{eq_tall}).
\item
Next-to-leading order (NLO) 
includes the first two terms on the r.h.s.\
of Eq.~(\ref{eq_tall}) (1PE \& iterative 2PE)
plus 
$V_{2\pi}^{'(2)}$
[Section~\ref{sec_NLO}, Eqs.~(\ref{eq_2C}) and (\ref{eq_2T})].
\item
Next-to-next-to-leading order (NNLO)
consists of NLO plus  
$V_{2\pi}^{'(3)}$
[Section~\ref{sec_NNLO},
Eqs.~(\ref{eq_3C})-(\ref{eq_3LS})].
\item
Next-to-next-to-next-to-leading order 
(denoted by N3LO in the figures)
is made up of NNLO plus 
$V_{2\pi}^{'(4)}$
[\ref{app_N3LO},
Eqs.~(\ref{eq_4c2C})-(\ref{eq_4M2sL}) 
and (\ref{eq_42lC})-(\ref{eq_42lT})]. 
\end{itemize}
It is clearly seen in 
Figs.~\ref{fig_f} and \ref{fig_g}
that the leading $2\pi$ exchange (NLO)
is, in general, rather small, insufficient to explain
the empirical facts in most partial waves. 
In contrast, the next order (NNLO)
is very large; in some cases, several times NLO. This is due to the
$\pi\pi N N$ contact interactions proportional
to the LECs $c_i$ that are introduced in the
sub-leading Lagrangian 
${\cal L}^{\Delta=1}$, Eq.~(\ref{eq_LD1}). 
These contacts are supposed to simulate the contributions
from intermediate $\Delta$-isobars and correlated $2\pi$
exchange which are known
to be large 
and crucial for a realistic model for the $NN$ interaction
at intermediate ranges
(see, e.~g., Ref.~\cite{MHE87}).

\begin{figure}[t]\centering
\vspace*{-2.8cm}
%\hspace*{-0.7cm}
\scalebox{0.55}{\includegraphics{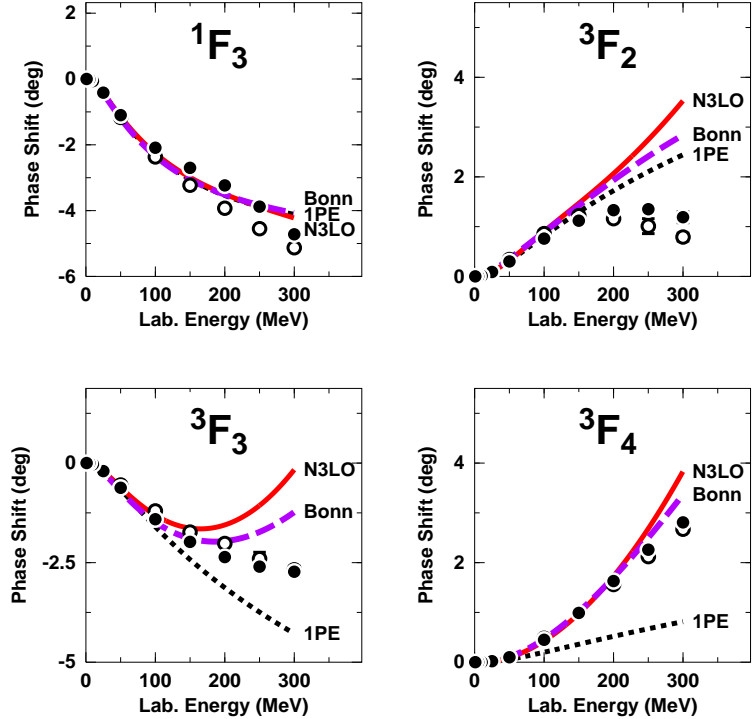}}
\vspace*{-3.0cm}
\caption{$F$-wave phase shifts of neutron-proton scattering
for laboratory kinetic energies below 300 MeV.
We show the results from one-pion-exchange (1PE),
and one- plus two-pion exchange as predicted by ChPT at
next-to-next-to-next-to-leading order (N3LO) and 
by the Bonn Full Model~\protect\cite{MHE87} (Bonn).
Note that the ``Bonn'' curve does not include the repulsive
$\omega$ and $\pi\rho$ exchanges of the full model, since 
this figure serves the purpose to compare just predictions
by different models/theories for the $\pi+2\pi$ contribution 
to the $NN$ interaction.
Empirical phase shifts (solid dots and open circles)
as in Fig.~\protect\ref{fig_f}.}
\label{fig_ff}
\end{figure}

\begin{figure}[t]\centering
\vspace*{-2.8cm}
%\hspace*{-0.7cm}
\scalebox{0.55}{\includegraphics{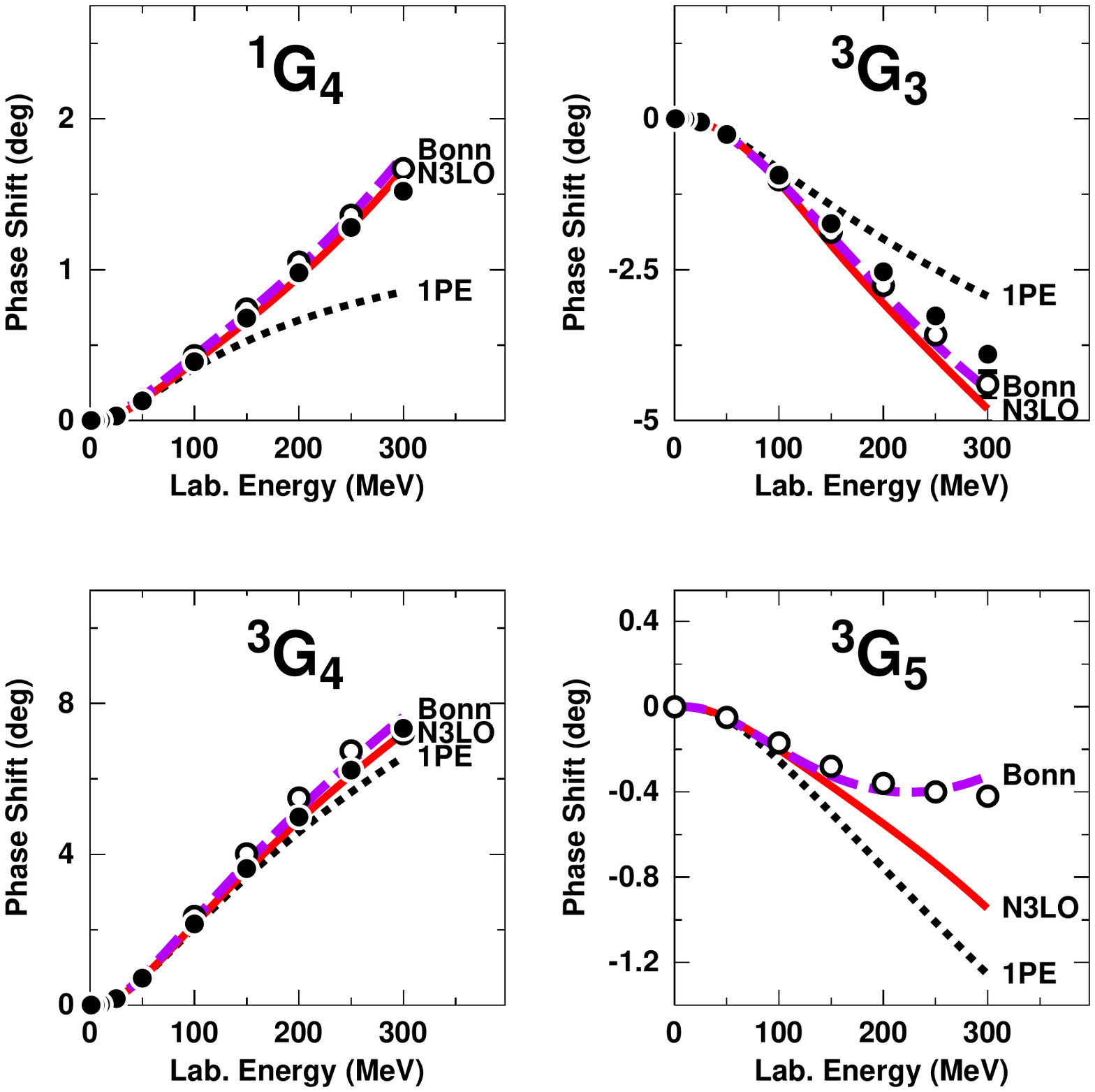}}
\vspace*{-3.0cm}
\caption{Same as Fig.~\ref{fig_ff}, but for $G$ waves.}
\label{fig_gg}
\end{figure}

At N$^3$LO a clearly identifiable
trend towards convergence emerges.
In $G$ waves (except for $^3G_5$, see \ref{app_N3LO_ph}
for a discussion of this issue),
N$^3$LO differs very little from NNLO
implying that we have reached convergence.
Also $^1F_3$ and $^3F_4$ appear fully converged.
However, in $^3F_2$ and $^3F_3$, N$^3$LO differs
noticeably from NNLO, but the difference is much smaller than the
one between NNLO and NLO. This is what we perceive as a trend towards
convergence.
Individual N$^3$LO contributions to peripheral phase shifts
are shown in \ref{app_N3LO_ph}.

In Figs.~\ref{fig_ff} and \ref{fig_gg}, 
we conduct a comparison between the predictions from chiral one- and
two-pion exchange at N$^3$LO and the corresponding 
predictions from conventional
meson theory (curve `Bonn'). 
As representative for conventional meson theory, we choose the Bonn
meson-exchange model for the $NN$ interaction~\cite{MHE87},
since it contains a comprehensive and thoughtfully constructed
model for $2\pi$ exchange.
This $2\pi$ model includes box and crossed box diagrams with
$NN$, $N\Delta$, and $\Delta\Delta$ intermediate states
as well as direct $\pi\pi$ interaction in $S$- and $P$-waves
(of the $\pi\pi$ system)
consistent with empirical information from $\pi N$ and $\pi\pi$
scattering.
Besides this the Bonn model also includes (repulsive) $\omega$-meson
exchange and irreducible diagrams of $\pi$ and $\rho$ exchange
(which are also repulsive).
However, note that
in the phase shift predictions displayed in 
Figs.~\ref{fig_ff} and \ref{fig_gg}, 
the ``Bonn'' curve includes only the $1\pi$ and $2\pi$ contributions
from the Bonn model; the short-range contributions are
left out since the purpose of the figure is to compare
different models/theories for $\pi+2\pi$.
In all waves shown (with the usual exception of $^3G_5$)
we see, in general, good agreement between N$^3$LO and Bonn.
In $^3F_2$ and $^3F_3$ above 150 MeV and in
$^3F_4$ above 250 MeV the chiral model to N$^3$LO is more
attractive than the Bonn $2\pi$ model. Note, however,
that the Bonn model is relativistic and, thus, includes
relativistic corrections up to infinite orders.
Thus, one may speculate that higher orders in 
ChPT may create some
repulsion, moving the Bonn and the chiral predictions even closer
together~\cite{foot2}.

The $2\pi$ exchange contribution to the $NN$ interaction
can also be derived
from {\it empirical} $\pi N$ and $\pi\pi$ input
using dispersion theory, which is based upon unitarity,
causality (analyticity), and crossing symmetry.
The amplitude $N\bar{N}\rightarrow \pi\pi$ is
constructed from $\pi N \rightarrow \pi N$
and $\pi N \rightarrow \pi\pi N$ data using 
crossing properties and analytic
continuation; this amplitude is then `squared' to yield 
the $N\bar{N}$ amplitude 
which is related to $NN$ by crossing symmetry~\cite{BJ76}.
The Paris group~\cite{Vin79,Lac80} pursued this path and 
calculated $NN$ phase shifts in peripheral partial waves.
Naively, the dispersion-theoretic approach is
the ideal one, since it is based exclusively on empirical
information. Unfortunately, in practice,
quite a few uncertainties enter.
First, there are ambiguities
in the analytic continuation and, second, 
the dispersion integrals have to be cut off at a certain
momentum to ensure reasonable results.
In Ref.~\cite{MHE87}, a thorough comparison was conducted
between the predictions by the Bonn model and the
Paris approach and it was demonstrated that the Bonn
predictions always lie comfortably within the range of uncertainty
of the dispersion-theoretic results.
Therefore, there is no need to perform a
separate comparison of our chiral N$^3$LO predictions with
dispersion theory, since it would not add anything that we
cannot conclude from Figs.~\ref{fig_ff} and \ref{fig_gg}.

Finally, we need to compare the
predictions with the empirical phase shifts.
In $F$ waves the N$^3$LO predictions above 200 MeV are,
in general, too attractive. Note, however, that this is
also true for the predictions by the Bonn $\pi + 2\pi$
model. In the {\it full} Bonn model, 
besides $\pi+2\pi$, 
(repulsive)
$\omega$ and $\pi\rho$ exchanges are included
which move
the predictions right on top of the data.
The exchange of a $\omega$ meson or  combined $\pi\rho$
exchange are $3\pi$ exchanges.
Three-pion exchange occurs first at chiral order four.
It has been investigated by Kaiser~\cite{Kai00a,Kai00b} and found to be
negligible, at this order.
However, $3\pi$ exchange at order five appears 
to be sizable~\cite{Kai01}
and may have impact on $F$ waves.
Besides this, there is the usual short-range phenomenology.
In ChPT, this short-range interaction
is parametrized in terms of four-nucleon contact terms
(since heavy mesons do not have a place in that theory).
Contact terms of order four (N$^3$LO) do not contribute to
$F$-waves, but order six does.
In summary, the remaining small 
discrepancies between the N$^3$LO predictions and the empirical
phase shifts may be straightened out in
fifth or sixth order of ChPT.

\subsection{$NN$ contact terms \label{sec_ct}}

The successful test of the chiral pion-exchange contributions
in peripheral partial waves in the previous section
has shown that we got the long- and intermediate-range
parts of the nuclear force right.
However, for a ``complete'' nuclear force, we have to 
describe correctly all partial waves, including the lower ones.
In fact, in calculations of $NN$ observables at low energies (cross
sections, analyzing powers, etc.),
the lower partial
waves with $L\leq 2$ are the most important ones, generating
the largest contributions.
The same is true for microscopic nuclear structure
calculations.
These lower partial waves are dominated by the
dynamics at short distances.
Therefore, we need to look now into the short-range part
of the $NN$ potential.

In conventional meson theory, the short-range nuclear force
is described by the exchange of heavy mesons, notably the
$\omega(782)$. 
Qualitatively, the short-distance behavior of the $NN$ 
potential is obtained
by Fourier transform of the propagator of
a heavy meson,
\begin{equation}
\int d^3q \frac{e^{i{\vec q} \cdot {\vec r}}}{m^2_\omega
+ {\vec q}^2} \;
\sim \;
 \frac{e^{-m_\omega r}}{r} \; .
\end{equation}

ChPT is an expansion in small momenta $Q$, too small
to resolve structures like a $\rho(770)$ or $\omega(782)$
meson, because $Q \ll \Lambda_\chi \approx m_{\rho,\omega}$.
But the latter relation allows us to expand the propagator 
of a heavy meson into a power series,
\begin{equation}
\frac{1}{m^2_\omega + Q^2} 
\approx 
\frac{1}{m^2_\omega} 
\left( 1 
- \frac{Q^2}{m^2_\omega}
+ \frac{Q^4}{m^4_\omega}
-+ \ldots
\right)
,
\end{equation}
where the $\omega$ is representative
for any heavy meson of interest.
The above expansion suggests that it should be 
possible to describe the short distance part of
the nuclear force simply in terms of powers of
$Q/m_\omega$, which fits in well
with our over-all 
power expansion since $Q/m_\omega \approx Q/\Lambda_\chi$.

A second purpose of contact terms is renormalization.
Dimensional regularization
of the loop integrals that occur in multi-pion exchange diagrams
typically generates polynomial terms
with coefficients that are, in part, infinite or scale
dependent (cf.\ \ref{app_NLO}). Contact terms pick up infinities
and remove scale dependences, which is why they are also known as counter terms.

The partial-wave decomposition of a power
$Q^\nu$ has an interesting property.
First note that $Q$ can only be either
the momentum transfer between the two interacting
nucleons $q$ or the average momentum $k$
[cf.\ Eq.~(\ref{eq_defqk}) for their definitions].
In any case, for even $\nu$,
\begin{equation}
Q^\nu = 
f_{\frac{\nu}{2}}(\cos \theta) 
\, ,
\end{equation}
where $f_m$ stands for a polynomial of degree $m$
and $\theta$ is the CMS scattering angle.
The partial-wave decomposition of $Q^\nu$ for a state
of orbital-angular momentum $L$
involves the integral
\begin{equation}
I^{(\nu)}_L  
=\int_{-1}^{+1} Q^\nu P_L(\cos \theta) d\cos \theta 
=\int_{-1}^{+1}
f_{\frac{\nu}{2}}(\cos \theta) 
 P_L(\cos \theta) d\cos \theta 
\,,
\end{equation}
where $P_L$ is a Legendre polynomial.
Due to the orthogonality of the $P_L$, 
\begin{equation}
I^{(\nu)}_L = 0  
\hspace*{.5cm}
\mbox{for}
\hspace*{.5cm}
L > \frac{\nu}{2} \, .
\end{equation}
Consequently, contact terms of order zero contribute only
in $S$-waves, while order-two terms contribute up to 
$P$-waves, order-four terms up to $D$-waves,
etc..

Due to parity, only even powers of $Q$
are allowed.
Thus, the expansion of the contact potential is
formally given by
\begin{equation}
V_{\rm ct} =
V_{\rm ct}^{(0)} + 
V_{\rm ct}^{(2)} + 
V_{\rm ct}^{(4)} 
+ \ldots \; ,
\end{equation}
where the superscript denotes the power or order.

We will now present, one by one, the various orders of 
$NN$ contact terms which result from the contact
Lagrangians presented in Section~\ref{sec_Lct}.

\subsubsection{Zeroth order (LO)}

The contact Lagrangian
$\widehat{\cal L}^{(0)}_{NN}$,
Eq.~(\ref{eq_LNN0}), 
which is part of 
$\widehat{\cal L}^{\Delta=0}$,
Eq.~(\ref{eq_LD0}), 
leads to the following $NN$ contact potential,
\begin{equation}
V_{\rm ct}^{(0)}(\vec{p'},\vec{p}) =
C_S +
C_T \, \vec{\sigma}_1 \cdot \vec{\sigma}_2 \, ,
\label{eq_ct0}
\end{equation}
and, in terms of partial waves, we have
\be
V_{\rm ct}^{(0)}(^1 S_0)          &=&  \widetilde{C}_{^1 S_0} =
4\pi\, ( C_S - 3 \, C_T )
\nonumber \\
V_{\rm ct}^{(0)}(^3 S_1)          &=&  \widetilde{C}_{^3 S_1} =
4\pi\, ( C_S + C_T ) \,.
\label{eq_ct0_pw}
\ee

\subsubsection{Second order (NLO)}

The contact Lagrangian
$\widehat{\cal L}^{(2)}_{NN}$,
Eq.~(\ref{eq_LNN2}), 
which is part of 
$\widehat{\cal L}^{\Delta=2}$,
Eq.~(\ref{eq_LD2}), 
generates the following $NN$ contact potential
\be
V_{\rm ct}^{(2)}(\vec{p'},\vec{p}) &=&
C_1 \, q^2 +
C_2 \, k^2 
\nonumber 
\\ &+& 
\left(
C_3 \, q^2 +
C_4 \, k^2 
\right) \vec{\sigma}_1 \cdot \vec{\sigma}_2 
\nonumber 
\\
&+& C_5 \left( -i \vec{S} \cdot (\vec{q} \times \vec{k}) \right)
\nonumber 
\\ &+& 
 C_6 \, ( \vec{\sigma}_1 \cdot \vec{q} )\,( \vec{\sigma}_2 \cdot 
\vec{q} )
\nonumber 
\\ &+& 
 C_7 \, ( \vec{\sigma}_1 \cdot \vec{k} )\,( \vec{\sigma}_2 \cdot 
\vec{k} ) \,.
\label{eq_ct2}
\ee
The coefficients $C_i$ used here in the contact potential
are, of course, related to the coefficients $C_i'$
that occur in the Lagrangian
$\widehat{\cal L}^{(2)}_{NN}$,
Eq.~(\ref{eq_LNN2}).
The relation, which is unimportant for our purposes,
can be found in Refs.~\cite{ORK96,EGM98}.

There are many ways to perform the partial-wave decomposition
of the above potential.
We perceive the method presented 
by Erkelenz, Alzetta, and Holinde~\cite{EAH71} as the
most elegant one.
Thus, one obtains
\be
V_{\rm ct}^{(2)}(^1 S_0)          &=&  C_{^1 S_0} ( p^2 + {p'}^2 ) 
\nonumber \\
V_{\rm ct}^{(2)}(^3 P_0)          &=&  C_{^3 P_0} \, p p'
\nonumber \\
V_{\rm ct}^{(2)}(^1 P_1)          &=&  C_{^1 P_1} \, p p' 
\nonumber \\
V_{\rm ct}^{(2)}(^3 P_1)          &=&  C_{^3 P_1} \, p p' 
\nonumber \\
V_{\rm ct}^{(2)}(^3 S_1)          &=&  C_{^3 S_1} ( p^2 + {p'}^2 ) 
\nonumber \\
V_{\rm ct}^{(2)}(^3 S_1- ^3 D_1)  &=&  C_{^3 S_1- ^3 D_1}  p^2 
\nonumber \\
V_{\rm ct}^{(2)}(^3 D_1- ^3 S_1)  &=&  C_{^3 S_1- ^3 D_1}  {p'}^2 
\nonumber \\
V_{\rm ct}^{(2)}(^3 P_2)          &=&  C_{^3 P_2} \, p p' 
\label{eq_ct2_pw}
\ee
with
\be
C_{^1 S_0} 
&=&
4\pi\, \left( C_1 + \frac{1}{4} C_2 - 3 C_3 - \frac{3}{4} C_4 - 
C_6 - \frac{1}{4} C_7 \right)
\nonumber 
\\
C_{^3 P_0} 
&=&
4\pi\, \left( -\frac{2}{3} C_1 + \frac{1}{6} C_2 - \frac{2}{3} C_3 
+ \frac{1}{6} C_4 - \frac{2}{3} C_5
+ 2 C_6 - \frac{1}{2} C_7 \right)
\nonumber 
\\
C_{^1 P_1} 
&=&
4\pi\, \left( -\frac{2}{3} C_1 + \frac{1}{6} C_2 + 2 C_3 
- \frac{1}{2} C_4 
+ \frac{2}{3} C_6 - \frac{1}{6} C_7 \right)
\nonumber 
\\
C_{^3 P_1} 
&=&
4\pi\, \left( -\frac{2}{3} C_1 + \frac{1}{6} C_2 - \frac{2}{3} C_3 
+ \frac{1}{6} C_4 - \frac{1}{3} C_5
- \frac{4}{3} C_6 + \frac{1}{3} C_7 \right)
\nonumber 
\\
C_{^3 S_1} 
&=&
4\pi\, \left( C_1 + \frac{1}{4} C_2 + C_3 + \frac{1}{4} C_4 + 
\frac{1}{3} C_6 + \frac{1}{12} C_7 \right)
\nonumber 
\\
C_{^3 S_1- ^3 D_1} 
&=&
4\pi\, \left( -\frac{2\sqrt{2}}{3} C_6 - \frac{\sqrt{2}}{6} 
C_7 \right)
\nonumber 
\\
C_{^3 P_2} 
&=&
4\pi\, \left( -\frac{2}{3} C_1 + \frac{1}{6} C_2 - \frac{2}{3} C_3 
+ \frac{1}{6} C_4 + \frac{1}{3} C_5 \right) \,.
\label{eq_ct2_pwcoef}
\ee

\subsubsection{Fourth order (N$^3$LO)}

The contact potential of order four reads
\be
V_{\rm ct}^{(4)}(\vec{p'},\vec{p}) &=&
D_1 \, q^4 +
D_2 \, k^4 +
D_3 \, q^2 k^2 +
D_4 \, (\vec{q} \times \vec{k})^2 
\nonumber 
\\ &+& 
\left(
D_5 \, q^4 +
D_6 \, k^4 +
D_7 \, q^2 k^2 +
D_8 \, (\vec{q} \times \vec{k})^2 
\right) \vec{\sigma}_1 \cdot \vec{\sigma}_2 
\nonumber 
\\ &+& 
\left(
D_9 \, q^2 +
D_{10} \, k^2 
\right) \left( -i \vec{S} \cdot (\vec{q} \times \vec{k}) \right)
\nonumber 
\\ &+& 
\left(
D_{11} \, q^2 +
D_{12} \, k^2 
\right) ( \vec{\sigma}_1 \cdot \vec{q} )\,( \vec{\sigma}_2 
\cdot \vec{q})
\nonumber 
\\ &+& 
\left(
D_{13} \, q^2 +
D_{14} \, k^2 
\right) ( \vec{\sigma}_1 \cdot \vec{k} )\,( \vec{\sigma}_2 
\cdot \vec{k})
\nonumber 
\\ &+& 
D_{15} \left( 
\vec{\sigma}_1 \cdot (\vec{q} \times \vec{k}) \, \,
\vec{\sigma}_2 \cdot (\vec{q} \times \vec{k}) 
\right) .
\label{eq_ct4}
\ee
The rather lengthy partial-wave expressions of this order
are relegated to \ref{app_ct}.

\subsection{Definition of $NN$ potential
\label{sec_pot}}

We have now rounded up everything needed for a realistic
nuclear force---long, intermediate, and short ranged 
components---and so we can finally proceed to the lower
partial waves. However, here we encounter another problem.
The two-nucleon system at low angular momentum, particularly,
in $S$ waves, is characterized by the
presence of a shallow bound state (the deuteron)
and large scattering lengths.
Thus, perturbation theory does not apply.
In contrast to $\pi$-$\pi$ and $\pi$-$N$,
the interaction between nucleons is not suppressed
in the chiral limit ($Q\rightarrow 0$).
Weinberg~\cite{Wei90,Wei91} showed that the strong enhancement of the
scattering amplitude arises from purely nucleonic intermediate
states (``infrared enhancement''). He therefore suggested to use perturbation theory to
calculate the $NN$ potential (i.e., the irreducible graphs) and to apply this potential
in a scattering equation 
to obtain the $NN$ amplitude. We will follow
this prescription and discuss potential problems in the next subsection.

Since the irreducible diagrams that make up the potential
are calculated using
covariant perturbation theory (cf.~Section~\ref{sec_pi}), 
it is consistent to
start from the covariant
Bethe-Salpeter (BS) equation~\cite{SB51} describing
two-nucleon scattering.
In operator notation, the BS equation reads
\begin{equation}
T = {\cal V+V\,G} \,T
\label{eq_BS}
\end{equation}
where $T$ denotes the invariant T-matrix ($T=i{\cal M}$ with ${\cal M}$ the invariant
amplitude) for the two-nucleon scattering process,
${\cal V}$ the sum of all connected two-particle irreducible diagrams, and 
${\cal G}$ is $(-i)$ times the relativistic two-nucleon propagator. 
The BS equation
is equivalent to a set of two equations
\begin{eqnarray}
{T}&=&{V}+{V} \, g \, {T}
\label{eq_bbs}
\\
{V}&=&{\cal V + V\,(G}-g)\,{V} 
\label{eq_BS2}
\\
   &=& {\cal V} + {\cal V}_{1\pi}\,({\cal G}-g)\,{\cal V}_{1\pi} 
    + \ldots \, ,
\label{eq_BS3}
\end{eqnarray}
where $g$ is a covariant three-dimensional propagator
which preserves relativistic elastic unitarity.
We choose the propagator $g$ proposed by 
Blankenbecler and Sugar (BbS)~\cite{BS66}
(for more details on relativistic three-dimensional
reductions of the BS equation, see Ref.~\cite{Mac89}). 
The ellipsis in Eq.~(\ref{eq_BS3}) stands for 
terms of irreducible $3\pi$ and higher pion exchanges
which we neglect.

Note that when we speak of covariance in conjunction with
(heavy baryon) ChPT, we are not referring to manifest
covariance. Relativity and relativistic off-shell
effects are accounted for in terms of
a $Q/M_N$ expansion up to the given order
and up to the number of pions we take into consideration.
Thus, Eq.~(\ref{eq_BS3}) is evaluated in the following way,
\begin{eqnarray}
V  & \approx &
 \:  {\cal V} (\mbox{\rm on-shell}) \: + \:
 {\cal V}_{1\pi}\,{\cal G}\,{\cal V}_{1\pi}-{V}_{1\pi}\,{g}\,{V}_{1\pi} 
\\
 & \approx & 
 \; V_{1\pi} \; + \; \widetilde{V}_{2\pi} \; + \:
 {\cal V}_{1\pi}\,{\cal G}\,{\cal V}_{1\pi}-{V}_{1\pi}\,{g}\,{V}_{1\pi} 
\\
 & = & 
 \; V_{1\pi} \; + \; V_{2\pi}'
\,,
\label{eq_BS4}
\end{eqnarray}
where
${\cal V}_{1\pi}$
denotes the relativistic (off-shell) 1PE, while
$V_{1\pi}$ is the on-shell 1PE given 
in Eq.~(\ref{eq_1pe}).
$\widetilde{V}_{2\pi}$  stands for the
irreducible $2\pi$ exchanges calculated (on-shell) in Section~\ref{sec_2pi},
{\it but without the covariant planar box diagram}, 
which in the above equations is given by
${\cal V}_{1\pi}\,{\cal G}\,{\cal V}_{1\pi}$.
Furthermore,
${V}_{1\pi}\,{g}\,{V}_{1\pi}$
is the iterated 1PE discussed in 
Section~\ref{sec_2piit}.
Thus, the term
$({\cal V}_{1\pi}\,{\cal G}\,{\cal V}
_{1\pi}-{V}_{1\pi}\,{g}\,{V}_{1\pi})$
represents the irreducible part of the
box diagram contribution. Finally,
$V_{2\pi}'$ subsumes all $2\pi$ exchanges without the iterated
$V_{1\pi}$ and is, therefore, also known as the irreducible 2PE.
Since all contributions are calculated on shell, the potential
has no energy dependence.

Adding the short-range contact terms $V_{\rm ct}$ to the above,
yields the full $NN$ potential, 
\begin{equation}
V = V_{1\pi} + V_{2\pi}' + V_{\rm ct} 
\,.
\label{eq_pot1}
\end{equation}
Notice that the pion-exchange part of this potential
differs from the perturbative amplitude, Eq.~(\ref{eq_tall}),
by the absence of the iterative 2PE. The latter is generated
automatically when the potential is inserted into a Schroedinger
or Lippmann-Schwinger equation (see below).
We also note that adding the contact interactions to the
pion interactions will generate loop corrections
of the contact terms. This leads to a $m_\pi$-dependent renormalization
of the contact parameters and can be ignored if $m_\pi$-dependence
is not an issue.

As discussed, the irreducible 2PE, $V_{2\pi}'$, is organized
according to increasing orders,
\begin{equation}
V_{2\pi}' = 
V_{2\pi}^{'(2)} +
V_{2\pi}^{'(3)} +
V_{2\pi}^{'(4)} 
+ \ldots 
\,,
\end{equation}
and was calculated in Section~\ref{sec_2pi}:
$V_{2\pi}^{'(2)}$
is given by the contributions of
Eqs.~(\ref{eq_2C}) and (\ref{eq_2T}),
$V_{2\pi}^{'(3)}$
is made up from
Eqs.~(\ref{eq_3C})-(\ref{eq_3EM4}), and
$V_{2\pi}^{'(4)}$
is contained in \ref{app_N3LO}.

The contact potentials come in even orders,
\begin{equation}
V_{\rm ct} =
V_{\rm ct}^{(0)} + 
V_{\rm ct}^{(2)} + 
V_{\rm ct}^{(4)} 
+ \ldots
\,,
\end{equation}
and were presented in Section~\ref{sec_ct}.

In summary, the $NN$ potential $V$, calculated 
to certain orders,
is given by:
\begin{eqnarray}
V_{\rm LO} &=& V_{1\pi} + V_{\rm ct}^{(0)} 
\label{eq_Vlo} \\
V_{\rm NLO} &=& V_{\rm LO} + V_{2\pi}^{'(2)} + V_{\rm ct}^{(2)} 
\label{eq_Vnlo} \\
V_{\rm NNLO} &=& V_{\rm NLO} + V_{2\pi}^{'(3)} 
\label{eq_Vnnlo} \\
V_{{\rm N}^3{\rm LO}} &=&
 V_{\rm NNLO} + V_{2\pi}^{'(4)} + V_{\rm ct}^{(4)}
\label{eq_Vn3lo}
\end{eqnarray}

The potential $V$ satisfies the relativistic
BbS equation, Eq.~(\ref{eq_bbs}), which reads explicitly,
\begin{equation}
{T}({\vec p}~',{\vec p})= {V}({\vec p}~',{\vec p})+
\int \frac{d^3p''}{(2\pi)^3} \:
{V}({\vec p}~',{\vec p}~'') \:
\frac{M_N^2}{E_{p''}} \:  
\frac{1}
{{ p}^{2}-{p''}^{2}+i\epsilon} \:
{T}({\vec p}~'',{\vec p}) 
\label{eq_bbs2}
\end{equation}
with $E_{p''}\equiv \sqrt{M_N^2 + {p''}^2}$.
The advantage of using a relativistic scattering equation is that it automatically
includes relativistic corrections to all orders. Thus, in the scattering equation,
no propagator modifications are necessary when raising the order to which the
calculation is conducted.

Defining
\begin{equation}
\widehat{V}({\vec p}~',{\vec p})
\equiv 
\frac{1}{(2\pi)^3}
\sqrt{\frac{M_N}{E_{p'}}}\:  
{V}({\vec p}~',{\vec p})\:
 \sqrt{\frac{M_N}{E_{p}}}
\label{eq_minrel1}
\end{equation}
and
\begin{equation}
\widehat{T}({\vec p}~',{\vec p})
\equiv 
\frac{1}{(2\pi)^3}
\sqrt{\frac{M_N}{E_{p'}}}\:  
{T}({\vec p}~',{\vec p})\:
 \sqrt{\frac{M_N}{E_{p}}}
\,,
\label{eq_minrel2}
\end{equation}
where the factor $1/(2\pi)^3$ is added for convenience,
the BbS equation collapses into the usual, nonrelativistic
Lippmann-Schwinger (LS) equation,
\begin{equation}
 \widehat{T}({\vec p}~',{\vec p})= \widehat{V}({\vec p}~',{\vec p})+
\int d^3p''\:
\widehat{V}({\vec p}~',{\vec p}~'')\:
\frac{M_N}
{{ p}^{2}-{p''}^{2}+i\epsilon}\:
\widehat{T}({\vec p}~'',{\vec p}) \, .
\label{eq_LS}
\end{equation}
Since 
$\widehat V$ 
satisfies Eq.~(\ref{eq_LS}), 
it can be used like a usual nonrelativistic potential, and 
$\widehat{T}$ 
may be perceived as the conventional nonrelativistic 
T-matrix.
In applications, it is more convenient to use the K-matrix instead
of the T-matrix and to have the LS equation
decomposed into partial waves: all these technical issues
are explained in detail in Appendix A of Ref.~\cite{Mac01} where also 
the formulas for the calculation of $np$ and $pp$ (the latter with Coulomb) phase shifts
are provided.
The partial wave decomposition of the operators by which the potential
is represented can be found in section~4 of Ref.~\cite{EAH71}, and numerical methods
for solving the LS equation are explained in Ref.~\cite{Mac93}.

\subsection{Renormalization}
\label{sec_reno}

\subsubsection{Regularization and non-perturbative renormalization}
Iteration of $\widehat V$ in the LS equation, Eq.~(\ref{eq_LS}),
requires cutting $\widehat V$ off for high momenta to avoid infinities.
This is consistent with the fact that ChPT
is a low-momentum expansion which
is valid only for momenta $Q \ll \Lambda_\chi \approx 1$ GeV.
Therefore, the potential $\widehat V$
is multiplied
with the regulator function $f(p',p)$,
\begin{equation}
{\widehat V}(\vec{ p}~',{\vec p})
\longmapsto
{\widehat V}(\vec{ p}~',{\vec p}) \, f(p',p) 
\end{equation}
with
\begin{equation}
f(p',p) = \exp[-(p'/\Lambda)^{2n}-(p/\Lambda)^{2n}] \,,
\label{eq_f}
\end{equation}
such that
\begin{equation}
{\widehat V}(\vec{ p}~',{\vec p}) \, f(p',p) 
\approx
{\widehat V}(\vec{ p}~',{\vec p})
\left\{1-\left[\left(\frac{p'}{\Lambda}\right)^{2n}
+\left(\frac{p}{\Lambda}\right)^{2n}\right]+ \ldots \right\} 
\,.
\label{eq_reg_exp}
\end{equation}
Typical choices for the cutoff parameter $\Lambda$ that
appears in the regulator are 
$\Lambda \approx 0.5 \mbox{ GeV} \ll \Lambda_\chi \approx 1$ GeV.

Equation~(\ref{eq_reg_exp}) provides an indication of the fact that
the exponential cutoff does not necessarily
affect the given order at which 
the calculation is conducted.
For sufficiently large $n$, the regulator introduces contributions that 
are beyond the given order. Assuming a good rate
of convergence of the chiral expansion, such orders are small 
as compared to the given order and, thus, do not
affect the accuracy at the given order.
In calculations, one uses, of course,
the exponential form, Eq.~(\ref{eq_f}),
and not the expansion Eq.~(\ref{eq_reg_exp}). On a similar note, we also
do not expand the square-root factors
in Eqs.~(\ref{eq_minrel1}-\ref{eq_minrel2})
because they are kinematical factors which guarantee
relativistic elastic unitarity.

It is pretty obvious that results for the $T$-matrix may
depend sensitively on the regulator and its cutoff parameter.
This is acceptable if one wishes to build models.
For example, the meson models of the past~\cite{Mac89,MHE87}
always depended sensitively on the choices for the
cutoff parameters which, in fact,
were important for the fit of the $NN$ data.
However, the EFT approach wishes to be fundamental
in nature and not just another model.

In field theories, divergent integrals are not uncommon and methods have
been developed for how to deal with them.
One regulates the integrals and then removes the dependence
on the regularization parameters (scales, cutoffs)
by renormalization. In the end, the theory and its
predictions do not depend on cutoffs
or renormalization scales.

So-called renormalizable quantum field theories, like QED,
have essentially one set of prescriptions 
that takes care of renormalization through all orders. 
In contrast, 
EFTs are renormalized order by order. 

The renormalization of {\it perturbative}
EFT calculations is not a problem. {\it The problem
is nonperturbative renormalization.}
This problem typically occurs in {\it nuclear} EFT because
nuclear physics is characterized by bound states which
are nonperturbative in nature.
EFT power counting may be different for nonperturbative processes as
compared to perturbative ones. Such difference may be caused by the infrared
enhancement of the reducible diagrams generated in the LS equation.

Weinberg's implicit assumption~\cite{Wei90,Wei09} was that the counterterms
introduced to renormalize the perturbatively calculated
potential, based upon naive dimensional analysis (``Weinberg counting''),
are also sufficient to renormalize the nonperturbative
resummation of the potential in the LS equation.
In 1996, Kaplan, Savage, and Wise (KSW)~\cite{KSW96}
pointed out that there are problems with the Weinberg scheme
if the LS equation is renormalized 
by minimally-subtracted dimensional regularization.
This criticism resulted in a flurry of publications on
the renormalization of the nonperturbative
$NN$ problem
\cite{FMS00,PBC98,FTT99,Bir06,Bea02,VA05-1,NTK05,VA05-2,VA07,EM06,VA08,Ent08,YEP07,LK08,BKV08,Val08,ME10}.
The literature is too comprehensive
to discuss all contributions.
Let us just mention some of the work that has particular relevance
for our present discussion.

If the potential $V$ consists of contact terms only (a.k.a.\
pion-less theory), then
the nonperturbative summation (\ref{eq_LS})
can be performed analytically and the power counting is explicit.
However, when pion exchange is included, then (\ref{eq_LS})
can be solved only numerically and the power counting
is less transparent.
Perturbative ladder diagrams of arbitrarily high order,
where the rungs of the ladder represent a potential made up from
irreducible pion exchange,
suggest that an infinite number of counterterms is needed to achieve
cutoff independence for all the terms of increasing order generated
by the iterations.
For that reason, 
Kaplan, Savage, and Wise (KSW)~\cite{KSW96} proposed 
to sum the leading-order contact interaction to all orders (analytically)
and to add higher-order contacts and
pion exchange perturbatively up to the given order. Unfortunately,
it turned out that the order by order convergence of 1PE 
is poor in the $^3S_1$-$^3D_1$ state~\cite{FMS00}. 
The failure was triggered by the $1/r^3$ singularity of the 1PE tensor
force when iterated to second order. Therefore, KSW counting is no
longer taken into consideration (see, however, \cite{BKV08}).  
A balanced discussion of possible
solutions is provided in \cite{Bea02}.

Some researchers decided to take
a second look at Weinberg's original proposal.
A systematic investigation of Weinberg counting in leading order
has been conducted by Nogga, Timmermans, and van Kolck~\cite{NTK05}
in momentum space, and by Valderrama and Arriola
at LO and higher orders in
configuration space~\cite{VA05-1,VA05-2,VA07}. A comprehensive
discussion of both approaches and their equivalence can be found
in Refs.~\cite{Ent08,Val08}.

The LO $NN$ potential is given in (\ref{eq_Vlo}) and consists
of 1PE plus two nonderivative contact terms that contribute
only in $S$ waves.
Nogga {\it et al.}~\cite{NTK05} find that the given counterterms renormalize
the $S$ waves (i.e., stable results are obtained for $\Lambda \rightarrow \infty$) and
the naively expected infinite number of counterterms
is not needed. This means that Weinberg power counting does actually work in
$S$ waves at LO (ignoring the $m_\pi$ dependence of the contact interaction
discussed in Refs.~\cite{KSW96,Bea02}).
However, there are problems with a particular class of higher partial waves,
namely those  
in which the tensor force from 1PE is attractive. The first few cases
of this kind  of low angular momentum are
$^3P_0$, $^3P_2$, and $^3D_2$, which need a counterterm for cutoff independence. 
The leading (nonderivative) counterterms do not contribute in
$P$ and higher waves, which is why Weinberg counting fails in these cases. 
But the second order contact potential provides counterterms
for $P$ waves. Therefore, the promotion
of, particularly, the $^3P_0$ and $^3P_2$ contacts from NLO to LO would
fix the problem in $P$ waves. To take care of the $^3D_2$ problem,
a N$^3$LO contact, i.e. a term from $V^{(4)}_{\rm ct}$, needs to be promoted to LO.
Partial waves with orbital angular momentum $L\geq 3$ may be calculated in Born
approximation with sufficient accuracy and, therefore, do not pose renormalization
problems.
In this way, one arrives at a scheme
of `modified Weinberg counting'~\cite{NTK05} for the leading order 
two-nucleon interaction.

\subsubsection{Renormalization beyond leading order}

As shown below, for a quantitative chiral $NN$ potential one needs to advance all the way
to N$^3$LO. Thus, the renormalization issue needs to be discussed beyond LO.
Naively, the most perfect renormalization procedure is the one where the cutoff
parameter $\Lambda$ is carried to infinity while stable results are maintained.
This was done successfully at LO in the work by Nogga {\it et al.}~\cite{NTK05} described above.
At NNLO, the infinite-cutoff renormalization procedure has been investigated 
in~\cite{YEP07} for partial waves with total angular momentum $J\leq 1$ and
in~\cite{VA07} for all partial waves with $J\leq 5$. At N$^3$LO, an investigation
of the $^1S_0$ state exists~\cite{Ent08}.
From all of these works, it is evident that no counter term is effective in partial-waves with
short-range repulsion and only a single counter term can effectively be used in
partial-waves with short-range attraction. Thus, for the $\Lambda \rightarrow \infty$
renormalization prescription, even at N$^3$LO, we have either one or no counter term
per partial-wave state. This is inconsistent with any reasonable power-counting scheme
and, therefore, defies the principals of an EFT.

A possible way out of this dilemma was proposed already in~\cite{NTK05}
and reiterated in a recent paper by Long and van Kolck~\cite{LK08}. In the latter
reference, the authors examine the renormalization of an attractive $1/r^2$ potential
perturbed by a $1/r^4$ correction.
Generalizing their findings, they come to
the conclusion that, for any attractive $1/r^n$ potential (with $n\geq 2$),
partial waves with low angular momentum $L$ must be summed to all orders
and one contact term is needed for each $L$ to renormalize the LO
contribution. However, there exists
an angular momentum $L_p$ ($L_p\approx 3$ for the nuclear case, cf.\ Ref. ~\cite{NTK05}), 
above which the leading order can be calculated perturbatively. 
In short, naive dimensional analysis (NDA) does not apply at LO below $L_p$.
However, once this failure of NDA is corrected at LO, higher order corrections
can be added in perturbation theory using counterterm that follow NDA~\cite{LK08}.

Reference~\cite{LK08} used just a toy model and, therefore, a full investigation 
using the chiral expansion is needed to answer the question 
if this renormalization approach will work for the realistic nuclear force.
A first calculation of this kind for the $S$ waves was recently performed by
Valderrama~\cite{Val09}. The author renormalizes the LO interaction nonperturbatively
and then uses the LO
distorted wave to calculate the 2PE contributions at NLO and NNLO
perturbatively. It turns out that perturbative renormalizability requires the introduction
of three counterterms in $^1S_0$ and six in the coupled $^3S_1-^3D_1$
channels. 
Thus, the number of counterterms required in this scheme is larger than in the
Weinberg scheme, which reduces the predictive power. For a final evaluation
of this approach, also the results for $P$ and $D$ waves are needed, which
are not yet published.

However, even if such a project turns out to be successful for $NN$ scattering,
there is doubt if the interaction generated in this approach is of any use
for applications in nuclear few- and many-body problems.
In applications, one would first have to solve the many-body problem
with the renormalized LO interaction, and then add higher order corrections in
perturbation theory.
However, it was shown in a recent paper~\cite{Mac09} that the renormalized LO
interaction
is characterized by a very large tensor force from 1PE. This is no surprise since
LO is renormalized with $\Lambda \rightarrow \infty$ implying that the 1PE,
particularly its tensor force, is totally uncut.
As a consequence of this, the wound integral in nuclear matter, $\kappa$,
comes out to be about 40\%. The hole-line and coupled cluster expansions
are know to converge $\propto \kappa^{n-1}$
with $n$ the number of hole-lines or particles per cluster~\cite{Bra67,Day67,Bet71}.
For conventional nuclear forces, the wound integral is typically between 5 and 10\%
and the inclusion of three-body clusters (or three hole-lines) are needed to
obtain converged results in the many-body system~\cite{FOS04,Hag08,DW85}.
Thus, if the wound integral is 40\%, probably, up to six hole-lines need to be
included for approximate convergence. Such calculations are not feasible even with
the most powerful computers of today and will not be feasible any time soon.
Therefore, even if the renormalization procedure proposed in~\cite{LK08} will work
for $NN$ scattering, the interaction produced will be highly impractical (to say
the least) in applications in few- and many-body problems because of convergence problems
with the many-body energy and wave functions.

\subsubsection{Back to the beginnings}
The various problems with the renormalization procedures discussed above
may have a simple common reason:
An EFT that has validity only for momenta $Q < \Lambda_\chi$ is applied such that
momenta $Q \gg \Lambda_\chi$ are heavily involved (because the regulator cutoff
$\Lambda \rightarrow \infty$).
A recent paper by Epelbaum and Gegelia~\cite{EG09} illustrates the point:
The authors construct an exactly solvable toy-model that simulates a pionful EFT 
and yields finite results for
$\Lambda \rightarrow \infty$.  However, as it turns out, these
finite results are incompatible with the underlying EFT, while
for cutoffs in the order of the hard scale consistency is maintained.
In simple terms, the point to realize is this: 
{\it If an EFT calculation produces (accidentally) a finite result for
$\Lambda \rightarrow \infty$, then that
does not automatically imply that this result is also meaningful.}

This matter is further elucidated in
the lectures by Lepage of 1997~\cite{Lep97}.
Lepage points out that it makes little sense to take the momentum cutoff beyond
the range of validity of the effective theory. By assumption, our data involves energies
that are too low---wave lengths that are too long---to probe the true structure of
the theory at very short distances. When one goes beyond the hard-scale of the theory,
structures are seen that are almost certainly wrong. Thus, results cannot improve
and, in fact, they may degrade or, in more extreme cases, the theory may become
unstable or untunable. In fact, in the $NN$ case, this is what is happening in 
several partial waves (as reported above). Therefore, Lepage
suggests to take the following three steps when building an effective theory:
\begin{enumerate}
\item
Incorporate the correct long-range behavior: The long-range behavior of the underlying
theory must be known, and it must be built into the effective theory. In the case of
nuclear forces, the long-range theory is, of course, well known and 
given by one- and multi-pion exchanges.
\item
Introduce an ultraviolet cutoff to exclude high-momentum states, or, equivalent, to soften the 
short-distance behavior: The cutoff has two effects: First it excludes high-momentum states,
which are sensitive to the unknown short-distance dynamics; only states that we understand
are retained. Second it makes all interactions regular at $r=0$, thereby avoiding the infinities
that beset the naive approach.
\item
Add local correction terms (also known as contact or counter terms) 
to the effective Hamiltonian. These mimic the effects of the 
high-momentum states excluded by the cutoff introduced in the previous step.
In the meson-exchange picture, the short-range nuclear force is described by
heavy meson exchange, like the $\rho(770)$ and $\omega(782)$. However, at low
energy, such structures are not resolved. Since we must include contact terms 
anyhow, it is most efficient to use them to account for any heavy-meson exchange
as well.
The correction terms systematically remove dependence on the cutoff.
\end{enumerate}

A first investigation in the above spirit has been
conducted by Epelbaum and Mei\ss ner~\cite{EM06} in 2006.
The authors stress that there is no point in taking the cutoff $\Lambda$ beyond the
breakdown scale of the EFT, $\Lambda_\chi \approx m_\rho \approx 1$ GeV,
since the error of the calculation is not expected to decrease
in that regime.
Any value for the cutoff parameter $\Lambda$
is acceptable if the error associated with its finite value is within
the theoretical uncertainty at the given order.
The authors conduct an investigation at LO (including only the counter terms
implied by Weinberg counting)
and find that,
starting from $\Lambda \approx 3$ fm$^{-1}$, the error
in the $NN$ phase shifts due to keeping $\Lambda$ finite
stays within the theoretical uncertainty at LO.

\begin{figure}[t]\centering
\vspace*{-1.5cm}
%\hspace*{-1.2cm}
\scalebox{0.55}{\includegraphics{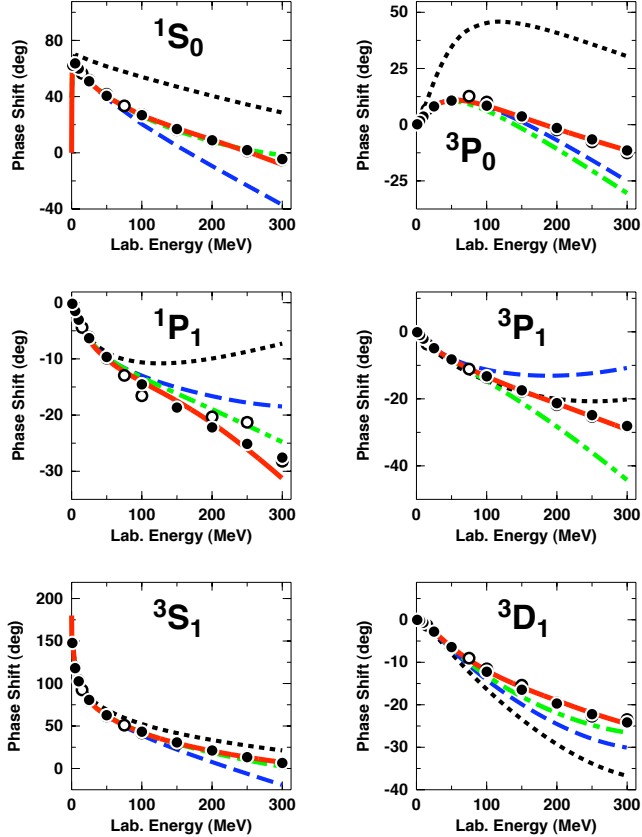}}
\vspace*{-2.5cm}
\caption{Phase shifts of $np$ scattering
as calculated from $NN$ potentials at different
orders of ChPT. The black dotted line is LO(500), 
the blue dashed is NLO(550/700)~\cite{EGM04},
the green dash-dotted NNLO(600/700)~\cite{EGM04},
and the red solid N$^3$LO(500)~\cite{EM03}, where the 
numbers in parentheses denote the cutoffs in MeV.
Partial waves with total angular momentum $J\leq 1$
are displayed.
Empirical phase shifts (solid dots and open circles)
as in Fig.~\ref{fig_f}.
\label{fig_phorders1}}
\end{figure}

\begin{figure}[t]\centering
\vspace*{-1.5cm}
%\hspace*{-1.2cm}
\scalebox{0.55}{\includegraphics{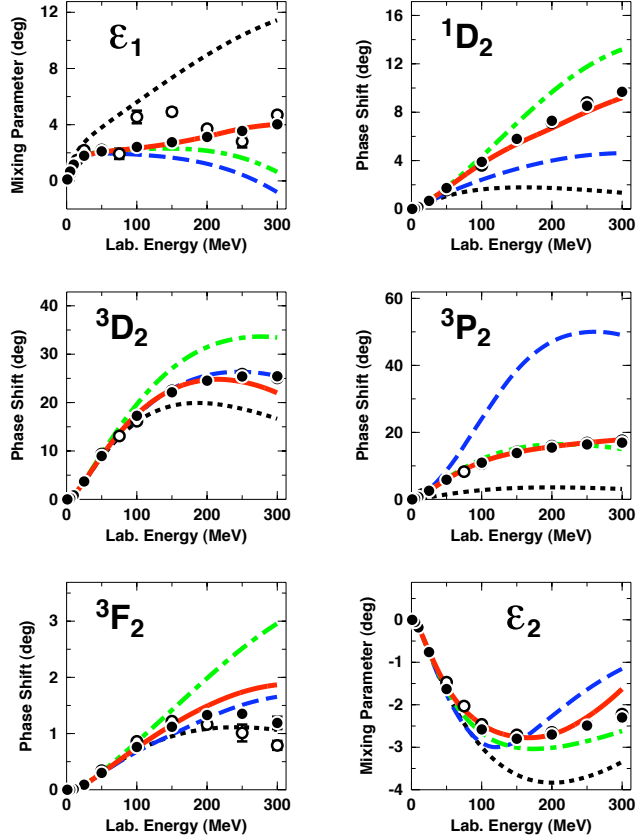}}
\vspace*{-2.5cm}
\caption{Same as Fig.~\ref{fig_phorders1}, but
$J=2$ phase shifts and 
$J\leq 2$ mixing parameters are shown.
\label{fig_phorders2}}
\end{figure}

\subsubsection{Concluding the renormalization issue}
Crucial for an EFT are regulator independence (within the range of validity
of the EFT) and a power counting scheme that allows for order-by-order
improvement with decreasing truncation error.
The purpose of renormalization is to achieve this regulator independence while maintaining
a functional power counting scheme.
After the comprehensive tries and errors of the past, it appears that there are two renormalization
schemes which have the potential to achieve the above goals and, therefore, should be investigated 
systematically in the near future.

In {\it scheme one}, the LO calculation is conducted nonperturbatively
(with $\Lambda \rightarrow \infty$ as in~\cite{NTK05})
and subleading orders are added perturbatively in distorted wave Born approximation.
As mentioned above,
Valderrama has started this in $S$ waves~\cite{Val09}, but results in higher
partial waves are needed to fully assess this approach.
Even though at this early stage any judgment is speculative, we take the liberty to predict
that this approach will be only of limited success and utility---for the following reasons.
First, it will probably require about twice as many counterterms as Weinberg counting
and, therefore, will have less predictive power. Second, this scheme may converge badly,
because the largest portion of the nuclear force, namely, the intermediate-range
attraction appears at NNLO. Third, as discussed in Ref.~\cite{Mac09}, this force may be problematic
(and, therefore, impractical) in applications in nuclear few- and many-body systems, 
because of a pathologically strong tensor force that will cause bad convergence of energy and wave functions. Finally, in the work that has been conducted so far within this
scheme by Valderrama,
it is found that only rather soft cutoffs can be used.

The latter point (namely, soft cutoffs) suggests that one may then as well
conduct the calculation nonperturbatively at all orders (up to N$^3$LO) using Weinberg counting, 
which is no problem with soft cutoffs. This is {\it scheme two} that we propose to
investigate systematically. 
In the spirit of Lepage, the cutoff independence should be examined
for cutoffs below the hard scale and not beyond. Ranges of cutoff independence within the
theoretical error are to be identified using `Lepage plots'~\cite{Lep97}.
A very systematic investigation of this kind does not exist at this time and is therefore needed.
However, there is comprehensive circumstantial
evidence from the numerous chiral $NN$ potentials constructed over the past 
decade~\cite{ORK94,EGM00,EM02a,EM02,EM03,ME05,EGM05}
(see Figs.~\ref{fig_phn3lo1} and \ref{fig_phn3lo2}, below)
indicating that this investigation will most likely be a success. 
The potentials discussed in the following section are all based upon
Weinberg counting.

\subsection{Constructing quantitative chiral $NN$ potentials
\label{sec_pot2}}

\subsubsection{What order?}

As discussed, the $NN$ potential can be calculated up to
various orders, cf.\ Eqs.~(\ref{eq_Vlo})-(\ref{eq_Vn3lo}),
and the accuracy increases as the order increases.
That triggers the obvious question:
To what order do we have to go for an accuracy that
we would perceive as necessary and sufficient
for, e.~g., reliable microscopic nuclear structure
calculations?
We had a first look at the order by order 
evolvement in Section~\ref{sec_peri}
where we compared phase shift predictions with empirical values
in peripheral partial waves. But that was only a small part
of the story. 
The most important partial waves are the lower ones,
since they generate the largest contributions
in most applications.
Now we have potentials at hand that
are defined through all partial waves, and so we can
investigate the issue of how well ChPT converges in the
important lower partial waves. 
This is demonstrated in Figs.~\ref{fig_phorders1} and \ref{fig_phorders2},
where we show the $J\leq 2$ phase parameters for potentials constructed
at LO, NLO, NNLO, and N$^3$LO. These figures clearly reveal
substantial improvements in the reproduction of the empirical
phase shifts as the orders go up.

There is an even better way to confront theory with
experiment.
We can calculate observables of
$NN$ scattering and compare to the experimental data.
It is customary to state the result of such a comparison
in terms of the $\chi^2$/datum where a value around unity
would signify a perfect fit.

\begin{table}[t]
\caption{Columns three and four show the
$\chi^2$/datum for the reproduction of the 1999 $np$ 
database~\cite{note2} (subdivided into energy intervals)
by families of $np$ potentials at NLO and NNLO constructed by the
Juelich group~\cite{EGM04}.
The $\chi^2$/datum is stated in terms of ranges
which result from a variation of the cutoff parameters
used in the regulator functions.
The values of these cutoff parameters in units of MeV are given in 
parentheses.
$T_{\rm lab}$ denotes the kinetic energy of the incident neutron
in the laboratory system.
\label{tab_chi2a}}
\smallskip
\begin{tabular*}{\textwidth}{@{\extracolsep{\fill}}cccc}
\hline 
\hline 
\noalign{\smallskip}
 $T_{\rm lab}$ bin &  \# of $np$ & 
\multicolumn{2}{c}{\it --- Juelich $np$ potentials --- }\\
 (MeV) 
 & data 
 & NLO
 & NNLO 
\\
 & & (550/700--400/500) & (600/700--450/500)
 \\
\hline 
\hline 
\noalign{\smallskip}
0--100&1058&4--5&1.4--1.9\\ 
100--190&501&77--121&12--32\\ 
190--290&843&140--220&25--69\\ 
\hline 
\noalign{\smallskip}
0--290&2402&67--105&12--27
\\ 
\hline 
\hline 
\end{tabular*}
\end{table}

\begin{table}[t]
\caption{Number of parameters needed for fitting the $np$ data
in phase-shift analysis and by a high-precision $NN$ potential
{\it versus} the total number of $NN$ contact terms of EFT based potentials 
to different orders. 
\label{tab_par}}
\smallskip
\begin{tabular*}{\textwidth}{@{\extracolsep{\fill}}ccc|ccc}
\hline 
\hline 
\noalign{\smallskip}
     & Nijmegen     & CD-Bonn        & 
               \multicolumn{3}{c}{\it --- Contact Potentials --- }\\
     & partial-wave & high-precision & $Q^0$ & $Q^2$  & $Q^4$ \\
     & analysis~\cite{Sto93} & potential~\cite{Mac01} & 
                                 LO & NLO/NNLO  & N$^3$LO \\
\hline 
\hline 
\noalign{\smallskip}
$^1S_0$         & 3 & 4 & 1&2 & 4 \\
$^3S_1$         & 3 & 4 & 1&2 & 4 \\
\hline
\noalign{\smallskip}
$^3S_1$-$^3D_1$ & 2 & 2 & 0&1 & 3 \\
\hline
\noalign{\smallskip}
$^1P_1$         & 3 & 3 & 0&1 & 2 \\
$^3P_0$         & 3 & 2 & 0&1 & 2 \\
$^3P_1$         & 2 & 2 & 0&1 & 2 \\
$^3P_2$         & 3 & 3 & 0&1 & 2 \\
\hline
\noalign{\smallskip}
$^3P_2$-$^3F_2$ & 2 & 1 & 0&0 & 1 \\
\hline
\noalign{\smallskip}
$^1D_2$         & 2 & 3 & 0&0 & 1 \\
$^3D_1$         & 2 & 1 & 0&0 & 1 \\
$^3D_2$         & 2 & 2 & 0&0 & 1 \\
$^3D_3$         & 1 & 2 & 0&0 & 1 \\
\hline
\noalign{\smallskip}
$^3D_3$-$^3G_3$ & 1 & 0 & 0&0 & 0 \\
\hline
\noalign{\smallskip}
$^1F_3$         & 1 & 1 & 0&0 & 0 \\
$^3F_2$         & 1 & 2 & 0&0 & 0 \\
$^3F_3$         & 1 & 2 & 0&0 & 0 \\
$^3F_4$         & 2 & 1 & 0&0 & 0 \\
\hline
\noalign{\smallskip}
$^3F_4$-$^3H_4$ & 0 & 0 & 0&0 & 0 \\
\hline
\noalign{\smallskip}
$^1G_4$         & 1 & 0 & 0&0 & 0 \\
$^3G_3$         & 0 & 1 & 0&0 & 0 \\
$^3G_4$         & 0 & 1 & 0&0 & 0 \\
$^3G_5$         & 0 & 1 & 0&0 & 0 \\
\hline
\hline
\noalign{\smallskip}
Total         & 35  & 38 & 2&9 & 24 \\
\hline
\hline
\noalign{\smallskip}
\end{tabular*}
\end{table}

Let's start with potentials developed to NLO and NNLO.
In Table~\ref{tab_chi2a}, we show
the $\chi^2$/datum for the fit of the world $np$ data
below 290 MeV for families of $np$ potentials at 
NLO and NNLO constructed by the Juelich group~\cite{EGM04}.
The NLO potentials produce the very large $\chi^2$/datum between 67 and 105,
and the NNLO are between 12 and 27.
The rate of improvement from one order to the other
is very encouraging, but the quality of the reproduction
of the $np$ data at NLO and NNLO is obviously
insufficient for reliable predictions.

Based upon these facts, it has been pointed out in 2002 by
Entem and Machleidt~\cite{EM02a,EM02} that one has
to proceed to N$^3$LO. Consequently, the first N$^3$LO  potential was
published in 2003~\cite{EM03}.

At N$^3$LO, there are a total of 24 contact terms (24 parameters)
which contribute to the partial waves with $L\leq 2$ 
(cf.~Section~\ref{sec_ct} and \ref{app_ct}).
These 24 LECs are essentially free constants
which parametrize
the short-ranged phenomenological part of the interaction.
In Table~\ref{tab_par}, column `$Q^4$/N$^3$LO', we show how these
terms are distributed over the partial waves.
Most important for the improved reproduction of the $NN$ 
phase shifts (and $NN$ observables) 
at N$^3$LO is the fact that 
contacts appear for the first time in $D$-waves. 
$D$-waves are not truly peripheral and, therefore,
1PE plus 2PE alone do not describe them well (Fig.~\ref{fig_phorders2}).
The $D$-wave contacts provide the necessary 
short-range corrections to get the $D$-phases right.
Besides this, at N$^3$LO, another contact is added to each
$P$-wave, which leads to substantial improvements,
particularly, in $^3P_0$ and $^3P_1$ above 100 MeV
(cf.\ Fig.~\ref{fig_phorders1}).

In Table~\ref{tab_par},
we also show the number of parameters
used in the Nijmegen partial wave analysis (PWA93)~\cite{Sto93}
and in the high-precision CD-Bonn potential~\cite{Mac01}.
The table reveals that, for $S$ and $P$ waves, 
the number of parameters
used in high-precision phenomenology and in EFT at N$^3$LO
are about the same.
{\bf Thus, the EFT approach provides retroactively a justification
for the phenomenology used in the 1990's to  obtain high-precision fits.}

At NLO and NNLO, the number of contact parameters is substantially
smaller than for PWA93 and CD-Bonn, which explains why
these orders are insufficient for a quantitative potential.
The 24 parameters of N$^3$LO are close to the 30+
used in PWA93 and high precision potentials. Consequently
(see following sections for details), at N$^3$LO,
a fit of the $NN$ data is possible
that is of about the
same quality as the one by the high-precision $NN$ 
potentials~\cite{Sto94,WSS95,MSS96,Mac01}.
Thus, one may perceive N$^3$LO as the order of ChPT that
is necessary and sufficient for a reliable $NN$ potential.

\subsubsection{Charge-dependence \label{sec_CD}}

So far we considered only neutron-proton scattering. However, as stressed repeatedly by
the Nijmegen group~\cite{Ber88,Sto93}, for a precise reproduction
of the $NN$ data, i.e., the $np$ and the $pp$ data,
it is crucial to take charge-dependence into account and to construct separate $np$, $pp$,
and $nn$ potentials which differ by the subtleties of charge-dependence.

By definition, {\it charge independence} (or isospin symmetry) is invariance under any 
rotation in isospin space. 
A violation of this symmetry is referred to as charge dependence or
charge independence breaking (CIB).
{\it Charge symmetry} is invariance under a rotation by 180$^0$ about the
$y$-axis in isospin space if the positive $z$-direction is associated
with the positive charge.
The violation of this symmetry is known as charge symmetry breaking (CSB).
Obviously, CSB is a special case of charge dependence.

CIB of the strong $NN$ interaction means that,
in the isospin $I=1$ state, the
proton-proton ($I_z=+1$), 
neutron-proton ($I_z=0$),
or neutron-neutron ($I_z=-1$)
interactions are (slightly) different,
after electromagnetic effects have been removed.
CSB of the $NN$ interaction refers to a difference
between proton-proton ($pp$) and neutron-neutron ($nn$)
interactions, only. 
For reviews, see Refs.~\cite{HM79,Mac89,MNS90,LM98a,LM98b,MS01,Gar09}.

CIB is seen most clearly in the $^1S_0$ 
$NN$ scattering lengths. 
The latest empirical values for the singlet scattering length $a$ 
and effective range $r$ are:
\begin{eqnarray}
a^N_{pp}&=&-17.3\pm 0.4 
\mbox{ fm \cite{MNS90}}, 
\quad \quad \quad \quad
r^N_{pp}=2.85\pm 0.04 \mbox{ fm \cite{MNS90}}; 
\\
%\end{eqnarray}
%\begin{eqnarray}
a^N_{nn}&=&-18.95\pm 0.40 \mbox{ fm \cite{Gon06,Che08}}, 
\quad \quad 
r^N_{nn} = 2.75\pm 0.11 \mbox{ fm \cite{MNS90}};
 \\
a_{np}&=&-23.740\pm 0.020 \mbox{ fm \cite{Mac01}},
\quad \quad \quad
r_{np}=2.77\pm 0.05 \mbox{ fm \cite{Mac01}}.
\end{eqnarray}
The values given for $pp$ and $nn$ 
scattering refer to the nuclear part of the interaction
as indicated by the superscript $N$; i.~e., 
electromagnetic effects have been removed from the experimental
values. 

The above values imply that
charge-symmetry is broken by
\begin{eqnarray}
\Delta a_{CSB}& \equiv& a_{pp}^N-a_{nn}^N = 1.65\pm 0.60 \mbox{ fm}, 
\label{eq_CSBlep1}
\\
\Delta r_{CSB}& \equiv& r_{pp}^N-r_{nn}^N = 0.10\pm 0.12 \mbox{ fm}; 
\label{eq_CSBlep2}
\end{eqnarray}
and
the following CIB is observed:
\begin{eqnarray}
\Delta a_{CIB} \equiv
\frac12 ( a_{pp}^N + a_{nn}^N )
 - 
 a_{np}
 &=& 5.6\pm 0.6 \mbox{ fm},
\label{eq_CIBlep1}
\\
\Delta r_{CIB} \equiv
 \frac12 ( r_{pp}^N + r_{nn}^N )
 - 
 r_{np}
 &=& 0.03\pm 0.13 \mbox{ fm}.
\label{eq_CIBlep2}
\end{eqnarray}
In summary, the $NN$ singlet scattering lengths show a small amount
of CSB and a clear signature of CIB.

The current understanding is that---on a fundamental level---charge-dependence
(`isospin violation') is due to a difference between the up and down quark masses
and electromagnetic interactions. 
As first discussed in Refs.~\cite{Kol93,Kol95,KFG96}, 
EFT is a suitable tool to also deal with isospin violations.
In the two-flavor case, the mass term in the QCD
Lagrangian, Eq.~(\ref{eq_LQCD}), can be written as [cf.\ Eq.~(\ref{eq_mmatr})]
\begin{eqnarray}
-
\bar{q} \,
 {\cal M} \, q
 & = & -\frac12 \bar{q} \, (m_u+m_d) \, I \, q - \frac12 \bar{q} \, (m_u-m_d) \, \tau_3 \,q
 \nonumber \\
 & = & -\frac12 \bar{q} \, (m_u+m_d) \, ( I -   \epsilon \, \tau_3 ) \, q
 \label{eq_IB}
\end{eqnarray}
with
\begin{equation}
\epsilon = \frac{m_d-m_u}{m_u+m_d} \sim  \frac13 \,,
\end{equation}
cf.\ Eqs.~(\ref{eq_umass}) and (\ref{eq_dmass}).
The first term of Eq.~(\ref{eq_IB}) conserves isospin but breaks chiral symmetry.
It is responsible for the nonvanishing pion mass and leads to terms $\propto (m_\pi^2)^n$
(with $n\geq 1$)
in the effective Lagrangian. The second term breaks isospin symmetry and generates
terms proportional to $(\epsilon m_\pi^2)^n$. Note that the isospin-breaking
effects are much smaller than the value for $\epsilon$ suggests, because the relevant scale
is $\Lambda_\chi$ rather than $(m_u+m_d)$. 

For isospin violating effects caused by the electromagnetic interaction,
the small parameter $e^2=4\pi\alpha \sim 1/10$ can be used
(where $\alpha=1/137.036$ denotes the fine-structure constant). 
It is then suggestive to consider
\begin{equation}
\epsilon \sim e \sim \frac{Q}{\Lambda_\chi}
\label{eq_IBpar1}
\end{equation}
as the general expansion parameter.
For photon loops, one may further assume
\begin{equation}
\frac{e^2}{(4\pi)^2} \sim \frac{Q^4}{\Lambda_\chi^4}
\,.
\label{eq_IBpar2}
\end{equation}
We will now briefly discuss various isospin violating contributions to the $NN$ interaction
listed in Table~\ref{tab_IBcontr}. For a more comprehensive and systematic study,
the interested reader is referred to Ref.~\cite{EM05}.

\begin{table}[t]
\caption{Isospin breaking contributions to the $NN$ interaction}
\label{tab_IBcontr}
\smallskip
\begin{tabular*}{\textwidth}{@{\extracolsep{\fill}}ll}
\hline 
\hline 
\noalign{\smallskip}
 {\bf Order} & {\bf Contributions} \\
\hline 
\hline 
\noalign{\smallskip}
NL\O \hspace*{.1cm} ($\nu=2$) 
                          & Pion-mass splitting in 1PE, \\
                          & Static Coulomb potential. \\  
\hline 
NNL\O \hspace*{.1cm} ($\nu=3$) 
                          & CSB contacts without derivatives, \\
                          & Charge-dependence of the pion-nucleon coupling constant in 1PE
                           ($\sim \epsilon \, m_\pi^2/\Lambda_\chi^2$). \\
\hline 
\noalign{\smallskip}
N$^3$L\O \hspace*{.1cm} ($\nu=4$)
                          & CIB contacts without derivatives, \\
                          & Charge-dependence of the pion-nucleon coupling constant in 1PE
                           [$\sim e^2/(4\pi)^2$], \\
                          & Pion-mass splitting in NLO 2PE, \\
                          & Nucleon-mass splitting in NLO 2PE and LS-equation, \\
                          & $\pi\gamma$-exchange, \\
                          & Relativistic corrections to the Coulomb potential ($\sim e^2 \, Q^2/M^2_N$), \\
                          & Further electromagnetic corrections.\\
\hline 
\hline 
\end{tabular*}
\end{table}

It is well known that the pion-mass difference is mainly due to electro-magnetic interactions
among the quarks and, thus, according to Eq.~(\ref{eq_IBpar2}), of order $\nu=4$. Therefore,
the effect of pion-mass difference on 1PE can be estimated to be
\begin{equation}
\frac{\Delta m_\pi^2}{m_\pi^2}  =
\frac{\Delta m_\pi^2}{\Lambda_\chi^2} \;
\frac{\Lambda_\chi^2}{m_\pi^2}
\sim \frac{e^2}{(4\pi)^2} \; \frac{\Lambda_\chi^2}{m_\pi^2}
\sim \frac{Q^2}{\Lambda_\chi^2}
\,.
\end{equation}
So, it is of order $\nu=2$ or NL\O \, (cf.\ Table~\ref{tab_IBcontr})
with the slash signifying that the extended power counting
scheme, Eqs.~(\ref{eq_IBpar1}) and (\ref{eq_IBpar2}), is applied.
This CIB effect is taken care of by replacing the $I=1$ isospin-symmetric 1PE potential, Eq.~(\ref{eq_1pe}),
by
\begin{equation}
V^{(pp)}_{1\pi}=V^{(nn)}_{1\pi}=V_{1\pi}(m_{\pi^0})
\end{equation}
for $pp$ and $nn$ scattering and
\begin{equation}
V^{(np,I=1)}_{1\pi}=
-V_{1\pi}(m_{\pi^0}) +2\,V_{1\pi}(m_{\pi^\pm})
\end{equation}
for $I=1$ $np$ scattering, with $V_{1\pi}(m_\pi)$ defined in Eq.~(\ref{eq_1pe_bare}). 
If the pion masses were all the same, the above potentials would be identical.
However, due to their mass splitting (Table~\ref{tab_LEC}), the $I=1$ $np$ potential is more attractive
than the $pp$ one, and this effect is known to explain about one half of the CIB scattering-length
difference, Eq.~(\ref{eq_CIBlep1}).
For completeness, we also give the $I=0$ $np$ 1PE potential which is
\begin{equation}
V^{(np,I=0)}_{1\pi}=
-V_{1\pi}(m_{\pi^0}) -2\,V_{1\pi}(m_{\pi^\pm})
\,.
\end{equation}

Due to the smallness of the pion mass, 1PE is also a sizable
contribution in all partial waves with $L>0$;
and due to the pion's relatively large mass splitting (3.4\%),
1PE creates relatively large charge-dependent effects 
in all partial waves 
(cf.\ Tables~V and VI and Fig.~4 of Ref.~\cite{Mac01}).
Therefore,
all modern phase shift analyses~\cite{SM99,Sto93} and all
modern $NN$ potentials~\cite{Mac01,Sto94,WSS95}
include the CIB effect created by 1PE.

The other NL\O \, contribution (Table~\ref{tab_IBcontr}) is due to the 
static Coulomb interaction which is $\sim e^2$ and,
therefore, also of order two. The Coulomb potential has considerable
impact on the $^1S_0$ phase shifts at low energies and creates CIB effects of similar size as
1PE in $P$ and $D$ waves (cf.\ Tables~V and VI and Fig.~5 of Ref.~\cite{Mac01}).
The calculation of phase shifts in the presence of the Coulomb potential is explained
in Appendix A.3 of Ref.~\cite{Mac01}.

At NNL\O \, ($\nu=3$), CSB contributions to the charge-dependence of the
pion-nucleon coupling constant ($\sim \epsilon \, m_\pi^2/\Lambda_\chi^2$) occur.
However, since empirically there is no clear evidence for such 
charge-dependence~\cite{STS93,MS01}, this contribution is ignored.
Moreover, at NNL\O, there is
a non-derivative CSB contact term
\begin{equation}
\propto \epsilon \, m_\pi^2 ( \bar{N} \tau_3 N)( \bar{N} N ) \,,
\label{eq_LCSB}
\end{equation}
which is crucial for the fit of the CSB splitting of the $^1S_0$ scattering length,
Eq.~(\ref{eq_CSBlep1}).

The leading CIB contact interactions are of electromagnetic origin and have the structure
\begin{equation}
\propto \frac{e^2}{(4\pi)^2} \, ( \bar{N} \tau_3 N)( \bar{N} \tau_3 N ) \,.
\label{eq_LCIB}
\end{equation}
They are obviously of order $\nu=4$ (N$^3$L\O) and
needed for the fit of the CIB splitting of the $^1S_0$ scattering length,
Eq.~(\ref{eq_CIBlep1}). Besides this, there are numerous other N$^3$L\O \, contributions
(cf.\ Table~\ref{tab_IBcontr}).
One-loop electromagnetic corrections to the pion-nucleon coupling constant
create effects of order $e^2/(4\pi)^2$ in the 1PE, which will be ignored for
the reasons given above. Corrections due to pion-mass difference in the NLO 2PE
can be estimated to be
\begin{equation}
\frac{\Delta m_\pi^2}{m_\pi^2} \;
\frac{Q^2}{\Lambda_\chi^2}
\sim 
\frac{Q^4}{\Lambda_\chi^4}
\,.
\end{equation}
As explained in Refs.~\cite{FK99,WME01}, these corrections can be calculated most conveniently
by separating the NLO 2PE as given in Eqs.~(\ref{eq_2C}) and (\ref{eq_2T})
 into an isoscalar and  an isovector piece:
 \begin{equation}
 V_{2\pi}=V_{2\pi}^0 +
\bbox{\tau}_1 \cdot \bbox{\tau}_2 \;
V_{2\pi}^1
\,.
\end{equation}
The isoscalar part $V_{2\pi}^0$ is given by~\cite{FK99}
\begin{equation}
V_{2\pi}^0 = \frac23 \, V_{2\pi}^0 (m_{\pi^\pm},m_{\pi^\pm}) + 
 \frac13 \, V_{2\pi}^0 (m_{\pi^0},m_{\pi^0}) 
 \approx  V_{2\pi}^0 (\bar{m}_{\pi},\bar{m}_{\pi}) 
 \,,
 \end{equation}
 where the arguments represent the masses of the two exchanged pions, which are given in
 Table~\ref{tab_LEC}, and with
 $\bar{m}_\pi$ denoting the average pion masses defined by
 \begin{equation}
 \bar{m}_\pi \equiv \frac23 \, m_{\pi^\pm} + \frac13 \, m_{\pi^0}
 = 138.0390 \mbox{ MeV.}
 \end{equation}
 For the isovector part $V_{2\pi}^1$ one obtains~\cite{FK99,WME01}
 \begin{equation}
 V_{2\pi}^1 = \left\{ \begin{array}{ll}
                  V_{2\pi}^1 (m_{\pi^\pm},m_{\pi^\pm})   & \mbox{ for $pp$ and $nn$,}  \\
                 2 \, V_{2\pi}^1 (m_{\pi^\pm},m_{\pi^0}) - V_{2\pi}^1 (m_{\pi^\pm},m_{\pi^\pm}) 
                 \approx V_{2\pi}^1 (m_{\pi^0},m_{\pi^0}) 
                                                                                & \mbox{ for $np, I=1$;}
                                \end{array}  \right. 
 \end{equation}
 and for $np, I=0$, we have  $V_{2\pi}=V_{2\pi} (\bar{m}_{\pi},\bar{m}_{\pi})$. 
 In conjunction with the work performed in Ref.~\cite{EM03}, it was found that
 the CIB effects due to pion-mass splitting in the NLO 2PE
 potential are negligibly small in $P$ and higher partial waves
 when also the pion-mass dependence of the polynomial terms is included 
(cf.\ discussion in Appendix A of Ref.~\cite{WME01}).
However, in the $^1S_0$ state, this effect is non-negligible
(causing $\Delta a_{CIB} \approx -0.5$ fm), but can be absorbed by the non-derivative
CIB contact. Thus, from a procedural point of view, the CIB effect from pion-mass splitting
in the NLO 2PE can simply be ignored (which is the procedure applied in Ref.~\cite{EM03}).

Nucleon-mass splitting in intermediate states of 2PE diagrams creates CSB
as discussed in detail in Refs.~\cite{Fri03,Fri04}. When derived just  for the NLO 2PE diagrams,
this effect is N$^3$L\O. Accurate calculations
conducted within the framework of conventional meson theory~\cite{LM98a} have shown that this effect
is very small in $P$ and higher partial waves for diagrams with only nucleon intermediate states
and moderate when single $\Delta$ excitation is taken into account (which corresponds to order
N$^4$L\O \, in $\Delta$-less ChPT).
The effect in $^1S_0$ is non-negligible, but can be absorbed by the non-derivative CSB contact term.
All existing models for the $NN$ interaction at N$^3$LO ignore this CSB.
However, it would be worthwhile to calculate this effect accurately for chiral 2PE and compare to the
comprehensive results obtained from conventional meson models~\cite{LM98a} where it was found that the  
CSB splitting of the singlet scattering length can be explained by nucleon-mass difference alone.

Nucleon masses also enter the outer legs of $NN$ scattering diagrams
and the LS equation. Here, practitioners use, in general, the accurate proton and
neutron masses and  the proper relativistic formula
for the relation between the on-shell CM momentum and the kinetic energy 
of the incident nucleon in the laboratory system
(cf. Appendix A.3 of Ref.~\cite{Mac01}).

\begin{table}[t]
\caption{Columns three to five display the
$\chi^2$/datum for the reproduction of the 1999 
{\boldmath\bf $np$ database}~\cite{note2}
(subdivided into energy intervals)
by various $np$ potentials.
For the chiral potentials, 
the $\chi^2$/datum is stated in terms of ranges
which result from a variation of the cutoff parameters
used in the regulator functions.
The values of these cutoff parameters 
in units of MeV
are given in parentheses.
$T_{\rm lab}$ denotes the kinetic energy of the incident nucleon
in the laboratory system.
\label{tab_chi2b}}
\smallskip
\begin{tabular*}{\textwidth}{@{\extracolsep{\fill}}ccccc}
%\begin{tabular}{cc|c|c|c}
\hline 
\hline 
\noalign{\smallskip}
 $T_{\rm lab}$ bin
 & \# of {\boldmath $np$}
 & {\it Idaho}
 & {\it Juelich}
 & Argonne         
\\
 (MeV)
 & data
 & N$^3$LO~\cite{EM03}
 & N$^3$LO~\cite{EGM05} 
 & $V_{18}$~\cite{WSS95}
\\
 & 
 & (500--600)
 & (600/700--450/500)
 & 
\\
\hline 
\hline 
\noalign{\smallskip}
0--100&1058&1.0--1.1&1.0--1.1&0.95\\ 
100--190&501&1.1--1.2&1.3--1.8&1.10\\ 
190--290&843&1.2--1.4&2.8--20.0&1.11\\ 
\hline 
\noalign{\smallskip}
0--290&2402&1.1--1.3&1.7--7.9&1.04
\\ 
\hline 
\hline 
\end{tabular*}
\end{table}

\begin{table}[b]
\caption{Same as Table~\ref{tab_chi2b} but for 
{\boldmath\bf $pp$}.
\label{tab_chi2c}}
\smallskip
\begin{tabular*}{\textwidth}{@{\extracolsep{\fill}}ccccc}
\hline 
\hline 
\noalign{\smallskip}
 $T_{\rm lab}$ bin
 & \# of {\boldmath $pp$}
 & {\it Idaho}
 & {\it Juelich}
 & Argonne         
\\
 (MeV)
 & data
 & N$^3$LO~\cite{EM03}
 & N$^3$LO~\cite{EGM05} 
 & $V_{18}$~\cite{WSS95}
\\
 &  
 & (500--600)
 & (600/700--450/500)
 & 
\\
\hline 
\hline 
\noalign{\smallskip}
0--100&795&1.0--1.7&1.0--3.8&1.0 \\ 
100--190&411&1.5--1.9&3.5--11.6&1.3 \\ 
190--290&851&1.9--2.7&4.3--44.4&1.8 \\ 
\hline 
\noalign{\smallskip}
0--290&2057&1.5--2.1&2.9--22.3&1.4 
\\ 
\hline 
\hline 
\end{tabular*}
\end{table}

Irreducible $\pi\gamma$ exchange~\cite{Kol98} causes CIB of order  N$^3$L\O \,
($\sim e^2/(4\pi)^2$). It is a moderate effect in $^1S_0$
($\Delta a_{CIB} \approx -0.35$ fm) and small in $P$ and higher partial waves
(cf.\ Ref.~\cite{Mac01}).
Some N$^3$LO potentials~\cite{EM03} do include this $\pi\gamma$ contribution.
Corrections to the $\pi\gamma$ graphs (which are N$^4$L\O) and even $2\pi\gamma$
diagrams (N$^6$L\O) have recently been calculated by Kaiser~\cite{Kai06a,Kai06b,Kai06c}
and found to be astonishingly large. No $NN$ calculation has yet included them.

Finally, at N$^3$L\O, several corrections to the long-range electromagnetic interaction occur. 
The leading relativistic correction to the static Coulomb potential~\cite{AS83,Ber88} is most conveniently included
by replacing the fine-structure constant $\alpha$ by 
\begin{equation}
\alpha' = \, \alpha \,\, \frac{E_p^2+p^2}{M_p \, E_p} \,.
\end{equation}
Other electromagnetic contributions to $pp$ scattering are
two-photon exchange, the Darwin-Foldy term, vacuum polarization,
and the magnetic moment interaction~\cite{Sto93,WSS95}.
In the case of $np$ scattering, the electromagnetic interaction consists
only of the magnetic moment contribution.
The electromagnetic interactions are important contributions to the scattering amplitude.
Note, however, that in the calculation of the strong nuclear phase shifts the electromagnetic interaction
is only of relevance when its distortion of the wave functions affects the nuclear phase shifts 
in a non-negligible way.
It is well known that this effect is large for the Coulomb potential in essentially all partial waves, 
but it is negligible for all
other electromagnetic interactions in all partial-waves, except $^1S_0$ 
below 30 MeV, where the effect can be calculated fairly model-independently and has
been tabulated in Ref.~\cite{Ber88}. Thus, in practice, it is sufficient to calculate $pp$ phase shifts
with only the Coulomb effect and $np$ phase shifts without any electromagnetic effects 
(which is the way the phase shifts published in Refs.~\cite{Sto94,Mac01}
and shown in the tables of \ref{app_par} are calculated).

\begin{figure}[t]\centering
\vspace*{-1.5cm}
%\hspace*{-1.2cm}
\scalebox{0.55}{\includegraphics{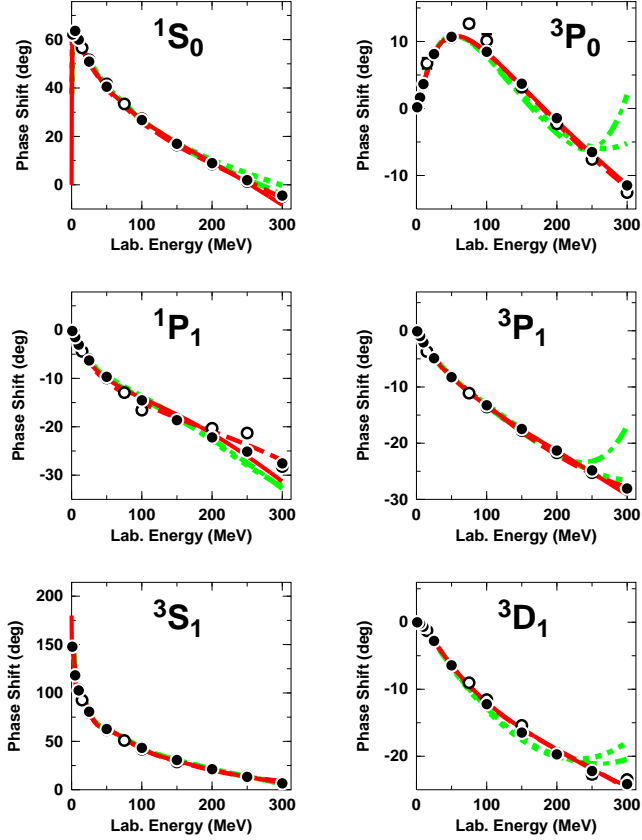}}
\vspace*{-2.5cm}
\caption{Neutron-proton phase parameters as described by
various chiral potentials at N$^3$LO.
The (red) solid and the dashed curves are calculated from
Idaho N$^3$LO potentials~\cite{EM03} with $\Lambda=$ 500 
and 600 MeV, respectively;
while the (green) dash-dotted and the dotted curves are based upon
Juelich N$^3$LO potentials~\cite{EGM05} with cutoff combinations
600/700 and 450/500 MeV, respectively.
Partial waves with total angular momentum $J\leq 1$ are displayed.
Empirical phase shifts (solid dots and open circles) 
as in Fig.~\ref{fig_f}.
\label{fig_phn3lo1}}
\end{figure}

\begin{figure}[t]\centering
\vspace*{-1.5cm}
%\hspace*{-1.2cm}
\scalebox{0.55}{\includegraphics{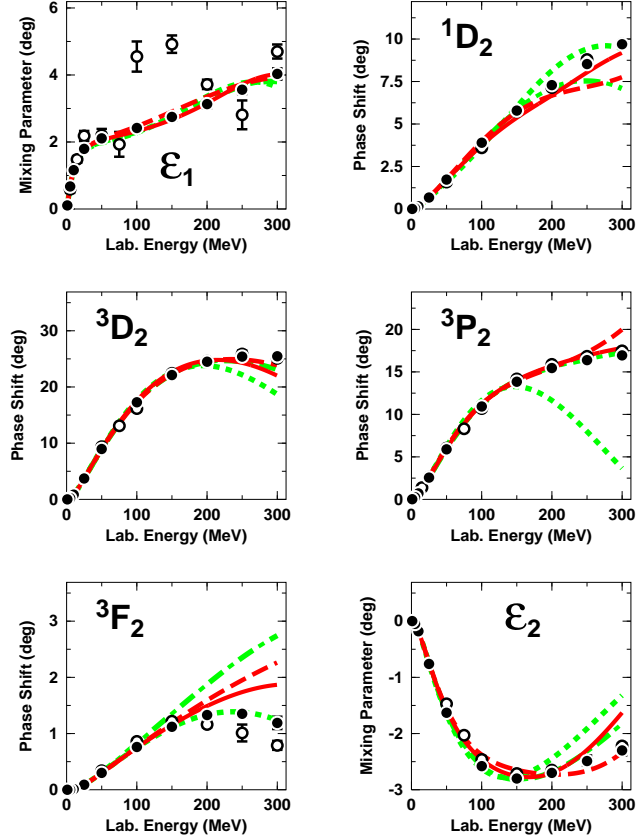}}
\vspace*{-2.5cm}
\caption{Same as Fig.~\ref{fig_phn3lo1} but
$J=2$ phase shifts and 
$J\leq 2$ mixing parameters are shown.
\label{fig_phn3lo2}}
\end{figure}

\subsubsection{A quantitative $NN$ potential at N$^3$LO
\label{sec_potn3lo}}

After previous sections have thoroughly prepared the
terrain, we will now present a quantitative $NN$ potential
at N$^3$LO, including details of construction and results.
We choose the Idaho N$^3$LO potential~\cite{EM03} as example.
The isospin symmetric part of the chiral $NN$ potential 
at N$^3$LO is defined in
Eq.~(\ref{eq_Vn3lo}) and the isospin violating terms
were discussed in the previous section. Numerous parameters
are involved
which can be subdivided into three groups: $\pi N$ LECs,
$NN$ contact parameters, and the cutoff parameter $\Lambda$
of the regulator Eq.~(\ref{eq_f}).
For $\Lambda$ we choose initially 500 MeV.
Within a certain reasonable range, results should not
depend sensitively on $\Lambda$ (cf.\ discussion
in Section~\ref{sec_reno}). Therefore, we have also
made a second fit for $\Lambda=600$ MeV.

\paragraph{Data fitting and results for $NN$ scattering}

The fitting procedure starts with 
the peripheral partial waves because they involve
fewer and more fundamental parameters.
Partial waves with $L\geq 3$
are exclusively determined by 1PE and 2PE because
the N$^3$LO contacts contribute to $L\leq 2$ only.
1PE and 2PE at N$^3$LO depend on
the axial-vector coupling constant, $g_A$ (we use $g_A=1.29$),
the pion decay constant, $f_\pi=92.4$ MeV,
and eight low-energy constants (LECs) that appear in the 
dimension-two and dimension-three $\pi N$ Lagrangians,
Eqs.~(\ref{eq_L2ct}) and (\ref{eq_L3rel}).
The LECs are listed in Table~\ref{tab_LEC},
where column `$NN$ Potential' shows the values
used for the present N$^3$LO potential.
In the fitting process,
we varied three of them, namely, $c_2$, $c_3$, and $c_4$.
We found that the other LECs are not very effective in the
$NN$ system and, therefore, we left them at their central
values as determined in $\pi N$ analysis.
The most influential constant is $c_3$, 
which---in terms of magnitude---has to be chosen
on the low side (slightly more than one standard deviation
below its $\pi N$ determination), otherwise there is
too much central attraction.
Concerning $c_4$, our choice $c_4 = 5.4$ GeV$^{-1}$ 
lowers the $^3F_2$ 
phase shift (and slightly the $^1F_3$) bringing it 
into closer agreement with the phase shift 
analysis---as compared to using the $\pi N$ value
$c_4=3.4$ GeV$^{-1}$. The other
$F$ waves and the higher partial waves are essentially unaffected
by this variation of $c_4$. 
Finally, the change of $c_2$ from its $\pi N$ value
of 3.28 GeV$^{-1}$ to 2.80 GeV$^{-1}$ (our choice)
brings about some subtle improvements of the fit, but
it is not essential.
Overall, the fit
of all $J\geq 3$ waves is very good.
The $F$-wave phase shifts are, in fact, described better
than in the perturbative calculation shown in Fig.~\ref{fig_f}
because the regulator moderates the attractive
surplus, thus, simulating correctly higher order
contributions beyond the present order.

\begin{table}[t]
\caption{Scattering lengths ($a$) and effective ranges ($r$) in units of fm.
($a_{pp}^C$ and $r_{pp}^C$ refer to the $pp$ parameters in the presence of
the Coulomb force. $a^N$ and $r^N$ denote parameters determined from the
nuclear force only and with all electromagnetic effects omitted.)
\label{tab_lep}}
\smallskip
\begin{tabular*}{\textwidth}{@{\extracolsep{\fill}}cccccc}
\hline 
\hline 
\noalign{\smallskip}
 & Idaho & Juelich \\
 & N$^3$LO~\cite{EM03} & N$^3$LO~\cite{EGM05} & CD-Bonn\cite{Mac01}
 & AV18\cite{WSS95} & Empirical \\
 & (500) & (550/600) \\
\hline
\noalign{\smallskip}
\multicolumn{6}{c}{\boldmath $^1S_0$} \\
$a_{pp}^C$  &--7.8188 & --7.8003& --7.8154 & --7.8138  
  &--7.8196(26)~\cite{Ber88} \\
  &&&&& --7.8149(29)~\cite{SES83} \\
$r_{pp}^C$  & 2.795   & 2.737& 2.773 & 2.787 
  & 2.790(14)~\cite{Ber88}  \\
  &&&&& 2.769(14)~\cite{SES83} \\
$a_{pp}^N$  &--17.083 & --16.423& --17.460 & --17.164 \\
$r_{pp}^N$  &  2.876  & 2.828& 2.845 & 2.865 \\
$a_{nn}^N$  &--18.900 & --18.900& --18.968 & --18.818 &--18.95(40)~\cite{Gon06,Che08} \\
$r_{nn}^N$  &  2.838  & 2.770& 2.819 & 2.834 &  2.75(11)~\cite{MNS90} \\
$a_{np}  $  &--23.732 & --23.613& --23.738 & --23.732 &--23.740(20)~\cite{Mac01} \\
$r_{np}  $  &  2.725  & 2.651& 2.671 & 2.697 & [2.77(5)]~\cite{Mac01}   \\
\multicolumn{6}{c}{\boldmath $^3S_1$} \\
$a_t$     &  5.417   & 5.417& 5.420 & 5.419 & 5.419(7)~\cite{Mac01}  \\
$r_t$     &  1.752   & 1.742& 1.751 & 1.753 & 1.753(8)~\cite{Mac01}  \\
\hline
\hline
\noalign{\smallskip}
\end{tabular*}
\end{table}

\begin{table}[b]
\small
\caption{Deuteron properties as predicted by various $NN$
potentials are compared to empirical information.
(Deuteron binding energy $B_d$, asymptotic $S$ state $A_S$,
asymptotic $D/S$ state $\eta$, deuteron radius $r_d$,
quadrupole moment $Q$, $D$-state probability $P_D$; the calculated
$r_d$ and $Q$ are without meson-exchange current contributions
and relativistic corrections.)
\label{tab_deu}}
\smallskip
\begin{tabular*}{\textwidth}{@{\extracolsep{\fill}}llllll}
\hline 
\hline 
\noalign{\smallskip}
 & Idaho & Juelich \\
 & N$^3$LO~\cite{EM03} & N$^3$LO~\cite{EGM05} & CD-Bonn\cite{Mac01}
 & AV18\cite{WSS95} & Empirical$^a$ \\
 & (500) & (550/600) \\
\hline
\noalign{\smallskip}
$B_d$ (MeV) &
 2.224575& 2.218279&
 2.224575 & 2.224575 & 2.224575(9) \\
$A_S$ (fm$^{-1/2}$) &
 0.8843& 0.8820 &
0.8846 & 0.8850 & 0.8846(9)  \\
$\eta$         & 
 0.0256& 0.0254&
0.0256& 0.0250&0.0256(4) \\
$r_d$ (fm)   &
 1.975& 1.977&
 1.966 &
 1.967 &
 1.97535(85) \\
$Q$ (fm$^2$) &
 0.275& 0.266&
 0.270 & 
 0.270 &
 0.2859(3)  \\
$P_D$ (\%)    & 
 4.51& 3.28&
4.85 & 5.76  \\
\hline
\hline
\noalign{\smallskip}
\end{tabular*}
\footnotesize
$^a$See Table XVIII of Ref.~\cite{Mac01} for references;
the empirical value for $r_d$ is from Ref.~\cite{Hub98}.\\
\end{table}

We turn now to the lower partial waves.
Here, the most important fit parameters are the ones associated
with the 24 contact terms that contribute to the partial waves
with $L\leq 2$ (cf.\ Section~\ref{sec_ct} and Table~\ref{tab_par}). 
In addition, we have two charge-dependent
contacts which are used to fit the
three different $^1S_0$ scattering 
lengths, $a_{pp}$, $a_{nn}$, and $a_{np}$.

In the optimization procedure, we fit first phase shifts,
and then we refine the fit by minimizing the
$\chi^2$ obtained from a direct comparison with the data.
We start with $pp$, since the $pp$ phase shifts and data are more
accurate than the $np$ ones. The $pp$ fit fixes essentially the $I=1$ potential.
The $I=1$ $np$ potential is just the $pp$ one modified
by charge-dependence due to nucleon-mass difference, pion-mass splitting in 1PE,
$\pi\gamma$ exchange, and omission of Coulomb (as discussed in Section~\ref{sec_CD}). 
In addition to this, the non-derivative contact in the $^1S_0$ state is changed such as to 
reproduce the $np$  scattering length.
The $nn$ potential is the $pp$ one without Coulomb, using neutron masses, and 
fitting the $nn$ scattering length in the $^1S_0$ state with the non-derivative contact.

The $\chi^2/$datum for the fit of the $np$ data below
290 MeV are shown in Table~\ref{tab_chi2b}, 
and the corresponding ones for $pp$
are given in Table~\ref{tab_chi2c}.
These tables reveal that at N$^3$LO
a $\chi^2$/datum comparable to the high-precision
Argonne $V_{18}$~\cite{WSS95} potential can, indeed, be achieved.
The Idaho N$^3$LO potential~\cite{EM03} with $\Lambda=500$ MeV produces
a $\chi^2$/datum = 1.1 
for the world $np$ data below 290 MeV
which compares well with the $\chi^2$/datum = 1.04
by the Argonne potential.
In 2005, also the Juelich group produced
several N$^3$LO $NN$ potentials~\cite{EGM05}, the best of which
fits the $np$ data with
a $\chi^2$/datum = 1.7 and the worse with 
7.9 (Table~\ref{tab_chi2b}).

Turning to $pp$,
the $\chi^2$ for $pp$ data are typically
larger than for $np$
because of the higher precision of $pp$ data.
Thus, the Argonne $V_{18}$ produces
a $\chi^2$/datum = 1.4 for the world $pp$ data
below 290 MeV and the best Idaho N$^3$LO $pp$ potential obtains
1.5. The fit by the best Juelich 
N$^3$LO $pp$ potential results in
a $\chi^2$/datum = 2.9 and the worst  produces 22.3.

Phase shifts of $np$ scattering from two Idaho 
(solid and dashed lines) and two Juelich (dash-dotted 
and dotted lines)
N$^3$LO $np$ potentials are shown in 
Figs.~\ref{fig_phn3lo1} and \ref{fig_phn3lo2}.
The phase shifts confirm what the corresponding
$\chi^2$ have already revealed.
Low-energy scattering parameters are listed in Table~\ref{tab_lep}.
A few more technical details and tables with numerical values for
contact and phase parameters are provided in \ref{app_par}.

\begin{figure}[t]\centering
\vspace*{-3.5cm}
%\hspace*{2.0cm}
\scalebox{0.45}{\includegraphics{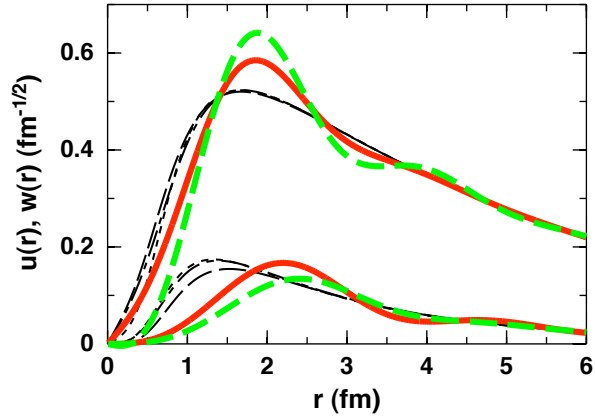}}
\vspace*{-3.5cm}
\caption{Deuteron wave functions: the family of larger curves are $S$-waves, 
the smaller ones $D$-waves. The thick colored lines represent 
the wave functions derived from chiral $NN$ potentials
at order N$^3$LO (red solid: Idaho(500)~\cite{EM03}, green dashed:
Juelich(550/600)~\cite{EGM05}). The thin dashed, dash-dotted, and dotted lines
refer to the wave functions of the CD-Bonn\protect~\cite{Mac01}, 
Nijm-I\protect~\cite{Sto94}, and AV18\protect~\cite{WSS95} potentials, 
respectively.
\label{fig_deu}}
\end{figure}

\paragraph{The deuteron}

The reproduction of the deuteron parameters is shown 
in Table~\ref{tab_deu}.
We present results for two N$^3$LO potentials, namely, Idaho~\cite{EM03}
with $\Lambda=500$ MeV and Juelich~\cite{EGM05} with cutoff combination
550/600 MeV.
Remarkable are the predictions by the chiral potentials
for the deuteron radius which are in good agreement with the latest
empirical value obtained by the isotope-shift method~\cite{Hub98}. 
All $NN$ potentials of the past 
(Table~\ref{tab_deu} includes two representative examples,
namely, CD-Bonn~\cite{Mac01} and AV18~\cite{WSS95})
fail to reproduce this very precise new value for the deuteron radius.

In Fig.~\ref{fig_deu}, we display the deuteron wave functions derived from
the N$^3$LO potentials and compare them
with wave functions based upon conventional $NN$ potentials from the
recent past. Characteristic differences are noticeable; in particular,
the chiral wave functions are shifted towards larger $r$ 
due to softer cutoffs and the effect of the contact terms,
which explains the larger deuteron radius.

\section{Nuclear many-body forces \label{sec_manyNF}}

The chiral 2NF discussed in the previous section has been applied in microscopic
calculations of nuclear structure with, in general, a great deal of 
success~\cite{Cor02,Cor05,Cor10,NC04,FNO05,Kow04,DH04,Wlo05,Dea05,Gou06,Hag08,Hag10,FOS04,FOS09}.
However, from high-precision studies conducted in the 1990s, it is well known that certain few-nucleon
reactions and nuclear structure issues require 3NFs for their microscopic explanation.
Outstanding examples are the $A_y$ puzzle of $N$-$d$ scattering~\cite{Glo96,EMW02}
and the ground state of $^{10}$B~\cite{Cau02}.
As noted before, 
an important advantage of the EFT approach to nuclear forces
is that it creates two- and many-nucleon forces on an equal
footing (cf.\ the overview given in Fig.~\ref{fig_hi}).
In this section, we will explain in some detail chiral three- and four-nucleon forces.
We will limit our presentation to the isospin-symmetric case;
isospin-violating 3NFs are discussed in Ref.~\cite{EMP05}.

\subsection{Three-nucleon forces}

Nuclear three-body forces in ChPT were initially discussed
by Weinberg~\cite{Wei92} and
the 3NF at NNLO was first derived by van Kolck~\cite{Kol94}.

For a 3NF, we have $A=3$ and $C=1$ and, thus, Eq.~(\ref{eq_nu})
implies
\begin{equation}
\nu = 2 + 2L + 
\sum_i \Delta_i \,.
\label{eq_nu3nf}
\end{equation}
We will use this equation to analyze 3NF contributions
order by order.

\subsubsection{Next-to-leading order}

\begin{figure}[t]\centering
\vspace*{-0.5cm}
\scalebox{1.0}{\includegraphics{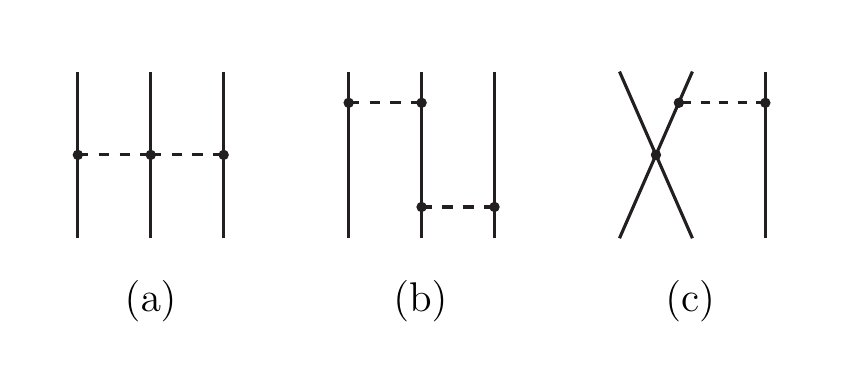}}
\vspace*{-0.75cm}
\caption{Three-nucleon force diagrams at NLO.
Notation as in Fig.~\ref{fig_hi}.}
\label{fig_3nf_nlo}
\end{figure}

The lowest possible power is obviously $\nu=2$ (NLO), which
is obtained for no loops ($L=0$) and 
only leading vertices
($\sum_i \Delta_i = 0$). We display typical graphs in Fig.~\ref{fig_3nf_nlo}.
As discussed by Weinberg~\cite{Wei92}, the contributions from these diagrams
vanish at NLO. To see this, let's first look at graph (a), which contains
a Weinberg-Tomozawa vertex, Eq.~(\ref{eq_LhatWT}), that includes a time-derivative of a pion field.
Since this diagram does not involve reducible topologies,
it can be treated as a Feynman diagram in which energy is conserved
at each vertex, so that the time-derivative yields a difference of nucleon
kinetic energies $\sim Q^2/M_N$ instead of $\sim Q$. Thus, the contribution from this
graph is suppressed by a factor $Q/M_N$ and demoted to NNLO.
Graphs (b) and (c) of Fig.~\ref{fig_3nf_nlo} are best discussed in terms of
time-ordered perturbation theory. Weinberg~\cite{Wei92} and van Kolck~\cite{Kol94} showed
that, at NLO, the irreducible topologies of these graphs cancel against the recoil corrections
from the reducible ones, leaving no net irreducible 3N contribution. 
What remains is just the iteration of the static 2N potentials.
In fact, this had been pointed out already
by Yang and Gl\"ockle~\cite{YG86} and Coon and Friar~\cite{CF86} in the 1980's. 

The bottom line is that there is no genuine 3NF contribution at NLO.
The first non-vanishing 3NF appears at NNLO.

\subsubsection{Next-to-next-to-leading order}

\begin{figure}[t]\centering
\vspace*{-4.0cm}
%\hspace*{-2.0cm}
\scalebox{0.55}{\includegraphics{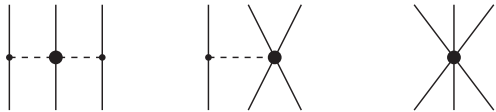}}
\vspace*{-10.0cm}
\caption{The three-nucleon force at NNLO.
From left to right: 2PE, 1PE, and contact diagrams.
Notation as in Fig.~\ref{fig_hi}.}
\label{fig_3nf_nnlo}
\end{figure}

The power $\nu=3$ (NNLO) is obtained when
there are no loops ($L=0$) and 
$\sum_i \Delta_i = 1$, i.e., 
$\Delta_i=1$ for one vertex 
while $\Delta_i=0$ for all other vertices.
There are three topologies which fulfill this condition,
known as the two-pion exchange (2PE), 1PE,
and contact graphs~\cite{Kol94,Epe02b}
(Fig.~\ref{fig_3nf_nnlo}).

Using the subleading vertices Eqs.~(\ref{eq_V2_3}) and (\ref{eq_V2_4}),
it  is straightforward to derive the 2PE 3N-potential to be
\begin{equation}
V^{\rm 3NF}_{\rm 2PE} = 
\left( \frac{g_A}{2f_\pi} \right)^2
\frac12 
\sum_{i \neq j \neq k}
\frac{
( \vec \sigma_i \cdot \vec q_i ) 
( \vec \sigma_j \cdot \vec q_j ) }{
( q^2_i + m^2_\pi )
( q^2_j + m^2_\pi ) } \;
F^{ab}_{ijk} \;
\tau^a_i \tau^b_j
\label{eq_3nf_nnloa}
\end{equation}
with $\vec q_i \equiv \vec{p_i}' - \vec p_i$, 
where 
$\vec p_i$ and $\vec{p_i}'$ are the initial
and final momenta of nucleon $i$, respectively, and
\begin{equation}
F^{ab}_{ijk} = \delta^{ab}
\left[ - \frac{4c_1 m^2_\pi}{f^2_\pi}
+ \frac{2c_3}{f^2_\pi} \; \vec q_i \cdot \vec q_j \right]
+ 
\frac{c_4}{f^2_\pi}  
\sum_{c} 
\epsilon^{abc} \;
\tau^c_k \; \vec \sigma_k \cdot [ \vec q_i \times \vec q_j] \; .
\label{eq_3nf_nnlob}
\end{equation}  
There are great similarities between this force and earlier derivations of
2PE 3NFs, notably the 50-year old Fujita-Miyazawa~\cite{FM57},
the Tucson-Melbourne (TM)~\cite{Coo79}, and the Brazil~\cite{CDR83}
forces. A thorough comparison between various 2PE 3NFs is conducted in
Ref.~\cite{FHK99} resulting in the recommendation to drop the so-called
``c-term'' from the TM force, since it does not have an equivalent in the
ChPT derived force, Eqs.~(\ref{eq_3nf_nnloa}), (\ref{eq_3nf_nnlob}), giving rise
to the construction of the TM' (or TM99) force~\cite{CH01}.

Notice that Eq.~(\ref{eq_3nf_nnlob}) does not include a $c_2$-term.
Due to two time-derivatives, the contribution from the $c_2$ vertex is $(Q/M_N)^2$ suppressed and
demoted by two orders. Note also that
the 2PE 3NF does not contain any new parameters,
because the LECs $c_1$, $c_3$, and $c_4$ appear already  in the 2PE 2NF (Section~\ref{sec_NNLO}) and
are fixed by $\pi N$ and/or $NN$ data.

The other two 3NF contributions shown in Fig.~\ref{fig_3nf_nnlo}
are easily derived by taking the last two terms of the $\Delta=1$ Lagrangian, Eq.~(\ref{eq_LD1}),
into account. The 1PE contribution is
\begin{equation}
V^{\rm 3NF}_{\rm 1PE} = 
-D \; \frac{g_A}{8f^2_\pi} 
\sum_{i \neq j \neq k}
\frac{\vec \sigma_j \cdot \vec q_j}{
 q^2_j + m^2_\pi }
( \mbox{\boldmath $\tau$}_i \cdot \mbox{\boldmath $\tau$}_j ) 
( \vec \sigma_i \cdot \vec q_j ) 
\label{eq_3nf_nnloc}
\end{equation}
and the 3N contact potential reads
\begin{equation}
V^{\rm 3NF}_{\rm ct} = E \; \frac12
\sum_{j \neq k} 
 \mbox{\boldmath $\tau$}_j \cdot \mbox{\boldmath $\tau$}_k  \; .
\label{eq_3nf_nnlod}
\end{equation}
These 3NF terms involve the two new parameters $D$ and $E$, 
which do not appear in the 2N problem.
There are many ways to pin these two parameters down.
In Ref.~\cite{Epe02b},
the triton binding energy and the $nd$ doublet scattering
length $^2a_{nd}$ were used.
One may also choose the binding
energies of $^3$H and $^4$He~\cite{Nog06} or
an optimal over-all fit of the properties of light nuclei~\cite{Nav07}.
Exploiting the consistency of interactions and currents in ChPT~\cite{GP06},
the parameter $D$ of the $\pi NNNN$ vertex involved in the
1PE 3NF can be constrained by $p$-wave pion-production data~\cite{HKM00}
or electro-weak processes like the tritium $\beta$-decay~\cite{GQN09}
or proton-proton fusion ($p\,p \rightarrow d \, e^+ \, \nu_e$)~\cite{Nak08}.
Once $D$ and $E$ are fixed, the results for other
3N, 4N, etc.\  observables are predictions.

The 3NF at NNLO has been applied in
calculations of few-nucleon 
reactions~\cite{Epe02b,Erm05,Kis05,Wit06,Ley06,Ste07,KE07,Mar09,Kie10,Viv10},
structure of light- and medium-mass 
nuclei~\cite{Nog06,Nav07,Hag07,Ots09},
and nuclear and neutron matter~\cite{Bog05,HS09}
with a good deal of success.
Yet, the famous `$A_y$ puzzle' of nucleon-deuteron scattering
is not resolved~\cite{Epe02b,KE07}.
When only 2NFs are applied,
the analyzing power in $p$-$^3$He scattering
is even more underpredicted than in $p$-$d$~\cite{Fis06,DF07}. 
However, when the
NNLO 3NF is added, the $p$-$^3$He $A_y$ substantially 
improves (more than in $p$-$d$)~\cite{Viv10}---but a discrepancy remains.
Furthermore, the spectra of light nuclei leave room for improvement~\cite{Nav07}.

We note that there are further 3NF contributions at NNLO, namely, the
$1/M_N$ corrections of the NLO 3NF diagrams (Fig.~\ref{fig_3nf_nlo}). Some of these terms involve
the vertices Eqs.~(\ref{eq_V2_1}) and (\ref{eq_V2_2}), and the $1/M_N$
correction of the $c_4$ vertex, Eq.~(\ref{eq_V2_4}); others are due to higher order recoil corrections.
Several of those contributions
have been calculated by Coon and Friar in 1986~\cite{CF86}.
These corrections are believed to be very small.

To summarize, the 3NF at NNLO is a remarkable contribution: It represents the leading many-body force
within the scheme of ChPT; it includes terms that were advocated already some 50 years ago; and it
produces noticeable improvements in few-nucleon reactions and the structure of
light nuclei.
But unresolved problems remain. Moreover, in the case of the
2NF, we have seen that one has to proceed to N$^3$LO to achieve sufficient accuracy.
Therefore, the 3NF at N$^3$LO is needed for at least two reasons: 
for consistency with the 2NF and to hopefully resolve outstanding problems in microscopic structure
and reactions.

\begin{figure}[t]\centering
\vspace*{-0.5cm}
%\hspace*{-1.5cm}
\scalebox{1.0}{\includegraphics{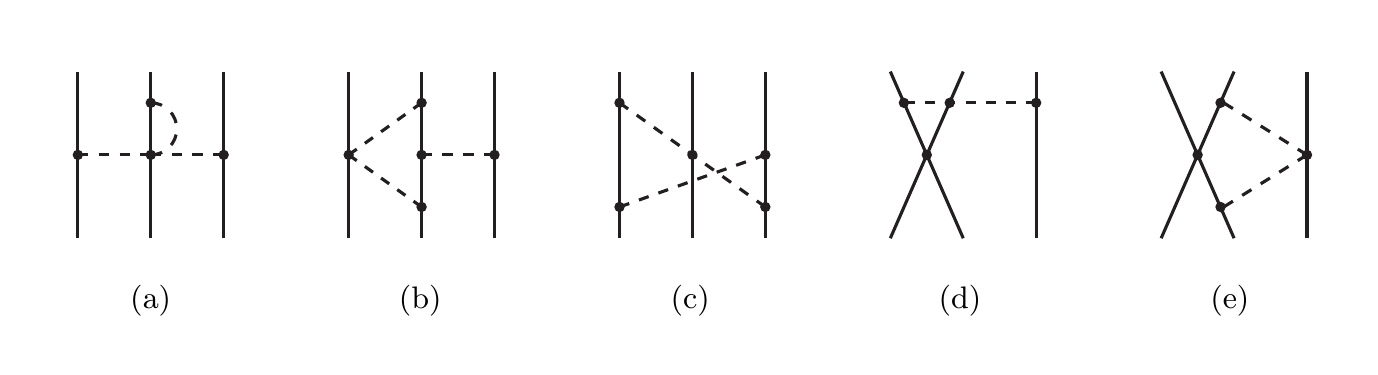}}
\vspace*{-0.75cm}
\caption{Leading one-loop 3NF diagrams at N$^3$LO.
We show one representative example for each of five topologies,
which are: (a) 2PE, (b) 1PE-2PE, (c) ring, (d) contact-1PE, (e) contact-2PE.
Notation as in Fig.~\ref{fig_hi}.}
\label{fig_3nf_n3lo}
\end{figure}

\subsubsection{Next-to-next-to-next-to-leading order}

According to Eq.~(\ref{eq_nu3nf}),
the value $\nu=4$, which corresponds to N$^3$LO, is obtained
for the following classes of diagrams.

\paragraph{3NF loop diagrams at N$^3$LO}
For this group of graphs, we have
$L=1$ and, therefore, all $\Delta_i$ have to be zero
to ensure $\nu=4$. 
Thus, these one-loop 3NF diagrams can include
only leading vertices, the parameters of which
are fixed from $\pi N$ and $NN$ analysis.
We show five representative examples
of this very large class of diagrams
in Fig.~\ref{fig_3nf_n3lo}. 
One sub-group of these diagrams (2PE graphs)
has been calculated by Ishikawa and Robilotta~\cite{IR07},
and two other topologies (1PE-2PE and ring diagrams)
have been evaluated by the Bonn-J\"ulich group~\cite{Ber08}.
The remaining topologies, which involve a leading four-nucleon
contact term [diagrams (d) and (e) of Fig.~\ref{fig_3nf_n3lo}],
are under construction by the Bonn-J\"ulich group.
The N$^3$LO 2PE 3NF has been applied in the calculation
of nucleon-deuteron observables in Ref.~\cite{IR07} 
producing very small effects.

The smallness of the 2PE loop 3NF at N$^3$LO is not unexpected.
It is consistent with
experience with corresponding 2NF diagrams: 
the NLO 2PE contribution to the $NN$ potential, which 
involves one loop and only leading vertices (Fig.~\ref{fig_nlo}), 
is also relatively small (Fig.~\ref{fig_f}).

By the same token, one may expect that also all the other N$^3$LO
3NF loop topologies will produce only small effects.

\paragraph{3NF tree diagrams at N$^3$LO}
The order $\nu=4$ is also obtained for the combination $L=0$ (no loops)
and $\sum_i \Delta_i = 2$.
Thus, either two vertices have to carry $\Delta_i=1$ or
one vertex has to be of the $\Delta_i=2$ kind,
while all other vertices are $\Delta_i=0$.
This is achieved if 
in the NNLO 3NF graphs of Fig.~\ref{fig_3nf_nnlo}
the power of one vertex is raised by one.
The latter happens if a relativistic
$1/M_N$ correction is applied.
A closer inspection reveals that all $1/M_N$ corrections of the
NNLO 3NF vanish and the first non-vanishing corrections
are proportional to $1/M_N^2$ and appear at N$^4$LO.
However, there are non-vanishing $1/M_N^2$ corrections of the NLO 3NF
and there are so-called drift corrections~\cite{Rob06} 
which contribute at N$^3$LO (some drift corrections are claimed to
contribute even at NLO~\cite{Rob06}). 
We do not expect these contributions to be sizable.
Moreover, there are contributions from the $\Delta_i =2$
Lagrangian~\cite{FMS98} proportional to the
low-energy constants $d_i$. As it turns out, these terms have
at least one time-derivative, which causes them to be
$Q/M_N$ suppressed and demoted to N$^4$LO.

Thus, besides some minor $1/M_N^2$ corrections, there are no tree
contributions to the 3NF at N$^3$LO.

{\it Summarizing the 3NF at N$^3$LO:}
For the reasons discussed, we anticipate that this 3NF is weak and will not solve
any of the outstanding problems. 
In view of this expectation, we have to look for
more sizable 3NF contributions elsewhere.

\subsubsection{The 3NF at N$^4$LO}

\begin{figure}[t]\centering
\vspace*{-0.5cm}
\scalebox{1.0}{\includegraphics{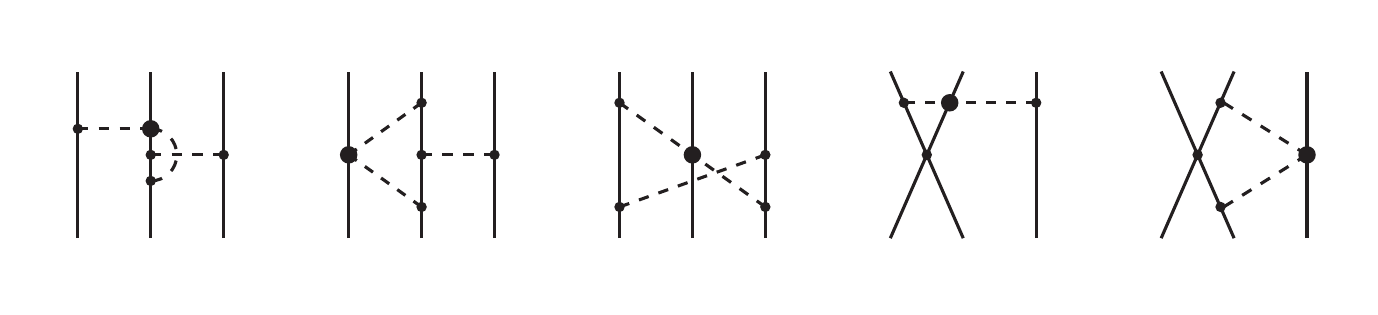}}
\vspace*{-0.75cm}
\caption{Sub-leading one-loop 3NF diagrams which appear at N$^4$LO.
Notation as in Fig.~\ref{fig_hi}.}
\label{fig_3nf_n4lo}
\end{figure}

The obvious step to take is to proceed to the next order,
N$^4$LO or $\nu=5$. Some of the tree diagrams that appear at this order were mentioned already:
the $1/M_N^2$ corrections of the NNLO 3NF and the trees with one $d_i$
vertex which are $1/M_N$ suppressed. Because of the suppression factors,
we do not expect sizable effects from these graphs.
Moreover, there are also tree diagrams with one vertex from the
$\,\Delta_i=3\,$ $\,\pi N$-Lagrangian~\cite{Fet00,FM00} proportional to the 
LECs $e_i$. Because of the high dimension of these vertices and assuming
reasonable convergence, we do not anticipate much from these trees either.

However, we believe that the loop contributions that occur at this order are truly important.
They are obtained by replacing in the N$^3$LO loops (Fig.~\ref{fig_3nf_n3lo})
one vertex by a $\Delta_i=1$ vertex. 
We show five examples of this large group of diagrams
in Fig.~\ref{fig_3nf_n4lo}.
This 3NF is presumably large and, thus, what we are looking for.

The reasons, why these graphs are large, can be argued as follows.
Corresponding 2NF diagrams are the three-pion exchange (3PE)
contributions to the $NN$ interaction. In analogy to 
Figs.~\ref{fig_3nf_n3lo} and \ref{fig_3nf_n4lo},
there are 3PE 2NF diagrams with only leading vertices (Fig.~\ref{fig_3pe}) 
and the ones with one (sub-leading) 
$c_i$ vertex (and the rest leading). These diagrams have been evaluated by Kaiser in 
Refs.~\cite{Kai00a,Kai00b} and \cite{Kai01}, respectively. 
Kaiser finds that the 3PE contributions with one sub-leading vertex are about
an order magnitude larger than the leading ones.

\subsubsection{3NF summary}

To make a complicated story short, this is the bottom line concerning 3NF~\cite{ME10}:
\begin{itemize}
\item
The leading chiral 3NF (that appears at NNLO) is sizable, improves predictions, but also leaves
unresolved problems. Therefore, additional {\it sizable}
3NF contributions are needed.
\item
The chiral 3NF at N$^3$LO involves only leading vertices and
most likely does {\it not} produce sizable contributions.
\item
Sizable contributions are expected from the subleading one-loop 3NF diagrams
that occur at N$^4$LO. {\it These 3NF contributions may
turn out to be the missing pieces in the 3NF puzzle and have the potential to solve
the outstanding problems in microscopic nuclear structure.}
\end{itemize}

\subsection{Four-nucleon forces}

\begin{figure}[t]\centering
\vspace*{-0.5cm}
\scalebox{1.1}{\includegraphics{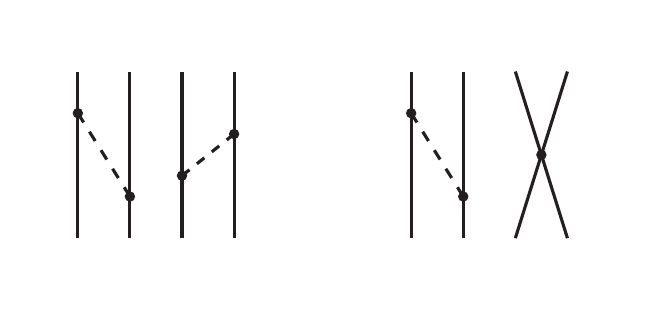}}
\vspace*{-0.75cm}
\caption{Disconnected four-nucleon force diagrams at NLO, which cancel against
the recoil corrections from corresponding iterative diagrams.}
\label{fig_4nf_nlo}
\end{figure}

\begin{figure}[t]\centering
%\vspace*{-1.5cm}
\scalebox{0.8}{\includegraphics{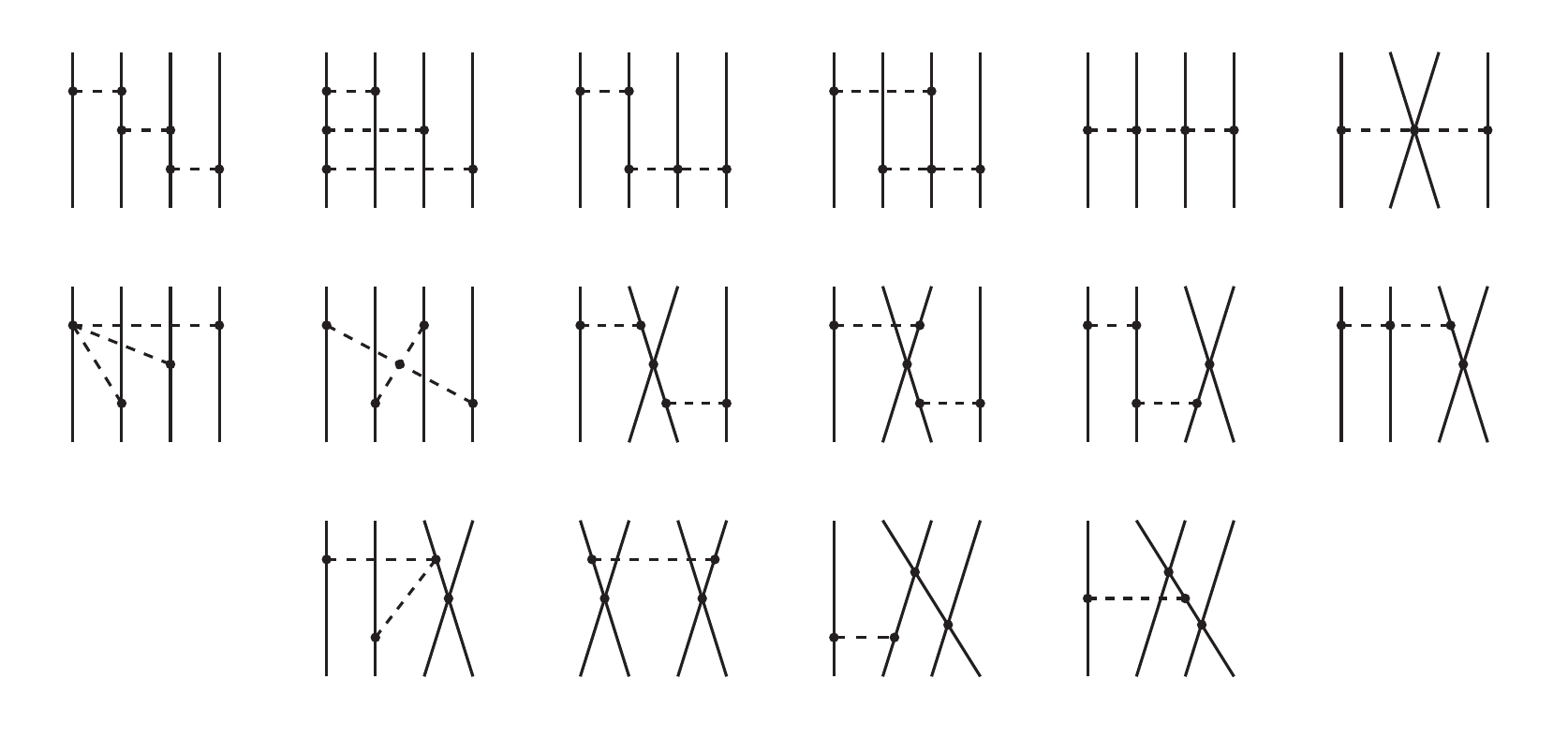}}
\vspace*{-0.5cm}
\caption{Leading four-nucleon force at N$^3$LO.}
\label{fig_4nf_n3lo}
\end{figure}

Formally, the lowest order four-nucleon force (4NF) occurs for two separately
interacting nucleon pairs with leading vertices and no loops, Fig.~\ref{fig_4nf_nlo}.
This scenario is characterized by $A=4$, $C=2$, $L=0$,
and $\sum_i \Delta_i = 0$ and therefore, according to Eq.~(\ref{eq_nu}),
has the power $\nu=2$ (NLO). However, similar to the NLO 3NF 
[Fig.~\ref{fig_3nf_nlo} (b) and (c)], the 4NF diagrams of Fig.~\ref{fig_4nf_nlo}
cancel against the recoil corrections from corresponding iterative diagrams~\cite{Kol94}.
The disconnected 4NF diagrams of order four, which are obtained for either 
$L=1$ or $\sum_i \Delta_i = 2$, also cancel~\cite{Epe07}.
Thus, we are left with just the connected ($C=1$) $A=4$ diagrams for which
Eq.~(\ref{eq_nu}) yields
\begin{equation}
\nu = 4 + 2L + 
\sum_i \Delta_i \,.
\label{eq_nu4nf}
\end{equation}
Therefore, a connected 4NF appears for the first time at $\nu = 4$ (N$^3$LO), with
no loops and only leading vertices, Fig.~\ref{fig_4nf_n3lo}. 
This 4NF  includes no new parameters and does not vanish~\cite{Epe07,Epe06a}.
Some graphs in Fig.~\ref{fig_4nf_n3lo} appear to be reducible (iterative).
Note, however, that these are Feynman diagrams, which are best analyzed
in terms of time-ordered perturbation theory. The various time-orderings include
also some irreducible topologies (which are, by definition, 4NFs). Or, in other words,
the Feynman diagram minus the reducible part of it yields the (irreducible)
contribution to the 4NF. The reducible part depends on the two-, three-, and four-body
scattering equations used.

Assuming a good rate of convergence, a contribution
of order $(Q/\Lambda_\chi)^4$ is expected to be rather small.
Thus, ChPT predicts 4NF to be essentially insignificant, 
consistent
with experience. Still, nothing is fully proven in physics
unless we have performed explicit calculations.
Recently, the leading 4NF (Fig.~\ref{fig_4nf_n3lo}) has been 
applied in a calculation of the $^4$He binding energy,
where it contributes a few 100 keV~\cite{Roz06}. It should be
noted that this preliminary
calculation involves many approximations,
but it certainly provides the right order of magnitude
of the result, which is
indeed very small
as compared to the full $^4$He binding energy
of 28.3 MeV.

\section{Introducing $\Delta$-isobar degrees of freedom \label{sec_delta}}

\begin{figure}[t]\centering
\vspace*{-0.5cm}
\scalebox{0.55}{\includegraphics{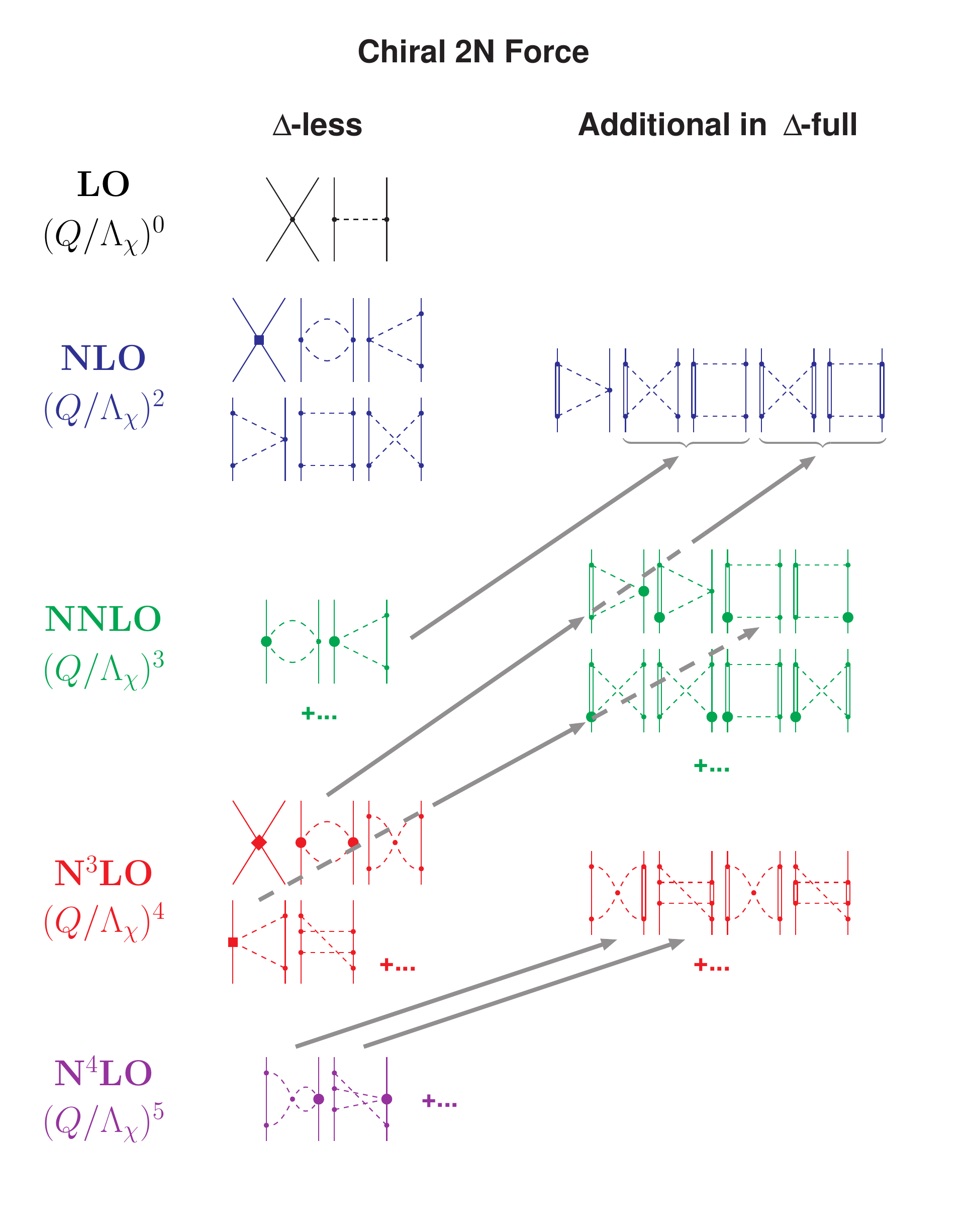}}
\vspace*{-1.0cm}
\caption{Chiral 2NF without and with $\Delta$-isobar degrees of freedom.
Arrows indicate the shift of strength when explicit $\Delta$'s are added to the theory.
Note that the $\Delta$-full theory consists of the diagrams involving $\Delta$'s
{\it plus} the $\Delta$-less ones. Double lines represent $\Delta$-isobars; remaining notation
as in Fig.~\ref{fig_hi}.}
\label{fig_delta_2nf}
\end{figure}

\begin{figure}[t]\centering
\vspace*{-0.5cm}
\scalebox{0.55}{\includegraphics{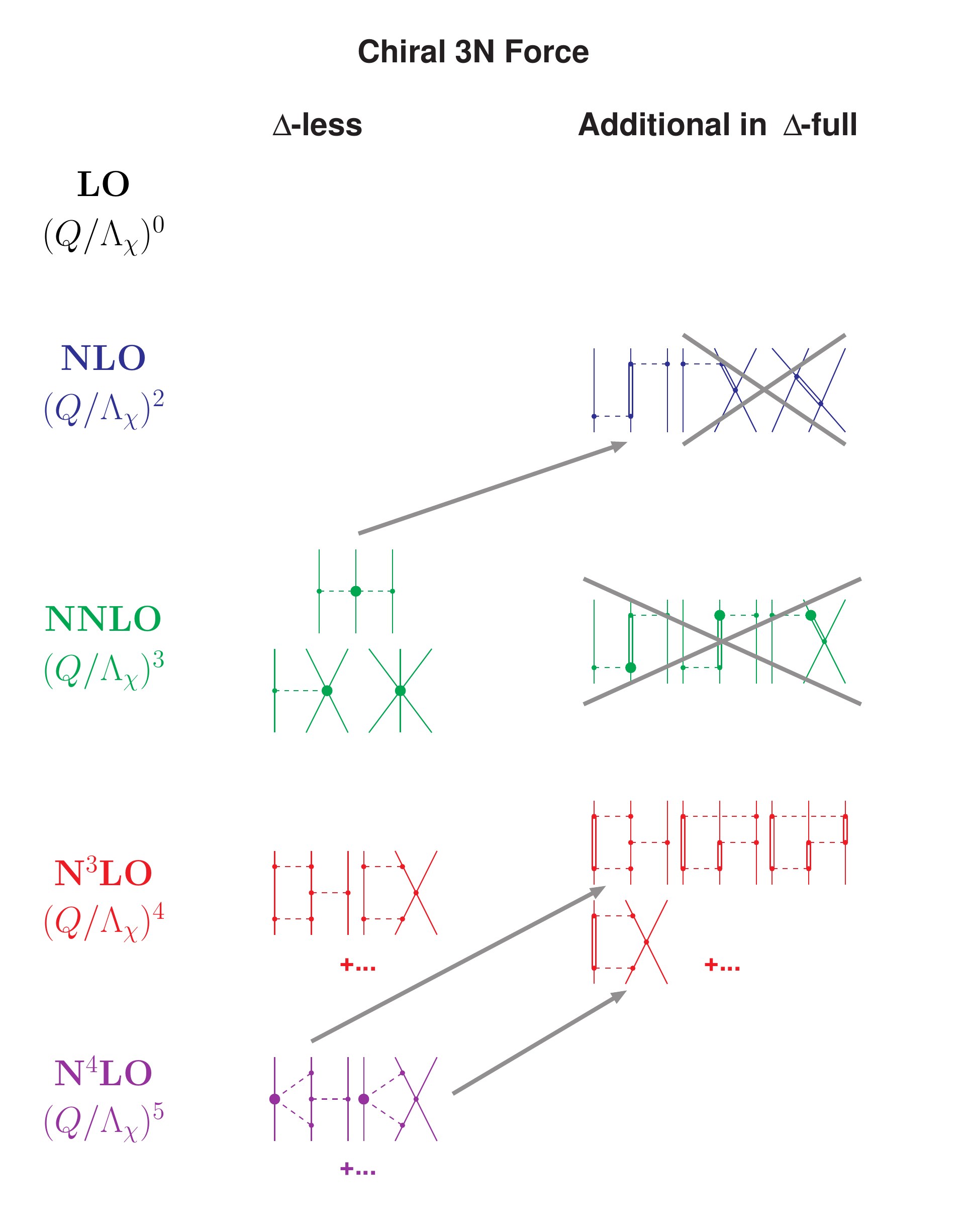}}
\vspace*{-0.5cm}
\caption{The 3NF without and with $\Delta$-isobar degrees of freedom.
Arrows indicate the shift of strength when explicit $\Delta$'s are added to the theory.
Note that the $\Delta$-full theory consists of the diagrams involving $\Delta$'s
{\it plus} the $\Delta$-less ones. Double lines represent $\Delta$-isobars; remaining notation
as in Fig.~\ref{fig_hi}.}
%\vspace*{-1.0cm}
\label{fig_delta_3nf}
\end{figure}

The lowest excited state of the nucleon is the
$\Delta(1232)$ resonance or isobar 
(a $\pi$-$N$ $P$-wave resonance with both spin and isospin 3/2)
with an excitation energy of $\Delta M=M_\Delta - M_N = 293$ MeV.
Because of its strong coupling to the $\pi$-$N$ system and low excitation energy,
it is an important ingredient for models of pion-nucleon scattering in the $\Delta$-region
and pion production from the
two-nucleon system at intermediate energies, where the particle production
proceeds prevailingly through the formation of $\Delta$ isobars~\cite{Mac89}.
At low energies, the more sophisticated conventional models for the 2$\pi$-exchange contribution to the
$NN$ interaction include the virtual excitation of $\Delta$'s, which in these models accounts
for about 50\% of the intermediate-range attraction of the nuclear force---as demonstrated
by the Bonn potential~\cite{MHE87,HM77}. 

Because of its relatively small excitation energy, it is not clear from the outset if, 
in an EFT, the $\Delta$ should be taken into account explicitly or integrated out as
a ``heavy'' degree of freedom. If it is included, then $\Delta M \sim m_\pi$ is considered
as another small expansion parameter, besides the pion mass and small external momenta.
This scheme has become known as the small scale expansion (SSE)~\cite{HHK98}.
Note, however, that this extension is of phenomenological character, since
$\Delta M$ does not vanish in the chiral limit.

In the chiral EFT discussed so far in this report 
(also known as the ``$\Delta$-less'' theory),
the effects due to $\Delta$ isobars
are taken into account implicitly. Note that the dimension-two LECs, the $c_i$,
have unnaturally large values (cf.\ Table~\ref{tab_LEC}). The reason for this is that
the $\Delta$-isobar (and some meson resonances) contribute considerably
to the $c_i$---a mechanism that has become known as
resonance saturation~\cite{BKM97}.
Therefore, the explicit inclusion of the $\Delta$ (``$\Delta$-full'' theory) will take strength out of these
LECs and move this strength to a lower order~\cite{OK92,ORK94,KGW98,KEM07,EKM08}. 
As a consequence, the convergence of the expansion improves, which
is another motivation for introducing explicit $\Delta$-degrees of freedom.
We observed that, in the $\Delta$-less theory, the subleading 2PE and 3PE contributions
to the 2NF are larger than the leading ones. The promotion of large contributions
by one order in the $\Delta$-full theory fixes this problem. 

In the heavy baryon formalism, the leading Lagrangian involving $\Delta$'s 
reads~\cite{Kol94,ORK96}
(listing only terms relevant to our present discussion)
\begin{equation}
\widehat{\cal L}^{\Delta_i=0}_\Delta = 
\bar{\Delta} (i \partial_0 - \Delta M) \Delta 
- \frac{h_A}{2f_\pi}  \left(\bar{N} {\bf T} {\vec S}  \Delta + \mbox{h.c.} \right) 
 \cdot \nabla \mbox{\boldmath ${\bf \pi}$}
-D_T \, \bar{N} \mbox{\boldmath ${\bf \tau}$} {\vec \sigma} N  \cdot
\left( \bar{N} {\bf T} {\vec S} \Delta + \mbox{h.c.} \right) ,
\end{equation}
where $\Delta$ is a four-component spinor in both spin and isospin space representing
the $\Delta$-isobar and $h_A$ and $D_T$ are LECs.\footnote{Our convention for $h_A$
is consistent with Refs.~\cite{Kol94,ORK96,KGW98,Epe06} and differs by a factor of two from
Refs.~\cite{HHK98,FM01,KEM07}.}
Moreover, $S^i$ are $2\times 4$ spin transition matrices
which satisfy $S^iS^{j\dagger}=(2\delta^{ij}-i\epsilon^{ijk}\sigma^k)/3$ and
$T^a$ are similar isospin matrices with
$T^aT^{b\dagger}=(2\delta^{ab}-i\epsilon^{abc}\tau^c)/3$. 
Notice that, due to the heavy baryon expansion, the mass of the $\Delta$-isobar, $M_\Delta$,
has disappeared and only the small mass difference $\Delta M$ enters. 

The LECs of the $\pi N$ Lagrangian are usually extracted in the analysis of $\pi$-$N$
scattering data and clearly come out differently in the $\Delta$-full theory as compared
to the $\Delta$-less one. While in the $\Delta$-less theory, the magnitude of the LECs
$c_3$ and $c_4$ is about 3-5 GeV$^{-1}$ (cf.\ Table~\ref{tab_LEC}),
they turn out to be around 1 GeV$^{-1}$ in the $\Delta$-full theory~\cite{KEM07}.

In the 2NF, the virtual excitation of $\Delta$-isobars requires at least one loop and, thus,
the contribution occurs first at $\nu=2$ (NLO), see Fig.~\ref{fig_delta_2nf}.
The $\Delta$ contributions to the 2PE were first evaluated in Refs.~\cite{OK92,ORK94,ORK96}
using time-ordered perturbation theory and later by Kaiser {\it et al.}~\cite{KGW98} in
covariant perturbation theory. Recently, also the NNLO contributions have been worked out~\cite{KEM07}.
Krebs {\it et al.}~\cite{KEM07} verified  the consistency between the $\Delta$-full and $\Delta$-less 
theories by showing that the contributions due to intermediate $\Delta$-excitations, expanded in 
powers of $1/\Delta M$, can be absorbed into a redefinition of the LECs of the $\Delta$-less
theory. The corresponding shift of the LECs $c_3,c_4$ is given by
\begin{equation}
c_3=-2c_4=-\frac{h_A^2}{9\Delta M}
\,.
\label{eq_c3c4}
\end{equation}
Using $h_A=3g_A/\sqrt{2}$ (large $N_c$ value), almost all of $c_3$
and an appreciable part of $c_4$ is explained by the $\Delta$ resonance.

The studies of Refs.~\cite{KGW98,KEM07} confirm that
a large amount of the intermediate-range attraction of the 2NF is shifted from NNLO to NLO
with the explicit introduction of the $\Delta$-isobar. 
However, it is also found that
the NNLO 2PE potential of the $\Delta$-less theory
provides a very good approximation to the NNLO potential in the $\Delta$-full theory.

The $\Delta$ isobar also changes the 3NF scenario, see Fig.~\ref{fig_delta_3nf}. 
The leading 2PE 3NF is promoted to NLO. In the $\Delta$-full theory, this term has
the same mathematical form as the corresponding term in the $\Delta$-less theory,
Eqs.~(\ref{eq_3nf_nnloa}) and (\ref{eq_3nf_nnlob}), provided one chooses $c_1=0$
and $c_3$, $c_4$ according to Eq.~(\ref{eq_c3c4}).
Note that the other two NLO 3NF terms involving $\Delta$'s vanish~\cite{EKM08}
as a consequence of the antisymmetrization of the 3N states.
The $\Delta$ contributions to the 3NF at NNLO~\cite{EKM08} vanish at this order,
because the subleading $N\Delta \pi$ vertex contains a time-derivative, which demotes
the contributions by one order. 
However, substantial 3NF contributions are expected at N$^3$LO from one-loop
diagrams with one, two, or three
intermediate $\Delta$-excitations,
which correspond to diagrams of order N$^4$LO, N$^5$LO, and N$^6$LO, respectively, 
in the $\Delta$-less theory. 
3NF loop-diagrams with one and two $\Delta$'s 
are included in the Illinois force~\cite{Pie01}
in a simplified way.

To summarize, the inclusion of explicit $\Delta$ degrees of freedom does
certainly improve the convergence of the chiral expansion by shifting sizable
contributions from NNLO to NLO. On the other hand, at NNLO
the results for the $\Delta$-full and $\Delta$-less theory are essentially the
same. Note that the $\Delta$-full theory consists of the diagrams involving
$\Delta$'s plus all diagrams of the $\Delta$-less theory. Thus, the $\Delta$-full
theory is much more involved. Moreover, in the $\Delta$-full theory, 
$1/M_N$ 2NF corrections appear at NNLO (not shown in Fig.~\ref{fig_delta_2nf}),
which were found to be uncomfortably large by Kaiser {\it et al.}~\cite{KGW98}. 
Thus, it appears that up to NNLO, the $\Delta$-less theory is more manageable.

The situation could, however, change at N$^3$LO where potentially large contributions
enter the picture. It may be more efficient to calculate these terms in the $\Delta$-full
theory, because in the $\Delta$-less theory they are spread out
over N$^3$LO, N$^4$LO and, in part, N$^5$LO. 
These higher order contributions are a crucial test for the convergence of
the chiral expansion of nuclear forces and represent a challenging topic for the
future.

\section{Conclusions
\label{sec_concl}}

The past 15 years have seen great progress in our understanding of nuclear forces
in terms of low-energy QCD. Key to this development was the realization that
low-energy QCD is equivalent to an effective field theory (EFT) which allows for 
a perturbative expansion that has become know as chiral perturbation theory (ChPT).
In this framework, two- and many-body forces emerge on an equal footing and the empirical fact
that nuclear many-body forces are substantially weaker than the two-nucleon force
is explained naturally.

In this review, we have shown in detail how the two-nucleon force is derived
from ChPT and demonstrated that,
at N$^3$LO, the accuracy can be achieved that
is necessary and sufficient for reliable microscopic nuclear 
structure predictions.
First calculations applying the N$^3$LO
$NN$ potential~\cite{EM03}
in the conventional shell model~\cite{Cor02,Cor05,Cor10},
the {\it ab initio} no-core shell model~\cite{NC04,FNO05},
the coupled cluster formalism~\cite{Kow04,DH04,Wlo05,Dea05,Gou06,Hag08,Hag10},
and the unitary-model-operator approach~\cite{FOS04,FOS09}
have produced promising results. 

We also discussed nuclear many-body forces based upon chiral EFT.
The 3NF at NNLO has been known for a while~\cite{Kol94,Epe02b} and
applied in 
few-nucleon reactions~\cite{Epe02b,Erm05,Kis05,Wit06,Ley06,Ste07,KE07,Mar09,Kie10,Viv10},
structure of light- and medium-mass nuclei~\cite{Nog06,Nav07,Hag07,Ots09},
and nuclear and neutron matter~\cite{Bog05,HS09} with some success.
However, the famous `$A_y$ puzzle' of nucleon-deuteron
scattering is not resolved by the 3NF at NNLO. 
Thus, one important open issue
are the few-nucleon forces beyond NNLO (``sub-leading
few-nucleon forces'') which, besides the $A_y$ puzzle, may also
resolve some important outstanding
nuclear structure problems. As explained, this may require going even beyond N$^3$LO.

Another open question is the convergence of the chiral expansion (of the two- as well as
the three-nucleon potentials) at orders beyond N$^3$LO,
for which the inclusion of $\Delta$-isobar degrees of freedom may be useful.
Furthermore, the non-perturbative renormalizations of the chiral potentials
require more work and a better understanding. 

Finally, we note that topics of interest we did not discuss 
include parity-violating nuclear forces and consistent electroweak
currents.

Having identified some of the open issues, we hope that this review will
be helpful towards future progress.

If the outstanding problems are resolved within the next few years,
then, after 80 years of desperate struggle, we
may finally claim that the nuclear force problem is essentially under control.
The greatest beneficiary of such progress will be the field of {\it ab initio} nuclear
structure physics. 

\section*{Acknowledgement}
The work by R. M. was supported in part by the U.S. Department of Energy
under Grant No.~DE-FG02-03ER41270.
The work of D. R. E. was funded by the Ministerio de Ciencia y
Tecnolog\'\i a under Contract No.~FPA2007-65748, the Junta de Castilla
y Le\'on under Contract No.~GR12,  and
the European Community-Research Infrastructure Integrating
Activity ``Study of Strongly Interacting Matter'' (HadronPhysics2
Grant No.~227431).

\appendix

\setcounter{figure}{0}
\setcounter{table}{0}

\section{Notation, conventions, and Feynman rules}

\subsection{Notation and conventions}
The contravariant space-time four-vector is given by
\begin{equation}
	x^\mu = ( t, \vec x ) 
\end{equation}
and the four-momentum vector reads
\begin{equation}
	p^\mu = ( E, \vec p ) \, .
\end{equation}
We use units such that $\hbar=c=1$.

Greek indices $\mu, \nu$, etc.\ run over the four space-time
coordinate labels $0,1,2,3$, with $x^0=t$ the time coordinate.
Latin indices $i,j,k$, and so on run over the three space
coordinate labels $1,2,3$.
The metric is diagonal with
\begin{equation}
g^{\mu\nu}=g_{\mu\nu}=
\left( \begin{array}{cccc}1 & & &0\\ & -1 & & \\ & & -1 & \\0& & & -1\end{array}\right)
\end{equation}
and the covariant versions of the above-mentioned vectors are
\begin{equation}
	x_\mu = g_{\mu\nu} x^\nu = ( t, -\vec x ) \,, 
	\quad \quad
	p_\mu = g_{\mu\nu} p^\nu = ( E, -\vec p ) \,,
\end{equation}
where summation over repeated indices is always understood; also
\begin{equation}
x^2 = x_\mu x^\mu = t^2-{\vec x}^2 \,. 
\end{equation}
While for an ordinary three-vector we have, in general,
$\vec x=(x^1,x^2,x^3)$, there is caution in place with the
(three-dimensional) nabla operator which is defined to be
\begin{equation}
\vec \nabla = (\nabla_1,\nabla_2,\nabla_3)
= \left(\frac{\partial}{\partial x^i}\right)
= (\partial_i)
= \left(-\frac{\partial}{\partial x_i}\right)
= (-\partial^i)
\end{equation}
The four-momentum operator reads
\begin{eqnarray}
 p^\mu &=&
  i \frac{\partial}{\partial x_\mu} 
= i \partial^\mu 
= \left(i\partial^0,-i\vec \nabla \right)
= \left(i\frac{\partial}{\partial t},\vec p\right)
\,, \\
 p_\mu &=&
  i \frac{\partial}{\partial x^\mu} 
= i \partial_\mu 
= \left(i\partial_0,i\vec \nabla \right) 
= \left(i\frac{\partial}{\partial t},-\vec p\right) \,.
\end{eqnarray}

The relativistic nucleon field satisfies the free
Dirac equation
\begin{equation}
\left(p\!\!\! \slash  - M_N \right) \Psi(x) \equiv
\left(\gamma^\mu p_\mu  - M_N \right) \Psi(x) 
= 0
\label{eq_Dirac}
\end{equation}
where $M_N$ denotes the nucleon mass and $\gamma_\mu$ the Dirac matrices which we apply
in Dirac-Pauli representation
\begin{equation}
\gamma^0 = \left( \begin{array}{cc} I & 0 \\ 0 & -I \end{array} \right) 
\quad \quad
\gamma^i = \left( \begin{array}{cc} 0 & \sigma^i \\ -\sigma^i & 0 \end{array} \right) 
\end{equation}
with $I$ the two-dimensional identity matrix and $\sigma^i$ the Pauli matrices
\begin{equation}
\sigma^1 = \left( \begin{array}{cc} 0 & 1 \\ 1 & 0 \end{array} \right) 
\quad \quad
\sigma^2 = \left( \begin{array}{cc} 0 &-i \\ i & 0 \end{array} \right) 
\quad \quad
\sigma^3 = \left( \begin{array}{cc} 1 & 0 \\ 0 &-1 \end{array} \right) 
\,. 
\end{equation}
\begin{eqnarray}
\left[ \sigma^i , \sigma^j \right] & = & 
\sigma^i \sigma^j - \sigma^j \sigma^i = 
2 i \epsilon^{ijk} \sigma^k \,,
\quad \quad \quad
\{\sigma^i,\sigma^j \} =
\sigma^i \sigma^j+ \sigma^j \sigma^i 
= 2\delta^{ij} \,,
\\
\sigma^i \sigma^j &=&
i \epsilon^{ijk} \sigma^k + \delta^{ij} \,,
\end{eqnarray}
with
\begin{equation}
\epsilon^{ijk} = \left\{ \begin{array}{ll}
                    +1 & \mbox{if $(i,j,k)$ even permutation of $(1,2,3)$}\\
                    -1 & \mbox{if odd permutation} \\
                     0 & \mbox{otherwise}
                         \end{array}
                 \right.
\end{equation}
Following convention, we denote the Pauli matrices
by $\tau^a$ ($a=1,2,3$) when operating in isospin space,
with
\begin{equation}
\tau^a \tau^b =
i \epsilon^{abc} \tau^c + \delta^{ab} \,.
\end{equation}
Notice that, for isospin components, it does not make sense to
distinguish between upper and lower indices and, therefore,
subscripts and superscripts have the same meaning. 

The Dirac matrices have the properties
\begin{equation}
\{\gamma^\mu,\gamma^\nu \}
= 2\,g^{\mu\nu} \,,
\quad \quad
{\gamma^0}^\dagger = \gamma^0 \gamma^0 \gamma^0 = \gamma^0 \,,
\quad \quad
{\gamma^i}^\dagger = \gamma^0 \gamma^i \gamma^0 = -\gamma^i \,.
\end{equation}
The $\gamma_5$-matrix is defined by
\begin{equation}
\gamma^5 = \gamma_5 = i \gamma^0 \gamma^1 \gamma^2 \gamma^3
= \left( \begin{array}{cc} 0 & I \\ I & 0 \end{array} \right) 
\,.
\end{equation}
\begin{equation}
\{\gamma^\mu,\gamma^5\}=0 \,,
\quad \quad
(\gamma^5)^2 = 1 \,,
\quad \quad
{\gamma^5}^\dagger = \gamma^5 \,.
\end{equation}
Commutator of $\gamma$ matrices:
\begin{equation}
\sigma^{\mu\nu} = \frac{i}{2} \left[\gamma^\mu,\gamma^\nu\right] \,, 
\quad \quad
\gamma^\mu\gamma^\nu = g^{\mu\nu} - i\sigma^{\mu\nu} \,,
\end{equation}
\begin{equation}
\sigma^{ij} = \epsilon^{ijk} 
     \left( \begin{array}{cc} \sigma^k & 0 \\ 0 & \sigma^k \end{array} \right) \,,
\quad \quad
\sigma^{0i} = i\left( \begin{array}{cc} 0 & \sigma^i \\ \sigma^i & 0 \end{array} \right) 
               = -\sigma^{i0}
\end{equation}

The relativistic Dirac field for positive-energy nucleons is
\begin{equation}
\Psi(x)=\sum_{s,t} 
\int \frac{d^3p}{(2\pi)^{3/2}}
\sqrt{\frac{M_N}{E_p}} \, u(\vec p,s) \, \xi_t \, e^{-ip\cdot x} \,
b(\vec p,s,t)
\end{equation}
with the Dirac spinor given by
\begin{equation}
	u(\vec p,s) = \sqrt{\frac{E_p+M_N}{2M_N}} \left(
	\begin{array}{c}
	  I   \\
	\frac{\vec \sigma \cdot \vec p}{E_p+M_N}
	\end{array}
	\right)
        \chi_s
\end{equation}
where $E_p=p^0=\sqrt{{\vec p}^2+M_N^2}$.
The Pauli spinors $\chi_s$ and $\xi_t$ 
describe, respectively, the spin and isospin of the nucleon.
Further, 
$b(\vec p,s,t)$
and
$b^\dagger(\vec p,s,t)$
are destruction and creation operators for a nucleon with
momentum $\vec p$, and spin and isospin quantum numbers
$s$ and $t$, respectively. They satisfy the anti-commutation
relations
\begin{equation}
\left\{b(\vec p,s,t), b^\dagger({\vec p}~',s',t')\right\}=
\delta_{s,s'}\delta_{t,t'}\delta^3(\vec p-{\vec p}~')
\,, 
\end{equation}
\begin{equation}
\left\{b(\vec p,s,t), b({\vec p}~',s',t')\right\}=
\left\{b^\dagger(\vec p,s,t), b^\dagger({\vec p}~',s',t')\right\}=0
\,.
\end{equation}
\begin{equation}
{\bar \Psi} \equiv \Psi^\dagger \gamma^0
\end{equation}

In the heavy-baryon formalism, the free field equation for nucleons
is, in leading order and using $v_\mu=(1,0,0,0)$,
\begin{equation}
i\partial_0 N(x) = 0
\label{eq_HBfree}
\end{equation}
and the nucleon field is
\begin{equation}
N(x)=\sum_{s,t} \int \frac{d^3l}{(2\pi)^{3/2}}
\, \chi_s \, \xi_t \, e^{-il\cdot x} \,
b(\vec l,s,t)
\end{equation}
where $l_0=0$ at leading order and $\vec l^2/2M_N$ at NLO.
\begin{equation}
{\bar N} \equiv N^\dagger \gamma^0 = N^\dagger
\end{equation}

The pion fields are in terms of their cartesian components 
($i=1,2,3$; no distinction between upper and lower index $i$)
\begin{equation}
\pi_i(x)=
\int \frac{d^3q}{(2\pi)^{3/2}} \frac{1}{\sqrt{2\omega}}
\left[ e^{-iq\cdot x}a_i(\vec q) + e^{iq\cdot x}a_i^\dagger(\vec q) \right]
\end{equation}
where $\omega=q^0=\sqrt{{\vec q}^2+m_\pi^2}$ and 
$a_i$ and $a_i^\dagger$ are, respectively, the destruction and creation
operators obeying the commutation relations
\begin{equation}
\left[a_i(\vec q),a_j^\dagger({\vec q}~') \right]
= \delta_{ij}\delta^3(\vec q - {\vec q}~')
\,,
\end{equation}
\begin{equation}
\left[a_i(\vec q),a_j({\vec q}~') \right] =
\left[a_i^\dagger(\vec q),a_j^\dagger({\vec q}~') \right] =0
\,.
\end{equation}
The charged and neutral pion fields are given by
\begin{eqnarray}
\pi_+ &=& \frac{1}{\sqrt{2}} \left(\pi_1+i\pi_2\right) \,, \\
\pi_- &=& \frac{1}{\sqrt{2}} \left(\pi_1-i\pi_2\right) \,, \\
\pi_0 &=& \pi_3 \,.
\end{eqnarray}

Throughout this article, we state amplitudes in terms of
the ``potential'' $V$ which is defined by
\begin{equation}
V \; = \; i \; {\cal M} 
\label{eq_V}
\end{equation}
where $\cal M$ is the invariant amplitude,
calculated according to Feynman rules.
The relation of $\cal M$ to the S-matrix
is
\begin{eqnarray}
\langle p_1\!' p_2\!' | S | p_1 p_2 \rangle &=&
\delta^3(\vec p_1 - \vec p_1\!') \,
\delta^3(\vec p_2 - \vec p_2\!') 
\nonumber \\ &&
 +\; (2\pi)^{4} \;
\delta^4(p_1+p_2-p_1\!'-p_2\!') \;
\frac{1}{(2\pi)^6} 
\left( \frac{M_N^4}{E_1' E_2' E_1 E_2} \right)^{\frac12}
\nonumber \\ &&
\times \; {\cal M}( p_1\!', p_2\!'; p_1, p_2 ) \,.
\label{eq_S}
\end{eqnarray}

\subsection{Feynman rules}

The basic formalism underlying the Feynman rules is the Dyson expansion
of the $S$-matrix [cf.\ e.~g., Eq.~(6.1.1) of Ref.~\cite{Wei95}].
Since we are applying {\it covariant} perturbation theory, we
assume ${\cal H}_I = - {\cal L}_I$ (where ${\cal H}_I$ denotes the interaction
Hamiltonian density and ${\cal L}_I$ the interaction Lagrangian density)
and use the usual covariant Feynman propagators.
From a procedural point of view, this is acceptable for the derivative couplings considered here
(see, however, Ref.~\cite{Ger71} for exceptions). For a discussion of the differences
between covariant perturbation theory and time-ordered perturbation theory (also
known as ``old fashioned'' perturbation theory), see the evaluation of the
football diagram, \ref{app_foot_topt} below.

In all {\it one-pion vertices} given below,
$q$ denotes the four-momentum of an outgoing pion 
of isospin component $a$. 
In all {\it two-pion vertices},
$q_1$ denotes the four-momentum of an ingoing pion with
isospin component $a$ and
$q_2$ the four-momentum of an outgoing pion with isospin
component $b$.
Nucleon momenta are, in general, denoted by $p$ and $p'$.

\subsubsection{Leading order}

The relativistic version of the leading order $\pi N$
Lagrangian has been given in Eq.~(\ref{eq_L1relalt}).
We consider the axial-vector (AV) and the Weinberg-Tomozawa (WT)
couplings, which involve one and two pion fields, respectively.
The relativistic AV interaction Lagrangian is given by
\begin{equation}
\mathcal{L}_{AV} = - \frac{g_A}{2 f_\pi} \, \bar{\Psi} \,\gamma^\mu \,
\gamma_5 \, \mbox{\boldmath $\tau$} 
\, \Psi \cdot \partial_\mu \mbox{\boldmath $\pi$} 
\,.
\end{equation}
For simple Lagrangians like this one,
the vertex is just
$i$ times the Lagrangian stripped off the fields which generates ($q$ out)
\begin{equation}
\frac{ g_A}{2 f_\pi} \, \gamma_\mu \gamma_5  \tau^a q^\mu
\,.
\label{eq_VAVrel}
\end{equation}
In the heavy baryon formalism, the AV Lagrangian reads
[cf.\ Eq.~(\ref{eq_L1})]
\begin{equation}
\widehat{\cal L}_{AV} = - \frac{g_A}{2 f_\pi} \, \bar{N}  \,
\mbox{\boldmath $\tau$} \cdot 
(\vec \sigma \cdot \vec \nabla) \mbox{\boldmath $\pi$} 
\, N 
\end{equation}
with vertex ($q$ out) 
\begin{equation}
-\frac{g_A}{2 f_\pi} \, \tau^a
\vec \sigma \cdot \vec{q}
\,.
\label{eq_VAV}
\end{equation}
The relativistic WT coupling term is
\begin{equation}
\mathcal{L}_{WT} = - \frac{1}{4 f_\pi^2} \, \bar{\Psi} \,\gamma^\mu \,
\mbox{\boldmath $\tau$} \cdot (\mbox{\boldmath $\pi$}\times 
\partial_\mu \mbox{\boldmath $\pi$})
\, \Psi 
\end{equation}
implying the vertex ($q_1$ in, $q_2$ out)
\begin{equation}
\frac{1}{4 f_\pi^2} \, \gamma_\mu \epsilon^{abc}  \tau^c
(q_1^\mu + q_2^\mu)
\,.
\label{eq_VWTrel}
\end{equation}
The Heavy Baryon version is 
\begin{equation}
\widehat{\cal L}_{WT} = - \frac{1}{4 f_\pi^2} \, \bar{N}  \,
\mbox{\boldmath $\tau$} \cdot (\mbox{\boldmath $\pi$}
\times \partial_0 \mbox{\boldmath $\pi$})
\, N 
\label{eq_LhatWT}
\end{equation}
with vertex ($q_1$ in, $q_2$ out)
\begin{equation}
\frac{1}{4 f_\pi^2} \, \epsilon^{abc}  \tau^c
(q_1^0 + q_2^0)
\,.
\label{eq_VWT}
\end{equation}
The relativistic nucleon propagator reads 
[cf.\ Eq.~(\ref{eq_Dirac})]
\begin{equation}
\frac{i}{\slash \! \! \!  p  - M_N + i\epsilon} =
\frac{i(\slash \! \! \!  p  + M_N)}{p^2 - M_N^2 + i\epsilon}
\end{equation}
and the leading order heavy baryon version of the nucleon propagator is
given by [cf.\ Eq.~(\ref{eq_HBfree})]
\begin{equation}
\frac{i}{l^0+i\epsilon} \,.
\end{equation}
The pion propagator is [cf.\ Eq.~(\ref{eq_Lpipi2a})]
\begin{equation}
\frac{i\delta^{ab}}{q^2-m_\pi^2+i\epsilon} \,.
\end{equation}

\subsubsection{Next-to-leading order \label{app_feyn_nlo}}
The ``fixed'' part of the dimension-two HB Lagrangian is given in Eq.~(\ref{eq_L2fixed}).
It leads to a nucleon kinetic energy correction 
\begin{equation}
-i \, \frac{{\vec p}~^2}{2M_N} 
\,,
\label{eq_V2_0}
\end{equation}
and produces the one-pion vertex ($q$ out)
\begin{equation}
\frac{g_A}{4M_Nf_\pi} \, \tau^a \, {\vec \sigma} \cdot (\vec p + {\vec p}~') \, q^0
\label{eq_V2_1}
\end{equation}
and the two-pion vertex ($q_1$ in, $q_2$ out)
\begin{equation}
-\frac{1}{8M_Nf_\pi^2} \, \epsilon^{abc} \, \tau^c \, (\vec p + {\vec p}~') \cdot ({\vec q}_1 + {\vec q}_2)
\,.
\label{eq_V2_2}
\end{equation} 

The vertices Eqs.~(\ref{eq_V2_1}) and (\ref{eq_V2_2}) are the first relativistic corrections
of the vertices Eqs.~(\ref{eq_VAV}) and (\ref{eq_VWT}), respectively. These corrections can also be
obtained by starting from the corresponding relativistic vertices Eqs.~(\ref{eq_VAVrel}) and
(\ref{eq_VWTrel}), sandwiching them between Dirac spinors, and making a $1/M_N$ expansion.

The ``contact'' part of the dimension-two Lagrangian, Eq.~(\ref{eq_L2ct}), gives rise to the following
two-pion vertices ($q_1$ in, $q_2$ out):
\begin{equation}
\frac{i\delta^{ab}}{f_\pi^2} \left[ -4c_1m_\pi^2 + \left(2c_2-\frac{g_A^2}{4M_N} \right)
q_1^0 q_2^0 +2c_3 q_{1\mu} q_2^{\mu} \right]
\label{eq_V2_3}
\end{equation}
and
\begin{equation}
-\frac{i}{f_\pi^2} \left(c_4+\frac{1}{4M_N} \right) \epsilon^{ijk} \epsilon^{abc} \sigma^i \tau^c  q_1^j q_2^k
\,.
\label{eq_V2_4}
\end{equation}
For more Feynman rules, see appendix A of Ref.~\cite{BKM95}.

\section{Two-pion exchange contributions to the 2NF at NLO}
\label{app_NLO}

\setcounter{figure}{0}
\setcounter{table}{0}

An overview of the 2PE diagrams that contribute at NLO was shown in
Fig.~\ref{fig_nlo}. We will now evaluate the various groups of diagrams
one by one.

\subsection{Triangle diagrams}

\subsubsection{The triangles in covariant perturbation theory}

\begin{figure}[t]\centering
\vspace*{-1cm}
\scalebox{0.8}{\includegraphics{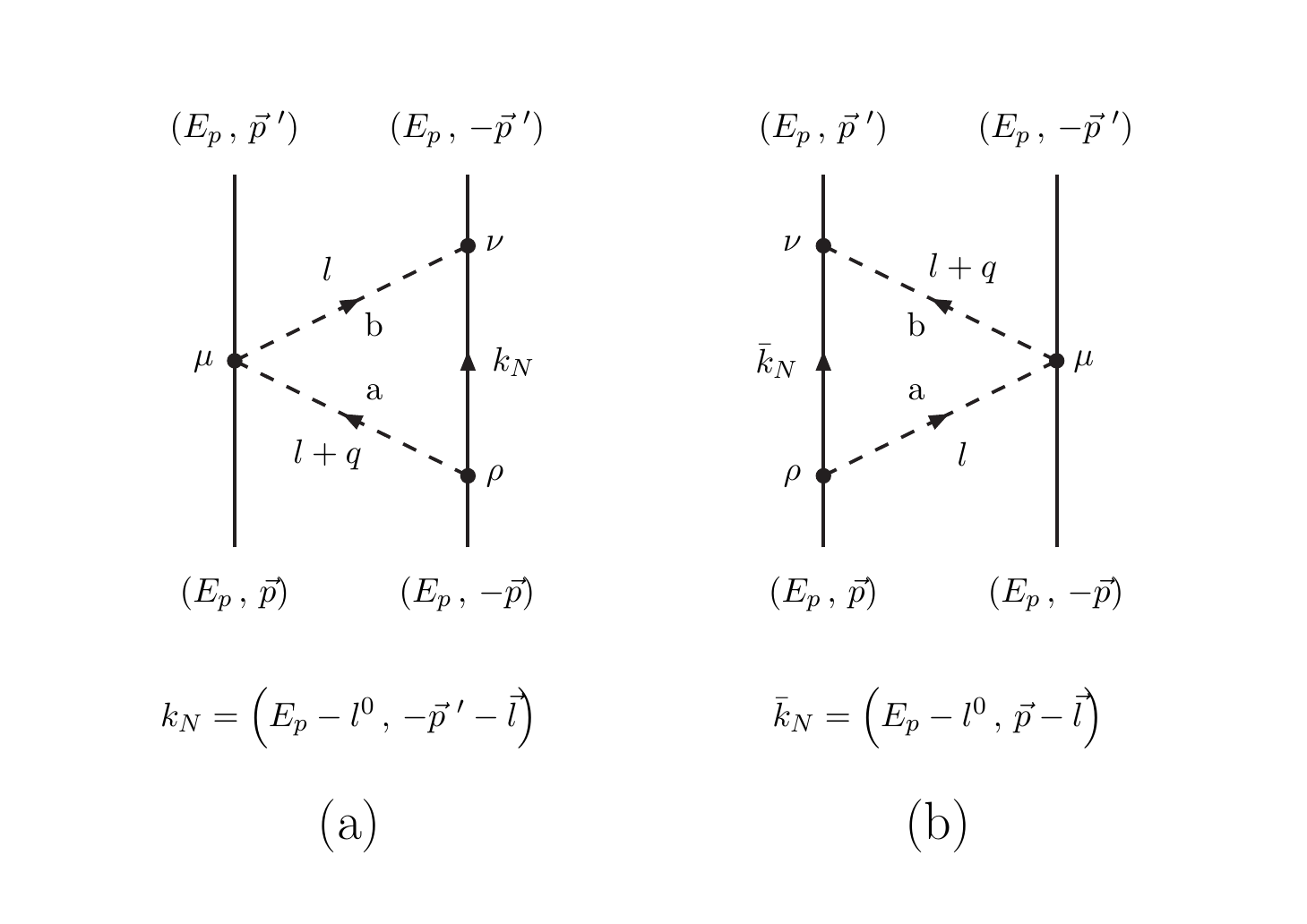}}
\vspace*{-0.75cm}
\caption{Two-pion exchange triangle diagrams at NLO.}
\label{fig_triangle}
\end{figure}

The triangle diagrams are shown in Fig.~\ref{fig_triangle}. 
The contributions from diagram (a) and (b) will be denoted by
$V_{\rm a}$ and $V_{\rm b}$, respectively.
According to relativistic Feynman rules, one obtains
\begin{eqnarray}
V_{\rm a} &=& \, i\, \frac{g_A^2}{16 f_\pi^4} \,
\int \frac{d^4 l}{(2\pi)^4} \,
(2 l^\mu + q^\mu) \epsilon^{abc} \tau^c_1 
\bar{u}_1 (\vec{p}~') \gamma_\mu u_1 (\vec{p})
\frac{i}{(l+q)^2-m_\pi^2+i\epsilon}
\times \nonumber \\ && \times
\frac{i}{l^2-m_\pi^2+i\epsilon}
\bar{u}_2 (-\vec{p}~') (-l^\nu) \tau^b_2 \gamma_\nu  \gamma_5
\frac{i(k \!\!\! \slash_{N} + M_N)}{k_{N}^2 - M_N^2 + i\epsilon}
(l^\rho+q^\rho) \tau^a_2 \gamma_\rho  \gamma_5
u_2 (-\vec{p})
  \,,     \\
\nonumber \\
V_{\rm b} &=& \, i\, \frac{g_A^2}{16 f_\pi^4} \,
\int \frac{d^4 l}{(2\pi)^4} \,
(2 l^\mu + q^\mu) \epsilon^{abc} \tau^c_2 
\bar{u}_2 (-\vec{p}~') \gamma_\mu u_2 (-\vec{p})
\frac{i}{(l+q)^2-m_\pi^2+i\epsilon}
\times \nonumber \\ && \times
\frac{i}{l^2-m_\pi^2+i\epsilon}
\bar{u}_1 (\vec{p}~') (-l^\nu-q^\nu) \tau^b_1 \gamma_\nu  \gamma_5
\frac{i(\bar{k} \!\!\!\slash_{N} + M_N)}{\bar{k}_{N}^2 - M_N^2 + i\epsilon}
(l^\rho) \tau^a_1 \gamma_\rho  \gamma_5
u_1 (\vec{p})
\,,
\end{eqnarray}
where the momenta are defined in Fig.~\ref{fig_triangle}
and $q=(0,{\vec p}~' - \vec p)$. 
The sum over the isospin operators is 
$\epsilon^{abc} \tau_i^c \tau_j^b \tau_j^a = 
- i \, 2 \: \bbox{\tau}_i \cdot \bbox{\tau}_j $
and, for the intermediate nucleon, we have 
$\gamma_\nu  \gamma_5 (k \!\!\!\slash_{N} + M_N) \gamma_\rho  \gamma_5 =
\gamma_\nu  (k\!\!\!\slash_{N} - M_N) \gamma_\rho$. Thus,
\begin{eqnarray}
V_{\rm a} &=& 
- \, i \: \bbox{\tau}_1 \cdot \bbox{\tau}_2 \:
\frac{g_A^2}{8 f_\pi^4} \,
\int \frac{d^4 l}{(2\pi)^4} \,
(2 l^\mu + q^\mu) 
\bar{u}_1 (\vec{p}~') \gamma_\mu u_1 (\vec{p})
\frac{1}{(l+q)^2-m_\pi^2+i\epsilon}
\times \nonumber \\ && \times
\frac{1}{l^2-m_\pi^2+i\epsilon}
\bar{u}_2 (-\vec{p}~') (-l^\nu) \gamma_\nu  
\frac{(k \!\!\! \slash_{N} - M_N)}{k_{N}^2 - M_N^2 + i\epsilon}
(l^\rho+q^\rho) \gamma_\rho 
u_2 (-\vec{p}) \,, \\ 
\nonumber \\
V_{\rm b} &=& 
-\, i \: \bbox{\tau}_1 \cdot \bbox{\tau}_2 \:
\frac{g_A^2}{8 f_\pi^4} \,
\int \frac{d^4 l}{(2\pi)^4} \,
(2 l^\mu + q^\mu) 
\bar{u}_2 (-\vec{p}~') \gamma_\mu u_2 (-\vec{p})
\frac{1}{(l+q)^2-m_\pi^2+i\epsilon}
\times \nonumber \\ && \times
\frac{1}{l^2-m_\pi^2+i\epsilon}
\bar{u}_1 (\vec{p}~') (-l^\nu-q^\nu) \gamma_\nu 
\frac{(\bar{k} \!\!\!\slash_{N} - M_N)}{\bar{k}_{N}^2 - M_N^2 + i\epsilon}
(l^\rho) \gamma_\rho  
u_1 (\vec{p})
\,.
\end{eqnarray}
At NLO, only the lowest order in the $1/M_N$ expansion
is included. The nucleon mass dependence comes from the Dirac spinors
and the nucleon propagator and performing such an expansion, 
we find at lowest order
\begin{eqnarray}
\bar{u}_i \gamma_\mu u_i & \approx &  \delta_{\mu 0}
\\
\bar{u}_i \gamma_\nu 
(k  \!\!\!\slash_{N} - M_N)
\gamma_\rho  u_i 
& \approx &
\delta_{\nu j} \delta_{\rho m} 2 M_N (\delta_{jm} 
+ i \epsilon^{jmn} \sigma_i^n)
\\
\frac{1}{k_{N}^2 - M_N^2 + i\epsilon} 
& \approx & 
\frac{1}{2M_N} \; \frac{1}{-l^0 + i\epsilon}
\end{eqnarray}
In this approximation, the amplitudes are
\begin{eqnarray}
V_{\rm a} &=& 
 -\, i\, \bbox{\tau}_1 \cdot \bbox{\tau}_2 \:
\frac{g_A^2}{4 f_\pi^4} \,
\int \frac{d^4 l}{(2\pi)^4} \,
\frac{1}{(l+q)^2-m_\pi^2+i\epsilon} \;
\frac{1}{l^2-m_\pi^2+i\epsilon} 
\times \nonumber \\ && \times \,
(\delta_{jm} + i \epsilon^{jmn} \sigma_2^n)
(l^j) (l^m+q^m) 
\label{eq_Va}
\\ 
\nonumber \\
V_{\rm b} &=& 
 -\, i\, \bbox{\tau}_1 \cdot \bbox{\tau}_2 \:
\frac{g_A^2}{4 f_\pi^4} \,
\int \frac{d^4 l}{(2\pi)^4} \,
\frac{1}{(l+q)^2-m_\pi^2+i\epsilon} \;
\frac{1}{l^2-m_\pi^2+i\epsilon} 
\times \nonumber \\ && \times \,
(\delta_{jm} + i \epsilon^{jmn} \sigma_1^n)
(l^j+q^j) (l^m) 
\label{eq_Vb}
\end{eqnarray}
where we used $q^0=0$. 

Alternatively, one can derive the NLO triangle diagram contributions 
by using the HB formalism from the outset.
Using the Feynman rules for the HB Lagrangians and propagators yields
\begin{eqnarray}
V_{\rm a}^{\rm HB} &=& \, i\, \frac{g_A^2}{16 f_\pi^4} \,
\int \frac{d^4 l}{(2\pi)^4} \,
(2 l^0) \epsilon^{abc} \tau^c_1 
\frac{i}{(l+q)^2-m_\pi^2+i\epsilon} \;
\frac{i}{l^2-m_\pi^2+i\epsilon} 
\times \nonumber \\ && \times \;
\tau^b_2 (\vec \sigma_2 \cdot \vec l)
\frac{i}{-l^0 + i\epsilon}
\tau^a_2 (-\vec \sigma_2 \cdot (\vec l + \vec q))
   \,,    \\
V_{\rm b}^{\rm HB} &=& \, i\, \frac{g_A^2}{16 f_\pi^4} \,
\int \frac{d^4 l}{(2\pi)^4} \,
(2 l^0) \epsilon^{abc} \tau^c_2 
\frac{i}{(l+q)^2-m_\pi^2+i\epsilon}
\frac{i}{l^2-m_\pi^2+i\epsilon} 
\times \nonumber \\ && \times \;
\tau^b_1 (\vec \sigma_1 \cdot (\vec l + \vec q))
\frac{i}{-l^0 + i\epsilon}
\tau^a_1 (-\vec \sigma_1 \cdot \vec l)
\,.
\end{eqnarray}

Performing the isospin sums and using the
identity
$(\vec \sigma \cdot \vec a)(\vec \sigma \cdot \vec b) =
\vec a \cdot \vec b + i \vec \sigma \cdot (\vec a \times \vec b)$,
we obtain
\begin{eqnarray}
V_{\rm a}^{\rm HB} &=& 
 -\, i\, \bbox{\tau}_1 \cdot \bbox{\tau}_2 \:
\frac{g_A^2}{4 f_\pi^4} \,
\int \frac{d^4 l}{(2\pi)^4} \,
\frac{1}{(l+q)^2-m_\pi^2+i\epsilon} \;
\frac{1}{l^2-m_\pi^2+i\epsilon}
\times \nonumber \\ && \times \;
\left[ \vec l \cdot (\vec l + \vec q) +
i\vec \sigma_2 \cdot (\vec l \times (\vec l + \vec q)) \right]
\label{eq_VHBa} \,, \\
\nonumber \\
V_{\rm b}^{\rm HB} &=& 
 -\, i\, \bbox{\tau}_1 \cdot \bbox{\tau}_2 \:
\frac{g_A^2}{4 f_\pi^4} \,
\int \frac{d^4 l}{(2\pi)^4} \,
\frac{1}{(l+q)^2-m_\pi^2+i\epsilon} \;
\frac{1}{l^2-m_\pi^2+i\epsilon}
\times \nonumber \\ && \times \;
\left[ (\vec l + \vec q) \cdot \vec l +
i\vec \sigma_1 \cdot ((\vec l + \vec q) \times \vec l) \right]
\,.
\label{eq_VHBb}
\end{eqnarray}
The two sets of Eqs.\ (\ref{eq_Va}), (\ref{eq_Vb})
and (\ref{eq_VHBa}), (\ref{eq_VHBb})
obviously agree---as they should.

Since $(\vec l +\vec q)\times \vec l = \vec q \times \vec l$
and since the integral over a term proportional to $\vec l$ will yield a result
$\propto \vec q$,  the spin dependent terms vanish and the sum
of both triangles is given by
\begin{eqnarray}
V = V_{\rm a} + V_{\rm b}
&=& 
 -\, i\, \bbox{\tau}_1 \cdot \bbox{\tau}_2 \:
\frac{g_A^2}{2 f_\pi^4} \,
\int \frac{d^4 l}{(2\pi)^4} \,
\frac{1}{(l+q)^2-m_\pi^2+i\epsilon} \;
%\times \nonumber \\ && \times
\frac{\vec l^2 + \vec l \cdot \vec q}{l^2-m_\pi^2+i\epsilon} 
\,.
\label{eq_tri4}
\end{eqnarray}

At this point, it might be of interest to compare
with what other authors obtain for the NLO triangle diagrams.
In the work of Ref.~\cite{ORK96}, time-ordered perturbation
theory (TOPT) (also known as old-fashioned PT)
is used in which loop integrals extend
only over the three space-dimensions.

Thus, we will perform the integration over the time
component in Eq.~(\ref{eq_tri4}) to make a comparison
with TOPT possible.
For convenience,
we first apply the transformation 
$l=\frac 1 2 (l'-q)$, which yields
\begin{eqnarray}
V &=& 
 -\, i\, \bbox{\tau}_1 \cdot \bbox{\tau}_2 \:
\frac{g_A^2}{8 f_\pi^4} \,
\int \frac{d^4 l'}{(2\pi)^4} \,
\frac{1}{(l'+q)^2-4m_\pi^2+i\epsilon} \;
\frac{\vec l'^2- \vec q^2}{(l'-q)^2-4m_\pi^2+i\epsilon} 
\,.
\end{eqnarray}
The poles in the lower half plane are at $l'^0=\omega_\pm-i\epsilon$ with
\begin{eqnarray}
\omega_\pm &=& \sqrt{(\vec l'\pm \vec q)^2+4m_\pi^2}
\end{eqnarray}
and, so, we can write the amplitude as
\begin{eqnarray}
V &=& 
 -\, i\, \bbox{\tau}_1 \cdot \bbox{\tau}_2 \:
\frac{g_A^2}{8 f_\pi^4} \,
\int \frac{d^4 l'}{(2\pi)^4} \,
\frac{1}{(l'^0)^2-\omega_+^2+i\epsilon} \;
\frac{\vec l'^2- \vec q^2}{(l'^0)^2-\omega_-^2+i\epsilon}
\,.
\end{eqnarray}
Now we perform the integration over $l'^0$ using the residue theorem
\begin{eqnarray}
V &=& 
 -\, \bbox{\tau}_1 \cdot \bbox{\tau}_2 \:
\frac{g_A^2}{8 f_\pi^4} \,
\int \frac{d^3 l'}{(2\pi)^3} \,
(\vec l'^2- \vec q^2) \left\{ 
\frac{1}{2\omega_+} \frac{1}{\omega_+^2-\omega_-^2} +
\frac{1}{\omega_-^2-\omega_+^2} \frac{1}{2\omega_-} \right\}
\nonumber \\
&=& 
 -\, \bbox{\tau}_1 \cdot \bbox{\tau}_2 \:
\frac{g_A^2}{8 f_\pi^4} \,
\int \frac{d^3 l'}{(2\pi)^3} \,
(\vec l'^2- \vec q^2) \left\{ 
\frac{1}{\omega_+^2-\omega_-^2} 
\frac{\omega_--\omega_+}{2\omega_+\omega_-} \right\}
\nonumber \\
&=& 
\: \bbox{\tau}_1 \cdot \bbox{\tau}_2 \:
\frac{g_A^2}{16 f_\pi^4} \,
\int \frac{d^3 l'}{(2\pi)^3} \,
\frac{\vec l'^2- \vec q^2}{\omega_+\omega_-(\omega_++\omega_-)} 
\,.
\end{eqnarray}
Except for a typo in Ref.~\cite{ORK96} and differences in notation, 
this agrees with the second term of Eq.~(20)
of Ref.~\cite{ORK96} as well as the first term of
Eq.~(4.30) of Ref.~\cite{EGM98}, which both  represent the triangle diagrams.

\subsubsection{Dimensional regularization of the triangles
\label{sec_dimreg}}

The integral is divergent and, therefore, requires
regularization. In Ref.~\cite{ORK96}, this is done by introducing
a Gaussian cutoff function $\exp(-\vec l^2/\Lambda^2)$ which, however, 
makes the results dependent on the cutoff parameter $\Lambda$.
To avoid such cutoff dependence,
we will use dimensional 
regularization which was first introduced by 't Hooft and Veltman
in 1972~\cite{HV72}.\footnote{Accessible introductions into
dimensional regularization can be found in Refs.~\cite{Col84,MS84}
and a very pedagogical presentation is provided in appendix A.2 of the
Scherer review~\cite{Sch03}.}

For this it is convenient to go back to 
Eq.~(\ref{eq_tri4}) and
apply a trick discovered by Feynman~\cite{Fey49},
\begin{equation}
\frac{1}{ab} = \int_0^1 \frac{dx}{[b+(a-b)x]^2}
\,,
\label{eq_trick}
\end{equation}
which allows to write the product of propagators
in terms of a linear combination.
Applying Feynman's trick in 
Eq.~(\ref{eq_tri4}) yields
\begin{equation}
V =
\: \bbox{\tau}_1 \cdot \bbox{\tau}_2 \:
\frac{g_A^2}{2 f_\pi^4} \,
 \int_0^1 dx \int \frac{d^{4} l}{i(2 \pi)^4}
\frac{\vec l^2 + \vec l \cdot \vec q}
{[l^2+(2lq+q^2)x-m_\pi^2 + i\epsilon]^2}
\,.
\label{eq_tri4a}
\end{equation}
The momentum integral is not yet quite ready for dimensional
regularization because 
the integrand mixes three- and four-dimensional versions
of the loop-momentum $l$.
The dimension of the integration variable $l$ must be consistent
throughout the integrand.

Formal (four-dimensional) covariance can be re-covered 
by re-introducing the four-vector
$v_\mu=(1,0,0,0)$ and re-writing the numerator as
$\vec l^2 + \vec l \cdot \vec q = (l\cdot v)^2 -l^2 - l\cdot q$.
After a shift of the integration variable $l \rightarrow l+xq$,
well-known formulae for the dimensional regularization of
covariant integrals can be applied [see, e.g., 
Eqs.~(10.23) and (10.32)-(10.34) of Ref.~\cite{MS84}].

Alternatively, one may go for consistency in three dimensions
and reduce the momentum integral
of Eq.~(\ref{eq_tri4a}) 
to the three space-dimensions---which is what we will do here.
This has the advantage that the integration variable observes
an Euclidean metric from the outset (and there is no need for a Wick rotation).
Moreover, the planar box diagram (see below) requires treatment in three dimensions.

The integrand of
Eq.~(\ref{eq_tri4a}) 
has two poles as evidenced through
\begin{equation}
l^2+(2lq+q^2)x-m_\pi^2 + i\epsilon =
\left[l^0 + \sqrt{\vec{l}^2+m_\pi^2+(2\vec{l}\cdot\vec{q}+\vec{q}^2)x} 
-i\epsilon\right] 
%\times \nonumber \\ && \times
\left[l^0 - \sqrt{\vec{l}^2+m_\pi^2+(2\vec{l}\cdot\vec{q}+\vec{q}^2)x} 
+i\epsilon\right]
\end{equation}
where 
we used $q^0=0$. The pole in the lower half-plane is
\begin{eqnarray}
l^0_1 &=& \sqrt{\vec{l}^2+m_\pi^2+(2\vec{l}\cdot\vec{q}+\vec{q}^2)x} 
-i\epsilon
\,,
\end{eqnarray}
and, using the residue theorem, one obtains
\begin{eqnarray}
V &=&
\: \bbox{\tau}_1 \cdot \bbox{\tau}_2 \:
\frac{g_A^2}{2 f_\pi^4} \,
\int_0^1 dx \int \frac{dl^0\,d^{3}l}{i(2 \pi)^4}
\, \frac{\vec l^2 + \vec l \cdot \vec q}
{({l^0}-{l^0_1})^2 ({l^0}+{l^0_1})^2} 
\nonumber \\ &=& 
\: \bbox{\tau}_1 \cdot \bbox{\tau}_2 \:
\frac{g_A^2}{2 f_\pi^4} \,
\int_0^1 dx \int \frac{d^{3}l}{(2 \pi)^{3}}
\,\, 
\,\, \frac{1}{4} \, 
\frac{\vec l^2 + \vec l \cdot \vec q}{{(l_1^0)}^{3}}
\,.
\end{eqnarray}
To proceed towards dimensional regularization, 
we now generalize the three-dimensional integral to
$(D-1)$ dimensions.
Moreover, we introduce the renormalization scale $\lambda$
and multiply the integral by $\lambda^{4-D}$ such that the
dimension of the amplitude stays the same for any $D$~\cite{Hoo73,Col84},
\begin{equation}
V = \: \bbox{\tau}_1 \cdot \bbox{\tau}_2 \:
\frac{g_A^2}{8 f_\pi^4} \,
\lambda^{4-D} 
\int_0^1 dx \int \frac{d^{D-1}l}{(2 \pi)^{D-1}}
\,\, 
\frac{\vec l^2 + \vec l \cdot \vec q}{{(l_1^0)}^{3}}
\,.
\end{equation}
For convenience, we now divide all dimension-full quantities involved in the integral
by $\lambda$ and denote all dimensionless quantities 
(except for the dimensionless momentum $\vec k$)
by a tilde; for example, $\tilde q = \vec q / \lambda$.
Furthermore,
we shift the integration variable by introducing the new variable $\vec k$ 
with $\lambda \vec k = \vec l + x \vec q$, so that
there are no terms linear in the integration variable in the denominator. 
Thus, we have now
\begin{eqnarray}
l^0_1 &=& \lambda \sqrt{\vec{k}^2+\tilde m_\pi^2+\tilde q^2 x (1-x) } 
-i\epsilon
\,,
\end{eqnarray}
and our integral reads
\begin{eqnarray}
V &=&
\: \bbox{\tau}_1 \cdot \bbox{\tau}_2 \:
\frac{g_A^2}{8 f_\pi^4} \,
\lambda^{2}
\int_0^1 dx \int \frac{d^{D-1}k}{(2 \pi)^{D-1}} \,\,
\frac{(k-x\tilde{q})^2 + (k-x\tilde{q})\cdot \tilde q}
{[k^2 + \tilde{m}_\pi^2 + \tilde{q}^2 x (1-x)]^{3/2}}
\,.
\end{eqnarray}
Terms with odd powers of $k$ vanish,
\begin{eqnarray}
V &=&
\: \bbox{\tau}_1 \cdot \bbox{\tau}_2 \:
\frac{g_A^2}{8 f_\pi^4} \,
\lambda^{2}
\int_0^1 dx \int \frac{d^{D-1}k}{(2 \pi)^{D-1}} \,\,
\frac{k^2-x(1-x)\tilde{q}^2}
{[k^2 + \tilde{m}_\pi^2 + \tilde{q}^2 x (1-x)]^{3/2}}
\,.
\end{eqnarray}
The volume element in $(D-1)$ dimensions is
\begin{eqnarray}
d^{D-1}k &=& k^{D-2}\,dk\,d\Omega_{D-1}
\label{eq_vol1}
\end{eqnarray}
with
\begin{eqnarray}
\int d\Omega_{D-1} &=& \frac{2 \pi^{(D-1)/2}}{\Gamma(\frac{D-1}{2})}
\,.
\label{eq_vol2}
\end{eqnarray}
Thus,
\begin{eqnarray}
V &=&
\: \bbox{\tau}_1 \cdot \bbox{\tau}_2 \:
\frac{g_A^2}{8 f_\pi^4} \,
\frac{2\lambda^{2}}{\Gamma(\frac{D-1}{2}) (4 \pi)^{(D-1)/2}} 
\int_0^1 dx 
\int_0^\infty dk \,\,
\frac{k^D-x(1-x)\tilde{q}^2k^{D-2}}
{[k^2 + \tilde{m}_\pi^2 + \tilde{q}^2 x (1-x)]^{3/2}}
\,.
\end{eqnarray}
Using [cf.\ for example, Eq. (A.35) of Ref.~\cite{Sch03}]
\begin{eqnarray}
\int_0^\infty \frac{k^{D-2+n_1}}{(k^2+s)^{n/2}} \,dk =
\frac{\Gamma(\frac{D+n_1-1}{2})\Gamma(\frac{n+1-D-n_1}{2})}{2\Gamma(\frac{n}{2})}
\frac{1}{s^{\frac{n+1-D-n_1}{2}}}
\,,
\label{eq_SchA35}
\end{eqnarray}
we obtain
\begin{eqnarray}
V &=&
\: \bbox{\tau}_1 \cdot \bbox{\tau}_2 \:
\frac{g_A^2}{8 f_\pi^4} \,
\frac{\lambda^{2}}{\Gamma(\frac 3 2)\Gamma(\frac{D-1}{2}) (4 \pi)^{(D-1)/2}} \,
\int_0^1 dx 
\left\{
\frac{\Gamma(\frac{D+1}{2})\Gamma(\frac{2-D}{2})}
{[\tilde{m}_\pi^2 + \tilde{q}^2 x (1-x)]^{(2-D)/2}}
\right. \nonumber \\ && \left.
-x(1-x)\tilde{q}^2
\frac{\Gamma(\frac{D-1}{2})\Gamma(\frac{4-D}{2})}
{[\tilde{m}_\pi^2 + \tilde{q}^2 x (1-x)]^{(4-D)/2}}
\right\}
\,.
\end{eqnarray}
We choose $D=4-\eta$ and then take the limit $\eta \to 0$. 
To prepare for the latter, we expand the various expressions involved
up to first order in $\eta$:
\begin{eqnarray}
\frac{\Gamma(\frac{D+1}{2})}{\Gamma(\frac{D-1}{2})}
&=& \frac{D-1}{2} = \frac{3-\eta}{2}
\nonumber \\
\Gamma\left(\frac{4-D}{2}\right) &=&
\Gamma\left(\frac{\eta}{2}\right)
=  \frac 2 \eta - \gamma + {\cal O}(\eta)
\nonumber \\
\Gamma\left(\frac{2-D}{2}\right) &=&
\frac{\Gamma(\frac{\eta}{2})}{\frac \eta 2 -1}
=  -\frac 2 \eta + \gamma -1 + {\cal O}(\eta)
\nonumber \\
\frac{1}{(4 \pi)^{\frac{D-1}{2}}} &=& \frac{1}{(4 \pi)^{\frac{3}{2}}} (4\pi)^{\frac{\eta}{2}}
\approx \frac{1}{(4 \pi)^{\frac{3}{2}}} \left(1+\frac{\eta}{2} \ln(4\pi)\right)
\nonumber \\
\frac{1}{(\tilde{m}_\pi^2 + \tilde{q}^2 x (1-x))^{\frac{4-D}{2}}} &\approx&
1-\frac{\eta}{2} \ln(\tilde{m}_\pi^2 + \tilde{q}^2 x (1-x))
\nonumber \\
\frac{1}{(\tilde{m}_\pi^2 + \tilde{q}^2 x (1-x))^{\frac{2-D}{2}}} &\approx&
\left[1-\frac{\eta}{2} \ln(\tilde{m}_\pi^2 + \tilde{q}^2 x (1-x))\right]
(\tilde{m}_\pi^2 + \tilde{q}^2 x (1-x))
\label{eq_eta}
\end{eqnarray}
Concerning the properties of the $\Gamma$-function, see, e.g.,
Ref.~\cite{AS70}; $\gamma = 0.5772\ldots$ is Euler's constant; 
and we used $a^\epsilon=\exp(\epsilon \ln a)\approx 1+\epsilon \ln a + \ldots$.

Hence
\begin{eqnarray}
V &=&
\: \bbox{\tau}_1 \cdot \bbox{\tau}_2 \:
\frac{g_A^2}{8 f_\pi^4} \,
\frac{\lambda^{2}}{\Gamma(\frac 3 2)} \,
\frac{1}{(4 \pi)^{\frac{3}{2}}} \left(1+\frac{\eta}{2} \ln(4\pi)\right)
\int_0^1 dx 
\nonumber \\ && 
\left\{ \frac{3-\eta}{2}
\left( -\frac 2 \eta + \gamma -1 + {\cal O}(\eta) \right)
\left[1-\frac{\eta}{2} \ln(\tilde{m}_\pi^2 + \tilde{q}^2 x (1-x))\right]
(\tilde{m}_\pi^2 + \tilde{q}^2 x (1-x))
\right. \nonumber \\ && \left.
-x(1-x)\tilde{q}^2
\left( \frac 2 \eta - \gamma + {\cal O}(\eta) \right)
\left[ 1-\frac{\eta}{2} \ln(\tilde{m}_\pi^2 + \tilde{q}^2 x (1-x)) \right]
\right\}
\end{eqnarray}
and neglecting terms of order $\eta$
\begin{eqnarray}
V &=&
\: \bbox{\tau}_1 \cdot \bbox{\tau}_2 \:
\frac{g_A^2}{8 f_\pi^4} \,
\frac{\lambda^{2}}{\Gamma(\frac 3 2)} \,
\frac{1}{(4 \pi)^{\frac{3}{2}}} 
\left[
\left( -\frac 2 \eta +\gamma -1 -\ln (4\pi) \right)
\left(\frac 3 2 \tilde{m}_\pi^2 +\frac{5}{12} \tilde{q}^2 \right)
\right. \nonumber \\ && \left.
+ \tilde m_\pi^2 +\frac 1 3 \tilde{q}^2  
+\int_0^1 dx 
\left (\frac 3 2 \tilde{m}_\pi^2 + \frac 5 2\tilde{q}^2 x (1-x) \right)
\ln(\tilde{m}_\pi^2 + \tilde{q}^2 x (1-x))
\right]
\,.
\end{eqnarray}
The integral over the logarithmic term is
\begin{eqnarray}
\int_0^1 dx 
\left (\frac 3 2 \tilde{m}_\pi^2 + \frac 5 2\tilde{q}^2 x (1-x) \right)
\ln(\tilde{m}_\pi^2 + \tilde{q}^2 x (1-x)) & = &
-\frac 4 3 \tilde{m}_\pi^2 - \frac{25}{36} \tilde q^2 +
\left(3 \tilde m_\pi^2+ \frac 5 6 \tilde q^2 \right) \ln (\tilde m_\pi) 
\nonumber \\ && 
 + \frac 1 6 (8 \tilde m_\pi^2+ 5 \tilde q^2 ) L(q)
\,,
\end{eqnarray}
where
\begin{equation} 
L(q)  \equiv  {w\over q} \ln {w+q \over 2m_\pi} \,,
\quad
 w  \equiv  \sqrt{4m_\pi^2+q^2} \,,
 \quad \mbox{and} \quad
 q=|\vec q| \,.
\label{eq_L}
\end{equation}
Introducing the abbreviation
\begin{eqnarray}
R &\equiv& 
-\frac{2}{\eta} + \gamma-1-\ln(4\pi)
\,,
\label{eq_R}
\end{eqnarray}
the final result for the {\it triangle diagrams at NLO} in dimensional
regularization reads
\begin{equation}
V =
\: \bbox{\tau}_1 \cdot \bbox{\tau}_2 \:
\frac{g_A^2}{384 \pi^2 f_\pi^4} \,
\left[
(18 m_\pi^2 +5 q^2 ) R
%\right. \nonumber \\ && \left.
-4 m_\pi^2 - \frac{13}{3} q^2 +
2(18 m_\pi^2 +5 q^2 ) \ln \left(\frac{m_\pi}{\lambda} \right) +
(16 m_\pi^2+10 q^2 ) L(q)
\right] .
\label{eq_triangles}
\end{equation}
Renormalization in a Modified Minimal Subtraction scheme 
($\overline{MS}$-scheme)~\cite{Col84}
amounts to omitting the $R$-term.

\subsection{Football diagram}

\subsubsection{The football in covariant perturbation theory}

\begin{figure}[t]\centering
\vspace*{-1cm}
\scalebox{0.9}{\includegraphics{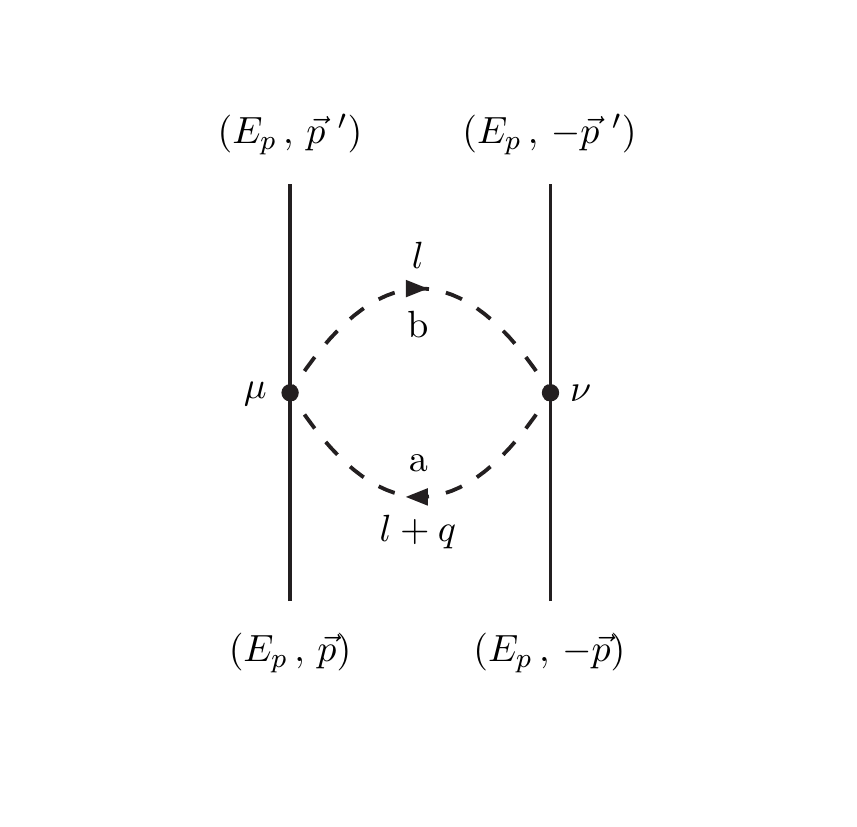}}
\vspace*{-1.5cm}
\caption{Two-pion exchange football diagram at NLO.}
\label{fig_football}
\end{figure}

The football diagram is shown in Fig.~\ref{fig_football}.
In the HB formalism, one obtains
\begin{eqnarray}
V &=& \, i \, \frac 1 2 \, \frac{1}{16 f_\pi^4} \,
\int \frac{d^4 l}{(2\pi)^4} \,
(2 l^0+q^0) \epsilon^{abc} \tau^c_1 
\frac{i}{(l+q)^2-m_\pi^2+i\epsilon} \;
%\times \nonumber \\ && \times
\frac{i}{l^2-m_\pi^2+i\epsilon}
(2 l^0+q^0) \epsilon^{bad} \tau^d_2 
\,,
\end{eqnarray}
where the factor $\frac 1 2$ is a combinatoric factor
and $q=(0,{\vec p}~' - \vec p)$.
Using 
$\epsilon^{abc} \tau^c_1 \epsilon^{bad} \tau^d_2 =
-2 \: \bbox{\tau}_1 \cdot \bbox{\tau}_2 \: $
and $q^0=0$,
\begin{eqnarray}
V &=&  i
\: \bbox{\tau}_1 \cdot \bbox{\tau}_2 \:
\frac{1}{4 f_\pi^4} 
\int \frac{d^4 l}{(2\pi)^4} \,
\frac{1}{(l+q)^2-m_\pi^2+i\epsilon}
\,\,
\frac{(l^0)^2}{l^2-m_\pi^2+i\epsilon}
\,.
\label{eq_foot1}
\end{eqnarray}
It is again of interest to compare
with what other authors obtain for the diagram under consideration.
As discussed in conjunction with the triangle diagrams,
in the work of Ref.~\cite{ORK96}, time-ordered perturbation
theory (TOPT) (also known as old-fashioned PT)
is used in which loop integrals extend
only over the three space-dimensions.
Therefore, we will perform the integration over the time
component in Eq.~(\ref{eq_foot1}) to make a comparison
with TOPT possible.
Substituting $l=\frac 1 2 (l'-q)$, 
\begin{eqnarray}
V &=& i
\: \bbox{\tau}_1 \cdot \bbox{\tau}_2 \:
\frac{1}{16 f_\pi^4} 
\int \frac{d^4 l'}{(2\pi)^4} \,
\frac{1}{(l'+q)^2-4m_\pi^2+i\epsilon} \,\,
\frac{(l'^0)^2}{(l'-q)^2-4m_\pi^2+i\epsilon}
\nonumber \\
&=& i
\: \bbox{\tau}_1 \cdot \bbox{\tau}_2 \:
\frac{1}{16 f_\pi^4} 
\int \frac{d^4 l'}{(2\pi)^4} \,
\frac{1}{(l'^0)^2-\omega_+^2+i\epsilon} \,\,
\frac{(l'^0)^2}{(l'^0)^2-\omega_-^2+i\epsilon}
\end{eqnarray}
with
\begin{equation}
\omega_\pm = \sqrt{(\vec l'\pm \vec q)^2+4m_\pi^2}
\,,
\end{equation}
and applying the residue theorem yields
\begin{eqnarray}
V &=& 
\: \bbox{\tau}_1 \cdot \bbox{\tau}_2 \:
\frac{1}{16 f_\pi^4} 
\int \frac{d^3 l'}{(2\pi)^3} \,
\left\{ 
\frac{1}{2\omega_+} \frac{\omega_+^2}{\omega_+^2-\omega_-^2} +
\frac{\omega_-^2}{\omega_-^2-\omega_+^2} \frac{1}{2\omega_-} \right\}
\nonumber \\ &=&
\: \bbox{\tau}_1 \cdot \bbox{\tau}_2 \:
\frac{1}{32 f_\pi^4} 
\int \frac{d^3 l'}{(2\pi)^3} \,
\frac{\omega_+-\omega_-}{\omega_+^2-\omega_-^2}
\nonumber \\ &=&
\: \bbox{\tau}_1 \cdot \bbox{\tau}_2 \:
\frac{1}{32 f_\pi^4} 
\int \frac{d^3 l'}{(2\pi)^3} \,
\frac{1}{\omega_++\omega_-}
\,.
\label{eq_foot2}
\end{eqnarray}
This amplitude for the football diagram does not agree with Ref.~\cite{ORK96},
where the result for the football is given by the first term
in Eq.~(20). In our notation, the result of Ref.~\cite{ORK96}
and also of Ref.~\cite{EGM98} [cf.\ Eq.~(4.30) therein] is
\begin{eqnarray}
V_{\rm ORK} &=& -
\: \bbox{\tau}_1 \cdot \bbox{\tau}_2 \:
\frac{1}{128 f_\pi^4} 
\int \frac{d^3 l'}{(2\pi)^3} \,
\frac{1}{\omega_+ \omega_-} \; 
\frac{(\omega_+-\omega_-)^2}{\omega_++\omega_-}
\,.
\label{eq_footORK}
\end{eqnarray}

\subsubsection{The proper calculation of the football in time-ordered perturbation theory
\label{app_foot_topt}}

We will explain now the reason for the discrepancy between 
Eqs.~(\ref{eq_foot2}) and (\ref{eq_footORK}).
In covariant PT theory one assumes for the interaction Hamiltonian density
${\cal H}_I = - {\cal L}_I$. 
This is strictly speaking not correct for
derivative coupling, because in that case one has
${\cal H}_I = - {\cal L}_I \, +$ {\it additional noncovariant terms}. 
For relatively simple cases of derivative coupling, the additional terms
are of the contact type. Now, in covariant PT, there are also
noncovariant contributions to the propagator, and it has been shown that 
for certain interactions with derivative coupling these
two groups of additional terms cancel. So, as a procedural matter,
in covariant PT, one can use
${\cal H}_I = - {\cal L}_I$ and the usual covariant Feynman propagators,
and everything comes out right (cf.\ e.~g., Ref.~\cite{Wei95},
pp.~318-323 therein).

In TOPT the story is different. Propagators are just non-covariant
energy denominators, which cancel nothing. Therefore,
the full interaction Hamiltonian 
including all additional terms has to be use to obtain
the same result as in covariant PT. However, in Ref.~\cite{ORK96}
${\cal H}_I = - {\cal L}_I$ is assumed in conjunction with TOPT. 
Thus, the contributions from the additional contact terms in the
Hamiltonian are missing in the calculation of Ref.~\cite{ORK96}
and that is the reason for the discrepancy with the covariant calculation,
as we will show now.

Using the notation of Ref.~\cite{Wei95},
the time-component of the current associated with
the Lagrangian density Eq.~(\ref{eq_LhatWT}) reads
\begin{eqnarray}
J^0_b &=& 
\frac{1}{4f_\pi^2} \bar N \tau^c N \epsilon^{abc} \pi_a 
\end{eqnarray}
and the additional term in the interaction Hamiltonian is
[cf.\ Eq.~(12) of Ref.~\cite{Wei91}]
\begin{eqnarray}
H^{\rm add}_I(t) &=& \int d^3 x\,\frac 1 2 \, (J^0_b)(J^0_b) =
\int d^3 x\,
\frac 1 2 \,
\frac{1}{16f_\pi^4} \,
\bar N \tau^c N \epsilon^{abc} \pi_a 
\bar N \tau^{c'} N \epsilon^{a'bc'} \pi_{a'} 
\,.
\end{eqnarray}
This generates the following leading contribution to the $S$-matrix
\begin{eqnarray}
\langle p'_1 p'_2 | S | p_1 p_2 \rangle &=&
(-i) \int dt\, H^{\rm add}_I(t) =
-i \int d^4 x\,
\frac{1}{16f_\pi^4} 
\frac 1 2 \epsilon^{abc} \epsilon^{a'bc'} 
\langle p'_1 p'_2 | \bar N \tau^c N \pi_a \bar N \tau^{c'} N \pi_{a'} | p_1 p_2 \rangle
\nonumber \\ &=&
\frac{-i}{(2\pi)^6} \int d^4 x\,
\, 2 \,
\frac{1}{16f_\pi^4} 
\frac 1 2 \epsilon^{abc} \epsilon^{a'bc'} \tau^c_1 \tau^{c'}_2 
e^{i(p'_1+p'_2-p_1-p_2) x}
\langle 0 | \pi_a \pi_{a'} | 0 \rangle
\,,
\label{eq_S1}
\end{eqnarray}
where the factor of 2 comes from the permutation of the nucleon fields
and the exchange term is left out.
For the pion fields, we obtain
\begin{eqnarray}
\langle 0 | \pi_a \pi_{a'} | 0 \rangle &=&
\frac{1}{(2\pi)^3} \int d^3 k d^3 k'\, \frac{1}{\sqrt{4\omega_k\omega_{k'}}} \;
%\times \nonumber \\ &\times& 
\langle 0 | \left(a_a(\vec k) e^{-ikx} + a^\dagger_a(\vec k)e^{ikx}\right)
\left(a_{a'}(\vec k') e^{-ik'x} + a^\dagger_{a'}(\vec k')e^{ik'x}\right) | 0 \rangle 
\nonumber \\ &=&
\frac{1}{(2\pi)^3} \int d^3 k d^3 k'\, \frac{1}{\sqrt{4\omega_k\omega_{k'}}}
\delta^3(\vec k - \vec k') \delta_{aa'} =
\int \frac{d^3 k}{(2\pi)^3} \, \frac{1}{2\omega_k}
\delta_{aa'}
\,.
\end{eqnarray}
Hence,
\begin{eqnarray}
\langle p'_1 p'_2 | S | p_1 p_2 \rangle &=&
\frac{-i}{(2\pi)^6} \int d^4 x\,
\frac{1}{16f_\pi^4} \,
\epsilon^{abc} \epsilon^{a'bc'} \tau^c_1 \tau^{c'}_2 
e^{i(p'_1+p'_2-p_1-p_2) x}
\int \frac{d^3 k}{(2\pi)^3} \, \frac{1}{2\omega_k}
\delta_{aa'}
\nonumber \\ &=&
-i (2\pi)^4 \delta^4 (p'_1+p'_2-p_1-p_2)
\, \frac{1}{(2\pi)^6} \,
\frac{1}{16f_\pi^4} \,
\epsilon^{abc} \epsilon^{a'bc'} \tau^c_1 \tau^{c'}_2 
\int \frac{d^3 k}{(2\pi)^3} \, \frac{1}{2\omega_k}
\delta_{aa'}
\nonumber \\ &=&
-i (2\pi)^4 \delta^4 (p'_1+p'_2-p_1-p_2)
\, \frac{1}{(2\pi)^6} \,
\frac{1}{8f_\pi^4} 
\: \bbox{\tau}_1 \cdot \bbox{\tau}_2 \:
\int \frac{d^3 k}{(2\pi)^3} \, \frac{1}{2\sqrt{\vec{k}^2+m_\pi^2}}
\nonumber \\ &=&
-i (2\pi)^4 \delta^4 (p'_1+p'_2-p_1-p_2)
\, \frac{1}{(2\pi)^6} \,
\frac{1}{32f_\pi^4} 
\: \bbox{\tau}_1 \cdot \bbox{\tau}_2 \:
\int \frac{d^3 l}{(2\pi)^3} \, \frac{1}{2\sqrt{{\vec l}^2+4m_\pi^2}}
\,,
\nonumber \\ 
\label{eq_S2}
\end{eqnarray}
where $\vec k=\vec l/2$ and
$\epsilon^{abc} \epsilon^{abc'} \tau^c_1 \tau^{c'}_2 = 2
\: \bbox{\tau}_1 \cdot \bbox{\tau}_2 \:$.
From the last equation, one can read off the contribution to the $T$-matrix
[cf.\ Eqs.~(\ref{eq_S}) and (\ref{eq_V})],
which is 
\begin{equation}
V^{\rm add} = 
\: \bbox{\tau}_1 \cdot \bbox{\tau}_2 \:
\frac{1}{32f_\pi^4} 
\int \frac{d^3 l}{(2\pi)^3} \, \frac{1}{2\sqrt{\vec l^2+4m_\pi^2}}
\label{eq_add}
\end{equation}
Obviously, this additional term
doesn't depend on external momenta, only on $m_\pi$.

For the amplitude, calculated in TOPT and including all terms, we now get
\begin{eqnarray}
V 
&=& V_{\rm ORK} + 
V^{\rm add}\\
&=& V_{\rm ORK} + 
\: \bbox{\tau}_1 \cdot \bbox{\tau}_2 \:
\frac{1}{32f_\pi^4} 
\int \frac{d^3 l}{(2\pi)^3} \, \frac{1}{2\sqrt{\vec l^2 + 4m_\pi^2}}
\\ &=& V_{\rm ORK} + 
\: \bbox{\tau}_1 \cdot \bbox{\tau}_2 \:
\frac{1}{32 f_\pi^4} 
\int \frac{d^3 l'}{(2\pi)^3} \,
\frac{1}{2\omega_+} 
\\ &=& V_{\rm ORK} + 
\: \bbox{\tau}_1 \cdot \bbox{\tau}_2 \:
\frac{1}{32 f_\pi^4} 
\int \frac{d^3 l'}{(2\pi)^3} \,
\left\{
\frac{1}{4\omega_+} +
\frac{1}{4\omega_-}
\right\}
\\ &=& V_{\rm ORK} + 
\: \bbox{\tau}_1 \cdot \bbox{\tau}_2 \:
\frac{1}{32 f_\pi^4} 
\int \frac{d^3 l'}{(2\pi)^3} \,
\frac{(\omega_++\omega_-)^2}{\omega_++\omega_-}
\frac{1}{4\omega_+\omega_-}
\\ &=&
\: \bbox{\tau}_1 \cdot \bbox{\tau}_2 \:
\frac{1}{32 f_\pi^4} 
\int \frac{d^3 l'}{(2\pi)^3} \,
\frac{1}{\omega_++\omega_-}
\end{eqnarray}
where we used the transformations 
$\vec l=\vec l' +\vec q$ 
and $\vec l' \mapsto -\vec l'$.
This result is now identical to the one obtained
using covariant PT, Eq.~(\ref{eq_foot2}), as it should.
We note that the term, Eq.~(\ref{eq_add}), 
left out in the calculation of Ref.~\cite{ORK96},
is a contact term $\propto m_\pi^2$, which affects the $m_\pi$-dependence of
the 2PE potential. Such terms are relevant for the charge-dependence of the nuclear force~\cite{WME01}
and for considerations of the chiral limit~\cite{EMG03}.
If these more subtle aspects are not of interest, then omitting the contact Eq.~(\ref{eq_add})
is acceptable,
since contact terms with adjustable parameters
are added to the theory anyhow (cf.\ Section~\ref{sec_Lct}).

\subsubsection{Dimensional regularization of the football}

We start from Eq.~(\ref{eq_foot1}),
\begin{eqnarray}
V &=& i
\: \bbox{\tau}_1 \cdot \bbox{\tau}_2 \:
\frac{1}{4 f_\pi^4} 
\int \frac{d^4 l}{(2\pi)^4} \,
\frac{1}{(l+q)^2-m_\pi^2+i\epsilon} \,\,
\frac{(l^0)^2}{l^2-m_\pi^2+i\epsilon}
\,,
\end{eqnarray}
and apply the Feynman trick Eq.~(\ref{eq_trick}),
\begin{equation}
V = -
\: \bbox{\tau}_1 \cdot \bbox{\tau}_2 \:
\frac{1}{4 f_\pi^4} \,
 \int_0^1 dx \int \frac{d^{4} l}{i(2 \pi)^4}
 \frac{(l^0)^2}
{[l^2+(2lq+q^2)x-m_\pi^2 + i\epsilon]^2}
\,.
\end{equation}
Using the residue theorem for the $l^0$ integration, we obtain
\begin{eqnarray}
V &=& -
\: \bbox{\tau}_1 \cdot \bbox{\tau}_2 \:
\frac{1}{4 f_\pi^4} \,
\int_0^1 dx \int \frac{dl^0\,d^{3}l}{i(2 \pi)^4}
\frac{(l^0)^2}
{({l^0}-{l^0_1})^2 ({l^0}+{l^0_1})^2} 
\nonumber \\ &=& 
\: \bbox{\tau}_1 \cdot \bbox{\tau}_2 \:
\frac{1}{4 f_\pi^4} \,
\int_0^1 dx \int \frac{d^{3}l}{(2 \pi)^{3}}
\,\, 
\,\, \frac{1}{4} \, 
\frac{1}{l_1^0}
\end{eqnarray}
with
$l^0_1 = \sqrt{\vec{l}^2+m_\pi^2+(2\vec{l}\cdot\vec{q}+\vec{q}^2)x} 
-i\epsilon$.
 
We treat this divergent integral by dimensional regularization 
(cf.\ \ref{sec_dimreg}).
For that purpose, we extend the integral to $(D-1)$ dimensions and shift the
integration variable to 
$\lambda \vec k = \vec l + x \vec q$ 
with $\lambda$ the renormalization scale:
\begin{eqnarray}
V &=&
\: \bbox{\tau}_1 \cdot \bbox{\tau}_2 \:
\frac{1}{16 f_\pi^4} \,
\lambda^{2}
\int_0^1 dx \int \frac{d^{D-1}k}{(2 \pi)^{D-1}} \,\,
\frac{1}
{[\vec{k}^2 + \tilde{m}_\pi^2 + \tilde{q}^2 x (1-x)]^{1/2}}
\,,
\end{eqnarray}
where quantities with a tilde are divided by $\lambda$ and, thus, dimensionless;
and $\vec k$ is dimensionless by definition. 
The angular integration yields
[cf.\ Eqs.~(\ref{eq_vol1}) and (\ref{eq_vol2})]
\begin{eqnarray}
V &=&
\: \bbox{\tau}_1 \cdot \bbox{\tau}_2 \:
\frac{1}{16 f_\pi^4} \,\,
\frac{2\lambda^2}{\Gamma(\frac{D-1}{2}) (4 \pi)^{(D-1)/2}} 
\int_0^1 dx \int_0^\infty dk \,\,
\frac{k^{D-2}}
{[\vec{k}^2 + \tilde{m}_\pi^2 + \tilde{q}^2 x (1-x)]^{1/2}}
\end{eqnarray}
and applying Eq.~(\ref{eq_SchA35}) we obtain
\begin{eqnarray}
V &=&
\: \bbox{\tau}_1 \cdot \bbox{\tau}_2 \:
\frac{1}{16 f_\pi^4} \,
\frac{\lambda^{2}\Gamma(\frac{2-D}{2})}
{\Gamma(\frac 1 2) (4 \pi)^{(D-1)/2}} \,
\int_0^1 dx 
\frac{1}
{[\tilde{m}_\pi^2 + \tilde{q}^2 x (1-x)]^{(2-D)/2}}
\,.
\end{eqnarray}
Choosing $D=4-\eta$ and using relations displayed in
Eq.~(\ref{eq_eta}) results in
\begin{eqnarray}
V &=&
\: \bbox{\tau}_1 \cdot \bbox{\tau}_2 \:
\frac{1}{16 f_\pi^4} \,
\frac{\lambda^{2}}{\Gamma(\frac 1 2)\,(4 \pi)^{\frac{3}{2}}} 
\left(1+\frac{\eta}{2} \ln(4\pi)\right)
\left( -\frac 2 \eta + \gamma -1 \right) \times
\nonumber \\ && \times
\int_0^1 dx 
\left[1-\frac{\eta}{2} \ln(\tilde{m}_\pi^2 + \tilde{q}^2 x (1-x))\right]
\left(\tilde{m}_\pi^2 + \tilde{q}^2 x (1-x)\right)
\,,
\end{eqnarray}
and, neglecting terms of order $\eta$, leads to
\begin{eqnarray}
V &=&
\: \bbox{\tau}_1 \cdot \bbox{\tau}_2 \:
\frac{1}{16 f_\pi^4} \,
\frac{\lambda^{2}}{\Gamma(\frac 1 2)\,(4 \pi)^{\frac{3}{2}}} 
\left[
\left( -\frac 2 \eta +\gamma -1 -\ln (4\pi) \right)
\left(\tilde{m}_\pi^2 +\frac{1}{6} \tilde{q}^2 \right)
\right. \nonumber \\ && \left.
+\int_0^1 dx 
\left (\tilde{m}_\pi^2 + \tilde{q}^2 x (1-x) \right)
\ln(\tilde{m}_\pi^2 + \tilde{q}^2 x (1-x))
\right]
\,.
\end{eqnarray}
The integral over the logarithmic term is
\begin{equation}
\int_0^1 dx 
\left (\tilde{m}_\pi^2 + \tilde{q}^2 x (1-x) \right)
\ln(\tilde{m}_\pi^2 + \tilde{q}^2 x (1-x)) =
%\nonumber \\ =
-\frac 4 3 \tilde{m}_\pi^2 - \frac{5}{18} \tilde q^2 +
\left(2\tilde m_\pi^2+ \frac 1 3 \tilde q^2 \right) \ln (\tilde m_\pi) +
\frac{\tilde q^2 + 4\tilde m_\pi^2}{3} L(q)
\end{equation}
with $L(q)$ given in Eq.~(\ref{eq_L}).

Thus, for the {\it football at NLO}, we finally get
\begin{eqnarray}
V &=&
\: \bbox{\tau}_1 \cdot \bbox{\tau}_2 \:
\frac{1}{384 \pi^2 f_\pi^4} \,
\left[
\frac12 R \, (6 m_\pi^2 + q^2 )
-4 m_\pi^2 - \frac{5}{6} q^2 +
%\right. \nonumber \\ && \left.
(6 m_\pi^2+ q^2 ) \ln \left(\frac{m_\pi}{\lambda}\right) +
w^2 L(q)
\right]
\label{eq_football}
\end{eqnarray}
with $R$ as defined in Eq.~(\ref{eq_R})
and $w$ given in Eq.~(\ref{eq_L}).

\subsection{Box and crossed box diagrams}

\begin{figure}[t]\centering
\vspace*{-1cm}
\scalebox{0.8}{\includegraphics{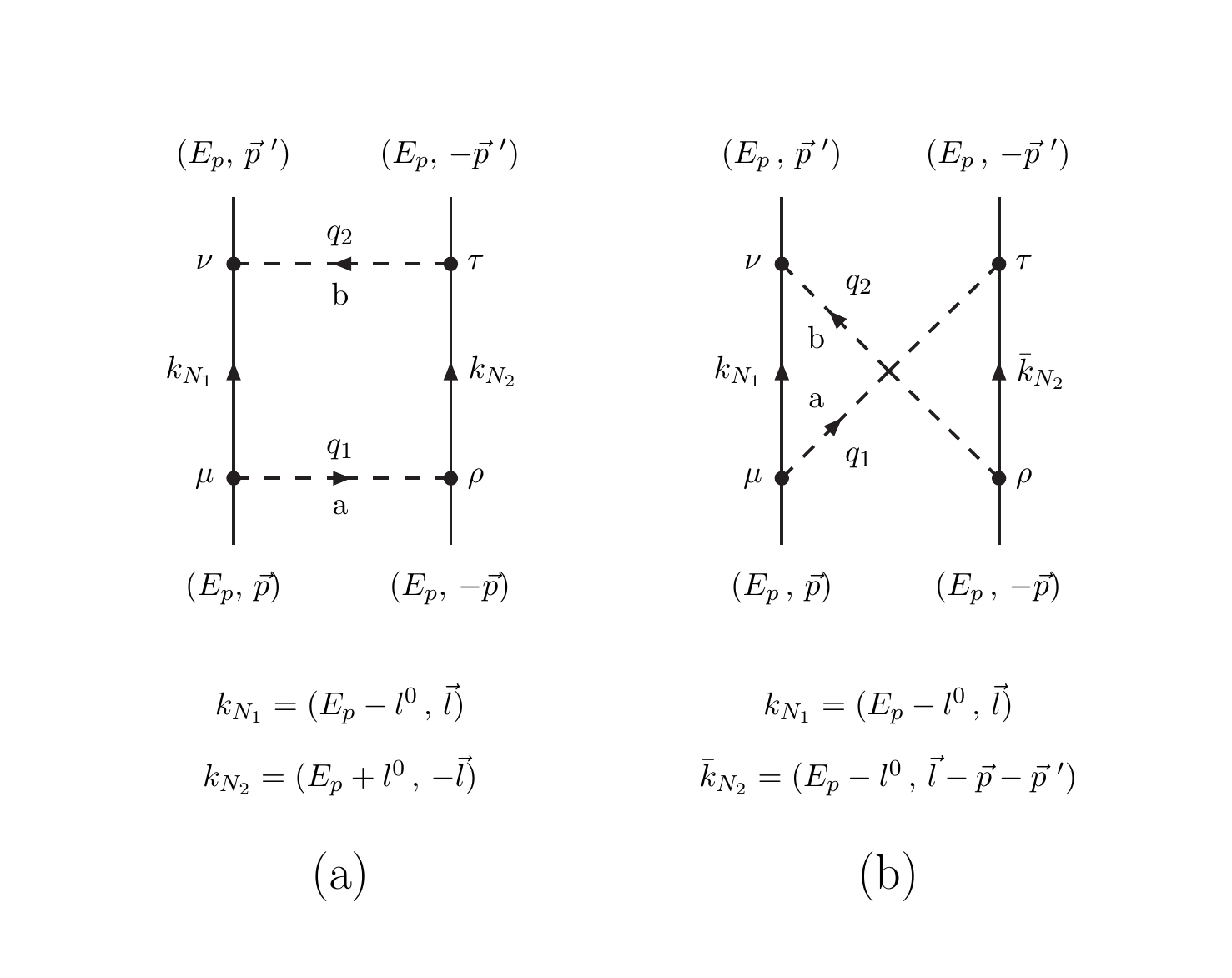}}
\vspace*{-1cm}
\caption{Two-pion exchange planar box (a) and crossed box (b) diagrams at NLO.}
\label{fig_boxes}
\end{figure}

The planar box (pb) and crossed box (cb) diagrams together with our notation are shown in Fig.~\ref{fig_boxes}
and the corresponding amplitudes are given by
\begin{eqnarray}
V^{\rm pb} &=& i \frac{g_A^4}{16 f_\pi^4} \,
\int \frac{d^4 l}{(2\pi)^4} \,\,
\frac{i}{q_1^2-m_\pi^2+i\epsilon} \,\,
\frac{i}{q_2^2-m_\pi^2+i\epsilon}
\times \nonumber \\ && \times
\bar{u}_1 ({\vec p}~') (-q^\nu_2) \tau^b_1 \gamma_\nu  \gamma_5
\frac{i(\slash \! \! \!  k_{N_1}  + M_N)}{k^2_{N_1} - M_N^2 + i\epsilon}
q_1^\mu \tau^a_1 \gamma_\mu  \gamma_5
u_1 (\vec{p})
\times \nonumber \\ && \times
\bar{u}_2 (-{\vec p}~') q_2^\tau \tau^b_2 \gamma_\tau  \gamma_5
\frac{i(\slash \! \! \!  k_{N_2}  + M_N)}{k^2_{N_2} - M_N^2 + i\epsilon}
(-q_1^\rho) \tau^a_2 \gamma_\rho  \gamma_5
u_2 (-\vec{p})
\,,
\\
V^{\rm cb} &=& i \frac{g_A^4}{16 f_\pi^4} \,
\int \frac{d^4 l}{(2\pi)^4} \,\,
\frac{i}{q_1^2-m_\pi^2+i\epsilon} \,\,
\frac{i}{q_2^2-m_\pi^2+i\epsilon}
\times \nonumber \\ && \times
\bar{u}_1 ({\vec p}~') (-q^\nu_2) \tau^b_1 \gamma_\nu  \gamma_5
\frac{i(\slash \! \! \!  k_{N_1}  + M_N)}{k^2_{N_1} - M_N^2 + i\epsilon}
q_1^\mu \tau^a_1 \gamma_\mu  \gamma_5
u_1 (\vec{p})
\times \nonumber \\ && \times
\bar{u}_2 (-{\vec p}~') (-q_1^\tau) \tau^a_2 \gamma_\tau  \gamma_5
\frac{i(\slash \! \! \!  {\bar k}_{N_2}  + M_N)}{ {\bar k}^2_{N_2} - M_N^2 + i\epsilon}
q_2^\rho \tau^b_2 \gamma_\rho  \gamma_5
u_2 (-\vec{p})
\,,
\end{eqnarray}
where we chose the loop momentum such that its time-component is
the energy transferred by the pions and its space-components are the 
ones of an intermediate nucleon;
$k_{N_1}=(E_p-l^0,\vec l)$, 
$k_{N_2}=(E_p+l^0,-\vec l)$, 
$q_1 =(l^0,\vec p - \vec l)$, and
$q_2 =(l^0,{\vec p}~' - \vec l)$, with
$E_p=\sqrt{{\vec p}~^2+M_N^2}$,
for the planar box diagram.
For the crossed box, only the momentum of nucleon~2 is different,
${\bar k}_{N_2}=(E_p-l^0,\vec l-\vec p -{\vec p}~')$.
The poles are located at
\begin{eqnarray}
l^0_1 &=& E_p \mp E_l \pm i \epsilon
\nonumber \\
l^0_2 &=& -E_p \mp E_l \pm i \epsilon
\nonumber \\
\bar l^0_2 &=& E_p \mp E_{\bar l} \pm i \epsilon
\nonumber \\
l^0_3 &=& \mp \omega_1 \pm i \epsilon
\nonumber \\
l^0_4 &=& \mp \omega_2 \pm i \epsilon
\end{eqnarray}
with $\omega_1=\sqrt{(\vec p-\vec l)^2+m_\pi^2}$,
$\omega_2=\sqrt{({\vec p}~'-\vec{l})^2+m_\pi^2}$,
$E_l=\sqrt{{\vec l}~^2+M_N^2}$, and
$E_{\bar l}=\sqrt{(\vec l-\vec p-{\vec p}~')^2+M_N^2}$.
The first three lines state the nucleon poles
and the last two lines describe the pion poles.
$l^0_2$ and $\bar l^0_2$ are the poles for nucleon $2$ in the box and crossed box
diagrams, respectively. 
We apply the Feynman trick Eq.~(\ref{eq_trick})
to the pion propagators and obtain
\begin{eqnarray}
V^{\rm pb} &=& i \frac{g_A^4}{16 f_\pi^4} \,
(3-2 \mbox{\boldmath $\tau$}_1 \cdot \mbox{\boldmath $\tau$}_2 )
\int_0^1\, dx\,\int \frac{d^4 l}{(2\pi)^4} \,
\frac{1}{[{l^0}^2 - \omega_1^2 +(\omega_1^2-\omega_2^2)x+i\epsilon]^2}
\times \nonumber \\ && \times
\bar{u}_1 ({\vec p}~') q^\nu_2 \gamma_\nu  
\frac{\slash \! \! \! k_{N_1}  - M_N}{(l^0-E_p)^2-(l^0_1-E_p)^2}
q_1^\mu \gamma_\mu  
u_1 (\vec{p})
\times \nonumber \\ && \times
\bar{u}_2 (-{\vec p}~') q_2^\tau \gamma_\tau  
\frac{\slash \! \! \! k_{N_2}  - M_N}{(l^0+E_p)^2-(l^0_2+E_p)^2}
q_1^\rho \gamma_\rho  
u_2 (-\vec{p})
\,,
\\
V^{\rm cb} &=& i \frac{g_A^4}{16 f_\pi^4} \,
(3+2 \mbox{\boldmath $\tau$}_1 \cdot \mbox{\boldmath $\tau$}_2 )
\int_0^1\, dx\,\int \frac{d^4 l}{(2\pi)^4} \,
\frac{1}{[{l^0}^2 - \omega_1^2 +(\omega_1^2-\omega_2^2)x+i\epsilon]^2}
\times \nonumber \\ && \times
\bar{u}_1 ({\vec p}~') q^\nu_2 \gamma_\nu  
\frac{\slash \! \! \! k_{N_1}  - M_N}{(l^0-E_p)^2-(l^0_1-E_p)^2}
q_1^\mu \gamma_\mu  
u_1 (\vec{p})
\times \nonumber \\ && \times
\bar{u}_2 (-{\vec p}~') q_1^\tau \gamma_\tau  
\frac{\slash \! \! \! \bar k_{N_2}  - M_N}{(l^0-E_p)^2-(\bar l^0_2-E_p)^2}
q_2^\rho \gamma_\rho  
u_2 (-\vec{p})
\,,
\end{eqnarray}
where we used 
$\tau_1^b\tau_1^a\tau_2^b\tau_2^a=
3-2 \mbox{\boldmath $\tau$}_1 \cdot \mbox{\boldmath $\tau$}_2$ and
$\tau_1^b\tau_1^a\tau_2^a\tau_2^b=
3+2 \mbox{\boldmath $\tau$}_1 \cdot \mbox{\boldmath $\tau$}_2$.
The pion poles are now given by
\begin{eqnarray}
	l^0_{34} &=& \mp \sqrt{(1-x)\omega_1^2+x\omega_2^2} \pm i \epsilon
\,.
\end{eqnarray}
Performing the $1/M_N$ expansion yields
\begin{eqnarray}
\bar{u}_1 \gamma_\nu
(k  \!\!\!\slash_{N_1} - M_N)
\gamma_\mu  u_1
& \approx &
\delta_{\nu i} \delta_{\mu j} 2 M_N (\delta_{ij}
+ i \epsilon^{ijr} \sigma_1^r)
\,,
\\
\bar{u}_2 \gamma_\tau
(k  \!\!\!\slash_{N_2} - M_N)
\gamma_\rho  u_2
& \approx &
\delta_{\tau m} \delta_{\rho n} 2 M_N (\delta_{mn}
+ i \epsilon^{mns} \sigma_2^s)
\,,
\\
\frac{1}{(l^0-E_p)^2-(l^0_1-E_p)^2} \times
\frac{1}{(l^0+E_p)^2-(l^0_2+E_p)^2} & \approx &
\frac{1}{4M_N^2} \frac{1}{(-l^0+i\zeta)(l^0+i\zeta)}
\,,
\\
\frac{1}{(l^0-E_p)^2-(l^0_1-E_p)^2} \times
\frac{1}{(l^0-E_p)^2-(\bar l^0_2-E_p)^2} & \approx &
\frac{1}{4M_N^2} \frac{1}{(l^0-i\epsilon)^2}
\,,
\end{eqnarray}
where 
$i\zeta=\frac{\vec p^2-\vec l^2}{2M_N}+i\epsilon$, 
which avoids the pinch singularity (cf. discussion on p.~7 of Ref.~\cite{Wei91}). 
Hence,
\begin{eqnarray}
V^{\rm pb} &=& i \frac{g_A^4}{16 f_\pi^4} \,
(3-2 \mbox{\boldmath $\tau$}_1 \cdot \mbox{\boldmath $\tau$}_2)
\int_0^1\, dx\,\int \frac{d^4 l}{(2\pi)^4} \,
\frac{q^i_2 q^j_1 (\delta_{ij} + i \epsilon^{ijr} \sigma_1^r)
q_2^m q_1^n (\delta_{mn} + i \epsilon^{mns} \sigma_2^s)}
{(l^0-l^0_{34})^2(l^0+l^0_{34})^2(l^0+i\zeta)(-l^0+i\zeta)}
\,,
\\
V^{\rm cb} &=& i \frac{g_A^4}{16 f_\pi^4} \,
(3+2 \mbox{\boldmath $\tau$}_1 \cdot \mbox{\boldmath $\tau$}_2)
\int_0^1\, dx\,\int \frac{d^4 l}{(2\pi)^4} \,
\frac{q^i_2 q^j_1 (\delta_{ij} + i \epsilon^{ijr} \sigma_1^r)
q_1^m q_2^n (\delta_{mn} + i \epsilon^{mns} \sigma_2^s)}
{(l^0-l^0_{34})^2(l^0+l^0_{34})^2(l^0-i\epsilon)^2}
\,.
\end{eqnarray}
In both diagrams we have a contribution from the pion poles, but in the box
diagram there is also a contribution from the nucleon poles which is enhanced
in the $1/M_N$ expansion and gives the iterated 1PE. We work this out by
closing the $l^0$ contour integral in the lower half-plane and, collecting the
contribution from the pole at $l^0=-i\zeta$, results in
\begin{eqnarray}
V^{\rm pb}_{\rm it} &=& \frac{g_A^4}{16 f_\pi^4} \,
(3-2 \mbox{\boldmath $\tau$}_1 \cdot \mbox{\boldmath $\tau$}_2)
\int_0^1\, dx\,\int \frac{d^3 l}{(2\pi)^3} \,
\frac{q^i_2 q^j_1 (\delta_{ij} + i \epsilon^{ijr} \sigma_1^r)
q_2^m q_1^n (\delta_{mn} + i \epsilon^{mns} \sigma_2^s)}
{(l^0_{34})^4(2i\zeta)}
\,.
\end{eqnarray}
Using
\begin{eqnarray} 
\int_0^1 dx\frac{1}{ (l^0_{34})^4 } &=&
\int_0^1 dx\frac{1}{ [\omega_1^2-(\omega_1^2-\omega_2^2)x]^2 } =
\frac{1}{ \omega_1^2 \omega_2^2} 
\,,
\\
q^i_2 q^j_1 (\delta_{ij} + i \epsilon^{ijr} \sigma_1^r)
q_2^m q_1^n (\delta_{mn} + i \epsilon^{mns} \sigma_2^s) &=&
\vec \sigma_1 \cdot (\vec p~' - \vec l) \vec \sigma_1 \cdot (\vec p - \vec l) 
\vec \sigma_2 \cdot (\vec p~' - \vec l) \vec \sigma_2 \cdot (\vec p - \vec l) 
\,,
\\
\frac{1}{2i\zeta} &=& \frac{M_N}{{\vec p}^2-{\vec l}^2+i\epsilon}
\,,
\end{eqnarray}
and changing $\vec l \to -\vec l$, we obtain
\begin{eqnarray}
	V_{\rm it}^{\rm pb} &=& \frac{g_A^4M_N}{16 f_\pi^4} \,
(3-2 \mbox{\boldmath $\tau$}_1 \cdot \mbox{\boldmath $\tau$}_2) \,\,
\int \frac{d^3 l}{(2\pi)^3} \,
\frac{
\vec \sigma_1 \cdot (\vec l+\vec p~') \, 
\vec \sigma_2 \cdot (\vec l+\vec p~') \, 
\vec \sigma_1 \cdot (\vec l+\vec p  ) \, 
\vec \sigma_2 \cdot (\vec l+\vec p  )}
{(\vec p^2-\vec{l}^2+i\epsilon)((\vec{l}+\vec{p})^2+m_\pi^2)((\vec{l}+{\vec p}~')^2+m_\pi^2)}
\end{eqnarray}
which is the LO iterated 1PE. 

Notice that the 
contribution from the nucleon pole in the crossed box diagram is zero.

Next, we calculate the contribution from the pion poles closing the $l^0$ contour
integral in the lower half-plane with the pole of interest located at
$l^0=l^0_{34}=\sqrt{(1-x)\omega_1^2+x\omega_2^2} - i \epsilon$:
\begin{eqnarray}
V^{\rm pb} &=& \frac{3g_A^4}{64 f_\pi^4} \,
(3-2 \mbox{\boldmath $\tau$}_1 \cdot \mbox{\boldmath $\tau$}_2) 
\int_0^1\, dx\,\int \frac{d^3 l}{(2\pi)^3} \,
\frac{(\vec q_1 \cdot \vec q_2 + i \vec \sigma_1 \cdot (\vec q_2 \times \vec q_1))
(\vec q_1 \cdot \vec q_2 + i \vec \sigma_2 \cdot (\vec q_2 \times \vec q_1))}
{(l^0_{34})^5} \,, 
\nonumber \\
\\
V^{\rm cb} &=& -\frac{3g_A^4}{64 f_\pi^4} \,
(3+2 \mbox{\boldmath $\tau$}_1 \cdot \mbox{\boldmath $\tau$}_2) 
\int_0^1\, dx\,\int \frac{d^3 l}{(2\pi)^3} \,
\frac{(\vec q_1 \cdot \vec q_2 + i \vec \sigma_1 \cdot (\vec q_2 \times \vec q_1))
(\vec q_1 \cdot \vec q_2 - i \vec \sigma_2 \cdot (\vec q_2 \times \vec q_1))}
{(l^0_{34})^5}  \,.
\nonumber \\
\end{eqnarray}
We treat the divergent integrals by dimensional regularization 
(cf.\ \ref{sec_dimreg}).
For that purpose, we extend the integrals to $(D-1)$ dimensions and shift the
integration variable to $\lambda \vec k = \vec p - \vec l + x \vec q$ with $\lambda$
the renormalization scale and $\vec q={\vec p}~' - \vec p$:
\begin{eqnarray}
V^{\rm pb} &=& \frac{3g_A^4}{64 f_\pi^4} \,
(3-2 \mbox{\boldmath $\tau$}_1 \cdot \mbox{\boldmath $\tau$}_2) 
\lambda^{2}
\int_0^1\, dx\,\int \frac{d^{D-1} k}{(2\pi)^{D-1}} \;
%\times \nonumber \\ &\times&
\frac{(\tilde q_1 \cdot \tilde q_2 + i \vec \sigma_1 \cdot (\tilde q_2 \times \tilde q_1))
(\tilde q_1 \cdot \tilde q_2 + i \vec \sigma_2 \cdot (\tilde q_2 \times \tilde q_1))}
{[k^2+\tilde m_\pi^2+\tilde q^2 x(1-x)]^{5/2}}
\,,
\nonumber \\
\\
V^{\rm cb} &=& -\frac{3g_A^4}{64 f_\pi^4} \,
(3+2 \mbox{\boldmath $\tau$}_1 \cdot \mbox{\boldmath $\tau$}_2) 
\lambda^{2}
\int_0^1\, dx\,\int \frac{d^{D-1} k}{(2\pi)^{D-1}} \;
%\times \nonumber \\ &\times&
\frac{(\tilde q_1 \cdot \tilde q_2 + i \vec \sigma_1 \cdot (\tilde q_2 \times \tilde q_1))
(\tilde q_1 \cdot \tilde q_2 - i \vec \sigma_2 \cdot (\tilde q_2 \times \tilde q_1))}
{[k^2+\tilde m_\pi^2+\tilde q^2 x(1-x)]^{5/2}}
\,,
\nonumber \\
\end{eqnarray}
where quantities with a tilde are divided by $\lambda$ and, thus, dimensionless;
and $\vec k$ is dimensionless by definition. Since
\begin{eqnarray}
\tilde q_1 &=& \vec k -x \tilde q
\,,
\\
\tilde q_2 &=& \vec k -(x-1) \tilde q
\,,
\end{eqnarray}
hence
\begin{eqnarray}
\tilde q_1 \cdot \tilde q_2 &=& k^2 + \tilde q^2 x(x-1) - (2x-1) \vec k \cdot \tilde q
\,,
\\
\tilde q_2 \times \tilde q_1 &=& \tilde q \times \vec k
\,.
\end{eqnarray}
Terms in odd powers of $\vec k$ vanish.
Furthermore, because of
\begin{eqnarray}
\int \frac{d^{D-1} k}{(2\pi)^{D-1}} \, 
\frac{k_i k_j}
{[k^2+\tilde m_\pi^2+\tilde q^2 x(1-x)]^{5/2}}
&=&
\frac{\delta_{ij}}{D-1} \int \frac{d^{D-1} k}{(2\pi)^{D-1}} \, 
\frac{k^2}
{[k^2+\tilde m_\pi^2+\tilde q^2 x(1-x)]^{5/2}}
\,,
\end{eqnarray}
we have
\begin{eqnarray}
\int \frac{d^{D-1} k}{(2\pi)^{D-1}} \, 
\frac{(\vec k\cdot \tilde q)^2}
{[k^2+\tilde m_\pi^2+\tilde q^2 x(1-x)]^{5/2}}
&=&
\frac{1}{D-1} \int \frac{d^{D-1} k}{(2\pi)^{D-1}} \, 
\frac{\tilde q^2 k^2}
{[k^2+\tilde m_\pi^2+\tilde q^2 x(1-x)]^{5/2}}
\,,
\nonumber \\
\int \frac{d^{D-1} k}{(2\pi)^{D-1}} \, 
\frac{(\vec k \cdot \tilde q)
i \vec \sigma_i \cdot (\tilde q \times \vec k)}
{[k^2+\tilde m_\pi^2+\tilde q^2 x(1-x)]^{5/2}}
&=& 0
\,,
\nonumber \\
\int \frac{d^{D-1} k}{(2\pi)^{D-1}} \, 
\frac{i \vec \sigma_1 \cdot (\tilde q \times \vec k)
i \vec \sigma_2 \cdot (\tilde q \times \vec k)}
{[k^2+\tilde m_\pi^2+\tilde q^2 x(1-x)]^{5/2}}
&=&
\frac{1}{D-1} \int \frac{d^{D-1} k}{(2\pi)^{D-1}} \, 
\frac{i^2 (\vec \sigma_1 \times \tilde q) \cdot (\vec \sigma_2 \times \tilde q) k^2}
{[k^2+\tilde m_\pi^2+\tilde q^2 x(1-x)]^{5/2}}
\,.
\nonumber \\
\label{eq_divers}
\end{eqnarray}
Whence
\begin{eqnarray}
V^{\rm pb} &=& \frac{3g_A^4}{64 f_\pi^4} \,
(3-2 \mbox{\boldmath $\tau$}_1 \cdot \mbox{\boldmath $\tau$}_2) 
\lambda^{2}
\int_0^1\, dx\,\int \frac{d^{D-1} k}{(2\pi)^{D-1}} \, 
\times \nonumber \\ &\times&
\frac{[k^2+\tilde q^2x(x-1)]^2+\frac{k^2}{D-1} \left[ (2x-1)^2\tilde q^2
- (\vec \sigma_1 \times \tilde q) \cdot (\vec \sigma_2 \times \tilde q) \right] }
{[k^2+\tilde m_\pi^2+\tilde q^2 x(1-x)]^{5/2}}
\,,
\\
V^{\rm cb} &=& -\frac{3g_A^4}{64 f_\pi^4} \,
(3+2 \mbox{\boldmath $\tau$}_1 \cdot \mbox{\boldmath $\tau$}_2) 
\lambda^{2}
\int_0^1\, dx\,\int \frac{d^{D-1} k}{(2\pi)^{D-1}} \,
\times \nonumber \\ &\times&
\frac{[k^2+\tilde q^2x(x-1)]^2+\frac{k^2}{D-1} \left[ (2x-1)^2\tilde q^2
+ (\vec \sigma_1 \times \tilde q) \cdot (\vec \sigma_2 \times \tilde q) \right] }
{[k^2+\tilde m_\pi^2+\tilde q^2 x(1-x)]^{5/2}}
\,.
\end{eqnarray}
Using $(\vec \sigma_1 \times \tilde q) \cdot (\vec \sigma_2 \times \tilde q)=
\lambda^{-2} [\vec q~^2 \vec \sigma_1\cdot \vec \sigma_2 -
(\vec \sigma_1\cdot\vec q)(\vec \sigma_2\cdot\vec q)]$, the sum of the
two diagrams is given by
\begin{eqnarray}
V^{\rm pb} + V^{\rm cb} &=& W_C  
(\mbox{\boldmath $\tau$}_1 \cdot \mbox{\boldmath $\tau$}_2) +
V_S (\vec \sigma_1 \cdot \vec \sigma_2)  
+ V_T 
(\vec \sigma_1\cdot\vec q)(\vec \sigma_2\cdot\vec q) 
\end{eqnarray}
with 
\begin{eqnarray}
W_C &=& -\frac{3g_A^4}{16 f_\pi^4} \,
\lambda^{2}
\int_0^1\, dx\,\int \frac{d^{D-1} k}{(2\pi)^{D-1}} \, 
\frac{[k^2+\tilde q^2x(x-1)]^2+\frac{k^2}{D-1}(2x-1)^2\tilde q^2}
{[k^2+\tilde m_\pi^2+\tilde q^2 x(1-x)]^{5/2}}
\,,
\\
V_T &=& \frac{9g_A^4}{32 f_\pi^4} \,
\int_0^1\, dx\,\int \frac{d^{D-1} k}{(2\pi)^{D-1}} \,
\frac{\frac{k^2}{D-1}}
{[k^2+\tilde m_\pi^2+\tilde q^2 x(1-x)]^{5/2}}
\,,
\\
V_S &=& -q^2 V_T
\,.
\end{eqnarray}
Applying the volume element in $(D-1)$ dimensions, Eqs.~(\ref{eq_vol1})
and (\ref{eq_vol2}), we find
\begin{eqnarray}
W_C &=& -\frac{3g_A^4}{16 f_\pi^4} \,
\frac{2\lambda^{2}}{\Gamma(\frac{D-1}{2}) (4 \pi)^{(D-1)/2}}
\int_0^1\, dx\,\int_0^\infty dk \,\,
\times \nonumber \\ &\times&
\frac{k^{D+2}+k^D \tilde q^2 \left[ 2x(x-1)+ \frac{(2x-1)^2}{D-1} \right]+
k^{D-2}\tilde q^4x^2(x-1)^2}
{[k^2+\tilde m_\pi^2+\tilde q^2 x(1-x)]^{5/2}}
\,,
\\
V_T &=& \frac{9g_A^4}{32 f_\pi^4} \,
\frac{2}{\Gamma(\frac{D-1}{2}) (4 \pi)^{(D-1)/2}}
\, \frac{1}{D-1}
\int_0^1\, dx\,\int_0^\infty dk \, \,
\frac{k^D}
{[k^2+\tilde m_\pi^2+\tilde q^2 x(1-x)]^{5/2}}
\,,
\end{eqnarray}
and performing the $k$-integration with the help of Eq.~(\ref{eq_SchA35}),
we obtain
\begin{eqnarray}
W_C &=& -\frac{3g_A^4}{16 f_\pi^4} \,
\frac{\lambda^{2}}{\Gamma(\frac{5}{2})\Gamma(\frac{D-1}{2}) (4 \pi)^{(D-1)/2}}
\int_0^1\, dx\,\, \left\{
\frac{\Gamma(\frac{D+3}{2})\Gamma(\frac{2-D}{2})}
{[\tilde m_\pi^2+\tilde q^2 x(1-x)]^{(2-D)/2}}
\right . \nonumber \\ && \left.
+\frac{\Gamma(\frac{D+1}{2})\Gamma(\frac{4-D}{2})
\tilde q^2 \left[ 2x(x-1)+ \frac{(2x-1)^2}{D-1} \right]}
{[\tilde m_\pi^2+\tilde q^2 x(1-x)]^{(4-D)/2}}
+\frac{\Gamma(\frac{D-1}{2})\Gamma(\frac{6-D}{2})
\tilde q^4x^2(x-1)^2}
{[\tilde m_\pi^2+\tilde q^2 x(1-x)]^{(6-D)/2}}
\right\}
\,,
\\
V_T &=& \frac{9g_A^4}{32 f_\pi^4} \,
\frac{1}{\Gamma(\frac{5}{2})\Gamma(\frac{D-1}{2}) (4 \pi)^{(D-1)/2}}
\, \frac{1}{D-1}
\int_0^1\, dx\,\,
\frac{\Gamma(\frac{D+1}{2})\Gamma(\frac{4-D}{2})}
{[\tilde m_\pi^2+\tilde q^2 x(1-x)]^{(4-D)/2}}
\,.
\end{eqnarray}
We choose $D=4-\eta$ and take the limit $\eta \to 0$
[cf.~Eq.~(\ref{eq_eta})],
\begin{eqnarray}
W_C &=& -\frac{g_A^4}{384\pi^2 f_\pi^4} \lambda^{2}\,
\int_0^1\, dx\,\, \left\{
R\,\left[45 \tilde m_\pi^2 +\tilde q^2(-6+105x-105x^2 ) \right] 
\right. \nonumber \\ && \left.
+\, 48 \tilde m_\pi^2 + 6\tilde q^2(-1+22x-22x^2 ) +
\frac{12 \tilde q^4 x^2(x-1)^2}{\tilde m_\pi^2+\tilde q^2 x(1-x)}
\right. \nonumber \\ && \left.
+\left[ 45\tilde m_\pi^2+\tilde q^2(-6+105x-105x^2)\right] 
\ln (\tilde m_\pi^2+\tilde q^2 x(1-x))
\right\}
\,,
\\
V_T &=& -\frac{3g_A^4}{128 \pi^2f_\pi^4} \,
\int_0^1\, dx\,\,
\left\{  R + 1 + \ln (\tilde m_\pi^2+\tilde q^2 x(1-x)) \right\}
\end{eqnarray}
with $R$ as defined in Eq.~(\ref{eq_R}).
And performing the $x$-integration, we finally get
for the {\it planar box plus crossed box diagrams at NLO}
\begin{eqnarray}
W_C &=& -\frac{g_A^4}{384\pi^2 f_\pi^4} 
\left\{
\left[\frac{1}{2} R+\ln \left(\frac{m_\pi}{\lambda}\right)\right] (90 m_\pi^2 +23q^2)
+16 m_\pi^2 + \frac{5}{6} q^2 
\right. \nonumber \\ && \left.
+\left(20m_\pi^2+23q^2+\frac{48m_\pi^4}{w^2}\right) L(q)
\right\}
\,,
\label{eq_WCboxes}
\\
V_T &=& - \frac{1}{q^2} V_S =  
-\frac{3g_A^4}{64 \pi^2f_\pi^4} \,
\left[ \frac{1}{2} R+\ln \left(\frac{m_\pi}{\lambda}\right) 
-\frac{1}{2} +L(q) \right]
\label{eq_VTboxes}
\end{eqnarray}
with $w$ and $L(q)$ given in Eq.~(\ref{eq_L}).

\subsection{Summary of 2PE contributions at NLO}

Adding up all 2PE contributions at NLO, namely, the box and crossed boxes
Eqs.~(\ref{eq_WCboxes}) and (\ref{eq_VTboxes}), 
the triangles Eq.~(\ref{eq_triangles}), and the football Eq.~(\ref{eq_football}),
we obtain
\begin{eqnarray} 
W_C &=&-{1\over384\pi^2 f_\pi^4} 
\left\{
\left[ 4m_\pi^2(5g_A^4-4g_A^2-1)
+q^2(23g_A^4 -10g_A^2-1) 
+ {48g_A^4 m_\pi^4 \over w^2} \right] L(q)
\right. \nonumber \\ && \left.
+ \left[ 6m_\pi^2(15g_A^4 - 6g_A^2 -1) + q^2(23g_A^4 - 10 g_A^2 -1) \right]
\ln\left(\frac{m_\pi}{\lambda}\right)
\right. \nonumber \\ && \left.
+ 4m_\pi^2(4g_A^4 + g_A^2 +1) + \frac{q^2}{6}(5g_A^4 + 26 g_A^2 +5)
\right\}
\\   
V_T &=& -{1\over q^2} V_{S} =
-\frac{3g_A^4}{64 \pi^2f_\pi^4} \,
\left[ L(q) +\ln \left(\frac{m_\pi}{\lambda}\right) 
-\frac{1}{2} \right]
\end{eqnarray}  
with $w$ and $L(q)$ given in Eq.~(\ref{eq_L}), and
where we applied renormalization in a 
Modified Minimal Subtraction scheme 
($\overline{MS}$-scheme)~\cite{Col84},
i.~e., the (infinite) $R$-terms are subtracted. 
The results fully agree with Ref.~\cite{KBW97}.
If one is not interested in subtle aspects, like charge-dependence~\cite{WME01} or 
the chiral limit ($m_\pi \rightarrow 0$)~\cite{EMG03},
one may omit
the polynomial terms in the above expressions [cf. Eqs.~(\ref{eq_2C}) and (\ref{eq_2T})],
since contact terms are added anyhow (cf. Section~\ref{sec_ct}) 
which have the same mathematical structure and, therefore,
can absorb the polynomial terms.
Thus, the contacts are getting renormalized by
the polynomials that result from dimensional regularization.

\section{Two-pion exchange contributions to the 2NF at NNLO
\label{app_NNLO}}

\setcounter{figure}{0}
\setcounter{table}{0}

Two-pion exchange diagrams that contribute at NNLO were shown in Fig.~\ref{fig_nnlo}.
These diagrams differ from the NLO diagrams, Fig.~\ref{fig_nlo}, by one insertion from
the dimension-two Lagrangian, the vertices of which are given in \ref{app_feyn_nlo}.
There are two kinds of second order vertices: The relativistic
corrections of the leading order Lagrangian, which are proportional to $1/M_N$, 
Eqs.~(\ref{eq_V2_0})-(\ref{eq_V2_2}), and new contact interactions proportional to the
LECs $c_i$, Eqs.~(\ref{eq_V2_3}) and (\ref{eq_V2_4}).
Since the shapes of the NNLO diagrams are the same as in the NLO case,
the evaluation is very similar to what we presented in detail in \ref{app_NLO}
for the NLO diagrams. 
The results for the irreducible 2PE at third order (NNLO) as derived by the Munich group~\cite{KBW97}
were given in Sec.~\ref{sec_NNLO}, Eqs.~(\ref{eq_3C})--(\ref{eq_3LS}).  
Note that the irreducible 2PE depends on what is chosen for the (reducible) iterated 1PE.
For this, the Munich group applies the expression Eq.~(\ref{eq_2piitKBW}).
Alternatively, one may also use the form Eq.~(\ref{eq_2piitEM}) implied by the 
BbS scheme~\cite{BS66}, which differs in third and higher orders from the Munich expression.
Therefore, when the BbS formalism is used, a (irreducible) correction term
has to be added to the Munich irreducible 2PE in higher orders. In terms of the notation introduced in
Eqs.~(\ref{eq_2piitKBW}) and (\ref{eq_2piitEM}), this correction term is given by: 
\begin{eqnarray}
V_{2\pi, \rm it}^{\rm(KBW)} - V_{2\pi, \rm it}^{\rm(EM)} 
&=&
\int \frac{d^3p''}{(2\pi)^3}\:
\left(\frac{1}{E_p}-\frac{1}{E_{p''}}\right)\frac{M_N^2}{p^2-{p''}^2+i\epsilon}
V_{1\pi} ({\vec p}~',{\vec p}~'')\,
V_{1\pi} ({\vec p}~'',{\vec p}) \\
&=&
-\int \frac{d^3p''}{(2\pi)^3}\:
\frac{M_N^2}{E_{p''}E_{p}(E_{p''}+E_{p})} 
V_{1\pi} ({\vec p}~',{\vec p}~'')\,
V_{1\pi} ({\vec p}~'',{\vec p}) \\
& \approx &
-\int \frac{d^3p''}{(2\pi)^3}\:
\frac{1}{2M_N}
V_{1\pi} ({\vec p}~',{\vec p}~'')\,
V_{1\pi} ({\vec p}~'',{\vec p}) \,,
\end{eqnarray}
with the last line showing the correction in third order or NNLO, which is what we want to calculate here.
Using Eq.~(\ref{eq_1pe}) and the relation $ \: \bbox{\tau}_1 \cdot \bbox{\tau}_2 \: 
\: \bbox{\tau}_1 \cdot \bbox{\tau}_2 \: =3-2 \: \bbox{\tau}_1 \cdot \bbox{\tau}_2$, and defining
$\vec l=\vec p-{\vec p}~''$, yields
\begin{eqnarray}
V_{2\pi, \rm it}^{\rm(KBW)} - V_{2\pi, \rm it}^{\rm(EM)}\, &=&
- \frac{g_A^4}{32 f_\pi^4 M_N}
(3-2 \: \bbox{\tau}_1 \cdot \bbox{\tau}_2 )
\int \frac{d^3 l}{(2 \pi)^3}
\;
\frac{\vec \sigma_1 \cdot (\vec l + \vec q) \,\, 
\vec \sigma_2 \cdot (\vec l + \vec q)}{(\vec l + \vec q)^2 +m_\pi^2}
\;
\frac{ \vec \sigma_1 \cdot \vec l \,\, 
\vec \sigma_2 \cdot \vec l }{l^2 +m_\pi^2} \\
&=&
- \frac{g_A^4}{32 f_\pi^4 M_N}
(3-2 \: \bbox{\tau}_1 \cdot \bbox{\tau}_2)
\int _0^1\,dx \int \frac{d^3 l}{(2 \pi)^3}
\frac{\vec \sigma_1 \cdot (\vec l + \vec q) \,\, 
\vec \sigma_2 \cdot (\vec l + \vec q)
\vec \sigma_1 \cdot \vec l \,\, \vec \sigma_2 \cdot \vec l }
{[\vec l^2 +(2\vec l \cdot \vec q+\vec q^2)x +m_\pi^2]^2} \,,
\end{eqnarray}
where, in the last line, we introduced the Feynman trick.
We now extend the integral to $(D-1)$ dimensions and make the change of variables
$\lambda \vec k= \vec l + x\vec q$ (cf.\ \ref{sec_dimreg}),
\begin{eqnarray}
V_{2\pi, \rm it}^{\rm(KBW)} - V_{2\pi, \rm it}^{\rm(EM)}\, &=&
- \frac{g_A^4}{32 f_\pi^4 M_N}
(3-2 \: \bbox{\tau}_1 \cdot \bbox{\tau}_2) \lambda^3
\int _0^1\,dx \int \frac{d^{D-1} k}{(2 \pi)^{D-1}}
\times
\nonumber \\ & \times &
\frac{\vec \sigma_1 \cdot [\vec k + (1-x)\tilde q] \,\, 
\vec \sigma_2 \cdot [\vec k + (1-x)\tilde q] \,\,
\vec \sigma_1 \cdot (\vec k-x\tilde q) \,\, \vec \sigma_2 \cdot (\vec k-x\tilde q)}
{[k^2 + \tilde m_\pi^2+\tilde q^2 x(1-x)]^2} \,.
\end{eqnarray}
Since
\begin{eqnarray}
\vec \sigma_i \cdot [\vec k + (1-x)\tilde q] \,\, \vec \sigma_i \cdot (\vec k-x\tilde q) 
&=& k^2+\tilde q^2x(x-1) -(2x-1)\vec k\cdot \tilde q
+i\vec \sigma_i \cdot (\tilde q\times\vec k) \,,
\end{eqnarray}
 we obtain, making use of Eq. (\ref{eq_divers}),
\begin{eqnarray}
V_{2\pi, \rm it}^{\rm(KBW)} - V_{2\pi, \rm it}^{\rm(EM)}\, &=&
- \frac{g_A^4}{32 f_\pi^4 M_N}
(3-2 \: \bbox{\tau}_1 \cdot \bbox{\tau}_2) \lambda^3
\int _0^1\,dx \int \frac{d^{D-1} k}{(2 \pi)^{D-1}}  \times
\nonumber \\ & \times &
\frac{[k^2+\tilde q^2x(x-1)]^2 +\frac{k^2}{D-1}[(2x-1)^2\tilde q^2-(\vec \sigma_1 \times \tilde q)(\vec \sigma_2 \times \tilde q)]
}
{[k^2 + \tilde m_\pi^2+\tilde q^2 x(1-x)]^2}  \,.
\end{eqnarray}
Applying $(\vec \sigma_1 \times \tilde q)(\vec \sigma_2 \times \tilde q)=
\lambda^{-2}[q^2 \vec \sigma_1\cdot\vec \sigma_2-(\vec \sigma_1\cdot\vec q)(\vec \sigma_2\cdot\vec q)]$, leads to
\begin{eqnarray}
V_C &=&
- \frac{3g_A^4}{32 f_\pi^4 M_N}
\lambda^3
\int _0^1\,dx \int \frac{d^{D-1} k}{(2 \pi)^{D-1}}
\frac{[k^2+\tilde q^2x(x-1)]^2 +\frac{k^2}{D-1}(2x-1)^2\tilde q^2
}
{[k^2 + \tilde m_\pi^2+\tilde q^2 x(1-x)]^2}
%\nonumber 
\\ 
W_C &=& -\frac 2 3 V_C
%\nonumber 
\\ 
V_T &=& - \frac{3g_A^4}{32 f_\pi^4 M_N}
\lambda
\int _0^1\,dx \int \frac{d^{D-1} k}{(2 \pi)^{D-1}}
\frac{\frac{k^2}{D-1}
}
{[k^2 + \tilde m_\pi^2+\tilde q^2 x(1-x)]^2}
%\nonumber 
\\ 
V_S &=& -q^2 V_T
%\nonumber 
\\ 
W_T &=& -\frac 2 3V_T
%\nonumber 
\\ 
W_S &=& -q^2 W_T
\,.
\end{eqnarray}
Using the volume element in $(D-1)$ dimensions, Eqs.~(\ref{eq_vol1}) and (\ref{eq_vol2}), we find
\begin{eqnarray}
V_C &=&
- \frac{3g_A^4}{32 f_\pi^4 M_N}
\frac{2\lambda^3}{\Gamma(\frac{D-1}{2})(4 \pi)^{(D-1)/2}}
\int _0^1\,dx \int_0^\infty dk
\frac{k^{D+2} +k^D\tilde q^2\left[ 2x(x-1)+\frac{(2x-1)^2}{D-1} \right]+
k^{D-2}\tilde q^4x^2(x-1)^2 
}
{[k^2 + \tilde m_\pi^2+\tilde q^2 x(1-x)]^2}
\,,
\nonumber 
\\   \\
V_T &=& - \frac{3g_A^4}{32 f_\pi^4 M_N}
\frac{2\lambda}{\Gamma(\frac{D-1}{2})(4 \pi)^{(D-1)/2}}
\frac{1}{D-1}
\int _0^1\,dx \int_0^\infty dk
\frac{k^D}{[k^2 + \tilde m_\pi^2+\tilde q^2 x(1-x)]^2}
\,.
\end{eqnarray}
Performing the $k$-integration yields
\begin{eqnarray}
V_C &=&
- \frac{3g_A^4}{32 f_\pi^4 M_N}
\frac{\lambda^3}{\Gamma(2)\Gamma(\frac{D-1}{2})(4 \pi)^{(D-1)/2}}
\int _0^1\,dx \left\{
\frac{\Gamma(\frac{D+3}{2})\Gamma(\frac{1-D}{2})}{[\tilde m_\pi^2+\tilde q^2 x(1-x)]^{(1-D)/2}} 
\nonumber \right. \\ & + & \left .
\frac{\Gamma(\frac{D+1}{2})\Gamma(\frac{3-D}{2}) \tilde q^2\left[ 2x(x-1)+\frac{(2x-1)^2}{D-1} \right]}{(\tilde m_\pi^2+\tilde q^2 x(1-x))^{(3-D)/2}} +
\frac{\Gamma(\frac{D-1}{2})\Gamma(\frac{5-D}{2})\tilde q^4x^2(x-1)^2 }{[\tilde m_\pi^2+\tilde q^2 x(1-x)]^{(5-D)/2}} 
\right\}
\,,
%\nonumber 
\\ 
V_T &=& - \frac{3g_A^4}{32 f_\pi^4 M_N}
\frac{\lambda}{\Gamma(2)\Gamma(\frac{D-1}{2})(4 \pi)^{(D-1)/2}}
\;
\frac{1}{D-1}
\int _0^1\,dx 
\frac{\Gamma(\frac{D+1}{2})\Gamma(\frac{3-D}{2})}{[\tilde m_\pi^2+\tilde q^2 x(1-x)]^{(3-D)/2}} 
\,.
\end{eqnarray}
We choose $D=4-\eta$ and take the limit $\eta \to 0$,
\begin{eqnarray}
V_C &=& - \frac{3g_A^4}{256 \pi f_\pi^4 M_N} \int_0^1 
\frac{5m_\pi^4-m_\pi^2q^2(1-20x+20x^2)+q^4x(-1+17x-32x^2+16x^3)}{\sqrt{m_\pi^2+q^2x(1-x)}}
\,,
%\nonumber 
\\
V_T &=& \frac{3g_A^4}{256 \pi f_\pi^4 M_N}  \int_0^1 \sqrt{m_\pi^2+q^2x(1-x)}
\,,
\end{eqnarray}
and, performing the $x$-integration, we finally obtain
\begin{eqnarray}
V_C &=& - \frac{3g_A^4}{256 \pi f_\pi^4 M_N} (
m_\pi w^2 + \widetilde w^4 A(q) )
\,,
%\nonumber 
\\
V_T &=& \frac{3g_A^4}{512 \pi f_\pi^4 M_N}  (m_\pi + w^2 A(q) )
\,.
\end{eqnarray}

\section{Two-pion exchange contributions to the 2NF at N$^3$LO
\label{app_N3LO}}

\setcounter{figure}{0}
\setcounter{table}{0}

The fourth order 2PE contributions consist of two classes: 
the one-loop (Fig.~\ref{fig_n3lo1}) and 
the two-loop diagrams (Fig.~\ref{fig_n3lo2}).

\subsection{One-loop diagrams}

This large pool of diagrams can be analyzed in a systematic
way by introducing the following well-defined
subdivisions.

\subsubsection{$c_i^2$ contributions.}

The only contribution of this kind comes from the football diagram 
with both
vertices proportional to $c_i$ (first row of
Fig.~\ref{fig_n3lo1}). One obtains~\cite{Kai01a}:
\begin{eqnarray}  
V_C & = & {3 L(q) \over 16 \pi^2 f_\pi^4 } 
\left[
\left( {c_2 \over 6} w^2 +c_3 \widetilde{w}^2 -4c_1 m_\pi^2 \right)^2 
+{c_2^2 \over 45 } w^4 
\right] \,, 
\label{eq_4c2C}
\\
W_T  &=&  -{1\over q^2} W_S 
     = {c_4^2 w^2 L(q) \over 96 \pi^2 f_\pi^4 } 
\,,
\label{eq_4c2T}
\end{eqnarray}
with $L(q)$ and $w$ defined in Eq.~(\ref{eq_L}) and
\begin{equation} 
\widetilde{w} \equiv  \sqrt{2m_\pi^2+q^2} \,. 
\end{equation}

\subsubsection{$c_i/M_N$ contributions.}

This class consists of diagrams with one vertex proportional to $c_i$
and one $1/M_N$ correction.
A few graphs that are representative for this class are shown in the
second row of Fig.~\ref{fig_n3lo1}. 
Symbols with a large solid dot and an open circle denote 
$1/M_N$ corrections of vertices
proportional to $c_i$. They are part of
$\widehat{\cal L}^{\Delta=2}$, Eq.~(\ref{eq_LD2}).
The result for this group of diagrams is~\cite{Kai01a}:
\begin{eqnarray} 
V_C & = & - {g_A^2\, L(q) \over 32 \pi^2 M_N f_\pi^4 } \left[ 
(c_2-6c_3) q^4 +4(6c_1+c_2-3c_3)q^2 m_\pi^2 
%\right.  \nonumber \\ && \left.
+6(c_2-2c_3)m_\pi^4
+24(2c_1+c_3)m_\pi^6 w^{-2} \right] \,,
\nonumber \\
\label{eq_4cMC}
\\
W_C &=& 
-{c_4 q^2 L(q) \over 192 \pi^2 M_N f_\pi^4 } 
\left[ g_A^2 (8m_\pi^2+5q^2) + w^2 \right] 
\,, \\
W_T  &=&  -{1\over q^2} W_S 
     = -{c_4 L(q) \over 192 \pi^2 M_N f_\pi^4 } 
\left[ g_A^2 (16m_\pi^2+7q^2) - w^2 \right] 
\label{eq_4cMS}
\,,  \\
V_{LS}& = & {c_2 \, g_A^2 \over 8 \pi^2 M_N f_\pi^4 } 
\, w^2 L(q) 
\,, \\
W_{LS}  &=& 
-{c_4 L(q) \over 48 \pi^2 M_N f_\pi^4 } 
\left[ g_A^2 (8m_\pi^2+5q^2) + w^2 \right] 
\,.
\label{eq_4cMLS}
\end{eqnarray}

\subsubsection{$1/M_N^2$ corrections.}

These are relativistic $1/M_N^2$ corrections of the leading order
$2\pi$ exchange diagrams. Typical examples
for this large class are shown in row 3--6 of
Fig.~\ref{fig_n3lo1}. 
This time, there is no correction from the iterated 1PE, 
Eq.~(\ref{eq_2piitKBW}) or
Eq.~(\ref{eq_2piitEM}),
since the expansion of the factor $M^2_N/E_p$ does not create
a term proportional to $1/M^2_N$. 
The total result for this class is~\cite{Kai01b},
\begin{eqnarray} 
V_C &=& -{g_A^4 \over 32\pi^2 M_N^2 f_\pi^4}
\Bigg[ 
L(q) \, \Big(2m_\pi^8 w^{-4}+8m_\pi^6 w^{-2} -q^4 -2m_\pi^4\Big)
+{ m_\pi^6 \over 2 w^{2}}\, 
\Bigg] ,
\label{eq_4M2C}
\\
W_C           &=& -{1\over 768\pi^2 M_N^2 f_\pi^4} \Bigg\{ L(q) 
\, \bigg[
                8g_A^2 \, \bigg({3\over 2} q^4 +3m_\pi^2 q^2 
+3m_\pi^4
               -6m_\pi^6 w^{-2} 
               -k^2(8m_\pi^2 +5q^2) \bigg)
\nonumber \\   && 
                + 4g_A^4 
                \bigg(k^2\big(20m_\pi^2+7q^2-16m_\pi^4 w^{-2}\big) 
                +16m_\pi^8 w^{-4}
%\nonumber \\   && 
                +12 m_\pi^6 w^{-2} 
-4m_\pi^4q^2w^{-2} -5q^4 -6m_\pi^2 q^2-6m_\pi^4 \bigg) 
\nonumber \\   && 
               -4k^2 w^2 \, \bigg]
%\nonumber \\   && 
                \,+\, {16 g_A^4 m_\pi^6 \over w^{2}} \Bigg\} \,,
\\
V_T &=& -{1\over q^2} V_S
    \; = \; {g_A^4 \, L(q) \over 32\pi^2 M_N^2 f_\pi^4} 
        \bigg(k^2+{5\over 8} q^2 +m_\pi^4 w^{-2} \bigg) \,,
\\
W_T &=& -{1\over q^2} W_S 
    \; = \; { L(q) \over 1536\pi^2 M_N^2 f_\pi^4} 
\Bigg[\, 4 g_A^4\, \bigg( 7m_\pi^2+{17\over 4} q^2 +4m_\pi^4 
w^{-2} \bigg) 
%\nonumber \\ &&
                 -\, 32 g_A^2\, \bigg( m_\pi^2+{7\over 16}q^2 \bigg) 
                 + w^2 \, \Bigg] \,, 
\nonumber \\
\\
V_{LS} &=& {g_A^4 \, L(q) \over 4\pi^2 M_N^2 f_\pi^4} 
\bigg( {11 \over 32} q^2 +m_\pi^4 w^{-2}\bigg) \,,
\\
W_{LS} &=&  { L(q) \over 256 \pi^2 M_N^2 f_\pi^4}
\Bigg[\, 16 g_A^2\, \bigg( m_\pi^2+{3\over 8}q^2\bigg)
+ \,\frac43\, g_A^4\, \bigg( 4m_\pi^4 w^{-2}-{11\over 4}q^2 
-9m_\pi^2 \bigg) 
                      -w^2 \, \Bigg] \,,
\\
V_{\sigma L} &=& {g_A^4 \, L(q) \over 32\pi^2 M_N^2 f_\pi^4}\;.
\label{eq_4M2sL}
\end{eqnarray} 

\subsection{Two-loop contributions.}

The two-loop contributions are quite involved.
In Fig.~\ref{fig_n3lo2}, we attempt a graphical representation of
this class. The gray disk stands for all one-loop $\pi N$ graphs
which are shown in some detail in the lower part of the figure.
Not all of the numerous graphs are displayed. 
Some of the missing ones
are obtained by permutation of the vertices along the nucleon line,
others by inverting initial and final states.
Vertices denoted by a small dot are from the
leading order Lagrangian
$\widehat{\cal L}^{\Delta=0}$,
Eq.~(\protect\ref{eq_LD0}).
The solid square represents vertices proportional to the LECs
$d_i$ introduced in 
${\cal L}^{(3)}_{\pi N}$
which is part of
$\widehat{\cal L}^{\Delta=2}$,
Eq.~(\protect\ref{eq_LD2}).
The $d_i$ vertices occur actually in one-loop $NN$ diagrams, but
we list them among the two-loop $NN$ contributions because they
are needed to absorb divergences generated by 
one-loop $\pi N$ graphs.
Using techniques from dispersion theory,
Kaiser~\cite{Kai01a} calculated the imaginary parts of the
$NN$ amplitudes, Im $V_\alpha(i\mu)$ and Im $W_\alpha(i\mu)$,
which result from analytic continuation to time-like
momentum transfer $q=i\mu-0^+$ with $\mu\geq 2m_\pi$.
From this,
the momentum-space amplitudes $V_\alpha(q)$ and $W_\alpha(q)$
are obtained
via the subtracted dispersion relations:
\begin{eqnarray} 
V_{C,S}(q) &=& 
-{2 q^6 \over \pi} \int_{2m_\pi}^\infty d\mu \,
{{\rm Im\,}V_{C,S}(i \mu) \over \mu^5 (\mu^2+q^2) }\,, 
\\
V_T(q) &=& 
{2 q^4 \over \pi} \int_{2m_\pi}^\infty d\mu \,
{{\rm Im\,}V_T(i \mu) \over \mu^3 (\mu^2+q^2) }\,, 
\end{eqnarray}
and similarly for $W_{C,S,T}$.

In most cases, the dispersion integrals can be solved
analytically and the following expressions are 
obtained~\cite{EM02}:
\begin{eqnarray}
V_C (q) & = &
 \frac{3g_A^4 \widetilde{w}^2 A(q)}{1024 \pi^2 f_\pi^6}
\left[ ( m_\pi^2 + 2q^2 )
\left( 2m_\pi + \widetilde{w}^2 A(q) \right)
+ 4g_A^2 m_\pi \widetilde{w}^2 \right] ;
\nonumber \\
\label{eq_42lC}
\\
W_C(q) & = & W_C^{(a)}(q) + W_C^{(b)}(q) \,,
\\
\mbox{\rm with \hspace*{1.5cm}} 
\nonumber \\
W_C^{(a)}(q) & = &
\frac{L(q)}{18432 \pi^4 f_\pi^6}
\Bigg\{
192 \pi^2 f_\pi^2
w^2 \bar{d}_3
\left[2g_A^2\widetilde{w}^2-\frac35(g_A^2-1)w^2\right]
\nonumber \\
&&
+\left[6g_A^2\widetilde{w}^2-(g_A^2-1)w^2\right]
\Bigg[
 384\pi^2f_\pi^2
\left(\widetilde{w}^2(\bar{d}_1+\bar{d}_2)+4m_\pi^2\bar{d}_5\right)
\nonumber \\
&&
+L(q)
\left(4m_\pi^2(1+2g_A^2)+q^2(1+5g_A^2)\right)
%\nonumber \\ &&
-\left(\frac{q^2}{3}(5+13g_A^2)+8m_\pi^2(1+2g_A^2)\right)
\Bigg]
\Bigg\}
\nonumber \\
\label{eq_WC}
\\
\mbox{\rm and \hspace*{1.6cm}}
\nonumber \\
W_C^{(b)}(q) & = &
-{2 q^6 \over \pi} \int_{2m_\pi}^\infty d\mu \,
{{\rm Im\,}W_C^{(b)}(i \mu) \over \mu^5 (\mu^2+q^2) }\,, 
\\
\mbox{\rm where\hspace*{1.4cm}}
\nonumber \\
{\rm Im}\, W_C^{(b)}(i\mu)&=& 
-{2\kappa \over 3\mu (8\pi f_\pi^2)^3} 
\int_0^1 dx\, 
\Big[ g_A^2(2m_\pi^2-\mu^2) +2(g_A^2-1)\kappa^2x^2 \Big]
\nonumber \\
&& \times \left\{ 
-\,3\kappa^2x^2 
+6 \kappa x \sqrt{m_\pi^2 +\kappa^2 x^2}  \; \ln{ \kappa x 
+\sqrt{m_\pi^2 
+\kappa^2 x^2}\over  m_\pi}
\right.
\nonumber \\
&&
\left.
+g_A^4\left(\mu^2 -2\kappa^2 x^2 -2m_\pi^2\right) 
%\right.  \nonumber \\ && \left.  \times 
\left[ {5\over 6} +{m_\pi^2\over \kappa^2 x^2} 
-\left( 1 +{m_\pi^2\over \kappa^2 x^2} \right)^{3/2} 
%\right. \right. \nonumber \\ &&  \left. \left. \times
\ln{ \kappa x +\sqrt{m_\pi^2 +\kappa^2 x^2}\over  m_\pi} \right] 
\right\};   
\nonumber \\
\\
V_T(q)  &=& V_T^{(a)}(q) + V_T^{(b)}(q) 
\nonumber \\
        &=& - {1 \over q^2}V_S(q) 
 \; = \; - {1\over q^2}\left(V_S^{(a)}(q)+V_S^{(b)}(q)\right) ,
\\
\mbox{\rm with\hspace*{1.5cm}}
\nonumber \\
V_T^{(a)}(q)  &=& - {1\over q^2}V_S^{(a)}(q) 
              \; = \; -\frac{g_A^2 w^2 L(q)}{32 \pi^2 f_\pi^4}
(\bar{d}_{14} - \bar{d}_{15}) 
\label{eq_VT}
\\
\mbox{\rm and \hspace*{1.5cm}}
\nonumber \\
V_T^{(b)}(q)  &=& - {1\over q^2}V_S^{(b)}(q) 
              \; = \; {2 q^4 \over \pi} \int_{2m_\pi}^\infty d\mu \,
{{\rm Im\,}V_T^{(b)}(i \mu) \over \mu^3 (\mu^2+q^2) } \,, 
\\
\mbox{\rm where\hspace*{1.4cm}}
\nonumber \\ 
{\rm Im}\, V_T^{(b)}(i\mu) &=& 
-{2g_A^6 \kappa^3 \over \mu (8\pi f_\pi^2)^3} 
\int_0^1 dx(1-x^2)
\left[
-{1\over 6}+{m_\pi^2 \over \kappa^2x^2}
%\right.  \nonumber \\ && \left.  
-\left( 1+{m_\pi^2 \over \kappa^2x^2} \right)^{3/2} 
\ln{ \kappa x +\sqrt{m_\pi^2 +\kappa^2 x^2}\over  m_\pi}
\right] ; 
\nonumber \\
\\
W_T(q) &=& - {1 \over q^2}W_S(q) 
%\nonumber \\
       = \frac{g_A^4 w^2 A(q)}{2048 \pi^2 f_\pi^6}
\left[ w^2 A(q) + 2m_\pi (1 + 2 g_A^2) \right];
\label{eq_42lT}
\end{eqnarray}
where 
\begin{equation} 
A(q) \equiv {1\over 2q}\arctan{q \over 2m_\pi} 
\quad \mbox{and} \quad
\kappa \equiv \sqrt{\mu^2/4-m_\pi^2}
\,.
\end{equation}

Note that the analytic solutions hold modulo polynomials.
We have checked the importance of those contributions where 
we could not find an analytic solution and where, therefore, the
integrations have to be performed numerically. It turns out
that the combined effect on $NN$ phase shifts from
$W_C^{(b)}$, $V_T^{(b)}$, and $V_S^{(b)}$
is smaller than 0.1 deg in $F$ and $G$ waves
and smaller than 0.01 deg in $H$ waves,
at $T_{\rm lab} = 300$ MeV (and less at lower energies). 
This renders these contributions negligible. 
Therefore, we omit
$W_C^{(b)}$, $V_T^{(b)}$, and $V_S^{(b)}$
in the construction of chiral $NN$ potentials
at order N$^3$LO (Section~\ref{sec_potn3lo}).

In Eqs.~(\ref{eq_WC}) and (\ref{eq_VT}), we use the scale-independent
LECs, $\bar{d}_i$, which are obtained by combining the 
scale-dependent ones, $d_i^r (\lambda)$, 
with the chiral logarithm,
$\ln (m_\pi/\lambda)$, or equivalently $\bar{d}_i = d^r_i(m_\pi)$.
For more details about this issue, see Ref.~\cite{FMS98}.

\subsection{Impact of individual N$^3$LO contributions on peripheral
phase shifts \label{app_N3LO_ph}}

\begin{figure}[t]\centering
\vspace{-0.7cm}
\hspace*{-0.5cm}
\scalebox{0.25}{\includegraphics{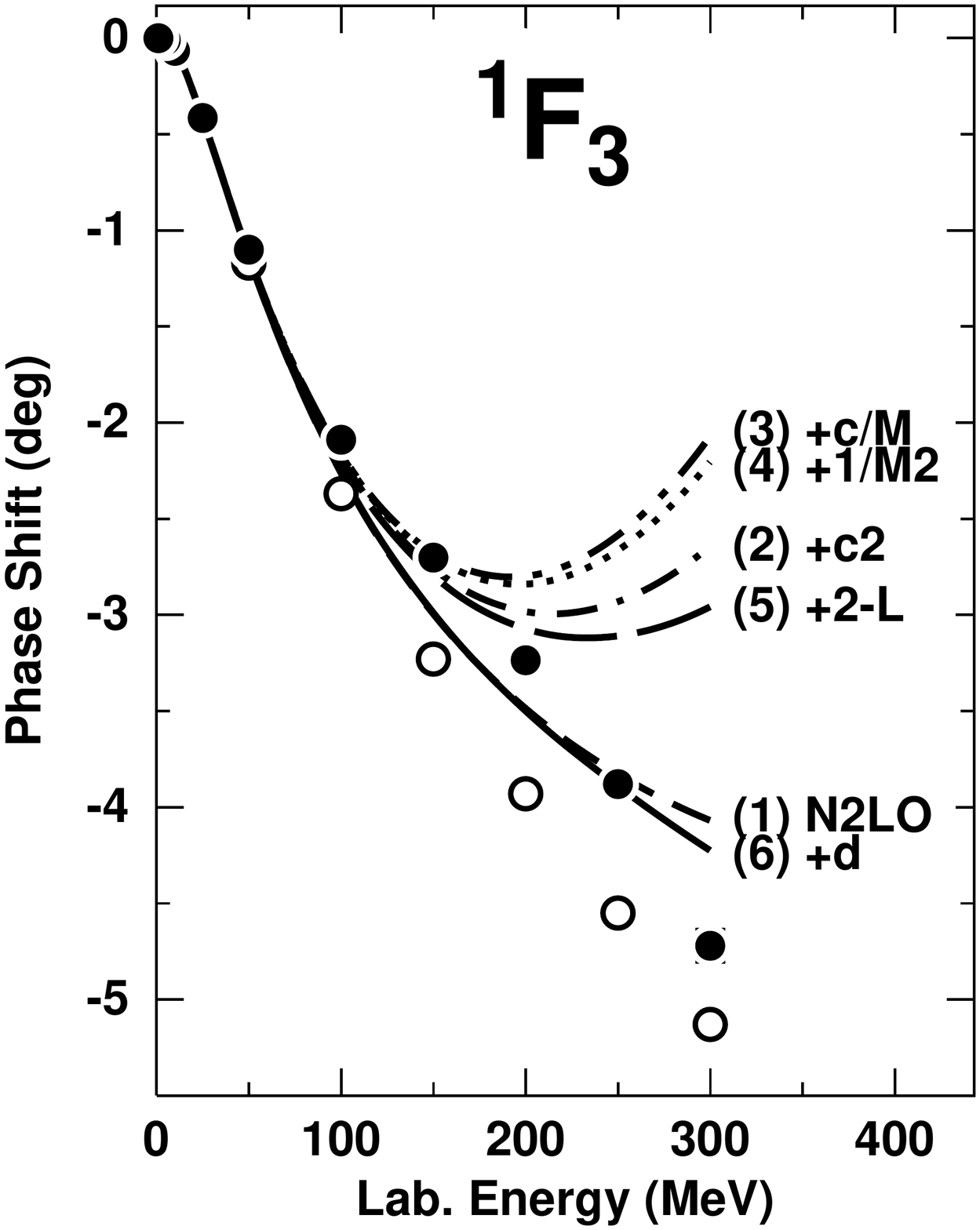}}
\hspace*{-0.5cm}
\scalebox{0.25}{\includegraphics{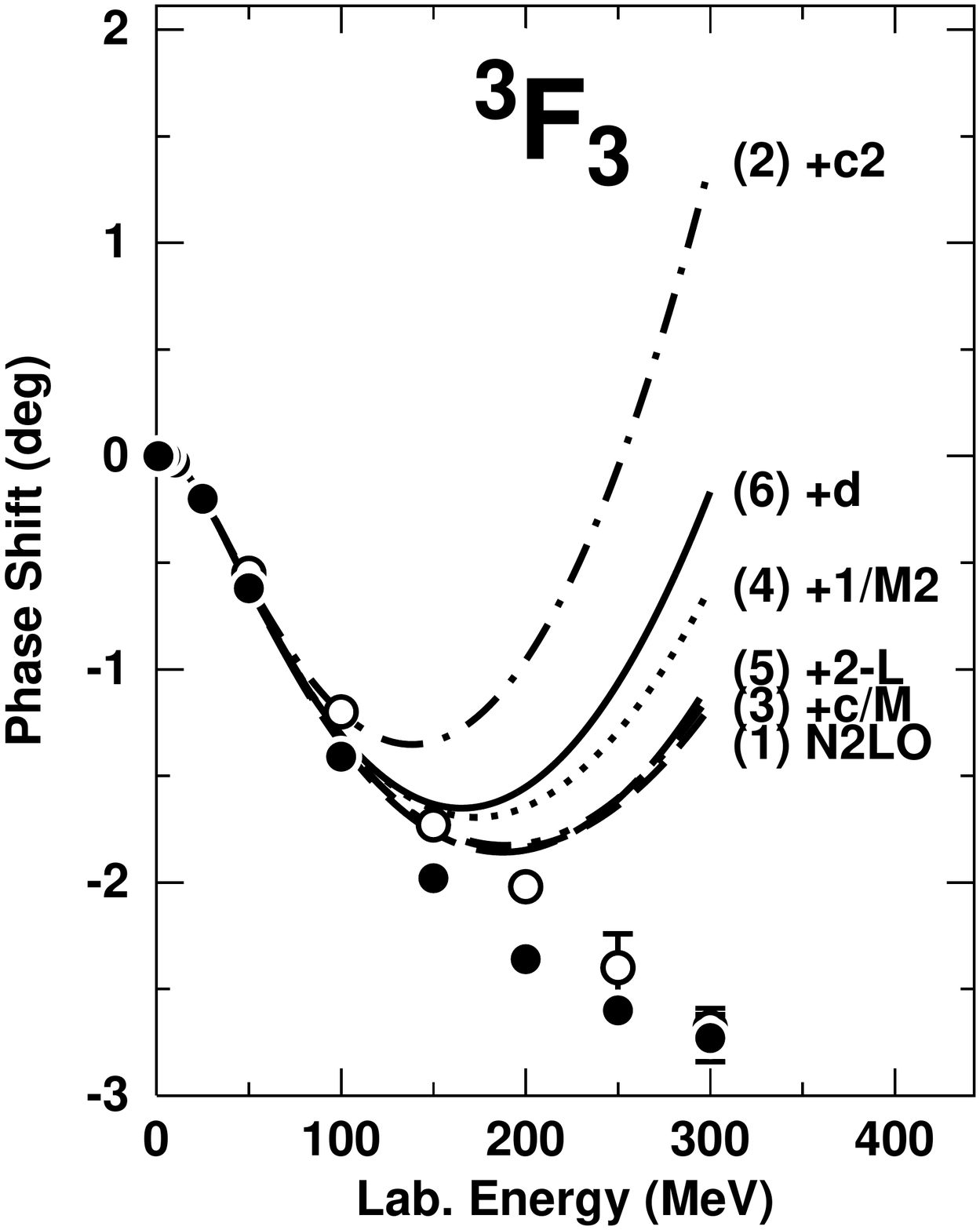}}

\vspace{-1.0cm}
\hspace*{-0.5cm}
\scalebox{0.25}{\includegraphics{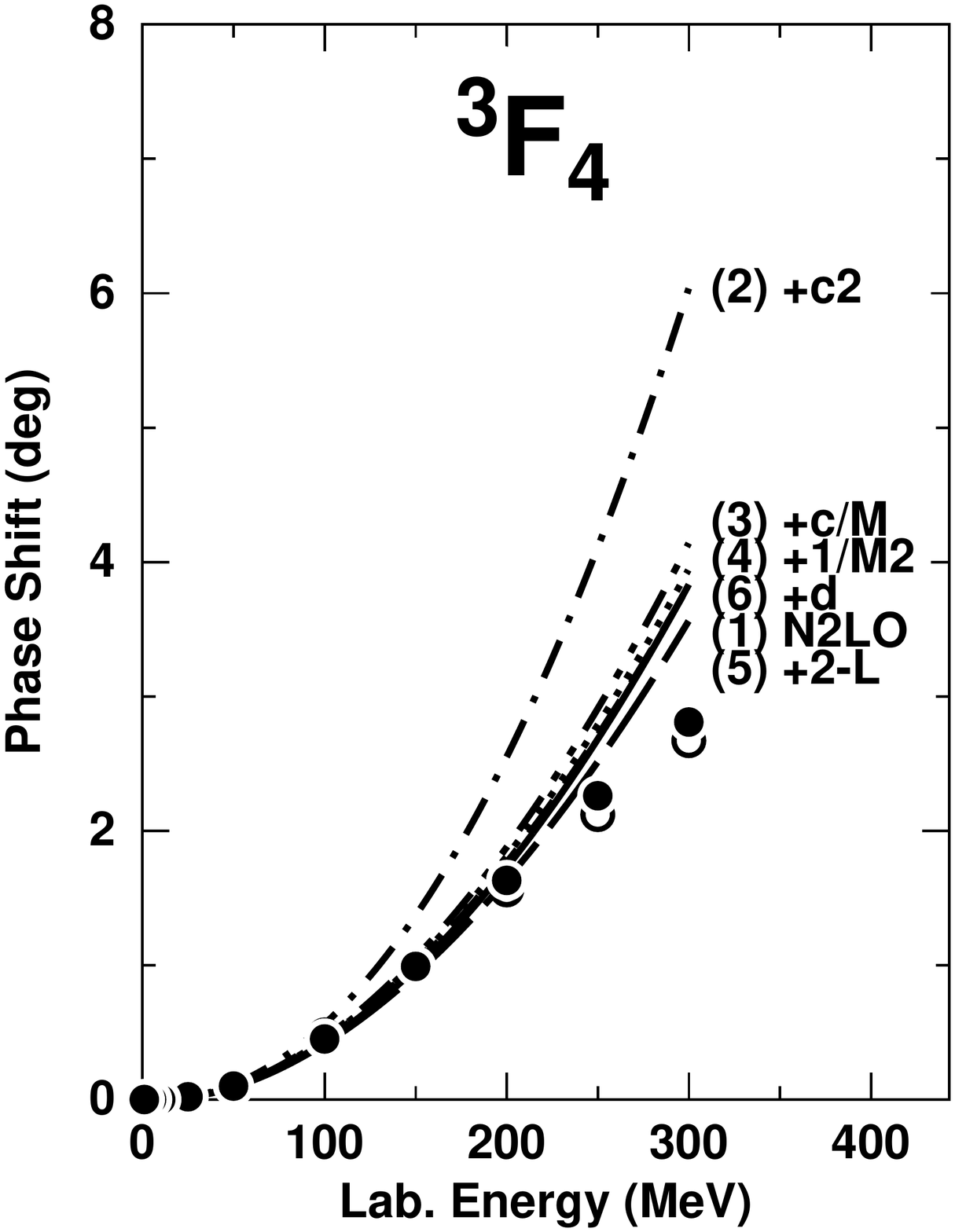}}
\hspace*{-0.5cm}
\scalebox{0.25}{\includegraphics{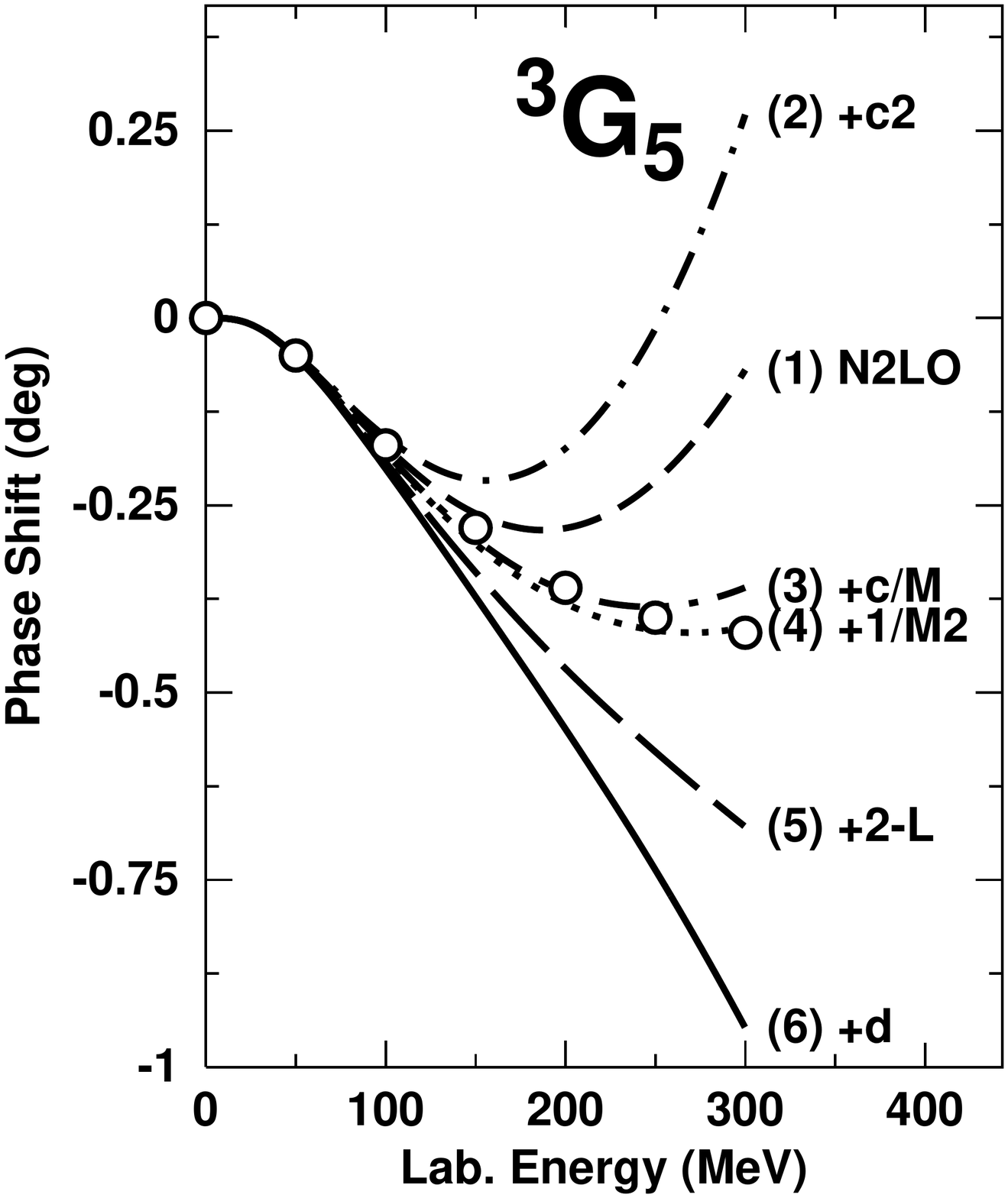}}

\vspace{-0.8cm}
\caption{The effect of individual fourth-order contributions
on the neutron-proton phase shifts in some selected peripheral
partial waves. The individual contributions
are added up successively in the order given in 
parentheses next to each curve.
Curve (1) is NNLO (``N2LO'') and curve (6) completes N$^3$LO.
For further explanations, see \ref{app_N3LO_ph}.
Empirical phase shifts (solid dots and open circles)
as in Fig.~\protect\ref{fig_f}.}
\label{fig_fff}
\end{figure}

The fourth order is obviously very diverse.
Here, we will show how the individual fourth-order contributions
impact $NN$ phase shifts in peripheral partial waves.
For this purpose, we display in 
Fig.~\ref{fig_fff} 
phase shifts
for four important peripheral partial waves, namely,
$^1F_3$, $^3F_3$, $^3F_4$, and $^3G_5$.
In each frame, the following curves
are shown:
\begin{description}
\item[(1)]
NNLO (``N2LO'').
\item[(2)] The previous curve plus
the $c_i^2$ graph, first row of 
Fig.~\ref{fig_n3lo1},
Eqs.~(\ref{eq_4c2C}) and (\ref{eq_4c2T}),
denoted by `c2' 
in the figure. 
\item[(3)] The previous curve plus
the $c_i/M_N$ contributions (denoted by `c/M'),
second row of Fig.~\ref{fig_n3lo1},
Eqs.~(\ref{eq_4cMC})-(\ref{eq_4cMLS}). 
\item[(4)] The previous curve plus
the $1/M^2_N$ corrections (`1/M2'),
row three to six of
Fig.~\ref{fig_n3lo1},
Eqs.~(\ref{eq_4M2C})-(\ref{eq_4M2sL}). 
\item[(5)] The previous curve plus
the two-loop contributions without the terms proportional
to $\bar{d}_i$ (`2-L'); i.e.
Fig.~\ref{fig_n3lo2}, but without the solid square;
Eqs.~(\ref{eq_42lC})-(\ref{eq_42lT}), but with
all $\bar{d}_i \equiv 0$. 
\item[(6)] The previous curve plus
the terms proportional
to $\bar{d}_i$ (denoted by `d' in the figure)
with the parameters given in Table~\ref{tab_LEC},
column `Peripheral perturbative $NN$'.
\end{description}
In summary, the various curves add up
successively
the individual N$^3$LO contributions 
in the order indicated in the curve label.
The last curve in this series, curve (6),
is the full N$^3$LO result.

The $c_i^2$ graph generates large attraction in all partial
waves [cf.\ differences between curves (1) and (2) in 
Fig.~\ref{fig_fff}].
This attraction is compensated by repulsion from the
$c_i/M_N$ diagrams, in most partial waves; the exception
is $^1F_3$ where $c_i/M_N$ adds more  attraction [curve (3)].
The $1/M_N^2$ corrections [difference between curves (3)
and (4)] are typically small.
Finally, the two-loop contributions create substantial repulsion
in $^1F_3$ and $^3G_5$ which brings $^1F_3$ into good agreement
with the data while causing a discrepancy for $^3G_5$. 
In $^3F_3$ and $^3F_4$, there are large cancelations
between the `pure' two-loop graphs and the $\bar{d}_i$ terms,
making the net two-loop contribution rather small.

A pivotal role in the above game is played by $W_S$, 
Eq.~(\ref{eq_4cMS}), from the $c_i/M_N$ group.
This attractive term receives a factor of 9 in $^1F_3$,
a factor $(-3)$ in $^3G_5$, and a factor of 1 in $^3F_3$
and $^3F_4$. Thus, this contribution is very attractive in
$^1F_3$ and repulsive in $^3G_5$. The latter is the reason for
the overcompensation of the $c_i^2$ graph by
the $c_i/M_N$ contribution in $^3G_5$ which is why the final
N$^3$LO result in this partial wave comes out too repulsive.
One can expect that $1/M_N$ corrections that occur at order
five or six will resolve this problem.

Before finishing this appendix, we like to point out that
the problem with the $^3G_5$ is not as dramatic as it may appear
from the phase shift plots---for two reasons.
First, the $^3G_5$ phase shifts are about one order of
magnitude smaller than the $F$ and most of the other $G$
phases. Thus, in absolute terms, the discrepancies seen
in $^3G_5$ are small.
In a certain sense, we are looking at `higher order noise'
under a magnifying glass.
Second, 
the $^3G_5$ partial wave contributes 0.06 MeV
to the energy per nucleon in nuclear matter,
the total of which is $-16$ MeV.
Consequently, small discrepancies in the reproduction
of $^3G_5$ by a $NN$ potential will have
negligible influence on the microscopic nuclear structure
predictions obtained with that potential.

\section{Fourth order $NN$ contact potential: Partial wave decomposition 
\label{app_ct}}

\setcounter{figure}{0}
\setcounter{table}{0}

The contact potential of order four,
Eq.~(\ref{eq_ct4}), decomposes into partial-waves as
follows
\be
V_{\rm ct}^{(4)}(^1 S_0)          &=&  \widehat{D}_{^1 S_0}          
({p'}^4 + p^4) +
                              D_{^1 S_0}          {p'}^2 p^2 
\nonumber 
\\
V_{\rm ct}^{(4)}(^3 P_0)          &=&        D_{^3 P_0}          
({p'}^3 p + p' p^3) 
\nonumber 
\\
V_{\rm ct}^{(4)}(^1 P_1)          &=&        D_{^1 P_1}          
({p'}^3 p + p' p^3) 
\nonumber 
\\
V_{\rm ct}^{(4)}(^3 P_1)          &=&        D_{^3 P_1}          
({p'}^3 p + p' p^3) 
\nonumber 
\\
V_{\rm ct}^{(4)}(^3 S_1)          &=&  \widehat{D}_{^3 S_1}          
({p'}^4 + p^4) +
                              D_{^3 S_1}          {p'}^2 p^2 
\nonumber 
\\
V_{\rm ct}^{(4)}(^3 D_1)          &=&        D_{^3 D_1}          
{p'}^2 p^2 
\nonumber 
\\
V_{\rm ct}^{(4)}(^3 S_1 - ^3 D_1) &=&  \widehat{D}_{^3 S_1 - ^3 D_1} 
p^4             +
                              D_{^3 S_1 - ^3 D_1} {p'}^2 p^2
\nonumber 
\\
V_{\rm ct}^{(4)}(^3 D_1 - ^3 S_1) &=&  \widehat{D}_{^3 S_1 - ^3 D_1} 
{p'}^4             +
                              D_{^3 S_1 - ^3 D_1} {p'}^2 p^2
\nonumber 
\\
V_{\rm ct}^{(4)}(^1 D_2)          &=&        D_{^1 D_2}          
{p'}^2 p^2 
\nonumber 
\\
V_{\rm ct}^{(4)}(^3 D_2)          &=&        D_{^3 D_2}          
{p'}^2 p^2 
\nonumber 
\\
V_{\rm ct}^{(4)}(^3 P_2)          &=&        D_{^3 P_2}          
({p'}^3 p + p' p^3) 
\nonumber 
\\
V_{\rm ct}^{(4)}(^3 P_2 - ^3 F_2) &=&        D_{^3 P_2 - ^3 F_2} {p'}p^3
\nonumber 
\\
V_{\rm ct}^{(4)}(^3 F_2 - ^3 P_2) &=&        D_{^3 P_2 - ^3 F_2} {p'}^3p
\nonumber 
\\
V_{\rm ct}^{(4)}(^3 D_3)          &=&        D_{^3 D_3}          
{p'}^2 p^2 
\label{eq_ct4_pw}
\ee
with the coefficients given by
\footnotesize
\be
\widehat{D}_{^1 S_0} &=& 
 4\pi \left(
             D_1    + \frac{1}{16} D_2    + \frac{1}{4}  D_3 - 
           3 D_5    - \frac{3}{16} D_6    - \frac{3}{4}  D_7 - 
             D_{11} - \frac{1}{4}  D_{12}  
 - \frac{1}{4} D_{13} 
%\right. \nonumber \\ && \left.
- \frac{1}{16} D_{14}
 \right)
\nonumber \\
D_{^1 S_0} &=& 
 4\pi \left(
\frac{10}{3} D_1    + \frac{5}{24} D_2    +  \frac{1}{6} D_3 + 
 \frac{2}{3} D_4    -           10 D_5    -  \frac{5}{8} D_6 - 
 \frac{1}{2} D_7    -            2 D_8    - \frac{10}{3} D_{11}  
 - \frac{1}{6} D_{12} 
 \right. \nonumber \\ && \left.
  -  \frac{1}{6} D_{13} - \frac{5}{24} D_{14} - 
 \frac{2}{3} D_{15}  
 \right)
\nonumber \\
D_{^3 P_0} &=& 
 4\pi \left(
-\frac{4}{3} D_1    + \frac{1}{12} D_2    -  \frac{4}{3} D_5     + 
\frac{1}{12} D_6    -  \frac{2}{3} D_9    -  \frac{1}{6} D_{10}  + 
 \frac{8}{3} D_{11} +  \frac{1}{3} D_{12} -  \frac{1}{3} D_{13}  
%\right. \nonumber \\ && \left.
 -  \frac{1}{6} D_{14}
 \right)
\nonumber \\
D_{^1 P_1} &=& 
 4\pi \left(
-\frac{4}{3} D_1    + \frac{1}{12} D_2    +            4 D_5     - 
 \frac{1}{4} D_6    +  \frac{4}{3} D_{11} - \frac{1}{12} D_{14}
 \right)
\nonumber \\
D_{^3 P_1} &=& 
 4\pi \left(
-\frac{4}{3} D_1    + \frac{1}{12} D_2    -  \frac{4}{3} D_5     + 
\frac{1}{12} D_6    -  \frac{1}{3} D_9    - \frac{1}{12} D_{10}  - 
           2 D_{11} -  \frac{1}{6} D_{12} +  \frac{1}{6} D_{13}   
%\right. \nonumber \\ && \left.
 + \frac{1}{8} D_{14}
 \right)
\nonumber \\
\widehat{D}_{^3 S_1} &=& 
 4\pi \left(
             D_1    + \frac{1}{16} D_2    +  \frac{1}{4} D_3     + 
	     D_5    + \frac{1}{16} D_6    +  \frac{1}{4} D_7     +  
 \frac{1}{3} D_{11} + \frac{1}{12} D_{12} + \frac{1}{12} D_{13}  
%\right. \nonumber \\ && \left.
 + \frac{1}{48} D_{14}
 \right)
\nonumber \\
D_{^3 S_1} &=& 
 4\pi \left(
\frac{10}{3} D_1    + \frac{5}{24} D_2    +  \frac{1}{6} D_3     + 
 \frac{2}{3} D_4    + \frac{10}{3} D_5    + \frac{5}{24} D_6     + 
 \frac{1}{6} D_7    +  \frac{2}{3} D_8    + \frac{10}{9} D_{11}  
+ \frac{1}{18} D_{12} 
\right. \nonumber \\ && \left.
 + \frac{1}{18} D_{13} + \frac{5}{72} D_{14}  + 
 \frac{2}{9} D_{15}
 \right)
\nonumber \\
D_{^3 D_1} &=& 
 4\pi \left(
\frac{8}{15} D_1    + \frac{1}{30} D_2    - \frac{2}{15} D_3- 
\frac{2}{15} D_4    + \frac{8}{15} D_5    + \frac{1}{30} D_6- 
\frac{2}{15} D_7    - \frac{2}{15} D_8    
+  \frac{2}{5} D_9 - \frac{1}{10} D_{10} 
\right. \nonumber \\ && \left.
   -  \frac{4}{9} D_{11} +  \frac{1}{9} D_{12}+ 
 \frac{1}{9} D_{13} - \frac{1}{36} D_{14} -\frac{16}{45} D_{15}
 \right)
\nonumber \\
\widehat{D}_{^3 S_1 - ^3 D_1} &=& 
 4\pi \left(
-\frac{2\,{\sqrt{2}}}{3} D_{11}  -  \frac{\sqrt{2}}{6} D_{12}- 
      \frac{\sqrt{2}}{6} D_{13}  - \frac{\sqrt{2}}{24} D_{14}
 \right)
\nonumber \\
D_{^3 S_1 - ^3 D_1} &=& 
 4\pi \left(
-\frac{14\,{\sqrt{2}}}{9} D_{11}  +      \frac{\sqrt{2}}{18} D_{12}+ 
      \frac{\sqrt{2}}{18} D_{13}  - \frac{7\,{\sqrt{2}}}{72} D_{14}+ 
  \frac{2\,{\sqrt{2}}}{9} D_{15}
 \right)
\nonumber \\
D_{^1 D_2} &=& 
 4\pi \left(
 \frac{8}{15} D_1    +  \frac{1}{30} D_2    -  \frac{2}{15} D_3    - 
 \frac{2}{15} D_4    -   \frac{8}{5} D_5    -  \frac{1}{10} D_6    + 
  \frac{2}{5} D_7    +   \frac{2}{5} D_8    -  \frac{8}{15} D_{11}  
 + \frac{2}{15} D_{12} 
 \right. \nonumber \\ && \left.
  +  \frac{2}{15} D_{13} -  \frac{1}{30} D_{14} + 
 \frac{2}{15} D_{15}
 \right)
\nonumber \\
D_{^3 D_2} &=& 
 4\pi \left(
 \frac{8}{15} D_1    +  \frac{1}{30} D_2    -  \frac{2}{15} D_3- 
 \frac{2}{15} D_4    +  \frac{8}{15} D_5    +  \frac{1}{30} D_6- 
 \frac{2}{15} D_7    -  \frac{2}{15} D_8    
+  \frac{2}{15} D_9 - \frac{1}{30} D_{10} 
\right. \nonumber \\ && \left.
  +   \frac{4}{5} D_{11} -   \frac{1}{5} D_{12}- 
  \frac{1}{5} D_{13} +  \frac{1}{20} D_{14} +  \frac{4}{15} D_{15}
 \right)
\nonumber \\
D_{^3 P_2} &=& 
 4\pi \left(
 -\frac{4}{3} D_1    +  \frac{1}{12} D_2    -   \frac{4}{3} D_5+ 
 \frac{1}{12} D_6    +   \frac{1}{3} D_9    +  \frac{1}{12} D_{10}- 
 \frac{2}{15} D_{11} +  \frac{1}{30} D_{12} 
%\right. \nonumber \\ && \left.
 -  \frac{1}{30} D_{13} + \frac{1}{120} D_{14}
 \right)
\nonumber \\
D_{^3 P_2 - ^3 F_2} &=& 
 4\pi \left(
   \frac{4\,{\sqrt{6}}}{15} D_{11} - 
      \frac{{\sqrt{6}}}{15} D_{12} + 
      \frac{{\sqrt{6}}}{15} D_{13} - 
        \frac{\sqrt{6}}{60} D_{14}
 \right)
\nonumber \\
D_{^3 D_3} &=& 
 4\pi \left(
 \frac{8}{15} D_1    +  \frac{1}{30} D_2    -  \frac{2}{15} D_3- 
 \frac{2}{15} D_4    +  \frac{8}{15} D_5    +  \frac{1}{30} D_6- 
 \frac{2}{15} D_7    -  \frac{2}{15} D_8    
%\right. \nonumber \\ && \left.
 -  \frac{4}{15} D_9 + 
 \frac{1}{15} D_{10} -  \frac{2}{15} D_{15}
 \right) .
 \nonumber \\
 \label{eq_ct4_pwcoef}
\ee
\normalsize

\section{Parameters of N$^3$LO $NN$ potentials
\label{app_par}}

\setcounter{figure}{0}
\setcounter{table}{0}

\begin{table}[t]
\caption{LECs 
for two N$^3$LO fits by the Idaho group~\cite{EM03} using $\Lambda=500$ MeV
and 600 MeV in the regulator function $f(p',p)$, Eq.~(\ref{eq_f}).
The  $C_S$ and $C_T$ of the zero-order counterterms given in Eq.~(\ref{eq_ct0}) 
are in units of $10^4$ GeV$^{-2}$;
the $C_i$, Eq.~(\ref{eq_ct2}), in $10^4$ GeV$^{-4}$; 
and the $D_i$, Eq.~(\ref{eq_ct4}),
 in $10^4$ GeV$^{-6}$.
 \label{tab_LEC_ct1}}
 \smallskip
\begin{tabular*}{\textwidth}{@{\extracolsep{\fill}}crr}
\hline 
\hline 
\noalign{\smallskip}
 LEC & $\Lambda = 500$ MeV & $\Lambda =600$ MeV 
 \\
 \hline
  $C_S^{pp}$ & -0.009991  &  -0.009943 \\
 $C_S^{nn}$ & -0.010011  &  -0.009949 \\
$C_S^{np}$ & -0.010028  &  -0.009955 \\
$C_T^{pp}$ & 0.000523   &  0.000695 \\
$C_T^{nn}$ & 0.000543   &  0.000701 \\
$C_T^{np}$ & 0.000561   &  0.000707 \\
$C_1$ &   0.051949 &   0.046433 \\
$C_2$ & 0.163034  &  0.174354 \\
$C_3$ & 0.003249  &  0.006562 \\
$C_4$ & -0.048954 & -0.050066 \\
$C_5$ & -0.075081 &      -0.086978 \\
$C_6$ &   -0.013343 &  -0.013639 \\
$C_7$ & -0.225500 &  -0.214188 \\
$D_1$ &   -0.016674 & -0.004441 \\
$D_2$ &  2.480231 & 2.751960 \\
$D_3$ &    0.915240 & 0.086650 \\
$D_4$ &     -0.811591 &  -0.061000 \\
 $D_5$ &      0.138064 & 0.126881 \\
$D_6$ &    1.249801 & 1.266717  \\
$D_7$ &  0.148074 & 0.268796 \\
$D_8$ &   -0.153405 &      -0.239967 \\
 $D_9$ &   -0.516607 &  -0.538043 \\
$D_{10}$ &      2.379565 & 2.429102 \\
$D_{11}$ &   -0.125289 &  -0.132228 \\
$D_{12}$ &  0.081807 & -0.053835 \\
 $D_{13}$ &   0.008037 & -0.081804 \\
$D_{14}$ &     -1.393312 &  -1.204164 \\
 $D_{15}$ &      0.164684 &   -0.005571 \\
\hline
\hline
\noalign{\smallskip}
\end{tabular*}
%\vspace{1.0cm}
\end{table}

\begin{table}[t]
\caption{Partial-wave LECs 
for two N$^3$LO fits by the Idaho group~\cite{EM03} using $\Lambda=500$ MeV
and 600 MeV in the regulator function $f(p',p)$, Eq.~(\ref{eq_f}).
The  $\widetilde{C}_i$ of the zero-order partial-wave counterterms given in Eq.~(\ref{eq_ct0_pw}) 
are in units of $10^4$ GeV$^{-2}$;
the $C_i$, Eq.~(\ref{eq_ct2_pw}), in $10^4$ GeV$^{-4}$; 
and the $D_i$, $\widehat{D}_i$, Eq.~(\ref{eq_ct4_pw}),
 in $10^4$ GeV$^{-6}$. The last column lists the exponent $n$ 
 of the regulator function, which is
 applied to the corresponding partial-wave counterterm.
 \label{tab_LEC_ct2}}
\smallskip
\begin{tabular*}{\textwidth}{@{\extracolsep{\fill}}cccc}
\hline 
\hline 
\noalign{\smallskip}
Partial-wave LEC & $\Lambda = 500$ MeV & $\Lambda =600$ MeV & $n$ 
 \\
 \hline
  $\widetilde{C}_{^1S_0}^{pp}$ &  -0.145286  & -0.151165 & 3 \\
  $\widetilde{C}_{^1S_0}^{nn}$ &  -0.146285  & -0.151467 & 3 \\
 $\widetilde{C}_{^1S_0}^{np}$ &  -0.147167  & -0.151745 & 3 \\
$C_{^1S_0}$  &  2.380    & 2.200 & 2 \\
$\widehat{D}_{^1S_0}$  & -2.545   & -4.890 & 2 \\
$D_{^1S_0}$  &-16.00 & -5.84  &     2 \\
$C_{^3P_0}$   & 1.487   & 1.548 & 2 \\ 
 $D_{^3P_0}$   & 0.245   & -0.215& 3 \\
  $C_{^1P_1}$   & 0.656    & 0.790 & 2 \\
  $D_{^1P_1}$   & 5.25     & 4.40 & 2 \\
  $C_{^3P_1}$   &-0.630     & -0.488 & 2 \\
  $D_{^3P_1}$   & 2.35     & 3.24 & 4 \\
 $\widetilde{C}_{^3S_1}$   &-0.118972496 & -0.116210 & 3 \\
  $C_{^3S_1}$   & 0.760    & 0.775 & 2 \\
 $\widehat{D}_{^3S_1}$    & 7.00    & 4.8004 &  2 \\
  $D_{^3S_1}$   & 6.55     & 10.8654 & 2 \\
  $D_{^3D_1}$   &-2.80    & -2.35 &  2 \\
  $C_{^3S_1-^3D_1}$  &0.826    & 0.796 & 2 \\
  $\widehat{D}_{^3S_1-^3D_1}$  &2.25    & 2.86 & 2 \\
  $D_{^3S_1-^3D_1}$  &6.61     & 5.58 & 2 \\
  $D_{^1D_2}$   &-1.770     & -1.764 & 4 \\
  $D_{^3D_2}$   &-1.46   & -1.27 &  2 \\
  $C_{^3P_2}$   &-0.538  & -0.548 &  2 \\
  $D_{^3P_2}$   & 2.295  & 2.554 &  2 \\
  $D_{^3P_2-^3F_2}$ &-0.465   & -0.525 & 4 \\
  $D_{^3D_3}$   & 5.66     &6.26 & 2, 3$^a$  \\ 
\hline
\hline
\noalign{\smallskip}
\end{tabular*}
\footnotesize
$^a$  $f(p',p) = 0.5 \{ \exp[-(p'/\Lambda)^{4}-(p/\Lambda)^{4}] 
+ \exp[-(p'/\Lambda)^{6}-(p/\Lambda)^{6}] \}$
is applied.
%\vspace{1.0cm}
\end{table}

This appendix provides detailed information on parameters involved
in the Idaho N$^3$LO $NN$ potentials~\cite{EM03}.
The $\pi N$ LECs were shown already in Table~\ref{tab_LEC} (column `$NN$ potential').
In Table~\ref{tab_LEC_ct1}, the LECs are listed, which are associated with the contact terms,
Eqs.~(\ref{eq_ct0}), (\ref{eq_ct2}), and (\ref{eq_ct4}), and,
in Table~\ref{tab_LEC_ct2}, the contact LECs are given in terms of partial-wave parameters.
Notice that through Eqs.~(\ref{eq_ct0_pw}), (\ref{eq_ct2_pwcoef}), and (\ref{eq_ct4_pwcoef})
there is a one-to-one correspondence between these two representations of the
contact LECs. 
The partial wave parameters appear one order of magnitude larger, because they are
multiplied by $(4\pi)$ in the partial-wave decomposition.
The partial-wave contact parameters of the J\"ulich N$^3$LO potentials
can be found in Ref.~\cite{EGM05}. There are similarities between the parameter sets.

At this point, one may raise the question of the {\it naturalness} of the LECs.
LECs may be perceived as {\it natural} if they are roughly of the following sizes:
\begin{eqnarray}
C_{S,T} & \sim & \frac{1}{f_\pi^2} \approx \; 0.01 \; \frac{10^4}{\rm GeV^2} \,,
\label{eq_CST} \\
C_i & \sim & \frac{1}{f_\pi^2 \, \Lambda^2} \approx \; 0.05 \; \frac{10^4}{\rm GeV^4} \,, \\
D_i & \sim & \frac{1}{f_\pi^2 \, \Lambda^4} \approx \; 0.2 \; \frac{10^4}{\rm GeV^6} \,, 
\end{eqnarray}
where  we used $\Lambda=0.5$ GeV for the numerical estimates.
Comparison with Table~\ref{tab_LEC_ct1} reveals that the values of the contact
LECs that result from a fit to the $NN$ data are in general natural, indeed.
Exceptions are $D_2$ and $D_{10}$ which are an order
of magnitude too large. 

The charge-dependent LO contact parameters 
 $\widetilde{C}_{^1S_0}^{pp}$, $\widetilde{C}_{^1S_0}^{nn}$, and $\widetilde{C}_{^1S_0}^{np}$
of Table~\ref{tab_LEC_ct2} can be analyzed as follows.
Based upon the isospin-symmetric Lagrangian Eq.~(\ref{eq_LNN0}) and the
isospin violating ones, Eqs.~(\ref{eq_LCSB}) and (\ref{eq_LCIB}),
the charge-dependent non-derivative contact parameters in the $^1S_0$ state are given by
\begin{eqnarray}
\widetilde{C}_{pp} & = &  \widetilde{C}_{\rm sym} + \widetilde{C}_{\rm CIB} + \widetilde{C}_{\rm CSB} \,, \\
\widetilde{C}_{nn} & = &  \widetilde{C}_{\rm sym} + \widetilde{C}_{\rm CIB} - \widetilde{C}_{\rm CSB} \,, \\
\widetilde{C}_{np} & = &  \widetilde{C}_{\rm sym} - \widetilde{C}_{\rm CIB} \,,
\end{eqnarray}
where   $\widetilde{C}_{\rm sym}$ is the same as $\widetilde{C}_{^1S_0}$ of Eq.~(\ref{eq_ct0_pw}).
The charge-symmetric and the CSB and CIB contributions can then
be obtained from
\begin{eqnarray}
\widetilde{C}_{\rm sym} & = & \frac12 \left[ \frac12 \left( \widetilde{C}_{ pp} + \widetilde{C}_{nn} \right)
 + \widetilde{C}_{np} \right] \,, \\
\widetilde{C}_{\rm CSB} & = & \frac12 \left( \widetilde{C}_{ pp} - \widetilde{C}_{nn} \right) \,, \\
\widetilde{C}_{\rm CIB} & = & \frac12 \left[ \frac12 \left( \widetilde{C}_{ pp} + \widetilde{C}_{nn} \right)
 - \widetilde{C}_{np} \right] \,. 
\end{eqnarray}
Using the numbers from the $\Lambda = 500$ MeV fit listed in Table~\ref{tab_LEC_ct2}, 
the following values are produced: 
\begin{eqnarray}
\widetilde{C}_{\rm sym} & = &  -0.1465 \; \frac{10^4}{\rm GeV^2} \,, \\
\widetilde{C}_{\rm CSB} & = & 0.0005 \; \frac{10^4}{\rm GeV^2} \,, \\
\widetilde{C}_{\rm CIB} & = & 0.0007 \; \frac{10^4}{\rm GeV^2}  \,. 
\end{eqnarray}
Note that these are partial-wave parameters, which carry a factor of $(4\pi)$ as compared to
the parameters of, e.g., Eq.~(\ref{eq_CST}).
As discussed in Section~\ref{sec_CD}, the Idaho N$^3$LO potential~\cite{EM03}
does not include explicitly the CIB effect from the pion-mass splitting in the NLO 2PE.
Thus, the above value for $\widetilde{C}_{\rm CIB}$ is not the pure contact contribution since it includes 
a simulation of this omitted (small) effect (causing $\Delta a_{CIB} \approx -0.5$ fm).
Moreover, also the CSB effect from nucleon-mass difference in the NLO 2PE
is not calculated explicitly and, thus, effectively contained in  $\widetilde{C}_{\rm CSB}$.
The exact size of this effect when evaluated in ChPT at NLO is not known, 
but could be a large fraction of the total CSB~\cite{LM98a}.

The regulator function $f(p',p)$, Eq.~(\ref{eq_f}), has two parameters, namely, the
`cutoff' $\Lambda$ and the power parameter $n$.
Fits were made for $\Lambda=500$ MeV and 600 MeV 
[denoted by N$^3$LO(500) and N$^3$LO(600)].
The parameter $n$ has to be chosen such that the higher powers
generated by the multiplication with the regulator [cf.\ Eq.~(\ref{eq_reg_exp})] are beyond the order
at which we are working, which is $\nu=4$.
In short, the requirement is 
\begin{equation}
2n+\nu_i>4 \,,
\label{eq_n}
\end{equation}
where $\nu_i$ denotes the order of a specific contribution to the potential. 
Thus, $n\geq 3$ for LO contributions,
$n\geq 2$ for NLO contributions, etc..
In the case of the Idaho N$^3$LO potentials,
$n=4$ is used for the 1PE and
$n=2$ for all 2PE contributions (which are NLO or higher).
For the contact contributions, $n$ is chosen in a partial-wave dependent
way and in accordance with Eq.~(\ref{eq_n}), see Table~\ref{tab_LEC_ct2}.

Phase shifts of $NN$ scattering 
as produced by the Idaho N$^3$LO(500) potential~\cite{EM03}
are shown in Tables~\ref{tab_pp}-\ref{tab_np0}
up to laboratory energies of 300 MeV.

\begin{table}
\caption{$pp$ phase shifts (in degrees) by the Idaho N$^3$LO(500) potential~\cite{EM03}.
\label{tab_pp}}
\smallskip
\begin{tabular*}{\textwidth}{@{\extracolsep{\fill}}crrrrrrrrrr}
\hline 
\hline 
\noalign{\smallskip}
 $T_{lab}$ (MeV)
 & $^1S_0$
 & $^3P_0$
 & $^3P_1$
 & $^1D_2$
 & $^3P_2$
 & $^3F_2$
 & $\epsilon_2$
 & $^3F_3$
 & $^1G_4$
 & $^3F_4$
\\
 \hline 
    1 &  32.76 &   0.13 &  -0.08 &   0.00 &   0.01 &   0.00 &   0.00 &   0.00 &   0.00 &   0.00 \\
    5 &  54.71 &   1.58 &  -0.90 &   0.04 &   0.21 &   0.00 &  -0.05 &   0.00 &   0.00 &   0.00 \\
   10 &  55.01 &   3.73 &  -2.05 &   0.17 &   0.64 &   0.01 &  -0.20 &  -0.03 &   0.00 &   0.00 \\
   25 &  48.37 &   8.57 &  -4.91 &   0.70 &   2.47 &   0.10 &  -0.80 &  -0.23 &   0.04 &   0.02 \\
   50 &  38.75 &  11.47 &  -8.31 &   1.70 &   5.87 &   0.33 &  -1.65 &  -0.69 &   0.15 &   0.10 \\
  100 &  25.74 &   9.30 & -13.39 &   3.68 &  11.11 &   0.80 &  -2.54 &  -1.48 &   0.41 &   0.42 \\
  150 &  16.66 &   4.34 & -17.63 &   5.28 &  14.01 &   1.22 &  -2.83 &  -1.96 &   0.66 &   0.87 \\
  200 &   8.73 &  -0.87 & -21.48 &   6.59 &  15.67 &   1.56 &  -2.76 &  -2.16 &   0.88 &   1.37 \\
  250 &   0.51 &  -5.78 & -25.43 &   7.95 &  16.85 &   1.81 &  -2.35 &  -2.11 &   1.05 &   1.83 \\
  300 &  -8.53 & -10.43 & -29.54 &   9.16 &  17.73 &   1.94 &  -1.67 &  -1.83 &   1.14 &   2.17 \\
\hline
\hline
\noalign{\smallskip}
\end{tabular*}
%\vspace{1.0cm}
\end{table}

\begin{table}
\caption{$nn$ phase shifts (in degrees) by the Idaho N$^3$LO(500) potential~\cite{EM03}.
\label{tab_nn}}
\smallskip
\begin{tabular*}{\textwidth}{@{\extracolsep{\fill}}crrrrrrrrrr}
\hline 
\hline 
\noalign{\smallskip}
 $T_{lab}$ (MeV)
 & $^1S_0$
 & $^3P_0$
 & $^3P_1$
 & $^1D_2$
 & $^3P_2$
 & $^3F_2$
 & $\epsilon_2$
 & $^3F_3$
 & $^1G_4$
 & $^3F_4$
\\
 \hline 
    1 &  57.51 &   0.21 &  -0.12 &   0.00 &   0.02 &   0.00 &   0.00 &   0.00 &   0.00 &   0.00 \\
    5 &  60.86 &   1.85 &  -1.04 &   0.05 &   0.26 &   0.00 &  -0.06 &  -0.01 &   0.00 &   0.00 \\
   10 &  57.62 &   4.10 &  -2.25 &   0.18 &   0.73 &   0.01 &  -0.22 &  -0.04 &   0.00 &   0.00 \\
   25 &  48.89 &   8.92 &  -5.15 &   0.73 &   2.66 &   0.11 &  -0.82 &  -0.24 &   0.04 &   0.02 \\
   50 &  38.67 &  11.60 &  -8.57 &   1.76 &   6.16 &   0.34 &  -1.68 &  -0.71 &   0.16 &   0.11 \\
  100 &  25.48 &   9.16 & -13.66 &   3.75 &  11.42 &   0.81 &  -2.55 &  -1.50 &   0.42 &   0.43 \\
  150 &  16.34 &   4.08 & -17.89 &   5.33 &  14.31 &   1.23 &  -2.82 &  -1.98 &   0.67 &   0.88 \\
  200 &   8.31 &  -1.18 & -21.69 &   6.63 &  15.99 &   1.58 &  -2.74 &  -2.17 &   0.89 &   1.39 \\
  250 &  -0.06 &  -6.11 & -25.56 &   7.97 &  17.19 &   1.83 &  -2.31 &  -2.11 &   1.06 &   1.86 \\
  300 &  -9.24 & -10.79 & -29.44 &   9.13 &  18.07 &   1.95 &  -1.63 &  -1.82 &   1.15 &   2.19 \\
\hline
\hline
\noalign{\smallskip}
\end{tabular*}
%\vspace{1.0cm}
\end{table}

\begin{table}
\caption{$I=1$ $np$ phase shifts (in degrees) by the Idaho N$^3$LO(500) potential~\cite{EM03}.
\label{tab_np1}}
\smallskip
\begin{tabular*}{\textwidth}{@{\extracolsep{\fill}}crrrrrrrrrr}
\hline 
\hline 
\noalign{\smallskip}
 $T_{lab}$ (MeV)
 & $^1S_0$
 & $^3P_0$
 & $^3P_1$
 & $^1D_2$
 & $^3P_2$
 & $^3F_2$
 & $\epsilon_2$
 & $^3F_3$
 & $^1G_4$
 & $^3F_4$
\\
 \hline 
    1 &  61.95 &   0.18 &  -0.11 &   0.00 &   0.02 &   0.00 &   0.00 &   0.00 &   0.00 &   0.00 \\
    5 &  63.34 &   1.63 &  -0.94 &   0.04 &   0.25 &   0.00 &  -0.05 &   0.00 &   0.00 &   0.00 \\
   10 &  59.55 &   3.66 &  -2.06 &   0.16 &   0.70 &   0.01 &  -0.18 &  -0.03 &   0.00 &   0.00 \\
   25 &  50.28 &   8.16 &  -4.86 &   0.68 &   2.56 &   0.09 &  -0.74 &  -0.20 &   0.03 &   0.02 \\
   50 &  39.82 &  10.75 &  -8.25 &   1.71 &   5.98 &   0.30 &  -1.57 &  -0.62 &   0.14 &   0.09 \\
  100 &  26.51 &   8.37 & -13.35 &   3.75 &  11.17 &   0.75 &  -2.46 &  -1.38 &   0.39 &   0.39 \\
  150 &  17.35 &   3.36 & -17.58 &   5.37 &  14.05 &   1.15 &  -2.76 &  -1.84 &   0.64 &   0.84 \\
  200 &   9.29 &  -1.86 & -21.39 &   6.68 &  15.72 &   1.49 &  -2.70 &  -2.03 &   0.87 &   1.33 \\
  250 &   0.89 &  -6.76 & -25.26 &   8.04 &  16.92 &   1.74 &  -2.29 &  -1.98 &   1.04 &   1.80 \\
  300 &  -8.31 & -11.40 & -29.17 &   9.19 &  17.81 &   1.87 &  -1.63 &  -1.70 &   1.14 &   2.13 \\
\hline
\hline
\noalign{\smallskip}
\end{tabular*}
%\vspace{1.0cm}
\end{table}

\begin{table}
\caption{$I=0$ $np$ phase shifts (in degrees) by the Idaho N$^3$LO(500) potential~\cite{EM03}.
\label{tab_np0}}
\smallskip
\begin{tabular*}{\textwidth}{@{\extracolsep{\fill}}crrrrrrrrrr}
\hline 
\hline 
\noalign{\smallskip}
 $T_{lab}$ (MeV)
 & $^1P_1$
 & $^3S_1$
 & $^3D_1$
 & $\epsilon_1$
 & $^3D_2$
 & $^1F_3$
 & $^3D_3$
 & $^3G_3$
 & $\epsilon_3$
 & $^3G_4$
\\
 \hline 
    1 &  -0.19 & 147.76 &  -0.01 &   0.11 &   0.01 &   0.00 &   0.00 &   0.00 &   0.00 &   0.00 \\
    5 &  -1.55 & 118.18 &  -0.19 &   0.67 &   0.22 &  -0.01 &   0.00 &   0.00 &   0.01 &   0.00 \\
   10 &  -3.19 & 102.58 &  -0.68 &   1.15 &   0.85 &  -0.07 &   0.00 &   0.00 &   0.08 &   0.01 \\
   25 &  -6.67 &  80.45 &  -2.81 &   1.76 &   3.72 &  -0.43 &  -0.03 &  -0.05 &   0.55 &   0.17 \\
   50 & -10.03 &  62.28 &  -6.41 &   2.05 &   8.99 &  -1.16 &   0.05 &  -0.26 &   1.61 &   0.72 \\
  100 & -14.05 &  42.31 & -11.94 &   2.32 &  17.56 &  -2.35 &   0.89 &  -0.95 &   3.50 &   2.14 \\
  150 & -17.49 &  29.79 & -15.83 &   2.67 &  22.68 &  -3.22 &   2.32 &  -1.77 &   4.83 &   3.51 \\
  200 & -21.43 &  20.63 & -19.01 &   3.16 &  24.71 &  -3.86 &   3.66 &  -2.56 &   5.66 &   4.62 \\
  250 & -26.08 &  13.33 & -21.94 &   3.72 &  24.29 &  -4.27 &   4.51 &  -3.19 &   6.02 &   5.35 \\
  300 & -31.28 &   7.24 & -24.48 &   4.02 &  22.00 &  -4.41 &   5.35 &  -3.53 &   5.91 &   5.57 \\
\hline
\hline
\noalign{\smallskip}
\end{tabular*}
%\vspace{1.0cm}
\end{table}

\pagebreak

\end{document}